\documentclass[12pt]{article}
\usepackage{epsfig}
\usepackage[koi8-r]{inputenc}
\usepackage[english,russian]{babel}
\begin{document}
\begin{otherlanguage}{english}
\title{Test problems in mechanics and special relativity}
\author{ Z.~K.~Silagadze \\
Budker Institute of Nuclear Physics  and \\ Novosibirsk State
University, 630 090, Novosibirsk, Russia }
\date{}

\maketitle

\begin{abstract}
These test problems were used by the author as weekly control works for 
the first year physics students at Novosibirsk State University in 2005. 
Solutions of the problems are also given. The problems were taken from or 
inspired by the sources listed at the end. 
\end{abstract}
\end{otherlanguage}

\begin{figure}[htb]
\centerline{\epsfxsize 80mm\epsfbox{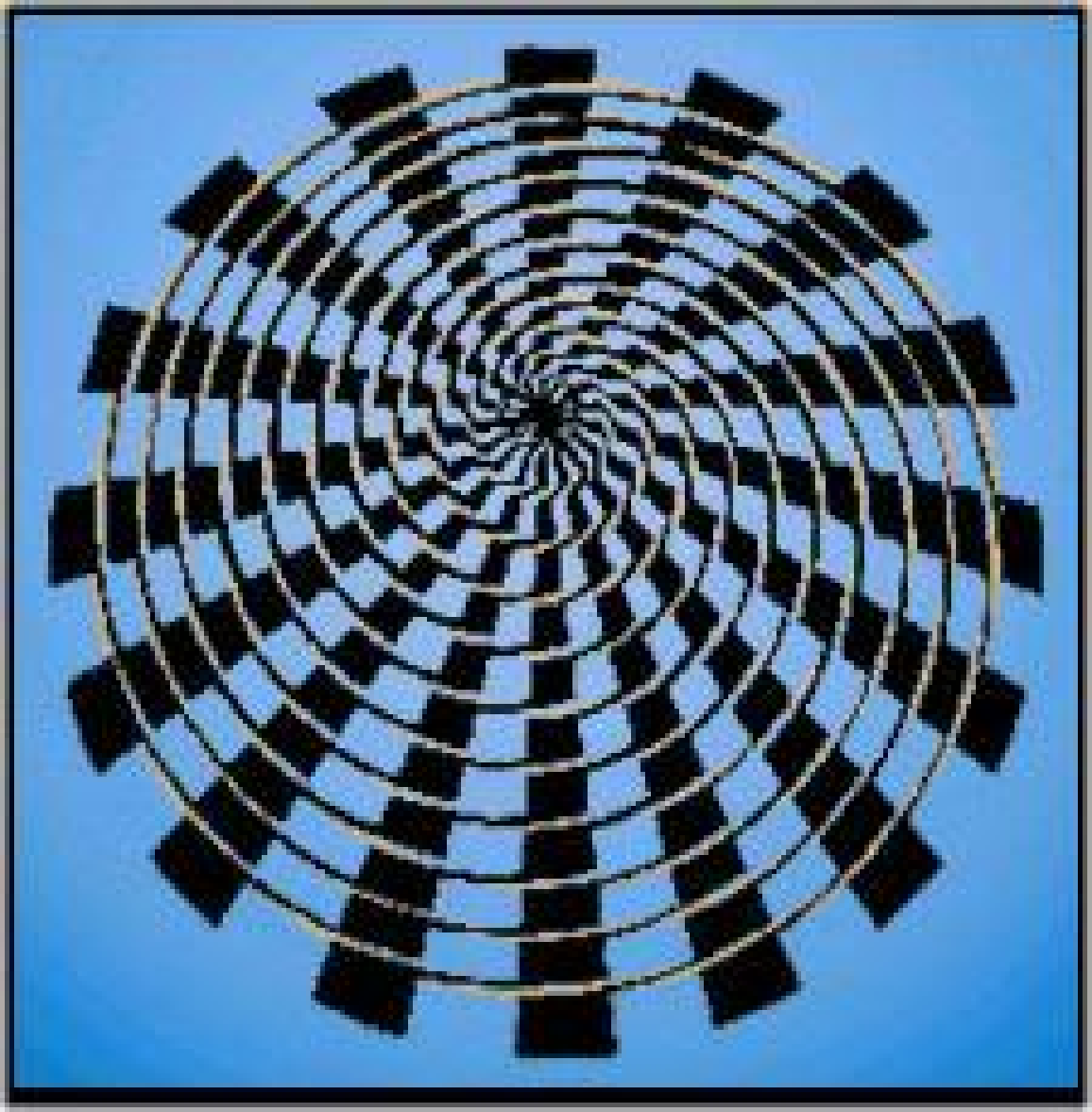}}
\end{figure}

\vspace*{5mm}
\centerline{{\LARGE Тестовые задачи по механике и теории относительности}}
\vspace*{3mm}
\centerline{{\Large З.~К.~Силагадзе}}
\vspace*{1mm}
\centerline{{\large Институт Ядерной Физики им. Будкера и}}
\vspace*{1mm}
\centerline{{\large Новосибирский Государственный Университет, 630 090, 
Новосибирск}}

\vspace*{5mm}
\begin{abstract}
Эти задачи были использованы автором для недельных контрольных работ по 
механике и теории относительности для студентов-первокурсников физического 
факультета Новосибирского Государственного Университета в 2005 году. 
Приводится также решения задач. Источники задач, или вдохновения, приведены 
в списке литературы.
\end{abstract}

\large

\newpage
\section*{Контрольная работа 1}
\setcounter{section}{1}
\setcounter{subsection}{1}
\subsection{}
Заяц бежит по прямой со скоростью $u$. Его начинает преследовать собака со 
скоростью $V$ и в ходе погони всегда бежит в направлении на зайца.
В начальный момент времени их положения показаны на рисунке. Найти уравнение 
траектории собаки в системе зайца.

\begin{figure}[htb]
\centerline{\epsfig{figure=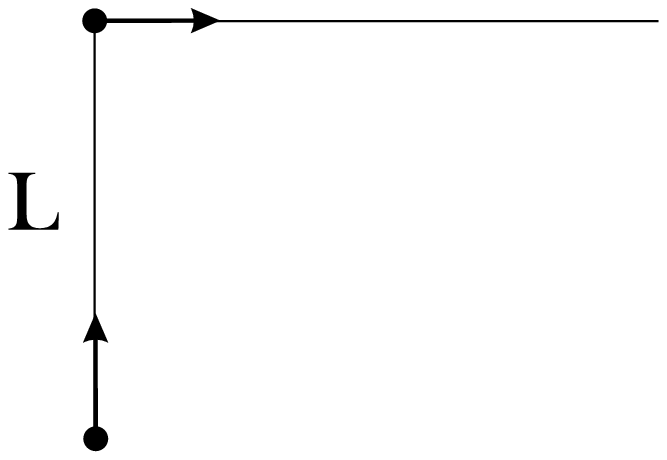,height=4cm}}
\end{figure}\noindent
A rabbit runs in a straight line with a speed $u$. A dog with a speed 
$V$ starts to pursuit it and during the pursuit always runs in the 
direction towards the rabbit. At the initial moment of time their 
positions are shown in the figure. Find the equation of the dog's 
trajectory in the rabbit's frame.

\subsection{}
Пусть скорости собаки и зайца в предыдущей задаче равны по величине и их 
начальное расположение таково (см. рисунок). Какому пределу стремится 
расстояние между ними со временем?

\begin{figure}[htb]
\centerline{\epsfig{figure=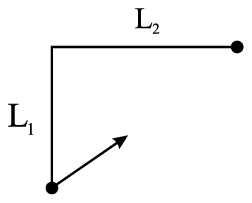,height=4cm}}
\end{figure}\noindent
Let speeds of the dog and the rabbit from the previous problem are equal in 
magnitude and their initial positions are as shown in the figure. To 
what limit converges the distance between them?

\subsection{}
Параболические координаты $\xi,\,\eta,\,\phi$ определяются формулами 
$$x=\sqrt{\xi\eta}\cos{\phi},\;y=\sqrt{\xi\eta}\sin{\phi},\;
z=\frac{1}{2}(\xi-\eta).$$
Выразить орты параболической системы через декартовые орты. Найти выражения 
для дифференциала длины дуги, площади, объема для этих координат.

\vspace*{5mm}\noindent
Parabolic coordinates $\xi,\,\eta,\,\phi $ are defined by formulas 
$$ x = \sqrt {\xi \eta} \cos {\phi}, \; y = \sqrt {\xi \eta} \sin 
{\phi}, \; z = \frac {1} {2} (\xi - \eta.) $$ 
Express coordinate orts of the parabolic system through Cartesian orts. 
Find elements of line, area and  volume for these coordinates.

\subsection{}
Электрон в постоянном магнитном поле движется по винтовой линии
$$x=a\cos{\omega t},\;y=a\sin{\omega t},\; z=bt.$$
Найти радиус кривизны его траектории.

\vspace*{5mm}\noindent
An Electron in a constant magnetic field moves in a helical orbit 
$$ x=a \cos {\omega t}, \; y=a \sin {\omega t}, \; z=bt. $$ 
Find the radius of curvature of its trajectory.

\subsection{}
Точка движется по плоскости так, что угол между вектором скорости и 
радиус-вектором во все время движения равен $\alpha$. Найти уравнение 
траектории точки, если в начальный момент $r(0)=r_0,\;\varphi(0)=\varphi_0$.

\vspace*{5mm}\noindent
A point moves on a plane so that the angle between its velocity and 
radius-vector is equal to $\alpha$ all the time. Find the trajectory 
of the point, if at the initial moment $r(0)=r_0,\;\varphi (0)=
\varphi_0$.

\section*{Контрольная работа 2}
\setcounter{section}{2}
\setcounter{subsection}{0}
\subsection{}
На баскетбольный мяч лежит мячик настольного тенниса. Мяч падает без начальной 
скорости с высоты 1~м. Оцените, на какую высоту подскочит мячик.

\begin{figure}[htb]
\centerline{\epsfig{figure=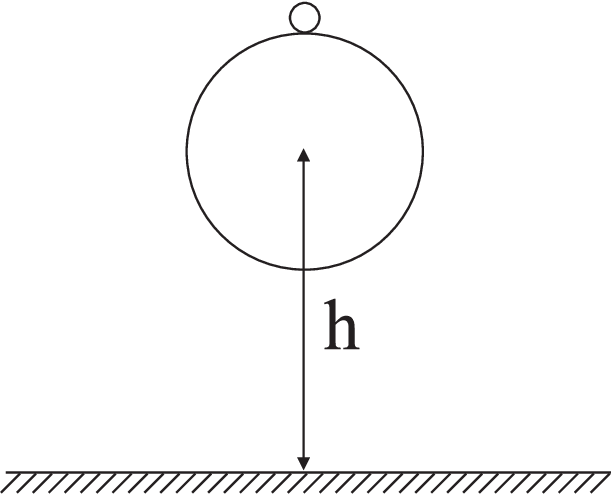,height=4cm}}
\end{figure}\noindent
A tennis ball sits atop a basketball as the two are dropped without 
initial speed from height 1~m. Estimate the height the tennis ball will 
rebound.

\subsection{}
У полого цилиндра один конец заклеен бумагой. Со стороны открытого конца к 
нему летит гвоздь с большой скоростью $V$. Шляпа гвоздя больше чем диаметр 
цилиндра, а их собственные длины равны. В системе цилиндра гвоздь сокращен 
$\gamma$-раз и поэтому не может пробить бумагу. Но в системе гвоздя сокращен 
цилиндр и поэтому бумага пробивается. Что происходит на самом деле в каждой 
из систем?

\begin{figure}[htb]
\centerline{\epsfig{figure=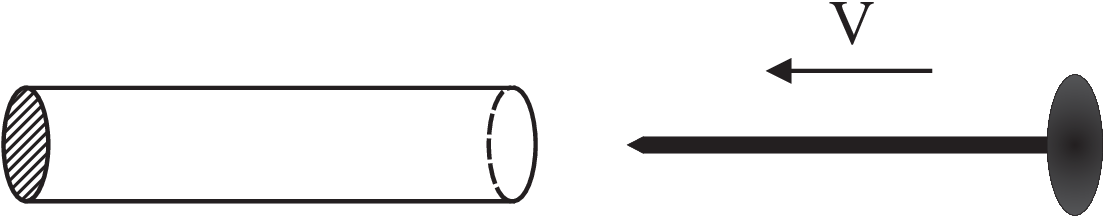,width=10cm}}
\end{figure}\noindent
One end of the hollow cylinder is sealed by a paper. A nail with the 
large velocity $V$ flies towards the open end. The hat of the nail is 
larger than the diameter of the cylinder, and their proper lengths are 
equal. In the cylinder's frame the nail is contracted $\gamma$-times 
and, consequently, cannot perforate the paper. But in the nail's frame 
it's the cylinder that's length-contracted and consequently the paper 
is pierced. What occurs actually in each of frames?

\subsection{}
Найти радиус кривизны $R$ траектории точки, если известны ее скорость 
$\vec{V}$ и ускорение $\vec{a}$.

\vspace*{5mm}\noindent
Assuming that the velocity $\vec{V}$ and acceleration $\vec{a}$ 
of a point are known find the radius of curvature $R$ of its trajectory.

\subsection{}
Параболические координаты на плоскости определяются так $x=\sigma \tau,$
$y=\frac{1}{2}(\tau^2-\sigma^2).$ Найти орты $\vec{e}_\sigma,\,
\vec{e}_\tau$ и элемент площади.

\vspace*{5mm}\noindent
Parabolic coordinates on a plane are defined as follows $x=\sigma\tau$, 
$y =\frac {1}{2}(\tau^2 -\sigma^2)$. Find the coordinate orts $\vec{e}_
\sigma,\,\vec {e} _\tau $ and the element of area.

\subsection{}
Точка движется на поверхности шара так, что ее скорость сохраняет постоянный 
угол $\alpha$ к меридиану. Найти уравнение траектории точки, если в начальный 
момент ее сферические координаты равны $\theta_0$ и $\varphi_0$.

\vspace*{5mm}\noindent
A point moves on the surface of a sphere so that its velocity vector
keeps a constant angle $\alpha$ to the meridian. Find the
trajectory of the point if at the initial moment its spherical 
coordinates were $\theta_0$ and $\varphi_0$.

\subsection{}
Пусть многомерное обобщение векторного произведения сохраняет следующие 
свойства обычного трехмерного векторного произведения:
$$\hspace*{-53mm}\vec{A}\times\vec{A}=0,$$ 
$$\hspace*{-10mm}(\vec{A}\times\vec{B})\cdot \vec{A}=(\vec{A}\times
\vec{B})\cdot \vec{B}=0,$$
$$|\vec{A}\times\vec{B}|=|\vec{A}|~|\vec{B}|,\;\; \mbox{если} \; \;
\vec{A}\cdot\vec{B}=0.$$ 
Доказать, что это возможно только в семимерном пространстве. 

\vspace*{5mm}\noindent
Let a multidimensional generalization of the cross product keeps the 
following properties of the usual three-dimensional vector product:
\clearpage
$$\hspace*{-53mm}\vec{A}\times\vec{A}=0,$$ 
$$\hspace*{-10mm}(\vec{A}\times\vec{B})\cdot \vec{A}=(\vec{A}\times
\vec{B})\cdot \vec{B}=0,$$
$$\hspace*{-5mm}|\vec{A}\times\vec{B}|=|\vec{A}|~|\vec{B}|,\;\; 
\mbox{if} \; \;\vec{A}\cdot\vec{B}=0.$$ 
Prove that this is possible only in seven-dimensional space.

\section*{Контрольная работа 3}
\setcounter{section}{3}
\setcounter{subsection}{0}
\subsection{}
В системе $S$ стержень имеет длину $L$ и движется со скоростью $u$ вдоль оси 
$x$. Какую длину имеет стержень в системе  $S^\prime$, которая движется вдоль 
оси $x$ со скоростью $V$?

\vspace*{5mm}\noindent
In the frame $S$ a rod has length $L$ and moves with a speed $u$ along 
the $x$ axis. What is the length of the rod in the system $S^\prime$ 
which moves with a speed $V$ along the $x$ axis?

\subsection{}
Если ввести координаты $x_+=ct+x,\, x_-=ct-x,$ то преобразование Лоренца 
можно записать так $x_+^\prime=e^{-\psi}x_+,\,x_-^\prime=e^{\psi}x_-$. 
Найти $\psi$ как функцию $V$ (параметр $\psi$ называется быстротой, 
а $x_+$ и $x_-$ -- координатами светового конуса).

\vspace*{5mm}\noindent
If one introduces the following coordinates $x_+=ct+x,\,x_-=ct-x,$ the
Lorentz transformations can be written in the form $x _+^\prime=
e^{-\psi}x_+,\,x_-^\prime=e^{\psi}x_-$. Find $\psi$ as a function of 
$V$ (the parameter $\psi$ is called rapidity, and $x_+$ and $x_-$ -- 
coordinates of the light cone).

\subsection{}
Найти результат релятивистского сложения $n$ коллинеарных скоростей $V_1$, 
$V_2$, $\ldots$,$V_n$. Что получится, если $V_1=V_2=\ldots=V_n$?

\vspace*{5mm}\noindent
Find the result of relativistic addition of $n$ collinear velocities 
$V_1$, $V_2$, $\ldots$, $V_n$. What does one get if $V_1=V_2=
\ldots=V_n$?

\subsection{}
Длинная люминисцентная лампа параллельна оси $x$ и движется вдоль этой оси со 
скоростью $V$. Лампа вспыхивает мгновенно в своей системе отсчета. С какой 
скоростью распространяется вспышка вдоль лампы с точки зрения неподвижного 
наблюдателя?

\vspace*{5mm}\noindent
The long luminescent lamp is parallel to the axis $x$ and moves along 
this axis with a speed $V$. The lamp flashes instantly in its proper 
frame. With what speed propagates the flash along the lamp from the point
of view of the motionless observer?

\subsection{}
Цилиндр вращается вокруг своей оси с угловой скоростью $\omega^\prime$. Показать, 
что для наблюдателя, который движется вдоль оси цилиндра со скоростью $V$, 
цилиндр окажется скрученным и найти величину кручения на единицу длины цилиндра.

\vspace*{5mm}\noindent
A cylinder rotates about its axis with angular velocity $\omega^\prime$. 
Show that for an observer who moves along the axis of the cylinder with 
a speed $V$ the cylinder will appear twisted and find the torsion 
magnitude per unit length of the cylinder.

\section*{Контрольная работа 4}
\setcounter{section}{4}
\setcounter{subsection}{0}
\subsection{}
Найти скорость наблюдателя, если далекий и маленький фонарь впереди ему 
кажется два раза больше, чем неподвижному наблюдателю рядом.

\vspace*{5mm}\noindent
Find the speed of an observer if a distant and small lantern ahead looks 
to him twice as large as to the motionless observer nearby.

\subsection{}
Массивное зеркало движется перпендикулярно своей плоскости со скоростью $V$. 
Найти угол отражения $\theta_2$ для луча света от такого зеркала, если угол 
падения равен $\theta_1$.

\begin{figure}[htb]
\centerline{\epsfig{figure=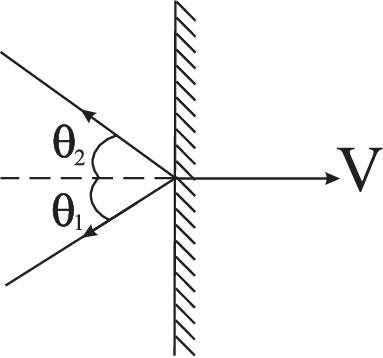,height=4cm}}
\end{figure}\noindent
A massive mirror moves perpendicularly to its plane with the speed $V$. 
Find the angle of reflection $\theta_2$ for a light ray from such mirror 
if its angle of incidence equals to $\theta_1$.

\subsection{}
Две системы отсчета движутся со скоростями $\vec{V}_1$ и $\vec{V}_2$. 
Доказать, что их относительная скорость $V$ удовлетворяет соотношению
$$\beta^2=\frac{(\vec{\beta}_1-\vec{\beta}_2)^2-(\vec{\beta}_1\times
\vec{\beta}_2)^2}{(1-\vec{\beta}_1\cdot\vec{\beta}_2)^2},$$
где $\vec{\beta}=\frac{\vec{V}}{c}$.

\vspace*{5mm}\noindent
Two reference frames move with speeds $\vec{V}_1$ and $\vec{V}_2$. Prove 
that their relative velocity $V$ satisfies the relation 
$$\beta^2=\frac{(\vec{\beta}_1-\vec{\beta}_2)^2-(\vec{\beta}_1\times 
\vec{\beta}_2)^2}{(1-\vec{\beta}_1\cdot\vec{\beta}_2)^2},$$ 
where $\vec{\beta}=\frac{\vec{V}}{c}$.

\subsection{}
Стержень движется со скоростью $V$. Какой угол $\theta$ составляет стержень с 
направлением движения, если в его системе покоя этот угол равен 
$\theta^\prime$?

\vspace*{5mm}\noindent
A rod moves with the speed $V$. What angle $\theta$ makes the rod with 
the direction of its motion if this angle is equal to $\theta^\prime$ 
in its rest frame?

\subsection{}
Вектор $\overrightarrow{AB\;}$ переносится по окружности с постоянной по 
величине скоростью $V$ так, что в каждой сопутствующей инерциальной системе 
отсчета вектор переносится параллельно самому себе. Найти минимальное значение
$V$ для которого после полного оборота вектор смотрит в противоположную к 
первоначальному сторону.

\begin{figure}[htb]
\centerline{\epsfig{figure=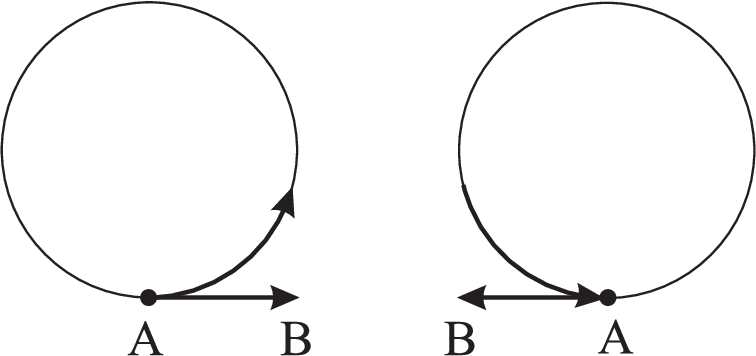,height=4cm}}
\end{figure}\noindent

A vector $\overrightarrow{AB\;}$ is transferred on a circle with the 
constant in magnitude speed $V$ so that in every instantaneous comoving 
inertial frame the vector is transferred parallel to itself. Find the 
minimal $V$ for which after a full turn the vector points opposite to its 
initial direction.

\section*{Контрольная работа 5}
\setcounter{section}{5}
\setcounter{subsection}{0}
\subsection{}
Частица налетает на такую же неподвижную частицу и происходит упругое 
столкновение. Доказать, что угол разлета частиц после столкновения всегда 
меньше, чем $90^\circ$.

\vspace*{5mm}\noindent  
A particle collides with another identical particle at rest. Prove that 
the opening angle between outgoing particle velocities after the elastic
scattering is always less than $90^\circ$.

\subsection{}
Под каким углом к оси $x$ летит фотон с частотой $\nu$ в системе $S$, если в 
системе $S^\prime$, движущегося вдоль оси $x$, фотон имеет частоту
$\nu^\prime$ и летит под углом $\theta^\prime$ к оси $x$.

\vspace*{5mm}\noindent
Under what angle to the $x$-axis moves the photon with frequency $\nu$ 
in frame $S$ if in frame $S^\prime$, moving along the $x$-axis, the 
photon has frequency $\nu^\prime$ and flies under the angle $\theta^
\prime$ to the $x$-axis.

\subsection{}
$K^0$-мезон распадается по схеме $K^0\to\pi^+\pi^-$. Найти энергию  $K$-мезона,
если максимально возможная энергия $\pi$-мезонов из этого распада $\alpha$-раз
больше минимальной. Для каких $\alpha$ не будет $\pi$-мезонов летящих в заднюю
полусферу?

\vspace*{5mm}\noindent
$K^0$-meson decays under the scheme $K^0\to\pi^+\pi^-$. Find the energy of 
the $K^0$-meson if the greatest possible energy of $\pi$-mesons from this 
decay is $\alpha$-times larger than the minimal one. For what $\alpha$ 
there will be no $\pi$-mesons flying into the backward hemisphere?

\subsection{}
Найти относительную скорость $\pi^+$ и $\pi^-$ мезонов из распада  
$K^0\to\pi^+\pi^-$.

\vspace*{5mm}\noindent
Find the relative velocity of the $\pi^+$ and $\pi^-$ mesons from the 
decay $K^0\to\pi^+\pi^-$.

\subsection{}
4-ускорение $a^\mu$ определяется выражением $a^\mu=\frac{du^\mu}{d\tau}$, где
$u^\mu$ - 4-скорость и $\tau$ - собственное время. Найдите скалярные 
произведения $a\cdot u$ и $a\cdot a$, если частица движется прямолинейно и 
зависимость ее скорости от лабораторной времени, $V(t)$, известна.

\vspace*{5mm}\noindent
4-acceleration $a^\mu$ is defined by the expression $a^\mu=\frac{du^\mu}
{d\tau}$, where $u^\mu$ is 4-velocity and $\tau$ -- proper time. Find 
scalar products $a\cdot u$ and $a\cdot a$ if the particle moves 
rectilinearly and the dependence of its speed on the laboratory time, 
$V(t)$, is known.

\section*{Контрольная работа 6}
\setcounter{section}{6}
\setcounter{subsection}{0}
\subsection{}
Два фотона с частотами $\nu_1$ и $\nu_2$ движутся в противоположных 
направлениях. Найти массу этой системы и скорость ее центра масс.

\vspace*{5mm}\noindent
Two photons with frequencies $\nu_1$ and $\nu_2$ are moving in opposite 
directions. Find the mass of this system and the speed of its center of 
mass.

\subsection{}
Неподвижный $\phi$-мезон распадается по схеме $\phi\to\pi^+\pi^-\pi^0$. найти
минимальный угол разлета $\gamma$-квантов от последующего распада $\pi^0\to
2\gamma$.

\vspace*{5mm}\noindent
$\phi$-meson at rest decays under the scheme $\phi\to\pi^+\pi^-\pi^0$. 
Find the minimal opening angle between the $\gamma$-quanta from the 
subsequent decay $\pi^0\to 2\gamma$.

\subsection{}
Быстрый электрон, имеющий скорость $V$, сталкивается с неподвижным ядром 
массы $M$ и испускает тормозный фотон. Какую максимальную энергию может иметь 
этот фотон?

\vspace*{5mm}\noindent
A fast electron with the speed $V$ collides with a motionless nucleus of
mass $M$ and emits a bremsstrahlung photon. What maximal energy this photon 
can have?

\subsection{}
К чему стремится отношение максимальных углов разлета в распадах 
$\phi\to K^+K^-$ и $\phi\to \pi^+\pi^-$ для очень больших энергии 
$\phi$-мезона?

\vspace*{5mm}\noindent
To what limit tends the ratio of the maximal opening angles in decays 
$\phi\to K^+ K^-$ and $\phi\to\pi^+\pi ^-$ for very energetic $\phi$-meson?

\subsection{}
Неподвижный $\omega$-мезон распадается по схеме $\omega\to\pi^+\pi^-\pi^0$.
В каких пределах может меняться энергия $\pi^-$-мезона, если энергия 
$\pi^+$-мезона равна $E_1=300~\mbox{МэВ}$?   

\vspace*{5mm}\noindent
$\omega$-meson at rest decays under the scheme $\omega\to\pi^+\pi^-\pi^0$. 
In what limits can vary the energy of the $\pi^-$-meson if the energy 
of the $\pi^+$-meson is equal to $E_1=300~\mbox{MeV}$?

\section*{Контрольная работа 7}
\setcounter{section}{7}
\setcounter{subsection}{0}
\subsection{}
В центре металлического желоба лежат два металлических шара, связанные нитью. 
Между шарами находится сжатая пружина. Нить пережигают и шары приобретают 
одинаковые по величине и противоположные скорости $u$. Пока шары не упали с 
желоба, лампочка в показанной на рисунке схеме не горит, т.к. элемент питания 
коротко замкнут. Но шары упадут одновременно и лампочка все равно не 
загорится. Однако, в системе наблюдателя, который движется со скоростью $V$ 
как на рисунке, правый шар падает раньше и лампочка должен загориться. 
Объясните парадокс.

\begin{figure}[htb]
\centerline{\epsfig{figure=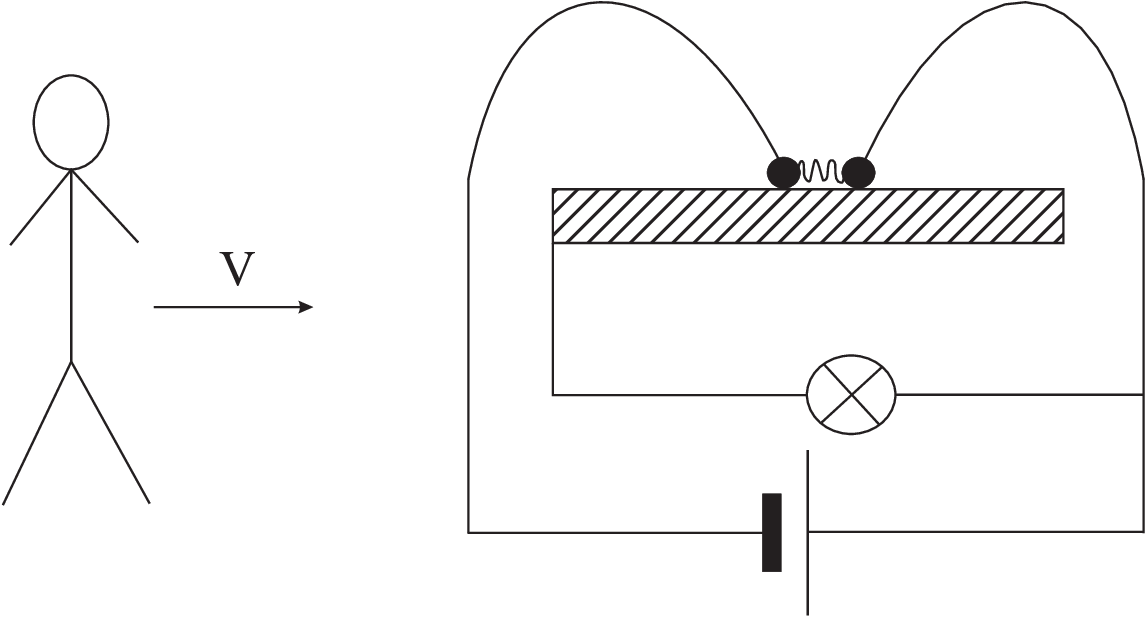,height=4cm}}
\end{figure}\noindent
In the center of a metal gutter two metal spheres connected by a strand 
lay. Between spheres there is a compressed spring. The strand is burnt 
through and the spheres acquire identical in magnitude and opposite 
speeds $u$. While spheres have not fallen from the gutter, the bulb in 
the electric scheme shown in the figure does not burn, since the electric
element is short circuited. But spheres will fall simultaneously and the 
bulb will not light up either. However, in the frame of an observer which 
moves with the speed $V$ as in the figure, the right sphere falls earlier 
and the bulb should light up. Explain the paradox.

\subsection{}
Ракета удаляется от Земли с постоянным собственным ускорением $g$. Через время 
$\tau$ после старта ракеты, радиолокатор посылает ей вдогонку короткий сигнал 
с частотой $\nu$. Какую частоту отраженного сигнала зафиксируют на Земле?

\vspace*{5mm}\noindent
A rocket leaves the Earth with constant proper acceleration $g$. After 
time  $\tau$ from its start, a radar sends to it a short signal with 
frequency $\nu$. What frequency of the reflected signal will be 
registered on the Earth?

\subsection{}
Электрон и протон находятся рядом в электрическом поле. Причем в начальный 
момент времени протон неподвижен, а электрон имеет энергию $E$ и движется в 
направлений поля. Для каких $E$ пойдет реакция $e^-+p\to e^-+p+e^-+e^+$ когда 
электрон и протон столкнутся?

\begin{figure}[htb]
\centerline{\epsfig{figure=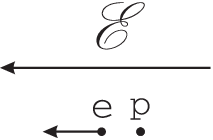,height=2cm}}
\end{figure}\noindent
An electron and a proton are close to each other in the electric field. 
At the initial moment of time the proton is motionless and the electron
has the energy $E$ and moves in the direction of the field. For what $E$ 
would the reaction $e^-+ p\to e^-+p+e^-+e^+$ be allowed when the electron
and the proton do collide?

\subsection{}
Ракета удаляется от Земли с постоянным в сопутствующей системе ускорением $g$.
Как связаны 4-координаты события $(ct^\prime,x^\prime,y^\prime,z^\prime)$ в 
системе ракеты с 4-координатами события $(ct,x,y,z)$ в системе Земли? 

\vspace*{5mm}\noindent
A rocket leaves the Earth with constant proper acceleration $g$. How are
related the 4-coordinates $(ct^\prime,x^\prime,y^\prime,z^\prime)$ 
of an event in the rocket's frame with 4-coordi\-nates of the event 
$(ct,x,y,z)$ in the Earth's frame?

\subsection{}
Идеальное зеркало массой 1~кг ускоряется лучом лазера мощности 
$9\cdot 10^9$~Вт, расположенного на Земле. За какое время зеркало улетит на 
расстояние $10^{15}$~м?

\vspace*{5mm}\noindent
An ideal mirror of mass 1~kg is accelerated by a beam of the laser of 
power $9\cdot 10^9$ watts, located on the Earth. For what time will the 
mirror depart on the distance $10^{15}$~m?

\section*{Контрольная работа 8}
\setcounter{section}{8}
\setcounter{subsection}{0}
\subsection{}
Однородный стержень длины $l$ опирается на гладкую горизонтальную поверхность 
и начинает падать на плоскость без начальной скорости из вертикального 
положения. найти траекторию верхнего конца стержня.

\vspace*{5mm}\noindent
A homogeneous rod of length $l$ leans on a smooth horizontal surface and 
starts to fall on the plane without initial speed from the vertical 
position. Find the trajectory of the top end of the rod.

\subsection{}
Электрон в начальный момент покоился на расстоянии $H$ от бесконечной 
проводящей незаряженной плоскости. Найти время, за которое электрон достигнет 
плоскости.

\vspace*{5mm}\noindent
At the initial moment an electron at rest was situated at a distance $H$ 
from the infinite, conducting, not charged plane. Find the time for which 
the electron will reach the plane.

\subsection{}
Точка с массой $m$ движется в поле с потенциалом $$U(x)=U_0\tg^2{\left (
\frac{x}{a}\right )}.$$ Найти закон движения точки.

\vspace*{5mm}\noindent
A point with mass $m$ moves in a field with potential $$U(x)=U_0\tg^2 
{\left (\frac{x}{a}\right)}.$$ Find the position $x(t)$ of the point
as a function of time.

\subsection{}
Шарик массы $m$ падает с высоты $H$ на горизонтальную плиту и упруго 
отскакивает. Найти высоту последующего подъема шарика, если сила сопротивления
воздуха $F=-\alpha V^2$.

\vspace*{5mm}\noindent
A ball of mass $m$ falls from the height $H$ on a horizontal plate and 
elastically rebounds. Find the height of the subsequent rise of the ball, 
if the air resistance force is $F=-\alpha V^2$.

\subsection{}
Электрон движется в поле магнитного монополя $$\vec{B}=g\frac{\vec{r}}{r^3}.$$
Как зависит момент импульса электрона $\vec{L}=\vec{r}\times \vec{p}$ от его
радиус-вектора?

\vspace*{5mm}\noindent
An electron moves in the field of a magnetic monopole $$\vec{B}=g\frac{
\vec {r}}{r^3}.$$ Find the angular momentum of the electron $\vec{L}
=\vec{r}\times\vec{p}$ as a function of its radius-vector.

\subsection{}
Однородное и постоянное электрическое поле $\cal{E}$ направлено вдоль оси $x$.
Из начала координат, вдоль оси $y$ вылетает релятивистский протон с энергией 
$E_0$. Найти его траекторию.

\vspace*{5mm}\noindent
A homogeneous and constant electric field $\cal{E}$ is directed along 
the $x$-axis. A relativistic proton with initial energy $E_0$ darts 
along the $y$-axis from the origin of coordinates. Find its trajectory.

\section*{Контрольная работа 9}
\setcounter{section}{9}
\setcounter{subsection}{0}
\subsection{}
Два поршня, способные двигаться без трения в герметично соединенных трубах, 
связаны жестким стержнем, как показано на рисунке. Площади поршней равны 
$S_1$ и $S_2$. В равновесии объем воздуха между поршнями равен $V$. Найти 
частоту малых колебаний поршней, если их суммарная масса вместе со стержнем 
равна $m$. Температура  поддерживается постоянной. Атмосферное давление равно
$p$.

\begin{figure}[htb]
\centerline{\epsfig{figure=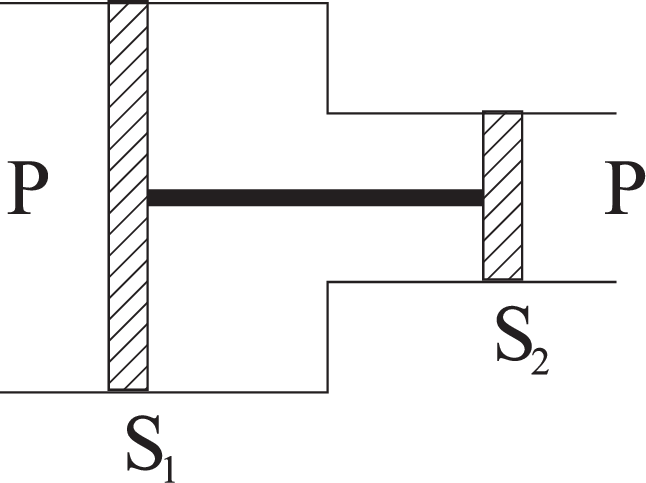,height=4cm}}
\end{figure}\noindent
Two pistons capable to move without friction in tightly connected pipes 
are connected by a rigid rod, as shown in the figure. The areas of pistons 
are equal to $S_1$ and $S_2$ respectively. At equilibrium the air volume 
between the pistons is equal to $V$. Find the frequency of small oscillations
of the pistons if their total mass together with the rod is $m$. 
The temperature is kept constant. The atmospheric pressure is equal to $p$.

\subsection{}
Тонкий и гибкий нерастяжимый канат длины $L$, сложенный пополам, уложен в 
тонкой горизонтальной гладкой трубке. Один конец нити закреплен, и примыкающая
к нему половина каната покоится, а второй половине сообщается начальная 
скорость $V_0$. Найти длину покоящейся части каната в тот момент, когда 
скорость точки изгиба станет равной скорости звука $u$.

\begin{figure}[htb]
\centerline{\epsfig{figure=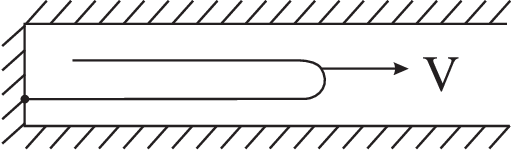,width=6cm}}
\end{figure}\noindent
The thin and flexible not extensible rope of length $L$, combined 
half-and-half, is laid in a thin horizontal smooth tube. One end of the 
rope is fixed, and the half of the rope connected to it is at rest, while
the initial velocity $V_0$ is given to the second half. Find the length 
of the motionless part of the rope at the moment when the speed of the
bending point becomes equal to the speed of sound $u$.

\subsection{}
Тело, масса которого равна $m$, может перемещаться по горизонтальной прямой. К
нему прикреплена пружина, коэффициент жесткости которой $k$. Второй конец 
пружины укреплен в неподвижной точке, как показано на рисунке. При угле 
$\alpha=\alpha_0$ пружина не деформирована. Найти частоту малых колебаний тела.

\begin{figure}[htb]
\centerline{\epsfig{figure=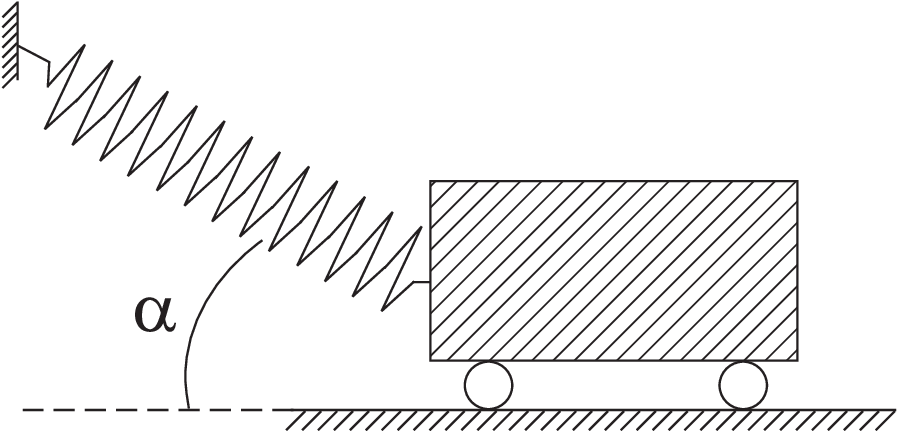,height=4cm}}
\end{figure}\noindent
A body of mass $m$ can move on a horizontal straight line with a spring 
of rigidity $k$ attached to it. The second end of the spring is fixed
as shown in the figure. When the angle $\alpha=\alpha_0$, the spring is not 
deformed. Find the frequency of small oscillations of the body.

\subsection{}
Найти изменение периода гармонического осциллятора при малой ангармонической 
добавке $\delta U=\frac{m\beta x^4}{4}\ll U$ к потенциалу 
$U(x)=\frac{m\omega^2 x^2}{2}$.

\vspace*{5mm}\noindent
Find the change of the oscillator period when a small anharmonic 
perturbation $\delta U=\frac{m\beta x^4}{4}\ll U $ is added to the harmonic
potential $U(x)=\frac{m\omega^2 x^2}{2}$.

\subsection{}
Фотон и электрон движутся навстречу. После лобового столкновения происходит 
реакция $\gamma +e^-\to e^-+e^-+e^+$. Найти энергию фотона, если все продукты 
реакции покоятся.

\vspace*{5mm}\noindent
Photon and electron move towards each other. After a head-on collision 
the reaction $\gamma+e^-\to e^-+e^-+e^+$ happens. Find the energy of the 
photon if all reaction products are at rest.

\section*{Контрольная работа 10}
\setcounter{section}{10}
\setcounter{subsection}{0}
\subsection{}
Шарик массы $m$ прикреплен к концам упругой безмассовой нити, которая 
перекинута через два горизонтальных гладких стержня. Стержни находятся на 
одном уровне и расстояние между ними $l$. Длина нити в ненапряженном состоянии
$2l$. Найти частоту малых вертикальных колебаний шарика, если в состоянии 
равновесия нить образует равносторонний треугольник. 

\begin{figure}[htb]
\centerline{\epsfig{figure=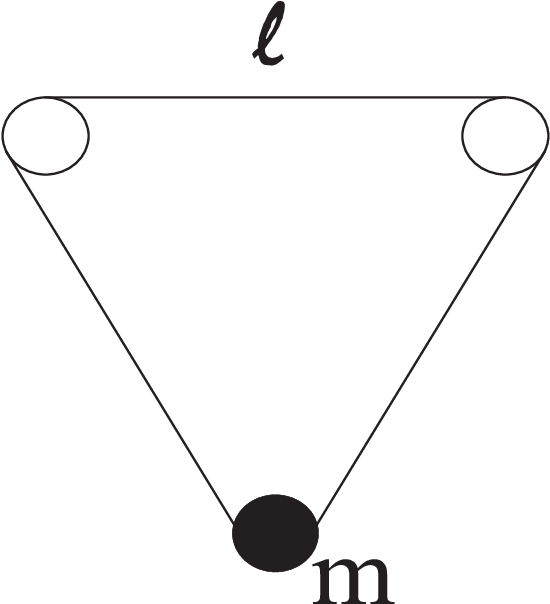,height=4cm}}
\end{figure}\noindent
A ball of mass $m$ is attached to the ends of an elastic massless string 
which is stretched over two horizontal smooth cores. The cores are at the 
same level and the distance between them is $l$. The length of not 
deformed string is $2l$. Find the frequency of small vertical oscillations
of the ball if at equilibrium the string forms a equilateral triangle.

\subsection{}
Линейному осциллятору частоты $\omega$ передается энергия под воздействием силы
$F(t)=F_0 e^{-t^2/\tau^2}$. Каков должен быть $\tau$, чтобы передача энергии
была максимально эффективной? В момент времени $t=-\infty$ осциллятор покоился
в положении равновесия.

\vspace*{5mm}\noindent
Energy is transferred to a linear oscillator of frequency $\omega$ under 
influence of the force $F(t)=F_0 e^{-t^2/\tau^2}$. For what $\tau$ is the
energy transfer maximally effective? At the time $t=-\infty$ the 
oscillator was at rest in the equilibrium point.

\subsection{}
Найти закон вынужденных колебаний линейного одномерного осциллятора под 
воздействием силы $F(t)$ при наличии силы трения $F=-\alpha V$.

\vspace*{5mm}\noindent
Find the law of forced oscillations of a linear one-dimensional 
oscillator under influence of the force $F(t)$ in presence of the
friction force $F=-\alpha V$.

\subsection{}
Песочные часы, в котором установился стационарный поток песка, поставили на 
весы. Доказать, что показания весов будет больше чем $Mg$, где $M$ масса 
песочных часов (вместе песком разумеется).

\vspace*{5mm}\noindent
An hourglass, in which the stationary stream of the sand was established, 
have been put on a sensitive balance. Prove that the display of the balance
will be more than $Mg$, where $M$ is the mass of the hourglass (with the 
sand of course).

\subsection{}
Из-за релятивистских эффектов период гармонического осциллятора зависит от 
амплитуды. Считая, что релятивистские эффекты маленькие, найти соответствующую
поправку.

\vspace*{5mm}\noindent
Because of the relativistic effects the period of harmonic oscillator depends
on its amplitude. Assuming the relativistic effects are small, find the 
corresponding correction.
 
\section*{Контрольная работа 11}
\setcounter{section}{11}
\setcounter{subsection}{0}
\subsection{}
Две одинаковые математические маятники, массы $m$ и частоты $\omega$, связаны 
слабой пружиной жесткости $k\ll m\omega^2$. Расстояние между точками подвеса
маятников равно длине не деформированной пружины. Один маятник отклонили от 
положения равновесия на маленькую величину. Через какое время колебания 
полностью перейдут ко второму маятнику?

\begin{figure}[htb]
\centerline{\epsfig{figure=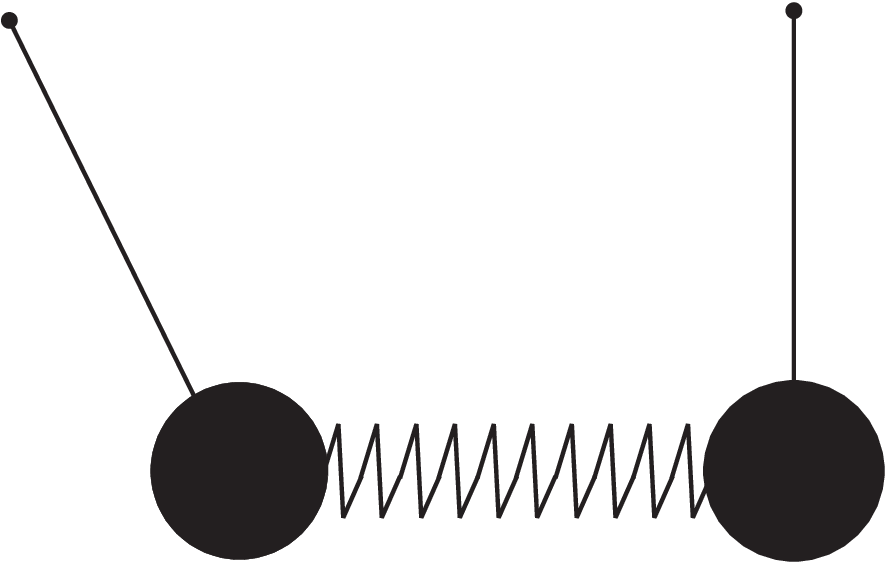,height=4cm}}
\end{figure}\noindent
Two identical mathematical pendula, with mass $m$ and frequency $\omega$, 
are connec\-ted by a weak spring of rigidity $k\ll m\omega^2$. The distance 
between the hanging points of the pendula equals to the length of the not 
deformed spring. One pendulum is set at a small distance from the 
equilibrium point and begins to swing. What time it takes the oscillations 
to pass completely to the second pendulum?

\subsection{}
Электрон движется по окружности радиуса $R$ в однородном магнитном поле. 
Величина поля медленно возрастает. Какой будет радиус орбиты электрона когда 
поле увеличится в два раза? Лагранжиан заряженной частицы в однородном 
магнитном поле, направленным вдоль оси $z$, имеет вид (в системе CGSE)
$${\cal{L}}=\frac{mV^2}{2}+\frac{qB}{2c}(x\dot y-y\dot x).$$  

\vspace*{5mm}\noindent
An electron moves on a circle of radius $R$ in a homogeneous magnetic 
field. The field slowly increases. What will be radius of the electron 
orbit when the magnetic field will increase twice? Lagrangian
of a charged particle in a homogeneous magnetic field, directed along 
the $z$-axis, looks like (in the CGSE system of units) 
$${\cal{L}}=\frac{mV^2}{2}+\frac {qB}{2c}(x\dot y-y\dot x).$$

\subsection{}
Маленькому кольцу, надетому на проволочную горизонтальную окружность радиуса 
$R$, сообщили начальную скорость $V_0\ll\sqrt{gR}$. Коэффициент трения кольца 
о проволоку равен $\mu$. Через какое время кольцо остановится?

\vspace*{5mm}\noindent
An  initial velocity $V_0 \ll \sqrt {gR}$ has been imparted to a small ring 
on a horizontal circular wire of radius $R$. The friction coefficient 
between the ring and the wire is equal to $\mu$. After what time will 
the ring stop?

\subsection{}
На крыше с углом наклона $\varphi$ лежит свинцовый лист длины $l$. Коэффициент
трения свинца о крышу $\mu>\tg{\varphi}$. Коэффициент линейного расширения 
свинца $\alpha$. В течении $N$ суток устойчивой погоды суточный перепад 
температуры составлял $\Delta t$. На какое расстояние сползет лист?

\vspace*{5mm}\noindent
A lead leaf of length $l$ lays on a roof with an angle of inclination 
$\varphi$. The friction coefficient between the lead and the roof 
$\mu>\tg{\varphi}$. Linear thermal expansion coefficient of lead is 
$\alpha$. During $N$ days of steady weather daily temperature drop was 
$\Delta t$. On what distance will the leaf slip?

\subsection{}
Неподвижный $\tau$ лептон испытывает радиационный распад $\tau^-\to\pi^-\pi^0
\nu_\tau\gamma$. В каких пределах может меняться энергия фотона, если массы 
подсистем $(\pi^-\pi^0)$ и $(\nu_\tau\gamma)$ равны друг другу и составляют
одну треть массы $\tau$ лептона $m_\tau$. Масса нейтрино $\nu_\tau$ равна 
нулю.  

\vspace*{5mm}\noindent
$\tau$ lepton at rest undergoes radiative decay $\tau^-\to\pi^-\pi^0
\nu_\tau\gamma$. In what limits does the energy of the photon change if 
invariant masses of subsystems $(\pi^-\pi^0)$ and $(\nu_\tau\gamma)$ are 
equal to each other and constitute one third of the $\tau$ lepton mass 
$m_\tau$. The mass of $\nu_\tau$ neutrino is equal to zero.

\section*{Контрольная работа 12}
\setcounter{section}{12}
\setcounter{subsection}{0}
\subsection{}
При каком направлении начальной скорости космический аппарат упадет на 
поверхность Земли вне зависимости от величины начальной скорости?

\vspace*{5mm}\noindent
At what direction of the initial speed a space vehicle will fall to 
Earth's surface irrespective of the initial speed magnitude?

\subsection{}
В каком центральном поле точка движется по траектории $$\frac{p}{r}=1+e\cos{
(\epsilon \varphi)}?$$ Здесь $p,\,e$ и $\epsilon$ -- некоторые константы.

\vspace*{5mm}\noindent
In what central force field does a point move on the trajectory 
$$\frac{p}{r}=1+e\cos{(\epsilon\varphi)}?$$ Here $p,\,e$ and $\epsilon$ 
are some constants.

\subsection{}
Найти наивыгоднейший (с точки зрения минимизации потребной начальной скорости)
угол запуска к горизонту баллистической ракеты на северном полюсе, чтобы она 
попала на экватор.

\vspace*{5mm}\noindent
Find the optimal launching angle to the horizon of a ballistic missile on 
the North Pole (from the point of view of minimization of the initial 
speed) for the missile to land on the equator.

\subsection{}
Как изменится третий закон Кеплера, если учесть движение Солнца, вызванное 
притяжением соответствующей планеты?

\vspace*{5mm}\noindent
How does Kepler's third law get modified if the movement of the Sun, 
caused by an attraction of the corresponding planet, is taken into 
account?

\subsection{}
Два спутника, имеющие равные массы, движутся в одном направлении по 
компланарным орбитам, одна из которых -- круговая радиуса $R$, а другая -- 
эллиптическая с расстояниями перигея и апогея $R$ и $8R$ соответственно. 
Полагая, что спутники путем непосредственной стыковки соединились друг с 
другом в точке соприкосновения их орбит
и дальнейшее движение продолжили вместе, найти апогей их новой орбиты.

\vspace*{5mm}\noindent
Two satellites, having equal masses, move in the same direction on 
coplanar orbits, one of which is circular of radius $R$, and the 
another is elliptic with the perigee and apogee distances $R$ and $8R$ 
respectively. Assuming that the satellites have been joined to each 
other at the contiguity point of their orbits and have their further motions 
continued together, find the apogee of their new orbit.

\subsection{}
Для каких значений момента импульса $L$ релятивистской частицы возможны 
круговые орбиты в потенциале $$U(r)=-\frac{\alpha}{r^\beta},\;\; \alpha>0,\,
\beta>0?$$ 

\vspace*{5mm}\noindent
For what values of the angular momentum $L$ are circular orbits possible 
in the potential $$U(r)=-\frac{\alpha}{r^\beta}, \;\; \alpha>0,\,
\beta>0,$$ for relativistic particles?

\section*{Контрольная работа 13}
\setcounter{section}{13}
\setcounter{subsection}{0}
\subsection{}
Найти зависимость угла рассеяния от прицельного параметра для быстрых 
электронов, пролетающих мимо очень тяжелого магнитного монополя. 

\vspace*{5mm}\noindent
Find the dependence of the scattering angle on the impact parameter for 
fast electrons flying past a very heavy magnetic monopole.

\subsection{}
Найти сечение рассеяния на угол, больший $90^\circ$, в центральном поле с 
потенциалом $$U(r)=\frac{\alpha}{r^2}, \;\;\alpha>0.$$ Налетающая частица имеет 
кинетическую энергию $T$.

\vspace*{5mm}\noindent
Find the cross section for scattering at angles greater than $90^\circ$ 
in the central force field with the potential $$U(r)=\frac{\alpha}{r^2}, 
\;\;\alpha>0.$$ The incoming particle has a kinetic energy $T$.

\subsection{}
Четыре звезды одинаковой массы $M$, находясь очень далеко друг от друга, 
расположены в вершинах квадрата и имеют одинаковые по величине скорости $V$, 
прицельные параметры которых, относительно центра квадрата, тоже одинаковы и 
равны $b$. Найти угол рассеяния звезд.

\begin{figure}[htb]
\centerline{\epsfig{figure=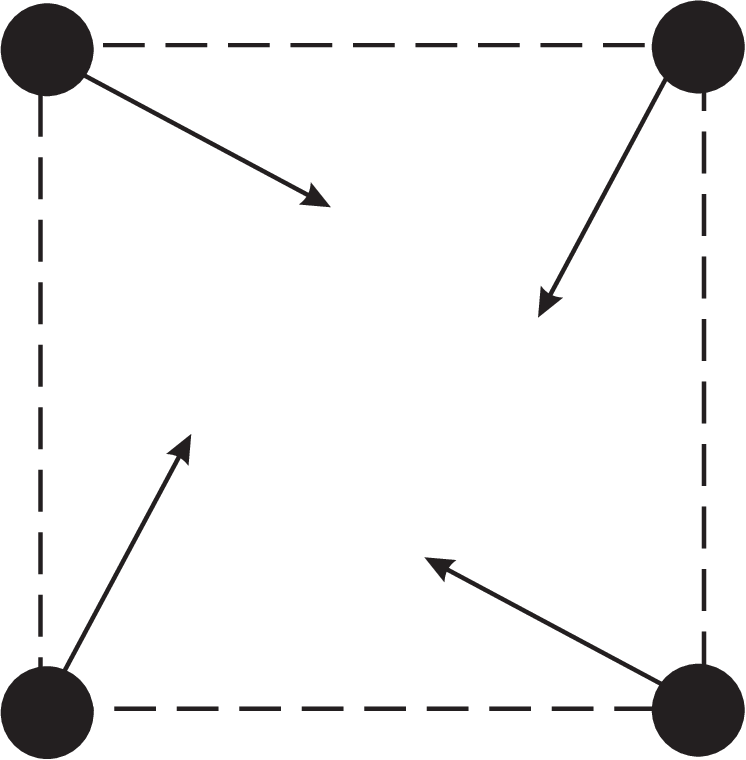,height=2.3cm}}
\end{figure}\noindent
Four stars of identical mass $M$, being very far from each other, are 
located at vertexes of a square and have identical in magnitude speeds 
$V$. Impact parameters of the stars, with respect to the center of 
the square, are also identical and equal $b$. Find the scattering angle 
of the stars.

\subsection{}
Одинаковые гладкие пластины длины $a$ укладываются одна на другой. Какое 
минимальное количество пластин понадобится, чтобы построить равновесную 
конструкцию с пролетом $L=3a$?

\begin{figure}[htb]
\centerline{\epsfig{figure=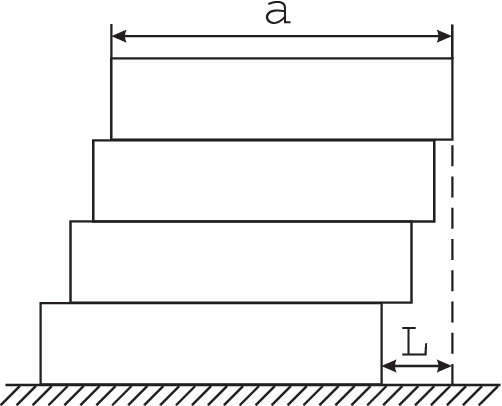,height=4cm}}
\end{figure}\noindent
Identical smooth plates of length $a$ are placed one on another. What 
minimum number of plates is required to construct an equilibrium 
construction with overhang $L=3a$?

\subsection{}
Однородный шар веса $Q$ и радиуса $R$ и гиря веса $P$ подвешены на веревках в 
точке $A$, как показано на рисунке. Расстояние от $A$ до центра шара $AO=l$. 
Определить, какой угол образует прямая $AO$ с вертикалью при равновесии.

\begin{figure}[htb]
\centerline{\epsfig{figure=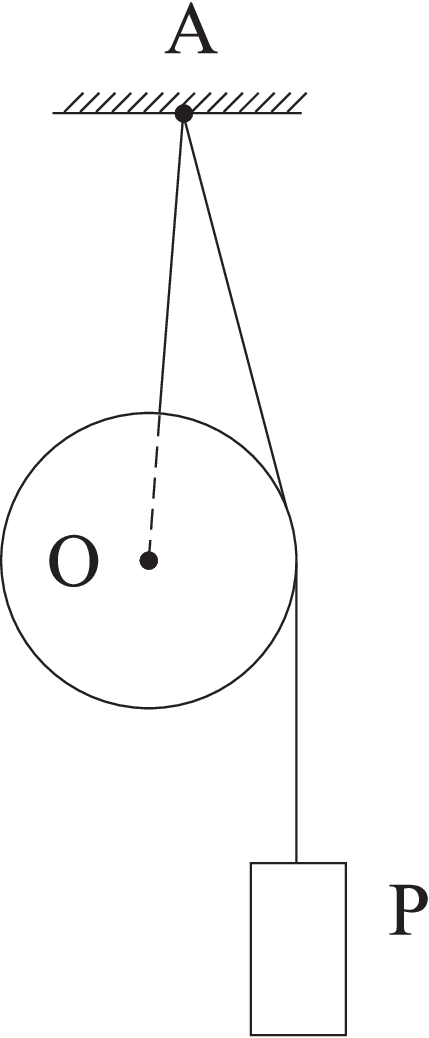,height=4cm}}
\end{figure}\noindent
A homogeneous sphere of weight $Q$ and radius $R$ and a weight $P$ 
are suspended on cords fixed in a point $A$, as shown in the figure. 
The distance from the point $A$ to the center of the sphere $AO=l$. 
Find the angle the straight line $AO$ makes with the vertical at 
balance.

\section*{Контрольная работа 14}
\setcounter{section}{14}
\setcounter{subsection}{0}
\subsection{}
Тело массы $M$ висит на гибком канате массы $m$. Найти частоту вертикальных 
колебании, если коэффициент жесткости каната равен $k$.

\begin{figure}[htb]
\centerline{\epsfig{figure=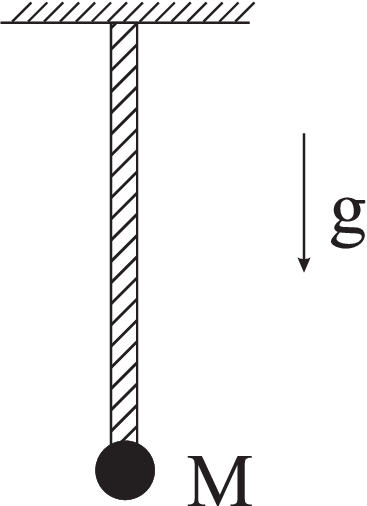,height=4cm}}
\end{figure}\noindent
A body of mass $M$ hangs on a flexible rope of mass $m$. Find the 
frequency of vertical oscillations if the coefficient of rigidity of the 
rope is equal to $k$.

\subsection{}
Кольцо массы $m$ подвешено симметрично на трех нерастяжимых нитях длины $l$ 
каждый. Найти частоту малых крутильных колебаний кольца относительно 
вертикальной оси проходящей через его центр.

\begin{figure}[htb]
\centerline{\epsfig{figure=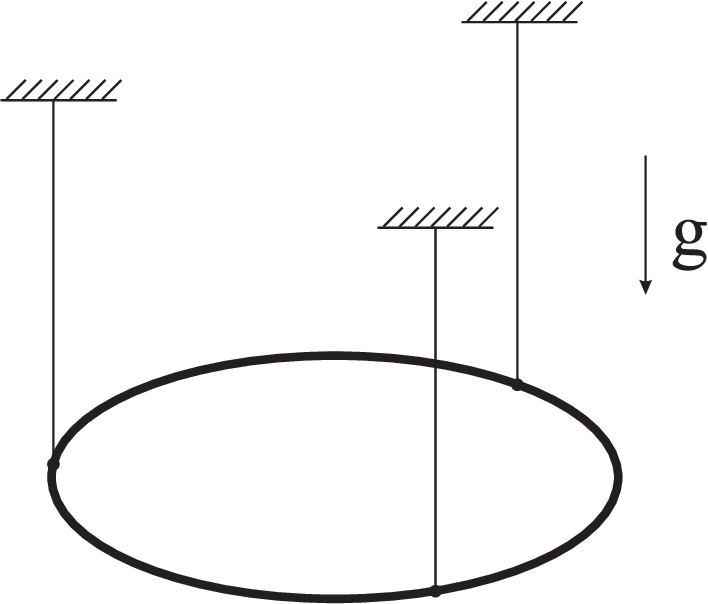,height=4cm}}
\end{figure}\noindent
A ring of mass $m$ is suspended symmetrically on three not extensible 
strings of length $l$ each. Find the frequency of small torsion 
oscillations of the ring around the vertical axis passing through its 
center.

\subsection{}
Частица движется внутри чашки по круговой горизонтальной орбите радиуса $r$. 
Поверхность чашки имеет форму гиперболоида вращения $\rho^2-z^2=R^2$ 
(в цилиндрических координатах). Для каких значении радиуса $r$ такое движение 
становится нестабильным?

\begin{figure}[htb]
\centerline{\epsfig{figure=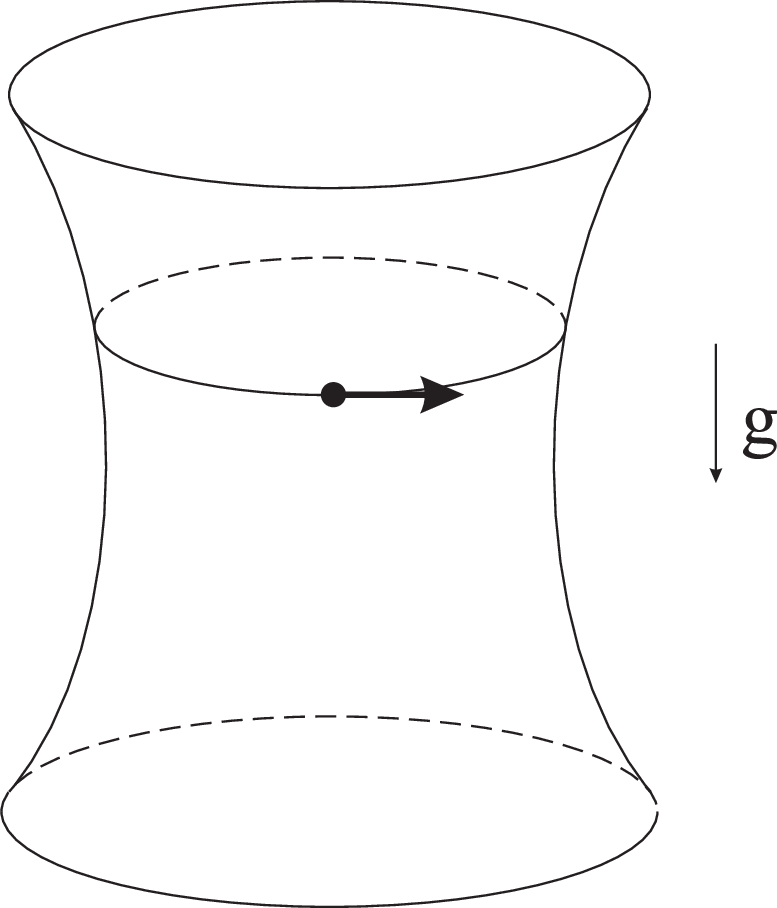,height=6cm}}
\end{figure}\noindent
A particle moves inside a cup on a circular horizontal orbit of 
radius $r$. The surface of the cup has the form of a hyperboloid of 
revolution $\rho^2-z^2=R^2$ (in cylindrical coordinates). For what values 
of radius $r$ does the orbit become unstable?

\subsection{}
Однородный стержень массы $M$ подвешен на двух вертикальных нерастяжимых нитях 
одинаковой длины, прикрепленных к концам стержня. Найти натяжение одной из 
нитей в момент обрыва другой.

\begin{figure}[htb]
\centerline{\epsfig{figure=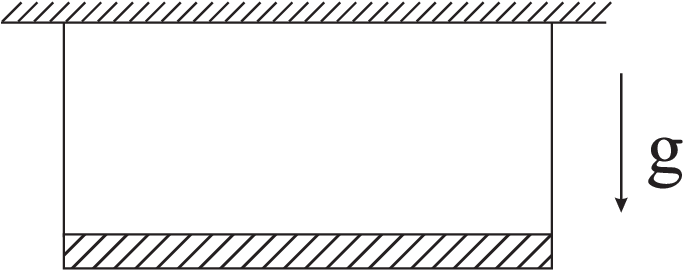,height=2cm}}
\end{figure}\noindent
A homogeneous core of mass $M$ is suspended on two vertical not extensible 
strings of identical lengths, attached to the ends of the core. Find the 
tension of one string at the moment of breakage of another.

\subsection{}
Однородный цилиндр с горизонтальной осью скатывается под действием силы 
тяжести со скольжением по наклонной плоскости при коэффициенте трения 
скольжения $\mu$. При каком угле наклона плоскости к горизонту это возможно?

\vspace*{5mm}\noindent
A homogeneous cylinder with a horizontal axis rolls down under gravity with 
sliding on an inclined plane. The sliding friction coefficient is $\mu$. 
For what inclination angles of the plane to the horizon is this possible?

\subsection{}
На концах трубки со сжатой легкой пружиной удерживаются нитью одинаковые 
шарики массой $m$. При разрыве нити шарики разлетаются с одинаковыми 
скоростями $u$. Объясните как сохраняется импульс при этом процессе с точки 
зрения наблюдателя, который движется вдоль оси трубки с релятивистской 
скоростью $V>u$.

\begin{figure}[htb]
\centerline{\epsfig{figure=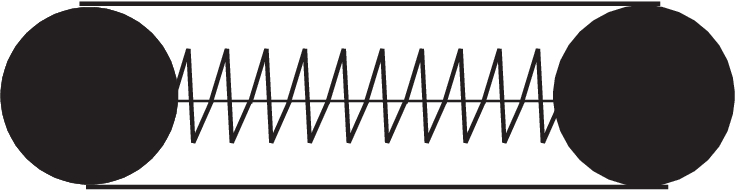,height=1cm}}
\end{figure}\noindent
On the ends of a tube with the compressed massless spring two identical 
balls of mass $m$ are kept by a string. At а breakage of the string the 
balls fly away with identical in magnitude speeds $u$. Explain how 
momentum is conserved in this process from the point of view of an observer 
who moves along the axis of the tube with relativistic speed $V>u$.

\section*{Контрольная работа 15}
\setcounter{section}{15}
\setcounter{subsection}{0}
\subsection{}
Полуцилиндр радиуса $R$ скользит по наклонной плоскости с углом наклона 
$\alpha$ к горизонту и совершает малые колебания вокруг своей оси, которая 
перпендикулярна к плоскости рисунка. Найти частоту этих колебаний. Трение 
отсутствует.

\begin{figure}[htb]
\centerline{\epsfig{figure=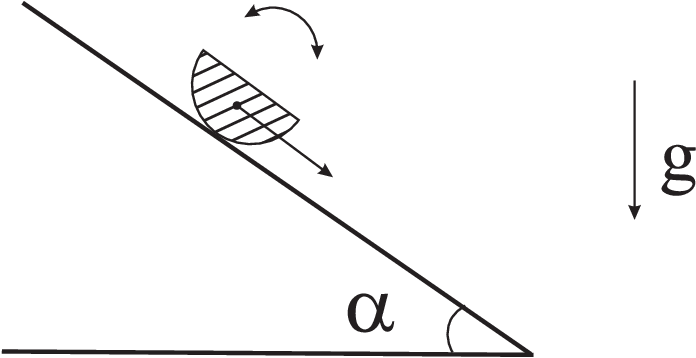,height=4cm}}
\end{figure}\noindent
A half cylinder of radius $R$ slides on an inclined plane with the angle
of inclination $\alpha$ to the horizon and performs small oscillations 
around its axis which is perpendicular to the plane of the figure. Find 
the frequency of these oscillations. Friction is absent.

\subsection{}
Песочные часы плавают в трубке, полностью заполненной водой, немного давя на 
верхнюю крышку трубки. Весь песок находится в нижней части часов. Начнут ли 
часы всплывать, если трубку перевернуть? Внутренний диаметр трубки лишь 
ненамного больше внешнего диаметра песочных часов.

\begin{figure}[htb]
\centerline{\epsfig{figure=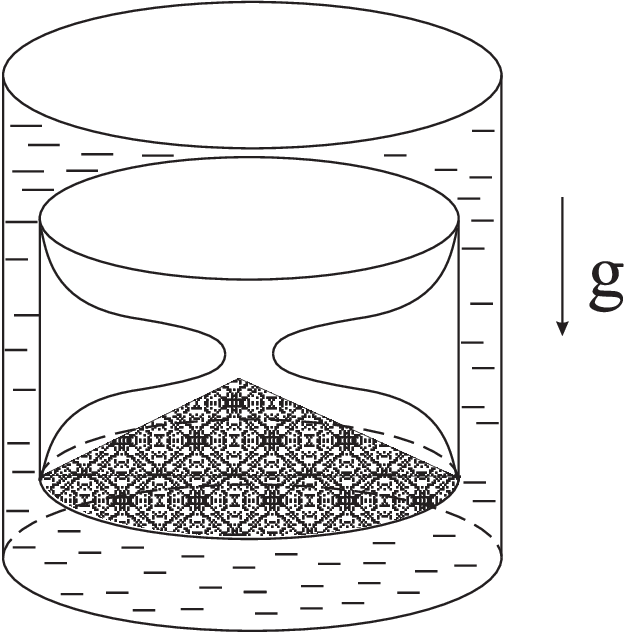,height=4cm}}
\end{figure}\noindent
In a sealed tube completely filled with water a run out hourglass floats
pressing a little on the top cover of the tube. The tube is flipped over. 
Does the hourglass start to float to the top? The internal diameter of 
the tube is only somewhat larger than the external diameter of the 
hourglass.

\subsection{}
Два бакала с вином стоят около краев двух соседних столов. На них опирается 
длинный и тонкий однородный деревянный стержень. Прольется ли вино, если с 
размаха ударить железной палкой ровно по середине стержня?

\begin{figure}[htb]
\centerline{\epsfig{figure=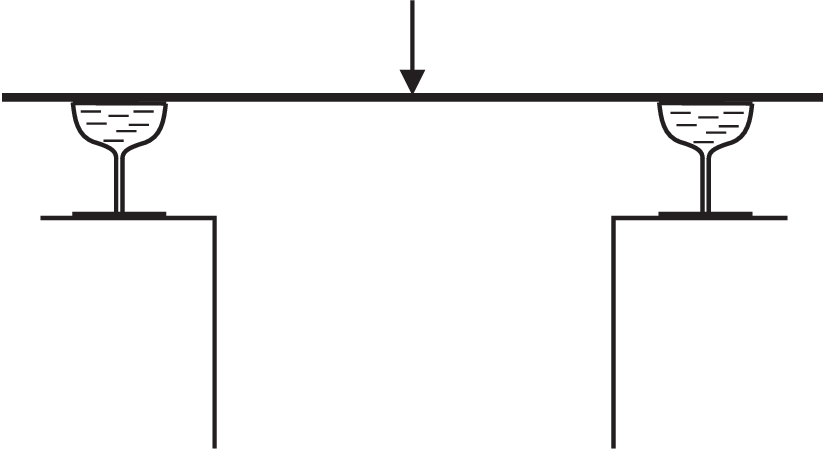,height=4cm}}
\end{figure}\noindent
A broomstick rests on two wine glasses as shown in the figure. Will the wine
spill out after a strong downward blow to the center of the broomstick with
an iron rod?

\subsection{}
Найти угол рассеяния электрона с прицельным параметром $\rho$ и со скоростью 
$V$ в поле магнитного монополя $$\vec{B}=g\frac{\vec{r}}{r^3}.$$

\vspace*{5mm}\noindent
Find  the scattering angle of an electron with the impact parameter $\rho$
and velocity $V$ in the field of a magnetic monopole $$\vec{B}=g\frac
{\vec{r}}{r^3}.$$

\subsection{}
Однородный шар катится без проскальзывания по поверхности поворотного стола, 
который вращается с угловой скоростью $\Omega$. Найти траекторию шара, если 
в начальный момент времени радиус-вектор центра шара равен $\vec{r}_0$, а его
скорость $\vec{V}_0$ (Относительно лабораторной системы).

\vspace*{5mm}\noindent
A homogeneous ball rolls without slipping on the surface of a turntable 
which rotates with angular velocity $\Omega$. Find the trajectory of the
ball if at the initial moment of time the radius vector of the ball equals 
to $\vec{r}_0$ and its velocity is $\vec{V}_0$ (with regard to the 
laboratory frame).

\subsection{}
Проволочка, концы которой закреплены, находится в неоднородном магнитном поле.
При этом натяжение проволочки равно $T$. Какой ток течет по проволочке, если 
ее форма совпадает с траекторией протона с импульсом $p$ в этом поле?

\vspace*{5mm}\noindent
A current-carrying wire with fixed ends is placed in a non-uniform 
magnetic field. The tension of the wire equals to $T$. What current does 
flow in the wire if its form coincides with the trajectory of a proton with 
momentum $p$ in this field?

\section*{\centerline{\LARGE Решения}}
\section*{Контрольная работа 1}
\setcounter{section}{1}
\setcounter{subsection}{0}
\subsection{}
\begin{figure}[htb]
\centerline{\epsfig{figure=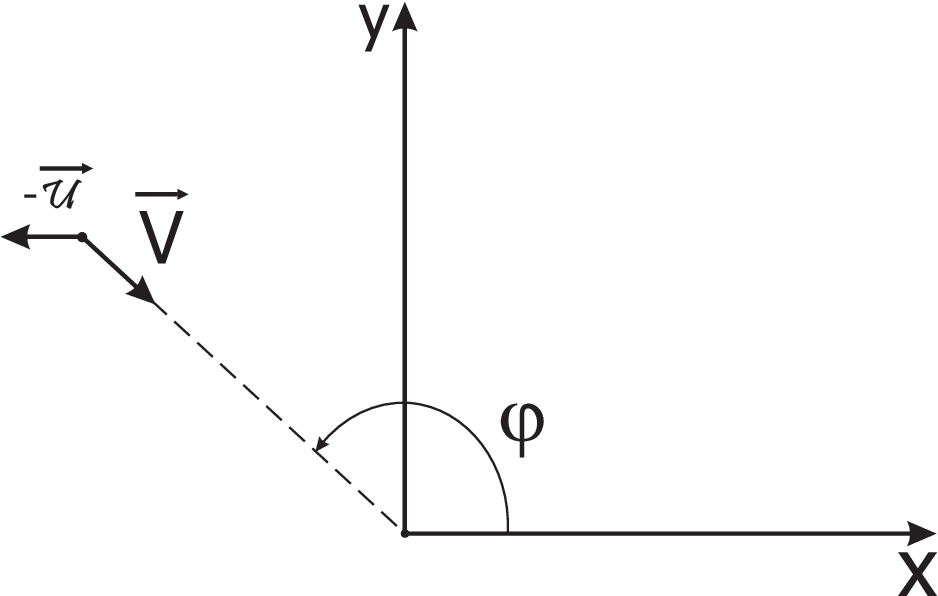,height=4cm}}
\end{figure}
Разложим скорость собаки (в системе зайца) на радиальную и тангенциальную 
компоненты
$$V_r=-V+u\cos{(\pi-\varphi)}=-V-u\cos{\varphi},\;
V_\varphi=u\sin{(\pi-\varphi)}=u\sin{\varphi}. $$
Но $V_r=\dot{r}$ и $V_\varphi=r\dot{\varphi}$. Поэтому
$$\dot{r}=-V-u\cos{\varphi},\; r\dot{\varphi}=u\sin{\varphi}.$$
Делим первое уравнение на второе и учтем 
$$\frac{\dot{r}}{\dot{\varphi}}=\frac{dr}{d\varphi},$$
получаем  
$$\frac{1}{r}\frac{dr}{d\varphi}=-\frac{V+u\cos{\varphi}}{u\sin{\varphi}}.$$
Поэтому
$$\ln{\frac{r}{L}}=-\int\limits_{\pi/2}^\varphi \frac{V+u\cos{\varphi}}
{u\sin{\varphi}}d\varphi.$$
Но
$$-\int\frac{V+u\cos{\varphi}}{u\sin{\varphi}}d\varphi=
\int\frac{\frac{V}{u}+z}{1-z^2}dz,$$
где $z=\cos{\varphi}$. Дальше разложим
$$\frac{\frac{V}{u}+z}{1-z^2}=\frac{A}{1-z}+\frac{B}{1+z},\;
A=\frac{1}{2}\left (1+\frac{V}{u}\right ),\; 
B=\frac{1}{2}\left (\frac{V}{u}-1 \right ).$$
Следовательно,
$$\int\frac{\frac{V}{u}+z}{1-z^2}dz=\ln{\left [ (1+z)^B(1-z)^{-A}\right ]}.$$
Поэтому
$$-\int\limits_{\pi/2}^\varphi \frac{V+u\cos{\varphi}}{u\sin{\varphi}}
d\varphi=\ln{\frac{(1+\cos{\varphi})^B}{(1-\cos{\varphi})^A}}=
\ln{\left [ \left (\ctg{\frac{\varphi}{2}}\right )^\frac{V}{u}\frac{1}
{\sin{\varphi}}\right ]}$$
и уравнение траектории будет
$$r=\frac{L}{\sin{\varphi}}\left (\ctg{\frac{\varphi}{2}}\right )
^\frac{V}{u}.$$

\vspace*{5mm}
\noindent $\bullet$ $\bullet$ $\bullet$

\begin{figure}[htb]
\centerline{\epsfig{figure=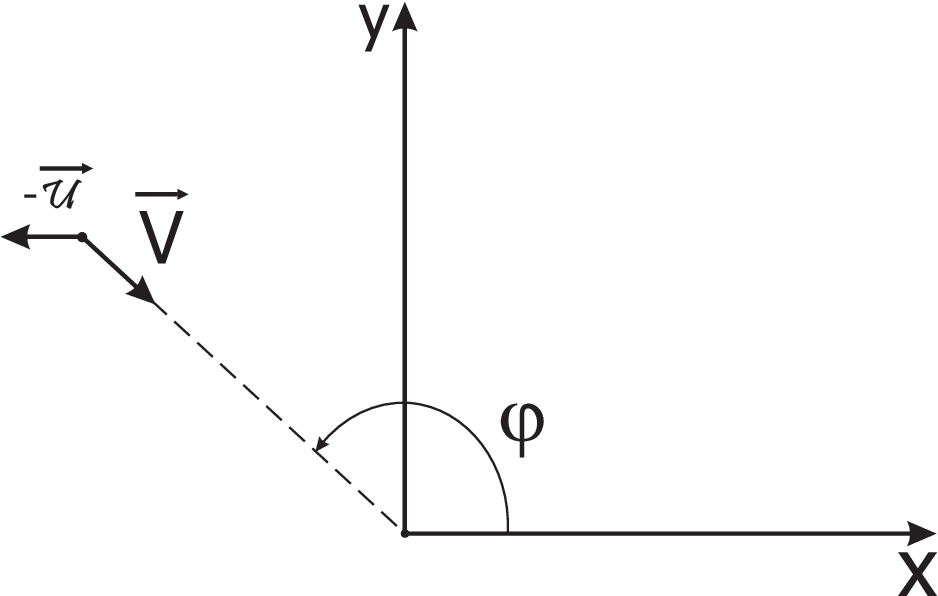,height=4cm}}
\end{figure}
Let us decompose the dog's velocity (in the rabbit's frame) into the radial 
and tangential components 
$$V_r=-V+u\cos{(\pi-\varphi)}=-V-u\cos{\varphi},\;
V_\varphi=u\sin{(\pi-\varphi)}=u\sin{\varphi}. $$
But $V_r=\dot{r}$ and $V_\varphi=r\dot{\varphi}$. Therefore,
$$\dot{r}=-V-u\cos{\varphi},\; r\dot{\varphi}=u\sin{\varphi}.$$
Dividing the first equation over the second one and taking into account that
$$\frac{\dot{r}}{\dot{\varphi}}=\frac{dr}{d\varphi},$$
we get  
$$\frac{1}{r}\frac{dr}{d\varphi}=-\frac{V+u\cos{\varphi}}{u\sin{\varphi}}.$$
Hence,
$$\ln{\frac{r}{L}}=-\int\limits_{\pi/2}^\varphi \frac{V+u\cos{\varphi}}
{u\sin{\varphi}}d\varphi.$$
However,
$$-\int\frac{V+u\cos{\varphi}}{u\sin{\varphi}}d\varphi=
\int\frac{\frac{V}{u}+z}{1-z^2}dz,$$
where $z=\cos{\varphi}$. Using the decomposition 
$$\frac{\frac{V}{u}+z}{1-z^2}=\frac{A}{1-z}+\frac{B}{1+z},\;
A=\frac{1}{2}\left (1+\frac{V}{u}\right ),\; 
B=\frac{1}{2}\left (\frac{V}{u}-1 \right ),$$
the integral is easily calculated
$$\int\frac{\frac{V}{u}+z}{1-z^2}dz=\ln{\left [ (1+z)^B(1-z)^{-A}\right ]}.$$
Hence,
$$-\int\limits_{\pi/2}^\varphi \frac{V+u\cos{\varphi}}{u\sin{\varphi}}
d\varphi=\ln{\frac{(1+\cos{\varphi})^B}{(1-\cos{\varphi})^A}}=
\ln{\left [ \left (\ctg{\frac{\varphi}{2}}\right )^\frac{V}{u}\frac{1}
{\sin{\varphi}}\right ]}$$
and we finally get the following equation for the trajectory
$$r=\frac{L}{\sin{\varphi}}\left (\ctg{\frac{\varphi}{2}}\right )
^\frac{V}{u}.$$

\subsection{}
\begin{figure}[htb]
\centerline{\epsfig{figure=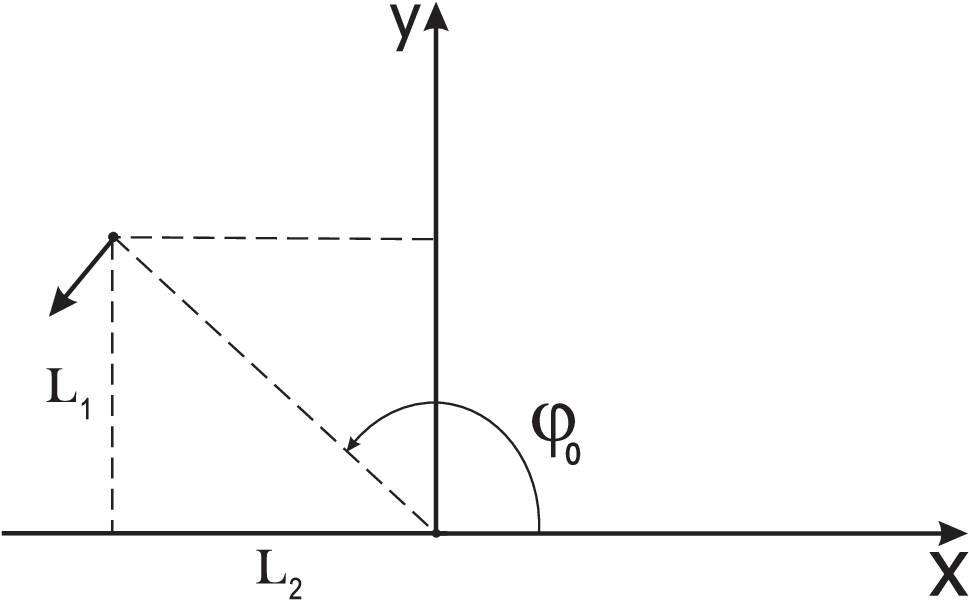,height=4cm}}
\end{figure}
\noindent Из
$$\dot{r}=-u(1+\cos{\varphi}),\; r\dot{\varphi}=u\sin{\varphi}$$
получаем
$$\dot{r}(1-\cos{\varphi})+r\dot{\varphi}\sin{\varphi}=-u(1-\cos^2{\varphi})+
u\sin^2{\varphi}=0.$$
Но
$$\dot{r}(1-\cos{\varphi})+r\dot{\varphi}\sin{\varphi}=
\frac{d}{dt}(r-r\cos{\varphi}).$$
Следовательно,
$$r(1-\cos{\varphi})=C,$$
где $C$ некоторая константа. При $t=0$ имеем (см. рисунок)
$$r_0=\sqrt{L_1^2+L_2^2},\; \cos{\varphi_0}=-\frac{L_2}{\sqrt{L_1^2+L_2^2}}.$$
Поэтому $C=L_2+\sqrt{L_1^2+L_2^2}$. 

В системе зайца $\varphi\to\pi$, когда $t\to\infty$. Действительно, 
$\dot{\varphi}=\frac{u}{r}\sin{\varphi}>0$, если $\varphi<\pi$, и $\varphi$ 
все время растет. Но при приближении к $\pi$, как $\dot{\varphi}$, так и 
$\dot{r}$ стремятся к нулю.

Следовательно, расстояние стремится к предельному значению
$$r_{min}=\frac{C}{1-\cos{\pi}}=\frac{L_2+\sqrt{L_1^2+L_2^2}}{2}.$$ 

\vspace*{5mm}
\noindent $\bullet$  $\bullet$  $\bullet$ 

\begin{figure}[htb]
\centerline{\epsfig{figure=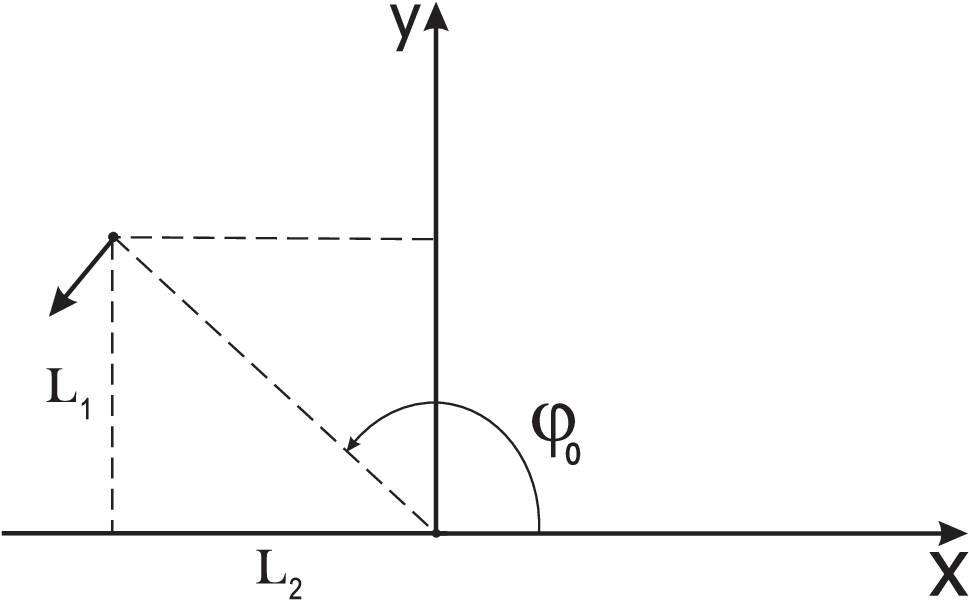,height=4cm}}
\end{figure}
\noindent From
$$\dot{r}=-u(1+\cos{\varphi}),\; r\dot{\varphi}=u\sin{\varphi}$$
we get
$$\dot{r}(1-\cos{\varphi})+r\dot{\varphi}\sin{\varphi}=-u(1-\cos^2{\varphi})+
u\sin^2{\varphi}=0.$$
But
$$\dot{r}(1-\cos{\varphi})+r\dot{\varphi}\sin{\varphi}=
\frac{d}{dt}(r-r\cos{\varphi}).$$
Therefore,
$$r(1-\cos{\varphi})=C,$$
where $C$ is a constant. At $t=0$ we have (see the figure)
$$r_0=\sqrt{L_1^2+L_2^2},\; \cos{\varphi_0}=-\frac{L_2}{\sqrt{L_1^2+L_2^2}}.$$
Hence, $C=L_2+\sqrt{L_1^2+L_2^2}$.

In the rabbits frame, $\varphi\to\pi$ when $t\to\infty$. Indeed,
$\dot{\varphi}=\frac{u}{r}\sin{\varphi}>0$, if $\varphi<\pi$, and, therefore,
$\varphi$ is increasing all the time. But when it approaches $\pi$, both 
$\dot{\varphi}$ and $\dot{r}$ approach zero.

Therefore, the limiting distance is
$$r_{min}=\frac{C}{1-\cos{\pi}}=\frac{L_2+\sqrt{L_1^2+L_2^2}}{2}.$$

\subsection{}
Найдем инфинитезимальные смещения вдоль координат $\xi,\,\eta$ и $\phi$:
$$d_{(\xi)}\vec{r}=\frac{1}{2} \left [ \sqrt{\frac
{\eta}{\xi}}\cos{\phi}~\vec{i}+\sqrt{\frac{\eta}{\xi}}\sin{\phi}~\vec{j}+
\vec{k}\right]d\xi,$$
$$d_{(\eta)}\vec{r}=\frac{1}{2} \left [ \sqrt{\frac{\xi}{\eta}}\cos{\phi}~
\vec{i}+\sqrt{\frac{\xi}{\eta}}\sin{\phi}~\vec{j}-\vec{k}\right]d\eta,$$
$$\hspace*{-15mm} d_{(\phi)}\vec{r}=\sqrt{\xi\eta}~(-\sin{\phi}~\vec{i}+
\cos{\phi}~\vec{j})~d\phi.$$
Где, например, $d_{(\xi)}\vec{r}=(dx)_{\eta,\phi=const}\vec{i}+
(dy)_{\eta,\phi=const}\vec{j}+(dz)_{\eta,\phi=const}\vec{k}$.
Орты - это единичные вектора по этим направлениям. Поэтому просто нормируем 
соответствующие инфинитезимальные смещения. Например
$$\vec{e}_\xi=\frac{d_{(\xi)}\vec{r}}{|d_{(\xi)}\vec{r}|},\;
|d_{(\xi)}\vec{r}|=\frac{1}{2}d\xi\sqrt{1+\frac{\eta}{\xi}}.$$
В результате получим
$$\vec{e}_\xi=\frac{1}{\sqrt{\xi+\eta}}~[\sqrt{\eta}\cos{\phi}~\vec{i}+
\sqrt{\eta}\sin{\phi}~\vec{j}+\sqrt{\xi}~\vec{k}],$$
$$\vec{e}_\eta=\frac{1}{\sqrt{\xi+\eta}}~[\sqrt{\xi}\cos{\phi}~\vec{i}+
\sqrt{\xi}\sin{\phi}~\vec{j}+\sqrt{\eta}~\vec{k}],$$
$$\hspace*{-47mm} \vec{e}_\phi=-\sin{\phi}~\vec{i}+\cos{\phi}~\vec{j}.$$
Элемент объема $dV=|d_{(\xi)}\vec{r}\cdot [d_{(\eta)}\vec{r} \times
d_{(\phi)}\vec{r}]|$. Замечая, что абсолютное значение детерминанта
$$\left | \begin{array}{ccc} \frac{1}{2}\sqrt{\frac{\eta}{\xi}}\cos \phi &
\frac{1}{2}\sqrt{\frac{\eta}{\xi}}\sin \phi & \frac{1}{2} \\
\frac{1}{2}\sqrt{\frac{\xi}{\eta}}\cos \phi &
\frac{1}{2}\sqrt{\frac{\xi}{\eta}}\sin \phi & -\frac{1}{2} \\
-\sqrt{\xi\eta}\sin \phi & \sqrt{\xi\eta}\cos \phi & 0 \end{array} 
\right | $$ 
равно $\frac{1}{4}(\xi+\eta)$, получаем
$$dV=\frac{\xi+\eta}{4}~d\xi ~d\eta ~d\phi.$$
Элементы площади равны
$$\hspace*{-13mm}
dS_{(\xi,\eta)}=|d_{(\xi)}\vec{r}\times d_{(\eta)}\vec{r}|=\frac{\xi+\eta}
{4\sqrt{\xi\eta}}\,d\xi d\eta,$$
$$dS_{(\xi,\phi)}=|d_{(\xi)}\vec{r}\times d_{(\phi)}\vec{r}|=\frac{1}{2}
\sqrt{\eta(\xi+\eta)}\,d\xi d\phi,$$
$$dS_{(\eta,\phi)}=|d_{(\eta)}\vec{r}\times d_{(\phi)}\vec{r}|=\frac{1}{2}
\sqrt{\xi(\xi+\eta)}\,d\eta d\phi.$$
Наконец, элемент дуги
$$(dl)^2=(dx)^2+(dy)^2+(dz)^2=\frac{1}{4}(\xi+\eta)\left [ \frac{(d\xi)^2}
{\xi}+\frac{(d\eta)^2}{\eta}\right ]+\xi\eta (d\phi)^2.$$

\vspace*{5mm}
\noindent $\bullet$  $\bullet$  $\bullet$
 
\noindent Let us find infinitesimal displacements along the coordinates 
$\xi,\,\eta$ and $\phi$:
$$d_{(\xi)}\vec{r}=\frac{1}{2} \left [ \sqrt{\frac
{\eta}{\xi}}\cos{\phi}~\vec{i}+\sqrt{\frac{\eta}{\xi}}\sin{\phi}~\vec{j}+
\vec{k}\right]d\xi,$$
$$d_{(\eta)}\vec{r}=\frac{1}{2} \left [ \sqrt{\frac{\xi}{\eta}}\cos{\phi}~
\vec{i}+\sqrt{\frac{\xi}{\eta}}\sin{\phi}~\vec{j}-\vec{k}\right]d\eta,$$
$$\hspace*{-15mm} d_{(\phi)}\vec{r}=\sqrt{\xi\eta}~(-\sin{\phi}~\vec{i}+
\cos{\phi}~\vec{j})~d\phi.$$
Where, for example, $d_{(\xi)}\vec{r}=(dx)_{\eta,\phi=const}\vec{i}+
(dy)_{\eta,\phi=const}\vec{j}+(dz)_{\eta,\phi=const}\vec{k}$.
Orts are unit vectors in these directions. Therefore, let us simply normalize
the corresponding infinitesimal displacements. For example,
$$\vec{e}_\xi=\frac{d_{(\xi)}\vec{r}}{|d_{(\xi)}\vec{r}|},\;
|d_{(\xi)}\vec{r}|=\frac{1}{2}d\xi\sqrt{1+\frac{\eta}{\xi}}.$$
As a result, we get
$$\vec{e}_\xi=\frac{1}{\sqrt{\xi+\eta}}~[\sqrt{\eta}\cos{\phi}~\vec{i}+
\sqrt{\eta}\sin{\phi}~\vec{j}+\sqrt{\xi}~\vec{k}],$$
$$\vec{e}_\eta=\frac{1}{\sqrt{\xi+\eta}}~[\sqrt{\xi}\cos{\phi}~\vec{i}+
\sqrt{\xi}\sin{\phi}~\vec{j}+\sqrt{\eta}~\vec{k}],$$
$$\hspace*{-47mm} \vec{e}_\phi=-\sin{\phi}~\vec{i}+\cos{\phi}~\vec{j}.$$
The volume element is $dV=|d_{(\xi)}\vec{r}\cdot [d_{(\eta)}\vec{r} \times
d_{(\phi)}\vec{r}]|$. Noting that the absolute value of the determinant
$$\left | \begin{array}{ccc} \frac{1}{2}\sqrt{\frac{\eta}{\xi}}\cos \phi &
\frac{1}{2}\sqrt{\frac{\eta}{\xi}}\sin \phi & \frac{1}{2} \\
\frac{1}{2}\sqrt{\frac{\xi}{\eta}}\cos \phi &
\frac{1}{2}\sqrt{\frac{\xi}{\eta}}\sin \phi & -\frac{1}{2} \\
-\sqrt{\xi\eta}\sin \phi & \sqrt{\xi\eta}\cos \phi & 0 \end{array} 
\right | $$ 
is $\frac{1}{4}(\xi+\eta)$, we get
$$dV=\frac{\xi+\eta}{4}~d\xi ~d\eta ~d\phi.$$
The area elements are
$$\hspace*{-13mm}
dS_{(\xi,\eta)}=|d_{(\xi)}\vec{r}\times d_{(\eta)}\vec{r}|=\frac{\xi+\eta}
{4\sqrt{\xi\eta}}\,d\xi d\eta,$$
$$dS_{(\xi,\phi)}=|d_{(\xi)}\vec{r}\times d_{(\phi)}\vec{r}|=\frac{1}{2}
\sqrt{\eta(\xi+\eta)}\,d\xi d\phi,$$
$$dS_{(\eta,\phi)}=|d_{(\eta)}\vec{r}\times d_{(\phi)}\vec{r}|=\frac{1}{2}
\sqrt{\xi(\xi+\eta)}\,d\eta d\phi.$$
At last, the line element is
$$(dl)^2=(dx)^2+(dy)^2+(dz)^2=\frac{1}{4}(\xi+\eta)\left [ \frac{(d\xi)^2}
{\xi}+\frac{(d\eta)^2}{\eta}\right ]+\xi\eta (d\phi)^2.$$

\subsection{}
Величина скорости электрона $V=\sqrt{\dot{x}^2+\dot{y}^2+\dot{z}^2}=
\sqrt{b^2+a^2\omega^2}$ не меняется со временем. Следовательно, тангенциальное
ускорение равно нулю и нормальное ускорение совпадает с величиной полного
ускорения:
$g_n=\sqrt{\ddot{x}^2+\ddot{y}^2+\ddot{z}^2}=a\omega^2$. Но $g_n=\frac{V^2}
{R}$ и радиус кривизны $$R=\frac{V^2}{g_n}=a+\frac{b^2}{a\omega^2}.$$

\vspace*{5mm}
\noindent $\bullet$  $\bullet$  $\bullet$
 
\noindent The magnitude of the electron's velocity $V=\sqrt{\dot{x}^2+
\dot{y}^2+\dot{z}^2}=\sqrt{b^2+a^2\omega^2}$ does not change over time. 
Therefore, the tangential acceleration is zero and the normal component of 
the acceleration coincides with the whole acceleration. Hence,
$g_n=\sqrt{\ddot{x}^2+\ddot{y}^2+\ddot{z}^2}=a\omega^2$. But $g_n=\frac{V^2}
{R}$ and the curvature radius is 
$$R=\frac{V^2}{g_n}=a+\frac{b^2}{a\omega^2}.$$

\subsection{}
\begin{figure}[htb]
\centerline{\epsfig{figure=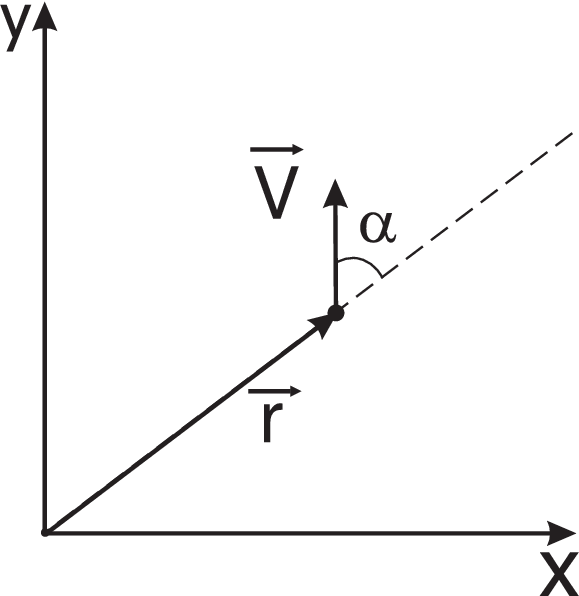,height=4cm}}
\end{figure}
Компоненты скорости в полярной системе координат $V_r=\dot{r}=V\cos\alpha$, и
$V_\varphi=r\dot{\phi}=V\sin\alpha$. Поэтому $$\frac{1}{r}\frac{dr}
{d\varphi}=\ctg{\alpha}.$$ Интегрируя, получаем уравнение траектории 
$$r=r_0e^{(\varphi-\varphi_0)\ctg{\alpha}}.$$

\vspace*{5mm}
\noindent $\bullet$  $\bullet$  $\bullet$
 
\begin{figure}[htb]
\centerline{\epsfig{figure=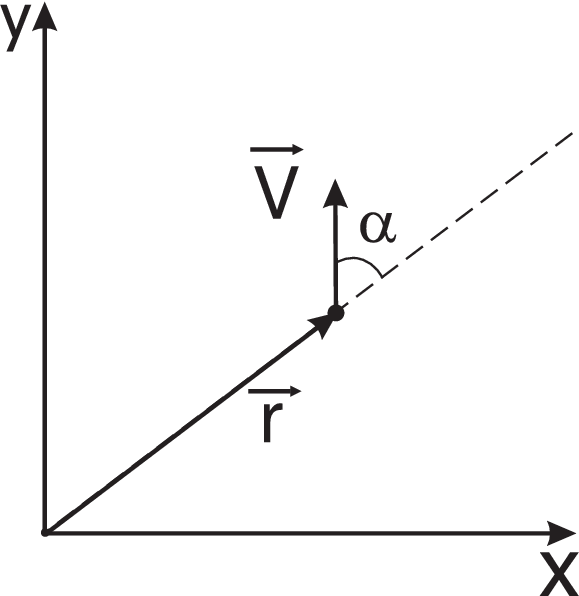,height=4cm}}
\end{figure}
The velocity components, in the polar coordinate system, are
$V_r=\dot{r}=V\cos\alpha$ and
$V_\varphi=r\dot{\phi}=V\sin\alpha$. Therefore, $$\frac{1}{r}\frac{dr}
{d\varphi}=\ctg{\alpha}.$$ Integrating, we get the trajectory equation
$$r=r_0e^{(\varphi-\varphi_0)\ctg{\alpha}}.$$

\clearpage
\section*{Контрольная работа 2}
\setcounter{section}{2}
\setcounter{subsection}{0}
\subsection{}  
\begin{figure}[htb]
\centerline{\epsfig{figure=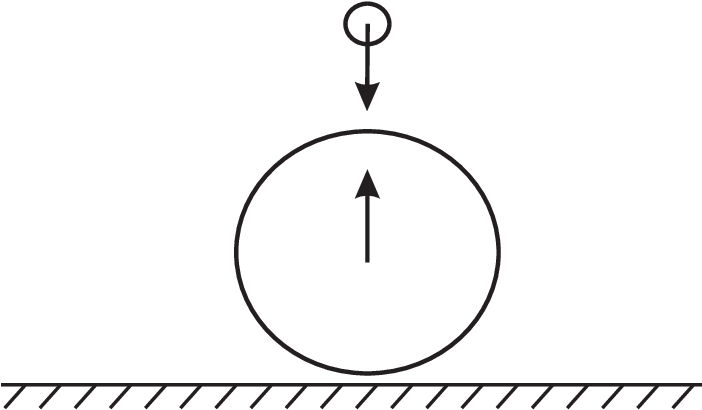,height=4cm}}
\end{figure}
Во время падения мяча из-за воздушного потока теннисный мячик оторвется и, 
следовательно столкнется уже отскочившим мячом. Для оценки можно считать, что 
и мяч, и мячик движутся на встречу со скоростью $V\approx \sqrt{2gh}$, где 
$h=1~\mbox{м}$. В системе покоя мяча, мячик движется со скоростью $2V$ и 
отскакивает с такой же скоростью, так как его масса много меньше массы 
баскетбольного мяча. Т.е. мячик отскакивает как бы от стенки. Следовательно, в 
системе земли скорость мячика после удара будет $2V+V=3V$. Высота подъема 
пропорциональна квадрату начальной скорости, поэтому теннисный мячик подскочит
на высоту $3^2h=9~\mbox{м}$. реальная высота подскока будет несколько меньше 
из-за сопротивления воздуха и потерь энергий при ударах, но все равно эта 
высота будет значительно больше, чем $1~\mbox{м}$. 

\vspace*{5mm}
\noindent $\bullet$  $\bullet$  $\bullet$
 
\begin{figure}[htb]
\centerline{\epsfig{figure=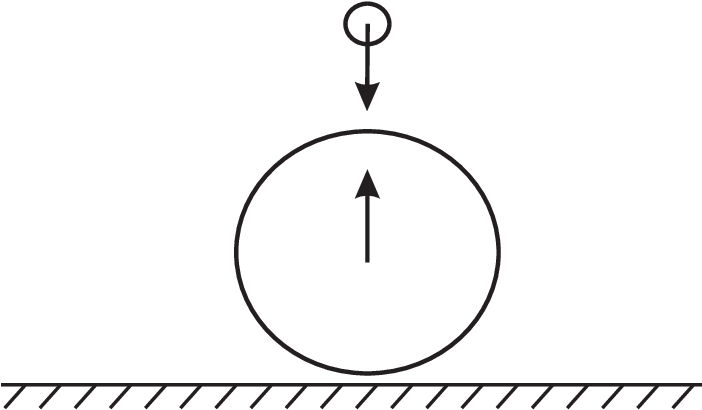,height=4cm}}
\end{figure}
During the fall of the ball, due to air flow, the tennis ball
will drag behind and consequently face a head-on collision with the 
basketball ball which already had bounced. For estimation purposes,
one can assume that upon the impact the balls are heading toward each other 
with the same velocity $V\approx \sqrt{2gh}$, where $h=1~\mbox{m}$.
In the rest frame of the basketball ball, the tennis ball is moving at a 
speed $2V$ and bounces at the same speed as its mass is much smaller than the 
mass of the basketball ball. That is, the tennis ball bounces as if on a wall.
Consequently, with regard to the ground, the tennis ball's speed after the 
impact will be  $2V+V=3V $. The maximum height is proportional to 
the square of initial velocity. Therefore, the tennis ball will jump
at the height of $3^2h=9~\mbox{m}$. In fact, the actual height will be 
somewhat less due to air resistance and energy loss upon the impact, but 
still this height will be considerably more than $1~\mbox{m}$.

\subsection{}
\begin{figure}[htb]
\centerline{\epsfig{figure=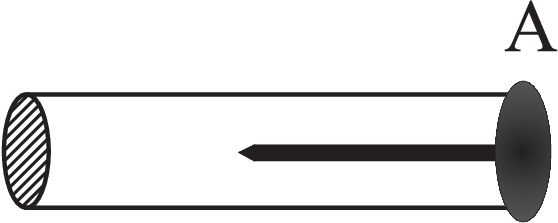,height=2cm}}
\end{figure}
В системе гвоздя ясно, что бумага пробивается, поэтому весь вопрос в том, 
как это достигается в системе цилиндра. В момент столкновения кончик гвоздя 
находится на расстоянии $\frac{L}{\gamma}$ от начала цилиндра. Но он не может 
мгновенно остановиться. Он вообще не ``знает'', что столкновение произошло, 
т.к. сигнал о столкновении может распространяться вдоль гвоздя со скоростью 
не больше, чем скорость света $c$. Так как в системе цилиндра кончик гвоздя 
убегает от этого сигнала со скоростью $V$, то он почувствует последствия 
столкновения не раньше, чем $t=\frac{L}{\gamma (c-V)}$. Все это время он 
движется со скоростью $V$ и, следовательно, к этому моменту времени окажется 
от начала цилиндра на расстоянии $\frac{L}{\gamma}+\frac{LV}{\gamma (c-V)}=
L\sqrt{\frac{1+\beta}{1-\beta}}$.
Так так $L\sqrt{\frac{1+\beta}{1-\beta}}>L$, то гвоздь не может не пробить 
бумагу.

\vspace*{5mm}
\noindent $\bullet$  $\bullet$  $\bullet$
 
\begin{figure}[htb]
\centerline{\epsfig{figure=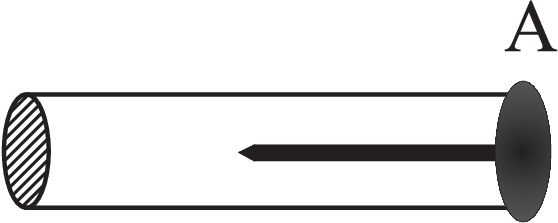,height=2cm}}
\end{figure}
In the nail's frame it is clear that the paper is punched, so the 
whole question is how this is achieved in the cylinder's rest frame. In this
frame, at the moment of collision, the tip of the nail is at a distance 
$\frac{L}{\gamma}$ from the beginning of the cylinder. But it cannot stop its 
flight instantly. It even does not ``know'' that the collision occurred, 
because the signal that the  collision happened can propagate along the nail 
with the speed no more than the speed of light $c$. In the cylinder's rest 
frame, the tip of the nail runs away from this signal at a rate $V$. 
Therefore, it will feel the effects of the collision not before  
$t=\frac{L}{\gamma (c-V)}$. All this time the tip moves at a speed $V$ and, 
therefore, when it stops it will be from the beginning of the cylinder at 
a distance $\frac{L}{\gamma}+\frac{LV}{\gamma (c-V)}=L\sqrt{\frac{1+\beta}
{1-\beta}}$. However, $L\sqrt{\frac{1+\beta}{1-\beta}}>L$ and, hence, the 
nail will punch the paper before its tip stops.

\subsection{}
Тангенциальное ускорение
$$a_\tau=\frac{dV}{dt}=\frac{d}{dt}\sqrt{\dot{x}^2+\dot{y}^2+\dot{z}^2}=
\frac{\dot{x}\ddot{x}+\dot{y}\ddot{y}+\dot{z}\ddot{z}}{\sqrt{\dot{x}^2+
\dot{y}^2+\dot{z}^2}}=\frac{\vec{a}\cdot\vec{V}}{V}.$$
Нормальное ускорение
$$a_n=\sqrt{a^2-a_\tau^2}=\sqrt{a^2-\frac{(\vec{a}\cdot\vec{V})^2}{V^2}}=
\frac{|\vec{a}\times \vec{V}|}{V}.$$
Но $a_n=\frac{V^2}{R}$, поэтому радиус кривизны $$R=\frac{V^2}{a_n}=\frac{
V^3}{|\vec{a}\times \vec{V}|}.$$

\vspace*{5mm}
\noindent $\bullet$  $\bullet$  $\bullet$
  
\noindent The tangential acceleration is
$$a_\tau=\frac{dV}{dt}=\frac{d}{dt}\sqrt{\dot{x}^2+\dot{y}^2+\dot{z}^2}=
\frac{\dot{x}\ddot{x}+\dot{y}\ddot{y}+\dot{z}\ddot{z}}{\sqrt{\dot{x}^2+
\dot{y}^2+\dot{z}^2}}=\frac{\vec{a}\cdot\vec{V}}{V}.$$
Hence, the normal acceleration equals to 
$$a_n=\sqrt{a^2-a_\tau^2}=\sqrt{a^2-\frac{(\vec{a}\cdot\vec{V})^2}{V^2}}=
\frac{|\vec{a}\times \vec{V}|}{V}.$$
But $a_n=\frac{V^2}{R}$, therefore the radius of curvature 
$$R=\frac{V^2}{a_n}=\frac{V^3}{|\vec{a}\times \vec{V}|}.$$

\subsection{}
Найдем инфинитезимальные смещения вдоль координат $\sigma$ и $\tau$: 
$$d_{(\sigma)}\vec{r}= (d\vec{r})_{\tau=const}=d\sigma [\tau \vec{i}-\sigma 
\vec{j}],\;\;  d_{(\tau)}\vec{r}=(d\vec{r})_{\sigma=const}=
d\tau [\sigma \vec{i}+\tau \vec{j}].$$ Орты $\vec{e}_\sigma$ и $\vec{e}_\tau$ 
отличаются от $d_{(\sigma)}\vec{r}$ и $d_{(\tau)}\vec{r}$ только нормировкой. 
Поэтому $$\vec{e}_\sigma=\frac{\tau \vec{i}-\sigma \vec{j}}{\sqrt{\tau^2+
\sigma^2}}\;\; \mbox{и} \;\; \vec{e}_\tau=\frac{\sigma \vec{i}+\tau \vec{j}}
{\sqrt{\tau^2+\sigma^2}}.$$ Элемент площади $dS=|d_{(\sigma)}\vec{r} \times 
d_{(\tau)}\vec{r}|=(\sigma^2+\tau^2)\,d\sigma d\tau$.

\vspace*{5mm}
\noindent $\bullet$  $\bullet$  $\bullet$
 
\noindent Let us find infinitesimal displacements along the coordinates
$\sigma$ and $\tau$:
$$d_{(\sigma)}\vec{r}= (d\vec{r})_{\tau=const}=d\sigma [\tau \vec{i}-\sigma
\vec{j}],\;\;  d_{(\tau)}\vec{r}=(d\vec{r})_{\sigma=const}=
d\tau [\sigma \vec{i}+\tau \vec{j}].$$ Orts $\vec{e}_\sigma$ and 
$\vec{e}_\tau$ differ from $d_{(\sigma)}\vec{r}$ and $d_{(\tau)}\vec{r}$
only by normalization. Therefore, $$\vec{e}_\sigma=\frac{\tau \vec{i}-\sigma
\vec{j}}{\sqrt{\tau^2+
\sigma^2}}\;\; \mbox{and} \;\; \vec{e}_\tau=\frac{\sigma \vec{i}+\tau \vec{j}}
{\sqrt{\tau^2+\sigma^2}}.$$ The area element is $dS=|d_{(\sigma)}\vec{r} 
\times d_{(\tau)}\vec{r}|=(\sigma^2+\tau^2)\,d\sigma d\tau$.

\subsection{}
Найдем орты сферической системы координат.
$$\vec{r}=r\sin{\theta}\cos{\varphi}\,
\vec{i}+r\sin{\theta}\sin{\varphi}\,\vec{j}+r\cos{\theta}\,\vec{j}.$$
Поэтому
инфинитезимальные смещения вдоль координат $r,\theta$ и $\varphi$ будут
$$d_{(r)}\vec{r}=dr[\sin{\theta}\cos{\varphi}\,\vec{i}+
\sin{\theta}\sin{\varphi}\,\vec{j}+ \cos{\theta}\,\vec{k}],$$
$$d_{(\theta)}\vec{r}=rd\theta [\cos{\theta}
\cos{\varphi}\, \vec{i}+\cos{\theta}\sin{\varphi}\,\vec{j}
-\sin{\theta}\,\vec{k}]$$
и
$$ d_{(\varphi)} \vec{r}=r\sin{\theta}d\varphi
[-\sin{\varphi}\,\vec{i}+\cos{\varphi}\,\vec{j}].$$
Нормируя, получаем орты:
$$\vec{e}_r=\sin{\theta}\cos{\varphi}\,\vec{i}+\sin{\theta}\sin{\varphi}\,
\vec{j}+\cos{\theta}\,\vec{k},$$
$$\vec{e}_\theta=\cos{\theta}\cos{\varphi}\,\vec{i}+
\cos{\theta}\sin{\varphi}\,\vec{j}-\sin{\theta}\,\vec{k}, \; \vec{e}_\varphi=
-\sin{\varphi}\,\vec{i}+\cos{\varphi}\,\vec{j}.$$
Поэтому $\vec{r}=r\vec{e}_r$ и скорость точки на поверхности шара (когда $r=
const$) будет $\vec{V}=\frac{d\vec{r}}{dt}=r\dot{\vec{e}}_r$. Но
$\dot{\vec{e}}_r=\dot{\theta}\vec{e}_\theta+\dot{\varphi}\sin{\theta}
\vec{e}_\varphi$ так как
$$\dot{\vec{e}}_r=\dot{\theta} [\cos{\theta}\cos{\varphi}\,\vec{i}+
\cos{\theta}\sin{\varphi}\,\vec{j}-\sin{\theta}\,\vec{k}]+
\dot{\varphi}\sin{\theta} [ -\sin{\varphi}\,\vec{i}+\cos{\varphi}\,\vec{i}]
.$$
Меридиан направлен вдоль $\vec{e}_\theta$. Поэтому, согласно условию, $\vec{V}
\cdot \vec{e}_\theta=V\cos{\alpha}$. Но тогда $\vec{V}\cdot \vec{e}_\varphi=
V\sin{\alpha}$, так как $\vec{e}_\theta$ и $\vec{e}_\varphi$
взаймно-перпендикулярны  и лежат в одной (касательной к шару) плоскости с
$\vec{V}$. Поставляя в эти равенства $\vec{V}=r\dot{\theta}\vec{e}_\theta+
r\dot{\varphi} \sin{\theta}\vec{e}_\varphi$, получаем
\begin{eqnarray}
&& r\dot{\theta}=V\cos{\alpha}, \nonumber \\ && r\dot{\varphi}\sin{\theta}=
V\sin{\alpha}. \nonumber \end{eqnarray}
Поделим одно на другое и учтем $\frac{\dot{\theta}}{\dot{\varphi}}=
\frac{d\theta}{d\varphi}$, получим
$$\frac{1}{\sin{\theta}}\frac{d\theta}{d\varphi}=\ctg{\alpha}.$$
Используя
$$\int\frac{d\theta}{\sin{\theta}}=\frac{-1}{2}\int\left [\frac{1}{1+
\cos{\theta}}+\frac{1}{1-\cos{\theta}}\right ]d\cos{\theta}=\ln{\sqrt{\frac{
1-\cos{\theta}}{1+\cos{\theta}}}}=\ln{\tg{\frac{\theta}{2}}},$$
получаем уравнение траектории
$$\tg{\frac{\theta}{2}}=\tg{\frac{\theta_0}{2}}~e^{(\varphi-\varphi_0)
\ctg{\alpha}}.$$
Эта кривая называется локсодромией.

\vspace*{5mm}
\noindent $\bullet$  $\bullet$  $\bullet$
  
\noindent Let us find the orts of the spherical coordinates.
$$\vec{r}=r\sin{\theta}\cos{\varphi}\,
\vec{i}+r\sin{\theta}\sin{\varphi}\,\vec{j}+r\cos{\theta}\,\vec{j}.$$ 
Therefore, the infinitesimal displacements along the coordinates $r,\theta$ 
and $\varphi$ will take the form 
$$d_{(r)}\vec{r}=dr[\sin{\theta}\cos{\varphi}\,\vec{i}+
\sin{\theta}\sin{\varphi}\,\vec{j}+ \cos{\theta}\,\vec{k}],$$ 
$$d_{(\theta)}\vec{r}=rd\theta [\cos{\theta}
\cos{\varphi}\, \vec{i}+\cos{\theta}\sin{\varphi}\,\vec{j}
-\sin{\theta}\,\vec{k}]$$
and
$$ d_{(\varphi)} \vec{r}=r\sin{\theta}d\varphi
[-\sin{\varphi}\,\vec{i}+\cos{\varphi}\,\vec{j}].$$ 
Normalizing, we get the orts:
$$\vec{e}_r=\sin{\theta}\cos{\varphi}\,\vec{i}+\sin{\theta}\sin{\varphi}\,
\vec{j}+\cos{\theta}\,\vec{k},$$ 
$$\vec{e}_\theta=\cos{\theta}\cos{\varphi}\,\vec{i}+
\cos{\theta}\sin{\varphi}\,\vec{j}-\sin{\theta}\,\vec{k}, \; \vec{e}_\varphi=
-\sin{\varphi}\,\vec{i}+\cos{\varphi}\,\vec{j}.$$
Therefore, $\vec{r}=r\vec{e}_r$ and the velocity of a point on the surface 
of the globe (when $r=const$) will be $\vec{V}=\frac{d\vec{r}}{dt}=
r\dot{\vec{e}}_r$. But 
$\dot{\vec{e}}_r=\dot{\theta}\vec{e}_\theta+\dot{\varphi}\sin{\theta}
\vec{e}_\varphi$, because
$$\dot{\vec{e}}_r=\dot{\theta} [\cos{\theta}\cos{\varphi}\,\vec{i}+
\cos{\theta}\sin{\varphi}\,\vec{j}-\sin{\theta}\,\vec{k}]+
\dot{\varphi}\sin{\theta} [ -\sin{\varphi}\,\vec{i}+\cos{\varphi}\,\vec{i}].$$
The meridian is directed along $\vec{e}_\theta$. Therefore, under the terms
of the problem, $\vec{V} \cdot \vec{e}_\theta=V\cos{\alpha}$. But then 
$\vec{V}\cdot \vec{e}_\varphi=V\sin{\alpha}$, because $\vec{e}_\theta$ and 
$\vec{e}_\varphi$ are mutually orthogonal and lay in the same (tangential
to the globe) plane together with $\vec{V}$. Substituting in these relations
$\vec{V}=r\dot{\theta}\vec{e}_\theta+r\dot{\varphi} \sin{\theta}
\vec{e}_\varphi$, we get
\begin{eqnarray}
&& r\dot{\theta}=V\cos{\alpha}, \nonumber \\ && r\dot{\varphi}\sin{\theta}=
V\sin{\alpha}. \nonumber \end{eqnarray}
Dividing the first equation over the second one and taking into account
$\frac{\dot{\theta}}{\dot{\varphi}}=
\frac{d\theta}{d\varphi}$, we get
$$\frac{1}{\sin{\theta}}\frac{d\theta}{d\varphi}=\ctg{\alpha}.$$
By using
$$\int\frac{d\theta}{\sin{\theta}}=\frac{-1}{2}\int\left [\frac{1}{1+
\cos{\theta}}+\frac{1}{1-\cos{\theta}}\right ]d\cos{\theta}=\ln{\sqrt{\frac{
1-\cos{\theta}}{1+\cos{\theta}}}}=\ln{\tg{\frac{\theta}{2}}},$$
we get the equation of the trajectory in the form
$$\tg{\frac{\theta}{2}}=\tg{\frac{\theta_0}{2}}~e^{(\varphi-\varphi_0)
\ctg{\alpha}}.$$
Such a curve is called a loxodrome. 

\subsection{}
Векторное произведение антикоммутативно, т.к.
$$0=(\vec{A}+\vec{B})\times
(\vec{A}+\vec{B})=\vec{A}\times\vec{B}+\vec{B}\times\vec{A}.$$
Также можно доказать, что смешанное произведение $(\vec{A}\times\vec{B})
\cdot\vec{C}$ меняет знак при перестановке любых двух сомножителей. Например,
$$0=((\vec{A}+\vec{C})\times\vec{B})
\cdot (\vec{A}+\vec{C})=(\vec{C}\times\vec{B})\cdot\vec{A}+
(\vec{A}\times\vec{B})\cdot\vec{C}.$$
Для любых двух векторов $\vec{A}$ и $\vec{B}$, норма $|\vec{A}\times\vec{B}|
^2$ равна
$$\left | \left (\vec{A}-\frac{\vec{A}\cdot\vec{B}}
{|\vec{B}|^2}~\vec{B}\right)\times\vec{B}\right |^2=
\left | \vec{A}-\frac{\vec{A}\cdot\vec{B}}{|\vec{B}|^2}~\vec{B}
\right |^2|\vec{B}|^2=|\vec{A}|^2 |\vec{B}|^2-(\vec{A}\cdot\vec{B})^2.$$
Следовательно, для любых двух векторов справедливо соотношение
$$(\vec{A}\times\vec{B})\cdot (\vec{A}\times\vec{B})=(\vec{A}\cdot\vec{A})
(\vec{B}\cdot\vec{B})-(\vec{A}\cdot\vec{B})^2.$$
Теперь рассмотрим
$$|\vec{A}\times (\vec{B}\times\vec{A})-(\vec{A}\cdot\vec{A})\vec{B}+
(\vec{A}\cdot\vec{B})\vec{A}|^2=$$
$$=|\vec{A}\times (\vec{B}\times\vec{A})|^2+
|\vec{A}|^4|\vec{B}|^2-(\vec{A}\cdot\vec{B})^2|\vec{A}|^2-
2|\vec{A}|^2(\vec{A}\times (\vec{B}\times\vec{A}))\cdot\vec{B}.$$
Но это равно нулю, так как
$$|\vec{A}\times (\vec{B}\times\vec{A})|^2=|\vec{A}|^2|\vec{B}\times\vec{A}|^2
=|\vec{A}|^4|\vec{B}|^2-(\vec{A}\cdot\vec{B})^2|\vec{A}|^2$$
и
$$(\vec{A}\times (\vec{B}\times\vec{A}))\cdot\vec{B}=(\vec{B}\times\vec{A})
\cdot (\vec{B}\times\vec{A})=|\vec{A}|^2|\vec{B}|^2-(\vec{A}\cdot\vec{B})^2.$$
Следовательно,
\begin{equation}
\vec{A}\times (\vec{B}\times\vec{A})=(\vec{A}\cdot\vec{A})\vec{B}-
(\vec{A}\cdot\vec{B})\vec{A}\;.
\label{eq5} \end{equation}
однако, хорошо известное для трехмерного векторного произведения соотношение
\begin{equation}
\vec{A}\times (\vec{B}\times\vec{C})=\vec{B}(\vec{A}\cdot\vec{C})-
\vec{C}(\vec{A}\cdot\vec{B})
\label{eq6} \end{equation}
в общем случае не выполняется. Чтобы доказать это, рассмотрим тернарное
произведение (которое равно нулю, если выполняется (\ref{eq6}))
$$\{\vec{A},\vec{B},\vec{C}\}=\vec{A}\times (\vec{B}\times\vec{C})-
\vec{B}(\vec{A}\cdot\vec{C})+\vec{C}(\vec{A}\cdot\vec{B}).$$
Тождество (\ref{eq5}) означает, что тернарное произведение, наподобие
смешанного произведения, тоже кососимметричная функция своих аргументов.
Например,
$$0=\{\vec{A}+\vec{B},\vec{A}+\vec{B},\vec{C}\}=\{\vec{A},\vec{B},\vec{C}\}+
\{\vec{B},\vec{A},\vec{C}\}.$$
Пусть $\vec{e}_i,\; i=1\div n$ некоторый ортонормированный базис в многомерном
векторном пространстве. Тогда
$$(\vec{e}_i\times\vec{A})\cdot(\vec{e}_i\times\vec{B})=
((\vec{e}_i\times\vec{B})\times\vec{e}_i)\cdot\vec{A}=
[\vec{B}-(\vec{B}\cdot\vec{e}_i)\vec{e}_i]\cdot\vec{A}$$
и, следовательно,
\begin{equation}
\sum\limits_{i=1}^n(\vec{e}_i\times\vec{A})\cdot(\vec{e}_i\times\vec{B})=
(n-1)\vec{A}\cdot\vec{B}.
\label{eq7} \end{equation}
Используя это тождество, получаем
$$\sum\limits_{i=1}^n\{\vec{e}_i,\vec{A},\vec{B}\}\cdot
\{\vec{e}_i,\vec{C},\vec{D}\}=$$
\begin{equation}
=(n-5)(\vec{A}\times \vec{B})\cdot
(\vec{C}\times \vec{D})+2(\vec{A}\cdot\vec{C})(\vec{B}\cdot\vec{D})-
2(\vec{A}\cdot\vec{D})(\vec{B}\cdot\vec{C}).
\label{eq8} \end{equation}
Поэтому
\begin{equation}
\sum\limits_{i,j=1}^n\{\vec{e}_i,\vec{e}_j,\vec{A}\}\cdot
\{\vec{e}_i,\vec{e}_j,\vec{B}\}=(n-1)(n-3)\vec{A}\cdot\vec{B}
\label{eq9} \end{equation}
и
\begin{equation}
\sum\limits_{i,j,k=1}^n\{\vec{e}_i,\vec{e}_j,\vec{e}_k\}\cdot
\{\vec{e}_i,\vec{e}_j,\vec{e}_k\}=n(n-1)(n-3).
\label{eq10} \end{equation}
Последнее уравнение показывает, что некоторые $\{\vec{e}_i,\vec{e}_j,
\vec{e}_k\}$ не равны нулю, если $n>3$. Таким образом тождество (\ref {eq6})
действительно справедливо только для обычного трехмерного векторного
произведения ($n=1$ случай, конечно, не интересен, потому что это
соответствует тождественно равному нулю  векторному произведению).

Далее заметим, что используя
$$\vec{A}\times (\vec{B}\times \vec{C})+(\vec{A}\times \vec{B})\times \vec{C}
=(\vec{A}+\vec{C})\times\vec{B}\times(\vec{A}+\vec{C})-
\vec{A}\times\vec{B}\times\vec{A}-\vec{C}\times\vec{B}\times\vec{C}=$$
$$=2\vec{A}\cdot\vec{C}~\vec{B}-\vec{A}\cdot\vec{B}~\vec{C}-
\vec{B}\cdot\vec{C}~\vec{A}$$
и
$$\vec{A}\times(\vec{B}\times(\vec{C}\times\vec{D}))=
\left . \frac{1}{2} \right [
\vec{A}\times (\vec{B}\times(\vec{C}\times\vec{D}))+
(\vec{A}\times \vec{B})\times(\vec{C}\times\vec{D})-$$
$$-(\vec{A}\times \vec{B})\times(\vec{C}\times\vec{D})-
((\vec{A}\times \vec{B})\times \vec{C})\times\vec{D}+
((\vec{A}\times \vec{B})\times\vec{C})\times\vec{D}+$$
$$+(\vec{A}\times (\vec{B}\times\vec{C}))\times\vec{D}-
(\vec{A}\times (\vec{B}\times\vec{C}))\times\vec{D}-
\vec{A}\times ((\vec{B}\times \vec{C})\times\vec{D})+ $$
$$+\vec{A}\times ((\vec{B}\times \vec{C})\times\vec{D})+
\vec{A}\times (\vec{B}\times ( \vec{C}\times\vec{D})) \left . \frac{}{}
\right ],$$
можно проверить следующее равенство
$$\vec{A}\times \{\vec{B},\vec{C},\vec{D}\}=-\{\vec{A},\vec{B},\vec{C}
\times \vec{D}\}+\vec{A}\times(\vec{B}\times(\vec{C}\times\vec{D}))-
\{\vec{A},\vec{C},\vec{D}\times \vec{B} \}+ $$ $$+
\vec{A}\times(\vec{C}\times(\vec{D}\times\vec{B}))-
\{\vec{A},\vec{D},\vec{B}\times \vec{C} \}+
\vec{A}\times(\vec{D}\times(\vec{B}\times\vec{C}))=$$
$$=-\{\vec{A},\vec{B},\vec{C}\times \vec{D} \}
-\{\vec{A},\vec{C},\vec{D}\times \vec{B} \}
-\{\vec{A},\vec{D},\vec{B}\times \vec{C} \}+
3\vec{A}\times\{\vec{B},\vec{C},\vec{D}\}.$$
Последнее равенство следует из
$$3\{\vec{B},\vec{C},\vec{D}\}=\{\vec{B},\vec{C},\vec{D}\}+
\{\vec{C},\vec{D},\vec{B}\}+\{\vec{D},\vec{B},\vec{C}\} =$$ $$=
\vec{B}\times(\vec{C}\times\vec{D})+
\vec{C}\times(\vec{D}\times\vec{B})+
\vec{D}\times(\vec{B}\times\vec{C}).$$
Следовательно, тернарное произведение удовлетворяет интересному соотношению
\begin{equation}
2~\vec{A}\times \{\vec{B},\vec{C},\vec{D}\}=\{\vec{A},\vec{B},\vec{C}
\times \vec{D}\}+\{\vec{A},\vec{C},\vec{D}\times \vec{B} \}+
\{\vec{A},\vec{D},\vec{B}\times \vec{C} \}.
\label{eq12}\end{equation}
Поэтому должно выполняться равенство
$$4\sum\limits_{i,j,k,l=1}^n|\vec{e}_i\times \{\vec{e}_j,\vec{e}_k,
\vec{e}_l\}|^2=$$ $$=
\sum\limits_{i,j,k,l=1}^n|\{\vec{e}_i,\vec{e}_j,
\vec{e}_k\times \vec{e}_l\}+\{\vec{e}_i,\vec{e}_k,\vec{e}_l\times
\vec{e}_j\}+\{\vec{e}_i,\vec{e}_l,\vec{e}_j\times\vec{e}_k\}|^2.$$
Левую сторону можно легко вычислить с помощью (\ref{eq7}) и (\ref{eq10}):
$$4\sum\limits_{i,j,k,l=1}^n|\vec{e}_i\times \{\vec{e}_j,\vec{e}_k,
\vec{e}_l\}|^2=4n(n-1)^2(n-3).$$
Чтобы вычислит правую сторону, можно использовать следующее равенство
\begin{equation}
\sum\limits_{i,j=1}^n\{\vec{e}_i,\vec{e}_j,\vec{A}\}\cdot
\{\vec{e}_i,\vec{e}_j\times\vec{B},\vec{C}\}=-(n-3)(n-6)\vec{A}\cdot
(\vec{B}\times\vec{C})
\label{eq13},\end{equation}
которое следует из (\ref{eq8}) и из равенства
$$\sum\limits_{i=1}^n(\vec{e}_i\times\vec{A})\cdot((\vec{e}_i\times\vec{B})
\times\vec{C})=$$ $$=\sum\limits_{i=1}^n(\vec{e}_i\times\vec{A})\cdot
[2\vec{e}_i\cdot\vec{C}~\vec{B}-\vec{B}\cdot\vec{C}~\vec{e}_i-
\vec{e}_i\cdot\vec{B}~\vec{C}-\vec{e}_i\times(\vec{B}\times\vec{C})]=$$
$$=-(n-4)\vec{A}\cdot(\vec{B}\times\vec{C}).$$
Теперь, имея (\ref{eq9}) и (\ref{eq13}) под рукой, можно довольно просто
вычислить
$$\sum\limits_{i,j,k,l=1}^n|\{\vec{e}_i,\vec{e}_j,
\vec{e}_k\times \vec{e}_l\}+\{\vec{e}_i,\vec{e}_k,\vec{e}_l\times
\vec{e}_j\}+\{\vec{e}_i,\vec{e}_l,\vec{e}_j\times\vec{e}_k\}|^2=$$
$$=3n(n-1)^2(n-3)+6n(n-1)(n-3)(n-6)=3n(n-1)(n-3)(3n-13).$$
Следовательно, должны иметь
$$4n(n-1)^2(n-3)=3n(n-1)(n-3)(3n-13)$$
и
$$3n(n-1)(n-3)(3n-13)-4n(n-1)^2(n-3)=5n(n-1)(n-3)(n-7)=0.$$
Это показывает, что нетривиальное векторное произведение определено только в
пространствах с размерностями три и семь.

\vspace*{5mm}
\noindent $\bullet$  $\bullet$  $\bullet$
 
\noindent It can be easily shown that the vector product is anti-commutative. 
Indeed,
$$0=(\vec{A}+\vec{B})\times
(\vec{A}+\vec{B})=\vec{A}\times\vec{B}+\vec{B}\times\vec{A}.$$ 
One can prove as well that the mixed product $(\vec{A}\times\vec{B})
\cdot\vec{C}$ is alternating in $\vec{A},\vec{B},\vec{C}$. For example,
$$0=((\vec{A}+\vec{C})\times\vec{B})
\cdot (\vec{A}+\vec{C})=(\vec{C}\times\vec{B})\cdot\vec{A}+
(\vec{A}\times\vec{B})\cdot\vec{C}.$$  
For any two vectors $\vec{A}$ and $\vec{B}$, the norm $|\vec{A}\times\vec{B}|
^2$ equals to
$$\left | \left (\vec{A}-\frac{\vec{A}\cdot\vec{B}}
{|\vec{B}|^2}~\vec{B}\right)\times\vec{B}\right |^2=
\left | \vec{A}-\frac{\vec{A}\cdot\vec{B}}{|\vec{B}|^2}~\vec{B}
\right |^2|\vec{B}|^2=|\vec{A}|^2 |\vec{B}|^2-(\vec{A}\cdot\vec{B})^2.$$
Therefore, the following relation holds
$$(\vec{A}\times\vec{B})\cdot (\vec{A}\times\vec{B})=(\vec{A}\cdot\vec{A})
(\vec{B}\cdot\vec{B})-(\vec{A}\cdot\vec{B})^2.$$
Now consider
$$|\vec{A}\times (\vec{B}\times\vec{A})-(\vec{A}\cdot\vec{A})\vec{B}+
(\vec{A}\cdot\vec{B})\vec{A}|^2=$$ 
$$=|\vec{A}\times (\vec{B}\times\vec{A})|^2+
|\vec{A}|^4|\vec{B}|^2-(\vec{A}\cdot\vec{B})^2|\vec{A}|^2-
2|\vec{A}|^2(\vec{A}\times (\vec{B}\times\vec{A}))\cdot\vec{B}.$$
But this equals zero because
$$|\vec{A}\times (\vec{B}\times\vec{A})|^2=|\vec{A}|^2|\vec{B}\times\vec{A}|^2
=|\vec{A}|^4|\vec{B}|^2-(\vec{A}\cdot\vec{B})^2|\vec{A}|^2$$
and
$$(\vec{A}\times (\vec{B}\times\vec{A}))\cdot\vec{B}=(\vec{B}\times\vec{A})
\cdot (\vec{B}\times\vec{A})=|\vec{A}|^2|\vec{B}|^2-(\vec{A}\cdot\vec{B})^2.$$ 
Therefore, 
\begin{equation}
\vec{A}\times (\vec{B}\times\vec{A})=(\vec{A}\cdot\vec{A})\vec{B}-
(\vec{A}\cdot\vec{B})\vec{A}\;.
\label{eqq5} \end{equation}
However, the well known identity for the three-dimensional vector product
\begin{equation}
\vec{A}\times (\vec{B}\times\vec{C})=\vec{B}(\vec{A}\cdot\vec{C})-
\vec{C}(\vec{A}\cdot\vec{B})
\label{eqq6} \end{equation}
is not valid in general case. To show this, let us consider the ternary 
product (which is zero if (\ref{eqq6}) is valid)
$$\{\vec{A},\vec{B},\vec{C}\}=\vec{A}\times (\vec{B}\times\vec{C})-
\vec{B}(\vec{A}\cdot\vec{C})+\vec{C}(\vec{A}\cdot\vec{B}).$$
Equation (\ref{eqq5}) implies that this ternary product is alternating in its 
arguments, like the mixed product. 
For example,
$$0=\{\vec{A}+\vec{B},\vec{A}+\vec{B},\vec{C}\}=\{\vec{A},\vec{B},\vec{C}\}+
\{\vec{B},\vec{A},\vec{C}\}.$$
Let $\vec{e}_i,\; i=1\div n$ be some orthonormal basis in the vector space. 
Then
$$(\vec{e}_i\times\vec{A})\cdot(\vec{e}_i\times\vec{B})=
((\vec{e}_i\times\vec{B})\times\vec{e}_i)\cdot\vec{A}=
[\vec{B}-(\vec{B}\cdot\vec{e}_i)\vec{e}_i]\cdot\vec{A}$$
and, therefore,
\begin{equation}
\sum\limits_{i=1}^n(\vec{e}_i\times\vec{A})\cdot(\vec{e}_i\times\vec{B})=
(n-1)\vec{A}\cdot\vec{B}.
\label{eqq7} \end{equation}
Using this identity, we obtain
$$\sum\limits_{i=1}^n\{\vec{e}_i,\vec{A},\vec{B}\}\cdot 
\{\vec{e}_i,\vec{C},\vec{D}\}=$$ 
\begin{equation}
=(n-5)(\vec{A}\times \vec{B})\cdot
(\vec{C}\times \vec{D})+2(\vec{A}\cdot\vec{C})(\vec{B}\cdot\vec{D})-
2(\vec{A}\cdot\vec{D})(\vec{B}\cdot\vec{C}).
\label{eqq8} \end{equation}
Hence
\begin{equation}
\sum\limits_{i,j=1}^n\{\vec{e}_i,\vec{e}_j,\vec{A}\}\cdot
\{\vec{e}_i,\vec{e}_j,\vec{B}\}=(n-1)(n-3)\vec{A}\cdot\vec{B}
\label{eqq9} \end{equation}
and
\begin{equation}
\sum\limits_{i,j,k=1}^n\{\vec{e}_i,\vec{e}_j,\vec{e}_k\}\cdot
\{\vec{e}_i,\vec{e}_j,\vec{e}_k\}=n(n-1)(n-3).
\label{eqq10} \end{equation}
It follows from the last equation that there exists some 
$\{\vec{e}_i,\vec{e}_j,\vec{e}_k\}$ that is not zero if $n>3$. So, really, 
equation (\ref {eqq6}) is valid only for the usual three-dimensional vector 
product (n = 1 case corresponds to identically vanishing vector product).
To continue, let us note that using
$$\vec{A}\times (\vec{B}\times \vec{C})+(\vec{A}\times \vec{B})\times \vec{C}
=(\vec{A}+\vec{C})\times\vec{B}\times(\vec{A}+\vec{C})-
\vec{A}\times\vec{B}\times\vec{A}-\vec{C}\times\vec{B}\times\vec{C}=$$
$$=2\vec{A}\cdot\vec{C}~\vec{B}-\vec{A}\cdot\vec{B}~\vec{C}-
\vec{B}\cdot\vec{C}~\vec{A}$$
and
$$\vec{A}\times(\vec{B}\times(\vec{C}\times\vec{D}))=
\left . \frac{1}{2} \right [
\vec{A}\times (\vec{B}\times(\vec{C}\times\vec{D}))+
(\vec{A}\times \vec{B})\times(\vec{C}\times\vec{D})-$$
$$-(\vec{A}\times \vec{B})\times(\vec{C}\times\vec{D})-
((\vec{A}\times \vec{B})\times \vec{C})\times\vec{D}+
((\vec{A}\times \vec{B})\times\vec{C})\times\vec{D}+$$
$$+(\vec{A}\times (\vec{B}\times\vec{C}))\times\vec{D}-
(\vec{A}\times (\vec{B}\times\vec{C}))\times\vec{D}-
\vec{A}\times ((\vec{B}\times \vec{C})\times\vec{D})+ $$ 
$$+\vec{A}\times ((\vec{B}\times \vec{C})\times\vec{D})+
\vec{A}\times (\vec{B}\times ( \vec{C}\times\vec{D})) \left . \frac{}{}
\right ],$$
we can check the equation
$$\vec{A}\times \{\vec{B},\vec{C},\vec{D}\}=-\{\vec{A},\vec{B},\vec{C}
\times \vec{D}\}+\vec{A}\times(\vec{B}\times(\vec{C}\times\vec{D}))-
\{\vec{A},\vec{C},\vec{D}\times \vec{B} \}+ $$ $$+
\vec{A}\times(\vec{C}\times(\vec{D}\times\vec{B}))-
\{\vec{A},\vec{D},\vec{B}\times \vec{C} \}+
\vec{A}\times(\vec{D}\times(\vec{B}\times\vec{C}))=$$
$$=-\{\vec{A},\vec{B},\vec{C}\times \vec{D} \}
-\{\vec{A},\vec{C},\vec{D}\times \vec{B} \}
-\{\vec{A},\vec{D},\vec{B}\times \vec{C} \}+
3\vec{A}\times\{\vec{B},\vec{C},\vec{D}\}.$$
The last identity follows from
$$3\{\vec{B},\vec{C},\vec{D}\}=\{\vec{B},\vec{C},\vec{D}\}+
\{\vec{C},\vec{D},\vec{B}\}+\{\vec{D},\vec{B},\vec{C}\} =$$ $$=
\vec{B}\times(\vec{C}\times\vec{D})+
\vec{C}\times(\vec{D}\times\vec{B})+
\vec{D}\times(\vec{B}\times\vec{C}).$$
As we see, the ternary product satisfies the following interesting identity
\begin{equation}
2~\vec{A}\times \{\vec{B},\vec{C},\vec{D}\}=\{\vec{A},\vec{B},\vec{C}
\times \vec{D}\}+\{\vec{A},\vec{C},\vec{D}\times \vec{B} \}+
\{\vec{A},\vec{D},\vec{B}\times \vec{C} \}.
\label{eqq12}\end{equation}
Hence we should have
$$4\sum\limits_{i,j,k,l=1}^n|\vec{e}_i\times \{\vec{e}_j,\vec{e}_k,
\vec{e}_l\}|^2=$$ $$=
\sum\limits_{i,j,k,l=1}^n|\{\vec{e}_i,\vec{e}_j,
\vec{e}_k\times \vec{e}_l\}+\{\vec{e}_i,\vec{e}_k,\vec{e}_l\times
\vec{e}_j\}+\{\vec{e}_i,\vec{e}_l,\vec{e}_j\times\vec{e}_k\}|^2.$$
Left-hand side is easily calculated by means of (\ref{eqq7}) and 
(\ref{eqq10}):
$$4\sum\limits_{i,j,k,l=1}^n|\vec{e}_i\times \{\vec{e}_j,\vec{e}_k,
\vec{e}_l\}|^2=4n(n-1)^2(n-3).$$
To calculate the right-hand side,  the following identity is useful
\begin{equation}
\sum\limits_{i,j=1}^n\{\vec{e}_i,\vec{e}_j,\vec{A}\}\cdot
\{\vec{e}_i,\vec{e}_j\times\vec{B},\vec{C}\}=-(n-3)(n-6)\vec{A}\cdot
(\vec{B}\times\vec{C}).
\label{eqq13}\end{equation}
This identity follows from (\ref{eqq8}) along with the relation
$$\sum\limits_{i=1}^n(\vec{e}_i\times\vec{A})\cdot((\vec{e}_i\times\vec{B})
\times\vec{C})=$$ $$=\sum\limits_{i=1}^n(\vec{e}_i\times\vec{A})\cdot
[2\vec{e}_i\cdot\vec{C}~\vec{B}-\vec{B}\cdot\vec{C}~\vec{e}_i-
\vec{e}_i\cdot\vec{B}~\vec{C}-\vec{e}_i\times(\vec{B}\times\vec{C})]=$$
$$=-(n-4)\vec{A}\cdot(\vec{B}\times\vec{C}).$$
Now, with (\ref{eqq9}) and (\ref{eqq13}) at hands, it is easy to calculate 
$$\sum\limits_{i,j,k,l=1}^n|\{\vec{e}_i,\vec{e}_j,
\vec{e}_k\times \vec{e}_l\}+\{\vec{e}_i,\vec{e}_k,\vec{e}_l\times
\vec{e}_j\}+\{\vec{e}_i,\vec{e}_l,\vec{e}_j\times\vec{e}_k\}|^2=$$
$$=3n(n-1)^2(n-3)+6n(n-1)(n-3)(n-6)=3n(n-1)(n-3)(3n-13).$$
Therefore we should have
$$4n(n-1)^2(n-3)=3n(n-1)(n-3)(3n-13).$$
Hence,
$$3n(n-1)(n-3)(3n-13)-4n(n-1)^2(n-3)=5n(n-1)(n-3)(n-7)=0.$$
This shows that a nontrivial vector product is defined only in
spaces with dimensions three and seven.

\section*{Контрольная работа 3}
\setcounter{section}{3}
\setcounter{subsection}{0}
\subsection{}
Пусть $L_0$ собственная длина стержня. Тогда $L=\frac{L_0}{\gamma_u}$ и
$L_0=\gamma_u L$. В системе $S^\prime$ стержень имеет скорость
$$u^\prime=u\ominus V=\frac{u-V}{1-\frac{uV}{c^2}}.$$
Поэтому длина стержня в этой системе $L^\prime=\frac{L_0}{\gamma_{u^\prime}}=
\frac{\gamma_u}{\gamma_{u^\prime}}L$. Но $$\gamma_{u\ominus V}=\gamma_u
\gamma_V \left (1-\frac{uV}{c^2} \right ).$$ Поэтому
$$L^\prime=\frac{L}{\gamma_V \left (1-\frac{uV}{c^2}\right )}.$$

\vspace*{5mm}
\noindent $\bullet$  $\bullet$  $\bullet$
 
\noindent
Let $L_0$ be a proper length of the rod. Then $L=\frac{L_0}{\gamma_u}$ and 
$L_0=\gamma_u L$. In the frame $S^\prime$, the rod moves with the velocity
$$u^\prime=u\ominus V=\frac{u-V}{1-\frac{uV}{c^2}}.$$
Therefore, the length of the rod in this frame is 
$L^\prime=\frac{L_0}{\gamma_{u^\prime}}=
\frac{\gamma_u}{\gamma_{u^\prime}}L$. But $$\gamma_{u\ominus V}=\gamma_u
\gamma_V \left (1-\frac{uV}{c^2} \right ).$$ Therefore,
$$L^\prime=\frac{L}{\gamma_V \left (1-\frac{uV}{c^2}\right )}.$$

\subsection{}
Согласно преобразованию Лоренца имеем $ct^\prime+x^\prime=\gamma [ct-\beta x+x
-\beta c t]=\gamma (1-\beta)(ct+x)$ и $ct^\prime-x^\prime=\gamma (1+\beta)
(ct-x)$. Но 
$$\gamma (1-\beta)=\sqrt{\frac{1-\beta}{1+\beta}}, \; {\mbox{и}} \;
\gamma (1+\beta)=\sqrt{\frac{1+\beta}{1-\beta}}=\frac{1}{\gamma (1-\beta)}.$$
следовательно, $x_+^\prime=e^{-\psi}x_+,\; x_-^\prime=e^{\psi}x_-$, где
$$e^{\psi}=\gamma (1+\beta)=\sqrt{\frac{1+\beta}{1-\beta}},\; {\mbox{или}}
\; \psi=\frac{1}{2}\ln{\frac{1+\beta}{1-\beta}}.$$
Заметим, что
$$\th{\psi}=\frac{e^\psi-e^{-\psi}}{e^\psi+e^{-\psi}}=\frac{\gamma (1+\beta)-
\gamma (1-\beta)}{\gamma (1+\beta)+\gamma (1-\beta)}=\beta.$$

\vspace*{5mm}
\noindent $\bullet$  $\bullet$  $\bullet$
 
\noindent 
According to the Lorentz transformations, $ct^\prime+x^\prime=\gamma 
[ct-\beta x+x -\beta c t]=\gamma (1-\beta)(ct+x)$ and $ct^\prime-x^\prime=
\gamma (1+\beta)(ct-x)$. But 
$$\gamma (1-\beta)=\sqrt{\frac{1-\beta}{1+\beta}}, \;\;\; {\mbox{and}} \;\;\;
\gamma (1+\beta)=\sqrt{\frac{1+\beta}{1-\beta}}=\frac{1}{\gamma (1-\beta)}.$$
Hence, $x_+^\prime=e^{-\psi}x_+,\; x_-^\prime=e^{\psi}x_-$, where
$$e^{\psi}=\gamma (1+\beta)=\sqrt{\frac{1+\beta}{1-\beta}},\;\;\; {\mbox{or}}
\;\;\; \psi=\frac{1}{2}\ln{\frac{1+\beta}{1-\beta}}.$$
Let us note that
$$\th{\psi}=\frac{e^\psi-e^{-\psi}}{e^\psi+e^{-\psi}}=\frac{\gamma (1+\beta)-
\gamma (1-\beta)}{\gamma (1+\beta)+\gamma (1-\beta)}=\beta.$$

\subsection{}
Пусть система $S_1$ движется со скоростью $V_1$ относительно $S$, 
а система $S_2$ со скоростью $V_2$ относительно $S_1$. Тогда 
$$x_-^{(1)}=\sqrt{\frac{1+\beta_1}{1-\beta_1}}\;x_- \; {\mbox{и}} \;
x_-^{(2)}=\sqrt{\frac{1+\beta_2}{1-\beta_2}}\;x_-^{(1)}.$$
Следовательно,
$$x_-^{(2)}=\sqrt{\frac{(1+\beta_1)(1+\beta_2)}{(1-\beta_1)(1-\beta_2)
}}\;x_-.$$
С другой стороны
$$x_-^{(2)}=\sqrt{\frac{1+\beta}{1-\beta}}\;x_-,$$
где
$$\beta=\beta_1\oplus\beta_2=\frac{\beta_1+\beta_2}{1+\beta_1\beta_2}$$
соответствует релятивистской сумме скоростей $V_1$ и $V_2$. В случае $n$ 
скоростей получим
$$\frac{1+\beta}{1-\beta}=\frac{(1+\beta_1)(1+\beta_2)\cdots (1+\beta_n)}
{(1-\beta_1)(1-\beta_2)\cdots (1-\beta_n)}. $$
Отсюда можно определить $\beta$. В частности, если $\beta_1=\beta_2=\cdots=
\beta_n$, то
$$\beta=\frac{(1+\beta_1)^n-(1-\beta_1)^n}{(1+\beta_1)^n+(1-\beta_1)^n}.$$

\vspace*{5mm}
\noindent $\bullet$  $\bullet$  $\bullet$

\noindent
Let the frame $S_1$ moves with velocity $V_1$ relative 
to $S$, and the frame $S_2$ -- with velocity $V_2$ relative to $S_1$. Then 
$$x_-^{(1)}=\sqrt{\frac{1+\beta_1}{1-\beta_1}}\;x_- \;\;\;{\mbox{and}}\;\;\;
x_-^{(2)}=\sqrt{\frac{1+\beta_2}{1-\beta_2}}\;x_-^{(1)}.$$
Hence,
$$x_-^{(2)}=\sqrt{\frac{(1+\beta_1)(1+\beta_2)}{(1-\beta_1)(1-\beta_2)
}}\;x_-.$$
On the other hand,
$$x_-^{(2)}=\sqrt{\frac{1+\beta}{1-\beta}}\;x_-,$$
where
$$\beta=\beta_1\oplus\beta_2=\frac{\beta_1+\beta_2}{1+\beta_1\beta_2}$$
is the relativistic sum of velocities $V_1$ and $V_2$. In the case of $n$ 
velocities, we get
$$\frac{1+\beta}{1-\beta}=\frac{(1+\beta_1)(1+\beta_2)\cdots (1+\beta_n)}
{(1-\beta_1)(1-\beta_2)\cdots (1-\beta_n)}. $$
From this equation one can determine $\beta$. In particular, if $\beta_1=
\beta_2=\cdots=\beta_n$, then
$$\beta=\frac{(1+\beta_1)^n-(1-\beta_1)^n}{(1+\beta_1)^n+(1-\beta_1)^n}.$$

\subsection{}
Рассмотрим точку $A$ лампы с координатой $x^\prime$ в системе лампы. 
Вспышка в точке $A$ в этой системе происходит при $t^\prime=0$, поэтому в системе 
неподвижного наблюдателя вспышка происходит при $t=\gamma(t^\prime+\frac{V}
{c^2} x^\prime)=\gamma\frac{V}{c^2}x^\prime$ и в точке с координатой $x=\gamma(
x^\prime+Vt^\prime)=\gamma x^\prime =\frac{c^2}{V} t$. Это показывает, что
в этой системе вспышка распространяется со скоростью $\frac{c^2}{V}$. То что 
эта скорость превосходит скорость света в вакууме $c$ связано с тем, что 
вспышки разных частей лампы причинно не связаны друг с другом, т.к. в системе 
лампы они происходят одновременно. Если причиной вспышки является подача 
напряжения, например, на левый конец, то вспышка в системе лампы будет 
распространяться с конечной скоростью $u^\prime < c$. Тогда в точке $x^\prime$
вспышка произойдет при $t^\prime=\frac{x^\prime}{u^\prime}$. Соответственно в 
Л-системе будем иметь
$$t=\gamma\left (\frac{x^\prime}{u^\prime}+\frac{V}{c^2}x^\prime\right )=
\gamma x^\prime\left (\frac{1}{u^\prime}+\frac{V}{c^2}\right ) \; {\mbox{и}}\; 
x=\gamma\left (x^\prime+V\frac{x^\prime}{u^\prime}\right )=\gamma x^\prime
\left (1+\frac{V}{u^\prime}\right ).$$
Т.е. скорость распространения вспышки в этой системе будет
$$u=\frac{x}{t}=\frac{1+\frac{V}{u^\prime}}{\frac{1}{u^\prime}+\frac{V}{c^2}}=
\frac{u^\prime+V}{1+\frac{u^\prime V}{c^2}}=u^\prime\oplus V<c.$$

\vspace*{5mm}
\noindent $\bullet$  $\bullet$  $\bullet$
  
\noindent Consider the point $A$ of the lamp with the coordinate 
$x^\prime$ in the lamp's rest frame. In this frame, the flash at this point 
occurs at $t^\prime=0$. Therefore, in the frame of stationary observer, 
the flash occurs at $t=\gamma(t^\prime+\frac{V} {c^2} x^\prime)=\gamma
\frac{V}{c^2}x^\prime$ and at a point with coordinate $x=\gamma(x^\prime+V
t^\prime)=\gamma x^\prime =\frac{c^2}{V} t$. This shows that, in this system, 
the flash propagates with speed $\frac{c^2}{V}$. The fact that this speed 
exceeds the speed of light in vacuum $c$ is a reflection of the fact that the 
flashes at different parts of the lamp are not causally related to each other, 
because in the lamp's rest frame they occur simultaneously. If the flash is 
caused by switching a voltage at, for example, left end side of the lamp, 
it will propagate in the lamp's rest frame with a finite speed $u^\prime < c$. 
Then at a point with coordinate $x^\prime$ the flash will occur at 
$t^\prime=\frac{x^\prime} {u^\prime}$. Correspondingly, in the laboratory 
frame, we will have
$$t=\gamma\left (\frac{x^\prime}{u^\prime}+\frac{V}{c^2}x^\prime\right )=
\gamma x^\prime\left (\frac{1}{u^\prime}+\frac{V}{c^2}\right ) \; 
{\mbox{and}}\;\; 
x=\gamma\left (x^\prime+V\frac{x^\prime}{u^\prime}\right )=\gamma x^\prime
\left (1+\frac{V}{u^\prime}\right ).$$ 
That is, the propagation speed of the flash in this frame is now 
$$u=\frac{x}{t}=\frac{1+\frac{V}{u^\prime}}{\frac{1}{u^\prime}+\frac{V}
{c^2}}=\frac{u^\prime+V}{1+\frac{u^\prime V}{c^2}}=u^\prime\oplus V<c.$$

\subsection{}
Представим себе, что вдоль образующей цилиндра поставлены точки. Каждый 
поперечный срез цилиндра представляет собой часы, где соответствующая точка 
играет роль конца стрелы. В системе $S^\prime$ эти часы синхронизованы, т.е. 
одновременно показывают, например, $12^{\underline{00}}$. В системе $S$ они 
уже не синхронизованы, т.е. когда часы с левого конца цилиндра показывают 
$12^{\underline{00}}$, часы с других частей цилиндра показывают другое время. 
Следовательно, наши точки в системе $S$ уже не лежат на одной линии в каждый 
данный момент времени. Это показывает, что цилиндр в системе $S$ должен быть 
скрученным. 

\begin{figure}[htb]
\centerline{\epsfig{figure=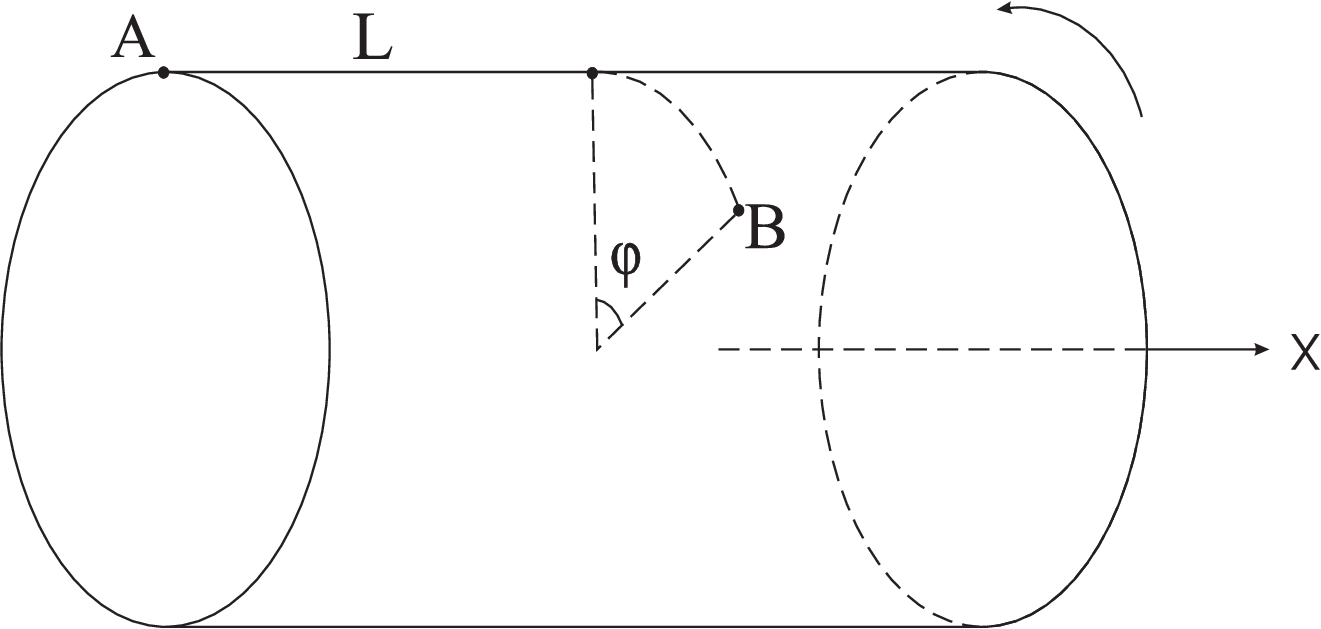,width=5cm}}
\end{figure}
\noindent
Посмотрим более внимательно. Пусть в момент $t=0$ точка $B$, 
которая находится на расстоянии $L$ вдоль цилиндра от точки $A$,
повернута на угол $\varphi$. Координатам $x_A=0,\,t_A=0$ и $x_B=L,\,t_B=0$ в
системе $S^\prime$ соответствует $x^\prime_A=0,\,t^\prime_A=0$ и $x^\prime_B=
\gamma (x_B-Vt_B)=\gamma L$, $t^\prime_B=\gamma (t_B-\frac{V}{c^2}x_B)=
-\gamma \frac{V}{c^2}L$. При этом, в этой системе точка $B$ в момент 
$t^\prime_B$ тоже повернута на угол $\varphi$, т.к. поперечные размеры при 
преобразовании Лоренца не меняются и, следовательно, поперечный угол тоже не 
меняется. За время  $t^\prime_A-t^\prime_B=\gamma \frac{V}{c^2}L$ точка $B$ 
должна успеть повернутся на угол $\varphi$, т.к. в системе $S^\prime$ часы 
синхронизированы и когда $A$  показывает $0$, $B$ тоже должен показывать 
такое-же время. Другими словами, в любой момент времени, в том числе и при 
$t^\prime=0$, все точки должны лежать на одной линии в системе $S^\prime$. 
Следовательно, $\varphi=\gamma\frac{V}{c^2}L \omega^\prime$, т.к. в системе 
$S^\prime$ цилиндр вращается с угловой скоростью 
$\omega^\prime$. Таким образом в системе $S$ имеем кручение на единицу длины
$\frac{\varphi}{L}=\gamma\frac{V}{c^2}\omega^\prime$. Из нашего рассуждения 
ясно, что кручение происходит в противоположном к вращению направлении 
(см. рисунок).

\vspace*{5mm}
\noindent $\bullet$  $\bullet$  $\bullet$

\noindent  
Let us imagine a set of points along the cylinder 
generatrix. Each cross-section of the cylinder is a clock where the 
corresponding point plays the role of the end of the clock's arrow. 
In the frame $S^\prime$, these clocks are synchronized, that is, all of them
show, for example, $12^{\underline{00}}$. In the frame $S$, however, they
are not synchronized, and when the clock on the far left end of the cylinder 
shows $12^{\underline{00}}$ other clocks along the cylinder show a different 
time. Consequently, our points no longer lie on one line in the frame $S$ 
at any given instant of this frame. This indicates that the cylinder in the 
frame $S$ must be twisted.

\begin{figure}[htp]
\centerline{\epsfig{figure=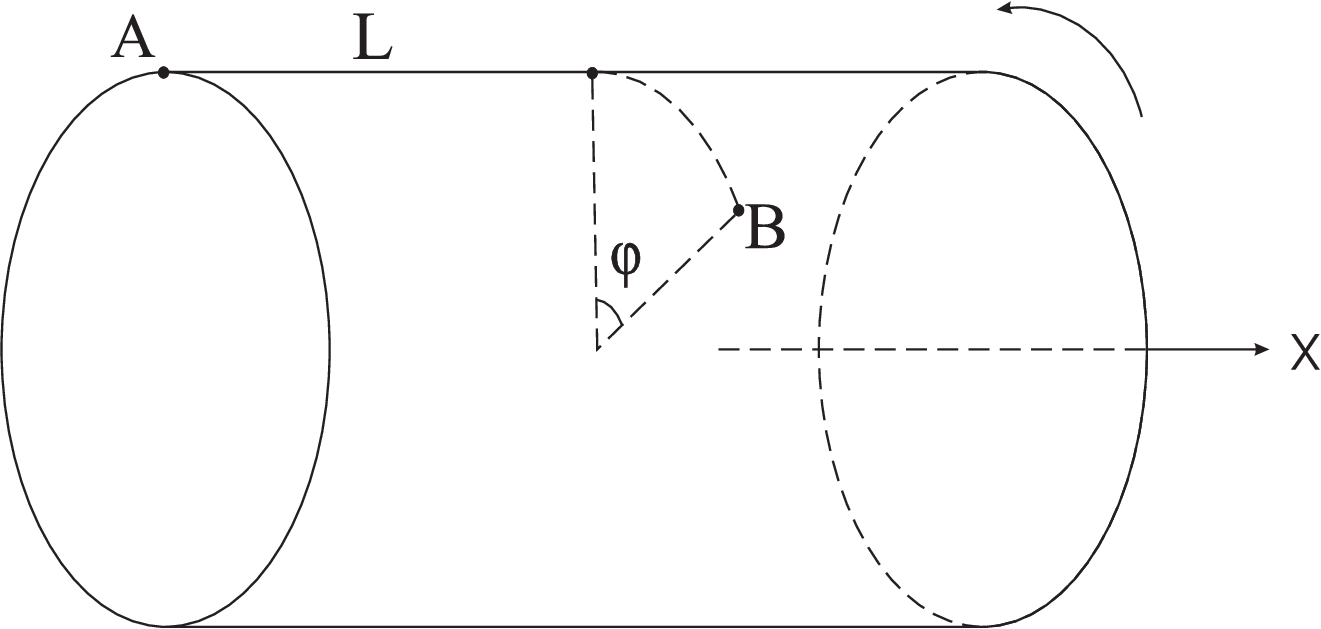,width=5cm}}
\end{figure}
\noindent
Let us look more closely. Suppose that at the time $t=0$ the point $B$, 
which is located at a distance $L$ along the cylinder from the point $A$, 
is rotated by an angle $\varphi$. Coordinates $x_A=0,\,t_A=0$ and $x_B=L,\,
t_B=0$ in the frame $S$ do correspond  to $x^\prime_A=0,\,t^\prime_A=0$ and 
$x^\prime_B=\gamma (x_B-Vt_B)=\gamma L$, $t^\prime_B=\gamma (t_B-\frac{V}
{c^2}x_B)=-\gamma \frac{V}{c^2}L$ in the frame $S^\prime$. 
At that, in this frame the point $B$ at the time $t^\prime_B$ is also rotated
by an angle $\varphi$, because transverse dimensions do not change under
Lorentz transformations and, therefore, a transverse angle is also not 
changed. During the time $t^\prime_A-t^\prime_B=\gamma \frac{V}{c^2}L$ 
the point $B$ should be able to turn by the angle $\varphi$, as in the frame 
$S^\prime$ the clocks are synchronized and when clock  $A$ shows $0$, the
clock $B$ also must show the same time. In other words, at any instant of 
time in $S^\prime$, including $t^\prime=0$, all the points considered above
should lie on one line. Consequently, $\varphi=\gamma\frac{V}{c^2}L 
\omega^\prime$, because in the frame $S^\prime$ the cylinder rotates with 
angular velocity $\omega^\prime$. Thus in the frame $S$ the torsion per unit 
length is $\frac{\varphi}{L}=\gamma\frac{V}{c^2}\omega^\prime$. 
It is clear from our discussion that the torsion occurs in the opposite 
to the rotation direction (see the figure). 

\section*{Контрольная работа 4}
\setcounter{section}{4}
\setcounter{subsection}{0}
\subsection{}
В системе $S$ компоненты скорости фотона $c_x=c\cos{\theta},\;c_y=c\sin
{\theta}$. В системе $S^\prime$ будем иметь
$$c^\prime_y=c\sin{2\theta}=\frac{c_y}{\gamma \left ( 1-\frac{c_xV}{c^2}
\right )},$$
или
$$\sin{2\theta}=\frac{\sin{\theta}}{\gamma (1-\beta \cos{\theta})}.$$
Но $\theta \ll 1$, поэтому $\sin{\theta}\approx \theta,\; \sin{2\theta}\approx 
2\theta$ и $\cos{\theta}\approx 1$. Следовательно, $$2=\frac{1}{\gamma
(1-\beta)}=\sqrt{\frac{1+\beta}{1-\beta}}$$ и $\beta=\frac{3}{5}$.

\vspace*{5mm}
\noindent $\bullet$  $\bullet$  $\bullet$

\noindent 
In the frame $S$, the velocity of the photon has the following components 
$c_x=c\cos{\theta},\;c_y=c\sin{\theta}$. Then, in the frame $S^\prime$,
we will have
$$c^\prime_y=c\sin{2\theta}=\frac{c_y}{\gamma \left ( 1-\frac{c_xV}{c^2}
\right )},$$
or
$$\sin{2\theta}=\frac{\sin{\theta}}{\gamma (1-\beta \cos{\theta})}.$$
But $\theta \ll 1$, therefore $\sin{\theta}\approx \theta,\; \sin{2\theta}
\approx 2\theta$ and $\cos{\theta}\approx 1$. Consequently, $$2=\frac{1}
{\gamma(1-\beta)}=\sqrt{\frac{1+\beta}{1-\beta}}$$ and $\beta=\frac{3}{5}$.

\subsection{}
В Л-системе компоненты скорости падающего фотона суть $c_x=c\cos{\theta_1}$,
$c_y=c\sin{\theta_1}$. В системе зеркала будем иметь 
$$c^\prime_x=\frac{c_x-V}{1-\frac{c_xV}{c^2}}=\frac{c\cos{\theta_1}-V}
{1-\beta\cos{\theta_1}},\;c^\prime_y=\frac{c_y}{\gamma\left (1-\frac{c_xV}
{c^2}\right )}=\frac{c\sin{\theta_1}}{\gamma(1-\beta\cos{\theta_1})}.$$
А после отражения $c^{\prime\prime}_x=-c^\prime_x$ и $c^{\prime\prime}_y=
c^\prime_y$. Поэтому в Л-системе $y$-компонента скорости отраженного фотона 
будет
$$c\sin{\theta_2}=\frac{c^{\prime\prime}_y}{\gamma\left(1+\frac{c^{\prime
\prime}_xV}{c^2}\right )}=\frac{c\sin{\theta_1}}{\gamma^2 \left (1-\beta
\frac{c^\prime_x}{c}\right )}\,\frac{1}{1-\beta\cos{\theta_1}}.$$
Но
$$1-\beta\frac{c^\prime_x}{c}=1-\beta\frac{\cos{\theta_1}-\beta}{1-\beta
\cos{\theta_1}}=\frac{1}{1-\beta\cos{\theta_1}}(1+\beta^2-2\beta
\cos{\theta_1}).$$
Окончательно
$$\sin{\theta_2}=\frac{\sin{\theta_1}}{\gamma^2(1+
\beta^2-2\beta \cos{\theta_1})}=\frac{(1-\beta^2)\,\sin{\theta_1}}{
1+\beta^2-2\beta\cos{\theta_1}}.$$

\vspace*{5mm}
\noindent $\bullet$  $\bullet$  $\bullet$
 
\noindent
In the laboratory frame, incident photon's velocity has the components
$c_x=c\cos{\theta_1}$, $c_y=c\sin{\theta_1}$. Then, in the rest frame of the 
mirror, we will have 
$$c^\prime_x=\frac{c_x-V}{1-\frac{c_xV}{c^2}}=\frac{c\cos{\theta_1}-V}
{1-\beta\cos{\theta_1}},\;c^\prime_y=\frac{c_y}{\gamma\left (1-\frac{c_xV}
{c^2}\right )}=\frac{c\sin{\theta_1}}{\gamma(1-\beta\cos{\theta_1})}.$$
Then after the reflection, in the rest frame of the mirror, 
$c^{\prime\prime}_x=-c^\prime_x$ and $c^{\prime\prime}_y=c^\prime_y$. 
Therefore, in the laboratory frame $y$-component of the reflected photon's 
velocity is
$$c\sin{\theta_2}=\frac{c^{\prime\prime}_y}{\gamma\left(1+\frac{c^{\prime
\prime}_xV}{c^2}\right )}=\frac{c\sin{\theta_1}}{\gamma^2 \left (1-\beta
\frac{c^\prime_x}{c}\right )}\,\frac{1}{1-\beta\cos{\theta_1}}.$$
But
$$1-\beta\frac{c^\prime_x}{c}=1-\beta\frac{\cos{\theta_1}-\beta}{1-\beta
\cos{\theta_1}}=\frac{1}{1-\beta\cos{\theta_1}}(1+\beta^2-2\beta
\cos{\theta_1}).$$
Hence, finally
$$\sin{\theta_2}=\frac{\sin{\theta_1}}{\gamma^2(1+
\beta^2-2\beta \cos{\theta_1})}=\frac{(1-\beta^2)\,\sin{\theta_1}}{
1+\beta^2-2\beta\cos{\theta_1}}.$$

\subsection{}
В Л-системе 4-скорости систем $S_1$ и $S_2$ имеют вид $u_1=(\gamma_1 c,\gamma_1
\vec{V}_1)$ и $u_2=(\gamma_2 c,\gamma_2\vec{V}_2)$, тогда как в системе $S_1$
будем иметь $u^\prime_1=(c,\vec{0})$, $u^\prime_2=(\gamma c,\gamma\vec{V})$,
где $\vec{V}$ есть относительная скорость $S_1$ и $S_2$ (скорость $S_2$ в 
системе $S_1$). Но скалярное произведение 4-векторов инвариантно: $u_1\cdot 
u_2=u^\prime_1\cdot u^\prime_2$. Следовательно, $\gamma_1\gamma_2 (c^2-
\vec{V_1}\cdot \vec{V_2})=\gamma c^2$. Отсюда $\gamma=\gamma_1\gamma_2 (1-
\vec{\beta_1}\cdot\vec{\beta_2})$. Но
$$\beta^2=1-\frac{1}{\gamma^2}=1-\frac{(1-\beta_1^2)(1-\beta_2^2)}
{(1-\vec{\beta_1}\cdot\vec{\beta_2})^2}=\frac{\beta_1^2+\beta_2^2-
2\vec{\beta_1}\cdot\vec{\beta_2}+(\vec{\beta_1}\cdot\vec{\beta_2})^2-
\beta_1^2\beta_2^2}{(1-\vec{\beta_1}\cdot\vec{\beta_2})^2}.$$
С другой стороны $(\vec{\beta_1}\times\vec{\beta_2})^2=\beta_1^2\beta_2^2
(1-\cos^2{\theta})=\beta_1^2\beta_2^2-(\vec{\beta_1}\cdot\vec{\beta_2})^2$,
и окончательно
$$\beta^2=\frac{(\vec{\beta_1}-\vec{\beta_2})^2-(\vec{\beta_1}\times
\vec{\beta_2})^2}{(1-\vec{\beta_1}\cdot\vec{\beta_2})^2}.$$

\vspace*{10mm}
\noindent $\bullet$  $\bullet$  $\bullet$
 
\noindent
In the laboratory frame, 4-velocities of $S_1$ and $S_2$ are
$u_1=(\gamma_1c,\gamma_1\vec{V}_1)$ and $u_2=(\gamma_2 c,\gamma_2\vec{V}_2)$, 
while in the frame $S_1$ we will have
$u^\prime_1=(c,\vec{0})$, $u^\prime_2=(\gamma c,\gamma\vec{V})$,
where $\vec{V}$ is the relative velocity of $S_1$ and $S_2$ (the velocity of
$S_2$ in the rest frame of $S_1$). But scalar product of four-vectors is 
invariant: $u_1\cdot u_2=u^\prime_1\cdot u^\prime_2$. Therefore, 
$\gamma_1\gamma_2 (c^2-\vec{V_1}\cdot \vec{V_2})=\gamma c^2$. 
It follows from this that
$\gamma=\gamma_1\gamma_2 (1-\vec{\beta_1}\cdot\vec{\beta_2})$. But
$$\beta^2=1-\frac{1}{\gamma^2}=1-\frac{(1-\beta_1^2)(1-\beta_2^2)}
{(1-\vec{\beta_1}\cdot\vec{\beta_2})^2}=\frac{\beta_1^2+\beta_2^2-
2\vec{\beta_1}\cdot\vec{\beta_2}+(\vec{\beta_1}\cdot\vec{\beta_2})^2-
\beta_1^2\beta_2^2}{(1-\vec{\beta_1}\cdot\vec{\beta_2})^2}.$$
On the other hand, $(\vec{\beta_1}\times\vec{\beta_2})^2=\beta_1^2\beta_2^2
(1-\cos^2{\theta})=\beta_1^2\beta_2^2-(\vec{\beta_1}\cdot\vec{\beta_2})^2$,
and finally
$$\beta^2=\frac{(\vec{\beta_1}-\vec{\beta_2})^2-(\vec{\beta_1}\times
\vec{\beta_2})^2}{(1-\vec{\beta_1}\cdot\vec{\beta_2})^2}.$$

\clearpage
\subsection{}
Пусть в момент времени $t=0$ стержень в $S$-системе занимает положение как на 
рисунке. 
\begin{figure}[htb]
\centerline{\epsfig{figure=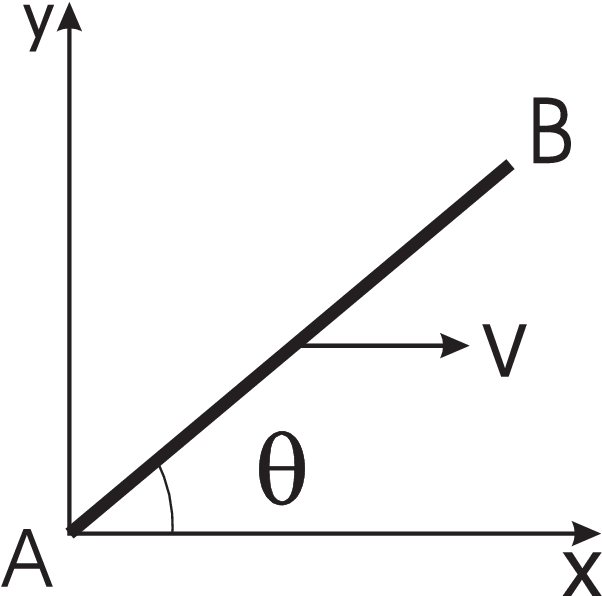,height=4cm}}
\end{figure}

\noindent Тогда в системе $S^\prime$ будем иметь $t^\prime_A=0,\,x^\prime_A=0,\,
y^\prime_A=0$ и $t^\prime_B=-\gamma\frac{V}{c^2}\;x_B=-\gamma\frac{V}{c^2}L
\cos{\theta}$, $x^\prime_B=\gamma x_B=\gamma L \cos{\theta}$, $y^\prime_B=
y_B=L \sin{\theta}$, где $L$ -- длина стержня в системе $S$. Несмотря на то, 
что $t^\prime_A\ne t^\prime_B$, точки $(x^\prime_A,y^\prime_A)$ и
$(x^\prime_B,y^\prime_B)$ определяют ориентацию стержня в системе $S^\prime$, 
так как в этой системе стержень неподвижен. 
\begin{figure}[htb]
\centerline{\epsfig{figure=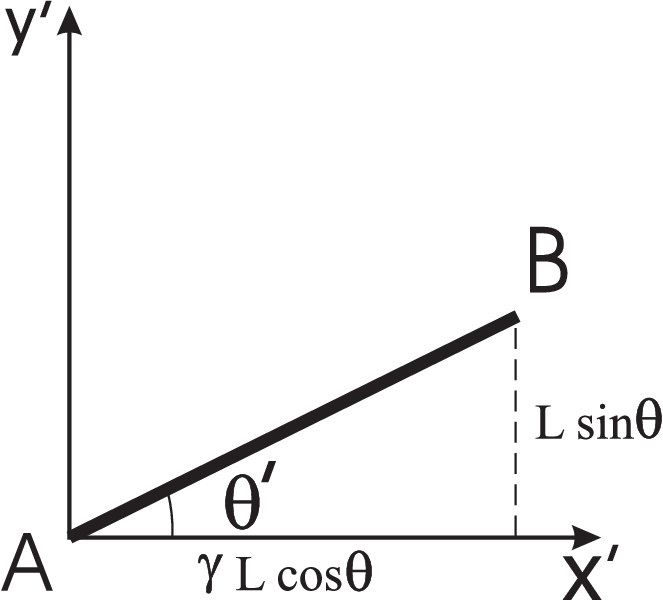,height=4cm}}
\end{figure}

\noindent Следовательно, $\tg{\theta^\prime}=\frac{\tg{\theta}}{\gamma}$ и 
$\tg{\theta}=\gamma\tg{\theta^\prime}$.

\vspace*{5mm}
\noindent $\bullet$  $\bullet$  $\bullet$
  
\noindent
Suppose that at the time $t=0$ the rod in the $S$-system 
occupies a position as indicated in the figure.
\begin{figure}[htb]
\centerline{\epsfig{figure=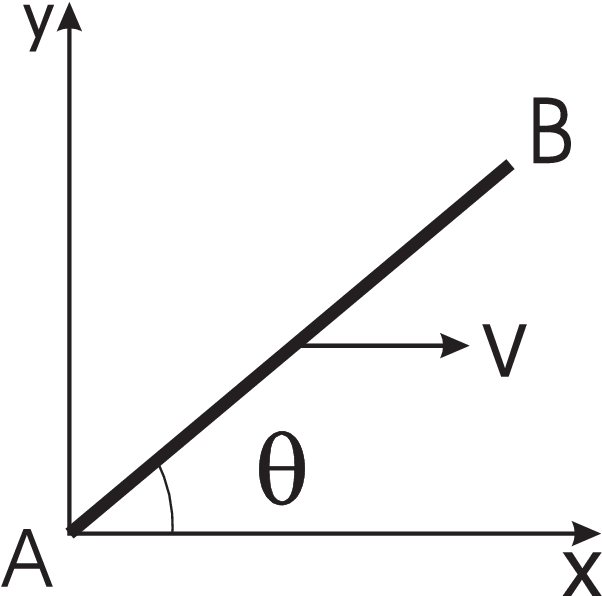,height=4cm}}
\end{figure}

\noindent In the system $S^\prime $ we will have
$t^\prime_A=0,\,x^\prime_A=0,\,
y^\prime_A=0$ and $t^\prime_B=-\gamma\frac{V}{c^2}\;x_B=-\gamma\frac{V}{c^2}L
\cos{\theta}$, $x^\prime_B=\gamma x_B=\gamma L \cos{\theta}$, $y^\prime_B=
y_B=L \sin{\theta}$, where $L$ is the length of the rod in the system $S$. 
Although $t^\prime_A\ne t^\prime_B$, the points $(x^\prime_A,y^\prime_A)$ and
$(x^\prime_B,y^\prime_B)$ still determine the orientation of the rod in the 
system $S^\prime$, because in this system the rod is stationary.
\begin{figure}[htb]
\centerline{\epsfig{figure=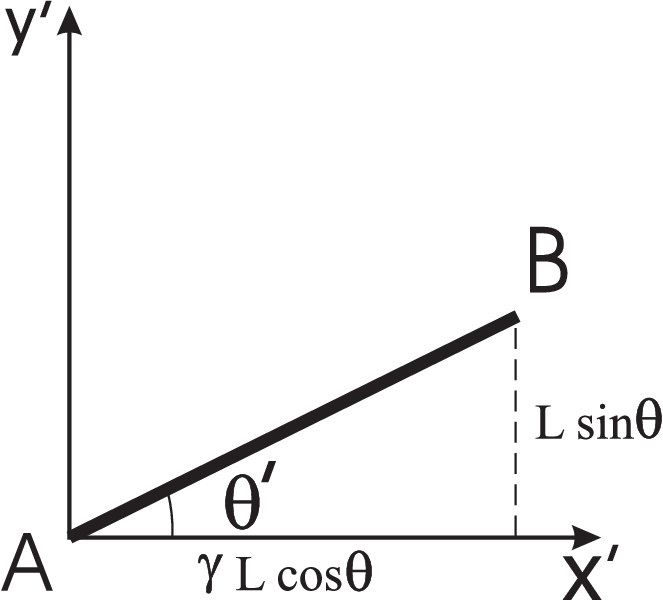,height=4cm}}
\end{figure}

\noindent Therefore, $\tg{\theta^\prime}=\frac{\tg{\theta}}{\gamma}$ and 
$\tg{\theta}=\gamma\tg{\theta^\prime}$.

\subsection{}
Пусть $S$ лабораторная система отсчета в которой в начальный момент 
$\overrightarrow{AB\;}$ составляет нулевой угол с осью $x$, а после полного 
оборота меняет ориентацию на 180$^\circ$. Заменим движение по окружности на 
движение  по правильному $N$-угольнику, где $N\to\infty$. Пусть двигаясь по 
$n$-ой грани многоугольника, которая наклонена под углом $\varphi$ к оси $x$
системы $S$, вектор $\overrightarrow{AB\;}$ составляет угол $\beta$ с осью 
$x$, а при переходе на следующую грань, с углом наклона $\varphi+\alpha$, 
$\alpha=\frac{2\pi}{N}$, угол становится $\beta+d\beta$. 

\begin{figure}[htb]
\centerline{\epsfig{figure=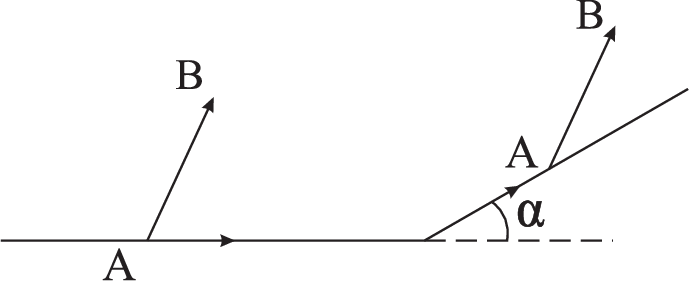,height=2cm,angle=20}}
\end{figure}\noindent
Рассмотрим систему $\tilde S$, которая
неподвижна относительно $S$, но ось $x$ которой повернута на угол $\varphi$ и
смотрит вдоль $n$-ой грани. Тогда в системе $\tilde S$ вектор 
$\overrightarrow{AB\;}$ составляет угол $\theta=\beta-\varphi$ с осью $x$
при движении по $n$-ой грани и угол $\theta+d\beta$ -- при движении по 
$(n+1)$-ой грани. Выберем начало отсчета времени так, что $t=0$ соответствует
переходу от $n$-ой грани к $(n+1)$-ой. Перед переходом, при $t=0_-$, координаты
точек $A$ и $B$ имеют вид $\tilde x_A=0,\tilde y_A=0,\tilde x_B=\tilde l 
\cos{\theta}, \tilde y_B=\tilde l \sin{\theta}$. Пусть $S^\prime$ -- система
в которой $\overrightarrow{AB\;}$ покоится при движении вдоль $n$-ой грани.
По преобразованиям Лоренца находим
$$t_A^\prime=0,\; x_A^\prime=0,\;y_A^\prime=0,\;t_B^\prime=-\gamma\frac{
V}{c^2}\tilde x_B,\; x_B^\prime=\gamma \tilde x_B,\; y_B^\prime=
\tilde y_B.$$
Несмотря на то, что $t_A^\prime\ne t_B^\prime$, угол наклона вектора 
$\overrightarrow{AB\;}$ к оси $x$ в системе $S^\prime$ определяется из 
$$\tg{\theta_-^{\,\prime}}=\frac{y_B^\prime}{x_B^\prime}=\frac{\tg{\theta}}
{\gamma},$$
т.к. в системе $S^\prime$ вектор $\overrightarrow{AB\;}$ неподвижен.

После перехода на $(n+1)$-ую грань, вектор $\overrightarrow{AB\;}$ уже движется
относительно $S^\prime$. Сразу после перехода, при $t=0_+$, будем иметь
$$\tilde x_A=0,\;\tilde y_A=0,\;\tilde x_B=\tilde l 
\cos{(\theta+d\beta)},\; \tilde y_B=\tilde l \sin{(\theta+d\beta)}.$$
Найдем момент времени $t=t_B$ такой, что событию $(t_B,\,\tilde x_B(t_B))$ 
отвечает время $t_B^\prime=0$ в системе $S^\prime$. Имеем
$$t_B^\prime=\gamma\left (t_B-\frac{V}{c^2}\,\tilde x_B(t_B)\right ).$$
Следовательно, $t_B=\frac{V}{c^2}\,\tilde x_B(t_B)$. Но $\tilde x_B(t_B)=
\tilde x_B+(V\cos{d\varphi})t_B\approx \tilde x_B+Vt_B$, где $d\varphi=
\frac{2\pi}{N}$. Отсюда
$$\tilde x_B(t_B)\approx\frac{\tilde x_B}{1-(V/c)^2}=\gamma^2 \tilde x_B
\;\;\mbox{и}\;\; t_B=\frac{V}{c^2}\gamma^2\tilde x_B.$$
Поэтому $x^\prime_B=\gamma(\tilde x_B(t_B)-Vt_B)\approx \gamma\tilde x_B=
\gamma\tilde l \cos{(\theta+d\beta)}$ и $y^\prime_B=\tilde y_B(t_B)=
\tilde y_B+(V\sin{d\varphi})t_B\approx \tilde y_B+Vt_Bd\varphi=
\tilde l \sin{(\theta+d\beta)}+(\gamma^2-1)\tilde l d\varphi\cos{(\theta+
d\beta)}$. Следовательно, угол наклона $\theta_+^\prime$, после перехода на 
$(n+1)$-ую грань, определяется из 
$$\tg{\theta_+^\prime}=\frac{1}{\gamma}\left [\tg{(\theta+d\beta)}+
(\gamma^2-1)d\varphi\right ].$$
Но в системе $S^\prime$ вектор $\overrightarrow{AB\;}$ переносится параллельно
самому себе, т.е.  $\theta_+^\prime=\theta_-^\prime$ и, следовательно, 
$\tg{\theta}=\tg{(\theta+d\beta)}+(\gamma^2-1)d\varphi$. Но
$$\tg{(\theta+d\beta)}=\frac{\tg{\theta}+\tg{(d\beta)}}{1-\tg{\theta}\tg
{(d\beta)}}\approx \frac{\tg{\theta}+d\beta}{1-d\beta\tg{\theta}}\approx
\tg{\theta}+(1+\tg^2{\theta})d\beta.$$
Поэтому получаем $$\frac{d\beta}{d\varphi}=-(\gamma^2-1)\cos^2{\theta},$$ 
или с учетом $\theta=\beta-\varphi$,
$$\frac{d\theta}{d\varphi}=-\left [1+(\gamma^2-1)\cos^2{\theta}\right].$$
Следовательно,
$$\varphi=-\int\limits_0^\theta\frac{d\theta}{1+(\gamma^2-1)\cos^2
{\theta}}.$$
Интеграл можно взять с помощью подстановки $\tg{\theta}=x$:
$$\int\frac{d\theta}{1+(\gamma^2-1)\cos^2{\theta}}=\int\frac{dx}{\gamma^2
+x^2}=\frac{1}{\gamma}\arctg{\frac{x}{\gamma}}=\frac{1}{\gamma}\arctg
{\frac{\tg{\theta}}{\gamma}}.$$
Окончательно, $\tg{\theta}=-\gamma\tg{(\gamma\varphi)}$. По условию задачи,
после полного оборота $\theta=\pi$. Поэтому должны иметь $\tg{(\gamma 2\pi)}
=0$, т.е. $\gamma$ полуцелое число (целым $\gamma$ соответствует $\theta=0$).
Минимальная скорость, $V=\frac{\sqrt{5}}{3}\,c$, получается при $\gamma=
\frac{3}{2}$.

\vspace*{5mm}
\noindent $\bullet$  $\bullet$  $\bullet$
 
\noindent
Let $S$ be the laboratory frame in which $\overrightarrow{AB\;}$ makes 
initially a zero angle with the $x$-axis but after the full turn changes its
orientation by  180$^\circ$. Let us replace the circle, on which the motion 
takes place, by a regular $N$-side polygon, where $N\to\infty$. 
Suppose the vector $\overrightarrow{AB\;}$ makes the angle $\beta$  with 
respect to the $x$-axis of the system $S$ while moving on the $n$-th side of 
the polygon tilted at an angle $\varphi$ with respect to the $x$-axis. This 
angle changes to $\beta+d\beta$ upon the transition to the next $(n+1)$-th
edge, with the tilting angle $\varphi+\alpha$,  $\alpha=\frac{2\pi}{N}$.
\begin{figure}[htb]
\centerline{\epsfig{figure=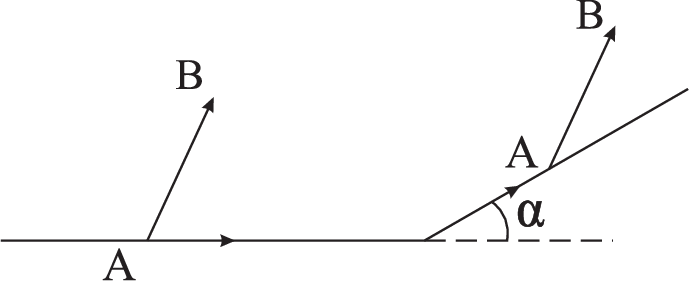,height=2cm,angle=20}}
\end{figure}

\noindent
Consider the system $\tilde S$, which is at rest with respect to $S$, but 
with the $x$-axis rotated by an angle $\varphi$, so that it is pointing 
along the $n$-th edge of the polygon. Then in the system $\tilde S$ the vector
$\overrightarrow{AB \;}$ makes an angle $\theta = \beta-\varphi $ with the
$x$-axis during the motion on the $n$-th edge and the angle $\theta + d\beta$ 
when driving on the $(n+1)$-th edge. We choose the origin of time so that 
$t=0$ corresponds to the transition from the $n$-th edge to the $(n +1)$-th. 
Before the transition, at $t=0_-$, coordinates of the points $A$ and $B$ are
$\tilde x_A=0,\tilde y_A=0,\tilde x_B=\tilde l \cos{\theta}, \tilde y_B=
\tilde l \sin{\theta}$. Let $S^\prime$ be the system in which
 $\overrightarrow{AB\;}$ is left at rest while moving along the $n$-th edge 
in the system $S$. According to the Lorentz transformations, we find
$$t_A^\prime=0,\; x_A^\prime=0,\;y_A^\prime=0,\;t_B^\prime=-\gamma\frac{
V}{c^2}\tilde x_B,\; x_B^\prime=\gamma \tilde x_B,\; y_B^\prime=
\tilde y_B.$$
Despite the fact that $t_A^\prime\ne t_B^\prime$, the angle of inclination of 
the vector $\overrightarrow{AB\;}$ to the $x$-axis is determined in the 
system $S^\prime$ by
$$\tg{\theta_-^{\,\prime}}=\frac{y_B^\prime}{x_B^\prime}=\frac{\tg{\theta}}
{\gamma},$$
because the vector $\overrightarrow{AB\;}$ remains stationary in this frame.

After the change to the $(n+1)$-th edge, the vector  $\overrightarrow{AB\;}$  
is now moving with respect to $S^\prime$. Immediately after the 
transition, at $t=0_+$, we have
$$\tilde x_A=0,\;\tilde y_A=0,\;\tilde x_B=\tilde l 
\cos{(\theta+d\beta)},\; \tilde y_B=\tilde l \sin{(\theta+d\beta)}.$$
Let us find a time $t=t_B$ such that the event  $(t_B,\,\tilde x_B(t_B))$ has
the time coordinate $t_B^\prime=0$ in the system $S^\prime$. We have
$$t_B^\prime=\gamma\left (t_B-\frac{V}{c^2}\,\tilde x_B(t_B)\right ).$$
Consequently, $t_B=\frac{V}{c^2}\,\tilde x_B(t_B)$. But $\tilde x_B(t_B)=
\tilde x_B+(V\cos{d\varphi})t_B\approx \tilde x_B+Vt_B$, where $d\varphi=
\frac{2\pi}{N}$. Hence
$$\tilde x_B(t_B)\approx\frac{\tilde x_B}{1-(V/c)^2}=\gamma^2 \tilde x_B
\;\;\mbox{and}\;\; t_B=\frac{V}{c^2}\gamma^2\tilde x_B.$$
Therefore $x^\prime_B=\gamma(\tilde x_B(t_B)-Vt_B)\approx \gamma\tilde x_B=
\gamma\tilde l \cos{(\theta+d\beta)}$ and $y^\prime_B=\tilde y_B(t_B)=
\tilde y_B+(V\sin{d\varphi})t_B\approx \tilde y_B+Vt_Bd\varphi=
\tilde l \sin{(\theta+d\beta)}+(\gamma^2-1)\tilde l d\varphi\cos{(\theta+
d\beta)}$. Consequently, the inclination angle $\theta_+^\prime$, after the 
transition on the $(n+1)$-th edge is determined from
$$\tg{\theta_+^\prime}=\frac{1}{\gamma}\left [\tg{(\theta+d\beta)}+
(\gamma^2-1)d\varphi\right ].$$
But in the system $S^\prime$ the vector $\overrightarrow{AB\;}$ is displaced 
parallel to itself, that is, $\theta_+^\prime=\theta_-^\prime$ and, 
consequently,
$\tg{\theta}=\tg{(\theta+d\beta)}+(\gamma^2-1)d\varphi$. But
$$\tg{(\theta+d\beta)}=\frac{\tg{\theta}+\tg{(d\beta)}}{1-\tg{\theta}\tg
{(d\beta)}}\approx \frac{\tg{\theta}+d\beta}{1-d\beta\tg{\theta}}\approx
\tg{\theta}+(1+\tg^2{\theta})d\beta.$$
Therefore, we get $$\frac{d\beta}{d\varphi}=-(\gamma^2-1)\cos^2{\theta},$$ 
or after taking into account that $\theta=\beta-\varphi$,
$$\frac{d\theta}{d\varphi}=-\left [1+(\gamma^2-1)\cos^2{\theta}\right].$$
Consequently,
$$\varphi=-\int\limits_0^\theta\frac{d\theta}{1+(\gamma^2-1)\cos^2
{\theta}}.$$
The integral can be calculated by substitution $\tg{\theta}=x$:
$$\int\frac{d\theta}{1+(\gamma^2-1)\cos^2{\theta}}=\int\frac{dx}{\gamma^2
+x^2}=\frac{1}{\gamma}\arctg{\frac{x}{\gamma}}=\frac{1}{\gamma}\arctg
{\frac{\tg{\theta}}{\gamma}}.$$
Finally, $\tg{\theta}=-\gamma\tg{(\gamma\varphi)}$. By the condition of the 
problem, after a full rotation $\theta=\pi$. Therefore, we must have 
$\tg{(\gamma 2\pi)}=0$ and $\gamma$ is half-integer (an integer $\gamma$ 
corresponds to $\theta=0$). The minimum speed, $V=\frac{\sqrt{5}}{3}\,c$,
is obtained when $\gamma=\frac{3}{2}$.

\section*{Контрольная работа 5}
\setcounter{section}{5}
\setcounter{subsection}{0}
\subsection{}
\begin{figure}[htb]
\centerline{\epsfig{figure=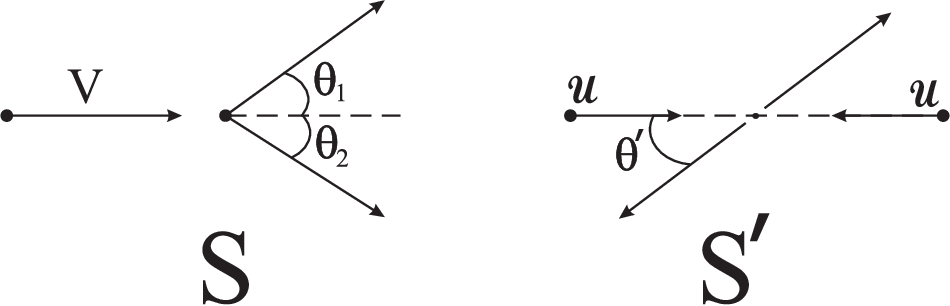,width=10cm}}
\end{figure}
Пусть $S^\prime$ система центра масс, который движется со скоростью $u$.
Вторая частица, которая неподвижна в системе $S$, в системе $S^\prime$
движется со скоростью $-u$. Следовательно, первая частица в этой
системе движется со скоростью $u$ (чтобы суммарный импульс
равнялся нулю). Пусть угол рассеяния в системе $S^\prime$ равен
$\theta^\prime$. Скорости частиц по величине не поменяются, т.к.
рассеяние упругое. Следовательно, после рассеяния
$V^\prime_{1,x}=u\cos{\theta^\prime}$, $V^\prime_{1,y}=u\sin{\theta^\prime}$,
$V^\prime_{2,x}=-u\cos{\theta^\prime}$ и $V^\prime_{2,y}=
-u\sin{\theta^\prime}$. 
По формуле сложения скоростей, в системе $S$ будем иметь $$
V_{1,x}=\frac{V^\prime_{1,x}+u}{1+\frac{V^\prime_{1,x}u}{c^2}}=
u\,\frac{1+\cos{\theta^\prime}}{1+\beta^2_u\cos{\theta^\prime}},\;
V_{1,y}=\frac{V^\prime_{1,y}}{\gamma_u\left(1+\frac{V^\prime_{1,x}u}{c^2}
\right)}=u\,\frac{\sin{\theta^\prime}}{\gamma_u\left(1+\beta_u^2\cos{\theta
^\prime}\right)},$$
$$ \hspace*{-46mm}
V_{2,x}=u\,\frac{1-\cos{\theta^\prime}}{1-\beta^2_u\cos{\theta^\prime}}
\;\;\;\; \mbox{и} \;\;\;\;
V_{2,y}=-u\,\frac{\sin{\theta^\prime}}{\gamma_u\left(1-\beta_u^2\cos{\theta
^\prime}\right)}.$$
Заметим, что $V_{1,x}>0$ и $V_{2,x}>0$. Поэтому в системе $S$ углы
рассеяния $\theta_1<\frac{\pi}{2}$ и $\theta_2<\frac{\pi}{2}$. Более того,
$$\tg{\theta_1}=\left|\frac{V_{1,y}}{V_{1,x}}\right|=
\frac{\sin{\theta^\prime}}{\gamma_u(1+\cos{\theta^\prime})}$$
и
$$\tg{\theta_2}=\left|\frac{V_{2,y}}{V_{2,x}}\right|=
\frac{\sin{\theta^\prime}}{\gamma_u(1-\cos{\theta^\prime})}.$$
Поэтому
$$\tg{\theta_1}\cdot\tg{\theta_2}=\frac{1}{\gamma_u^2}.$$
Или
$$\tg{\theta_1}=\frac{1}{\gamma_u^2}\,\ctg{\theta_2}=\frac{1}{\gamma_u^2}
\tg{\left (\frac{\pi}{2}-\theta_2\right )}<\tg{\left (\frac{\pi}{2}-\theta_2
\right )},$$ 
т.к. $\gamma_u>1$. Следовательно,
$\theta_1<\frac{\pi}{2}-\theta_2$ и
$\theta_1+\theta_2<\frac{\pi}{2}$.

\vspace*{5mm}
\noindent $\bullet$  $\bullet$  $\bullet$
 
\begin{figure}[htb]
\centerline{\epsfig{figure=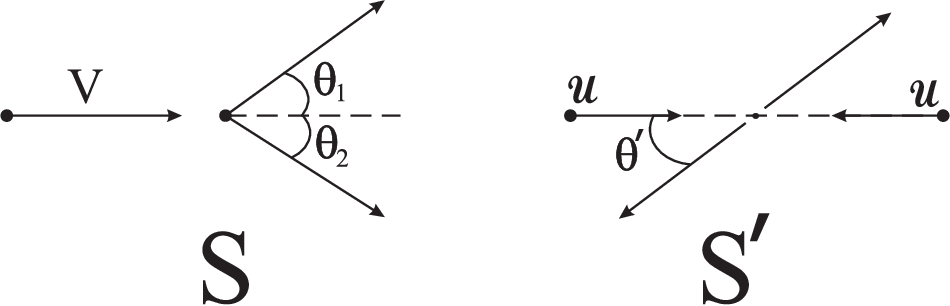,width=10cm}}
\end{figure}
Let $S^\prime$ be the center of mass system, which moves with velocity $u$.
the second particle, which is initially at rest in the system $S$, in the 
system $S^\prime$ moves with velocity $-u$. Consequently, the first particle 
in this system moves with velocity $u$ (so that the total momentum is zero). 
Let the scattering angle in the $S^\prime$ is $\theta^\prime$. Particle 
velocity magnitudes will not change after the scattering because the 
scattering is elastic. Consequently, after the scattering
$V^\prime_{1,x}=u\cos{\theta^\prime}$, $V^\prime_{1,y}=u\sin{\theta^\prime}$,
$V^\prime_{2,x}=-u\cos{\theta^\prime}$ and $V^\prime_{2,y}=
-u\sin{\theta^\prime}$. 
According to the velocity addition formula, in the system $S$ we will have
$$V_{1,x}=\frac{V^\prime_{1,x}+u}{1+\frac{V^\prime_{1,x}u}{c^2}}=
u\,\frac{1+\cos{\theta^\prime}}{1+\beta^2_u\cos{\theta^\prime}},\;
V_{1,y}=\frac{V^\prime_{1,y}}{\gamma_u\left(1+\frac{V^\prime_{1,x}u}{c^2}
\right)}=u\,\frac{\sin{\theta^\prime}}{\gamma_u\left(1+\beta_u^2\cos{\theta
^\prime}\right)},$$
$$ \hspace*{-46mm}
V_{2,x}=u\,\frac{1-\cos{\theta^\prime}}{1-\beta^2_u\cos{\theta^\prime}}
\;\;\;\; \mbox{and} \;\;\;\;
V_{2,y}=-u\,\frac{\sin{\theta^\prime}}{\gamma_u\left(1-\beta_u^2\cos{\theta
^\prime}\right)}.$$
Note that $V_{1,x}>0$ and $V_{2,x}>0$. Therefore, in the system $S$ the 
scattering angles $\theta_1<\frac{\pi}{2}$ and $\theta_2<\frac{\pi}{2}$.
Moreover,
$$\tg{\theta_1}=\left|\frac{V_{1,y}}{V_{1,x}}\right|=
\frac{\sin{\theta^\prime}}{\gamma_u(1+\cos{\theta^\prime})}$$
and
$$\tg{\theta_2}=\left|\frac{V_{2,y}}{V_{2,x}}\right|=
\frac{\sin{\theta^\prime}}{\gamma_u(1-\cos{\theta^\prime})}.$$
Therefore,
$$\tg{\theta_1}\cdot\tg{\theta_2}=\frac{1}{\gamma_u^2}.$$
Or
$$\tg{\theta_1}=\frac{1}{\gamma_u^2}\,\ctg{\theta_2}=\frac{1}{\gamma_u^2}
\tg{\left (\frac{\pi}{2}-\theta_2\right )}<\tg{\left (\frac{\pi}{2}-\theta_2
\right )},$$ 
because $\gamma_u>1$. Consequently,
$\theta_1<\frac{\pi}{2}-\theta_2$ and
$\theta_1+\theta_2<\frac{\pi}{2}$.

\subsection{}
В системе $S^\prime$ фотон имеет $y$-компоненту импульса $$p^\prime_y=\frac{h
\nu^\prime}{c}\sin{\theta^\prime}.$$ Но $p^\prime_y=p_y$ и в
системе $S$ $$p_y=\frac{h\nu}{c}\sin{\theta}.$$
Следовательно $\nu^\prime\sin{\theta^\prime}=\nu\sin{\theta}$ и
$$\sin{\theta}=\frac{\nu^\prime}{\nu}\sin{\theta^\prime}.$$

\vspace*{5mm}
\noindent $\bullet$  $\bullet$  $\bullet$
 
\noindent
In the system $S^\prime$, the photon's momentum has the following 
$y$-component 
$$p^\prime_y=\frac{h\nu^\prime}{c}\sin{\theta^\prime}.$$
But $p^\prime_y=p_y$ and in the
system $S$, $$p_y=\frac{h\nu}{c}\sin{\theta}.$$
Consequently,
$\nu^\prime\sin{\theta^\prime}=\nu\sin{\theta}$ and
$$\sin{\theta}=\frac{\nu^\prime}{\nu}\sin{\theta^\prime}.$$

\subsection{}
Рассмотрим распад в системе покоя $K^0$-мезона. В этой системе,
каждый пион имеет энергию
$\varepsilon^\prime=\frac{m_K}{2}$ (положили $c=1$) и
импульс $$p^\prime=\sqrt{\varepsilon^{\prime 2}-m_\pi^2}=\frac{m_K}{2}
\sqrt{1-\frac{4m_\pi^2}{m_K^2}}.$$
\noindent В Л-системе энергия $\pi$-мезона, который в системе покоя вылетел
под углом $\theta^\prime$, будет
$\varepsilon=\gamma(\varepsilon^\prime+\beta
p^\prime_x)=\gamma(\varepsilon^\prime+\beta p^\prime \cos{\theta^\prime})$. 
Поэтому
$$\alpha=\frac{\varepsilon_{max}}{\varepsilon_{min}}=
\frac{\varepsilon^\prime+\beta
p^\prime}{\varepsilon^\prime-\beta p^\prime}=\frac{1+\beta\beta^\prime_\pi}
{1-\beta\beta^\prime_\pi},$$ где
$$\beta^\prime_\pi=\frac{p^\prime}{\varepsilon^\prime}=\sqrt{1-
\frac{4m^2_\pi}{m^2_K}}$$
есть скорость $\pi$-мезона в системе покоя $K^0$-мезона. Отсюда
скорость $K^0$-мезона
$$\beta=\frac{\alpha-1}{\alpha+1}\,\frac{1}{\beta^\prime_\pi}.$$
Для энергии $E=\frac{mc^2}{\sqrt{1-\beta^2}}$ (восстановили $c$) 
после некоторой алгебры получим
$$E=m_Kc^2\;\frac{\alpha+1}{2}\,\sqrt{\frac{m_K^2-4m_\pi^2}{\alpha
m_K^2-(\alpha+1)^2\,m_\pi^2}}.$$ 
В Л-системе импульс пиона будет
$$p_x=\gamma\left (p^\prime_x+\beta\frac{\varepsilon^\prime}{c}\right )=\gamma
p^\prime\left (cos{\theta^\prime}+\frac{\beta}{\beta^\prime_\pi}\right ).$$ 
Видно, что $p_x\ge 0$,
если $\frac{\beta}{\beta^\prime_\pi}\ge 1$. Т.е. должны иметь
$$\frac{\alpha-1}{\alpha+1}\ge\beta^{\prime 2}_\pi=1-\frac{4m_\pi^2}
{m_K^2}.$$
Отсюда $$\alpha\ge \frac{m_K^2}{2m_\pi^2}-1.$$

\clearpage
\noindent $\bullet$  $\bullet$  $\bullet$
 
\noindent
Consider the decay in the rest system of the $K^0$ meson. In this system,
each pion has an energy $\varepsilon^\prime=\frac{m_K}{2}$ (we have assumed
$c=1$) and momentum
$$p^\prime=\sqrt{\varepsilon^{\prime 2}-m_\pi^2}=\frac{m_K}{2}
\sqrt{1-\frac{4m_\pi^2}{m_K^2}}.$$
\noindent In the L-system, the energy of the $\pi$-meson, which in the $K^0$ 
rest frame was emitted at angle $\theta^\prime$, will be
$\varepsilon=\gamma(\varepsilon^\prime+\beta
p^\prime_x)=\gamma(\varepsilon^\prime+\beta p^\prime \cos{\theta^\prime})$.
Therefore,
$$\alpha=\frac{\varepsilon_{max}}{\varepsilon_{min}}=
\frac{\varepsilon^\prime+\beta
p^\prime}{\varepsilon^\prime-\beta p^\prime}=\frac{1+\beta\beta^\prime_\pi}
{1-\beta\beta^\prime_\pi},$$ where
$$\beta^\prime_\pi=\frac{p^\prime}{\varepsilon^\prime}=\sqrt{1-
\frac{4m^2_\pi}{m^2_K}}$$
is the velocity of the $\pi$-meson in the $K^0$ rest frame. Hence for the 
velocity of the $K^0$-meson we get
$$\beta=\frac{\alpha-1}{\alpha+1}\,\frac{1}{\beta^\prime_\pi}.$$
After some algebra, for the energy $E=\frac{mc^2}{\sqrt{1-\beta^2}}$
(we have restored $c$), we get
$$E=m_Kc^2\;\frac{\alpha+1}{2}\,\sqrt{\frac{m_K^2-4m_\pi^2}{\alpha
m_K^2-(\alpha+1)^2\,m_\pi^2}}.$$
Pion's momentum in the L-frame is
$$p_x=\gamma\left (p^\prime_x+\beta\frac{\varepsilon^\prime}{c}\right )=\gamma
p^\prime\left (cos{\theta^\prime}+\frac{\beta}{\beta^\prime_\pi}\right ).$$ 
It is evident that $p_x\ge 0$, if $\frac{\beta}{\beta^\prime_\pi}\ge 1$.
Therefore we should have
$$\frac{\alpha-1}{\alpha+1}\ge\beta^{\prime 2}_\pi=1-\frac{4m_\pi^2}
{m_K^2}.$$
Hence
$$\alpha\ge \frac{m_K^2}{2m_\pi^2}-1.$$

\subsection{}
Для относительной скорости в задаче 4.3 получили
$$\beta_u^2=\frac{(\vec{\beta}_1-\vec{\beta}_2)^2-(\vec{\beta}_1
\times\vec{\beta}_2)^2}{(1-\vec{\beta}_1\cdot\vec{\beta}_2)^2}.$$
Так как распад симметричный, то $\beta_1=\beta_2$ и будет иметь
$$\beta_u^2=\frac{2\beta^2_1(1-\cos{\theta})-\beta_1^2\sin^2{\theta}}
{(1-\beta^2_1\cos{\theta})^2}.$$
Но
$2(1-cos{\theta})-\sin^2{\theta}=2(1-\cos{\theta})-(1-\cos{\theta})
(1+\cos{\theta})=(1-\cos{\theta})^2$. Следовательно
$$\beta_u^2=\beta^2_1\frac{(1-\cos{\theta})^2}{(1-\beta^2_1\cos
{\theta})^2}.$$
Энергия $\pi$-мезона после распада $E_1=\frac{E}{2}$, импульс $p_1=\sqrt{E_1^2-
m_\pi^2}=\frac{E}{2}\sqrt{1-\frac{4m_\pi^2}{E^2}}$ (положили $c=1$). Поэтому
$\beta_1=\frac{p_1}{E_1}=\sqrt{1-\frac{4m_\pi^2}{E^2}}$.
Но $\cos{\theta}$ однозначно определяется заданием $E$.
Действительно, из законов сохранения энергии
$m_K\gamma=2m_\pi\gamma_1$ и импульса
$m_K\beta\gamma=2m_\pi\beta_1\cos{\theta}$ следует
$$\cos{\theta}=\frac{\beta}{\beta_1}.$$ \noindent Поэтому получаем
$\beta_u^2=\frac{(\beta_1-\beta)^2}{(1-\beta_1\beta)^2}$.
Отсюда $1-\beta_u^2=\frac{(1-\beta_1^2)(1-\beta^2)}{(1-\beta_1\beta)^2}$,
или $\gamma_u=\gamma_1\gamma(1-\beta_1\beta)$. С другой стороны,
$$\gamma_1(1-\beta_1\beta)=\gamma_1(1-\beta_1^2\cos{\theta})=
\frac{1}{m_\pi^2\gamma_1}p_1\cdot
p_2,$$ где $p_1$, $p_2$ -- 4-импульсы пионов. Из закона сохранения
4-импульса $p_K=p_1+p_2$ следует
$m_K^2=(p_1+p_2)^2=2m_\pi^2+2p_1\cdot p_2$ и, следовательно, 
$$p_1\cdot p_2=\frac{m_K^2-2m_\pi^2}{2}.$$ 
\noindent Окончательно,
$$\gamma_u=\frac{1}{m_\pi^2}\,p_1\cdot
p_2=\frac{m_K^2}{2m_\pi^2}-1$$ и
$$\beta_u=\sqrt{1-\frac{1}{\gamma_u^2}}=\frac{\sqrt{1-
\frac{4m_\pi^2}{m_K^2}}}{1-\frac{2m_\pi^2}{m_K^2}}$$
не зависит от $E$.

Более просто этот результат можно получить, если заметить, что
$\gamma_u=\gamma_{\vec{V}_1\ominus\vec{V}_2}=\gamma_1\gamma_2(1-
\vec{\beta}_1\cdot\vec{\beta}_2)$
и $$\frac{m_K^2-2m_\pi^2}{2}=p_1\cdot
p_2=E_1E_2-\vec{p}_1\cdot\vec{p}_2=m_\pi^2\gamma_1\gamma_2
(1-\vec{\beta}_1\cdot\vec{\beta}_2).$$ Следовательно,
$\gamma_u=\frac{m_K^2-2m_\pi^2}{2m_\pi^2}$ даже для
несимметричного распада. 

Еще один простой способ: рассмотрим
распад в системе покоя одного из пионов. В этой системе до распада
$K^0$-мезон движется с некоторой скоростью $V^\prime$, а второй пион
после распада со скоростью $u$. Закон сохранения энергии будет
иметь вид $m_K\gamma^\prime=m_\pi+m_\pi\gamma_u$ (напомним, что $c=1$),
а закон сохранения импульса 
$m_K\beta^\prime\gamma^\prime=m_\pi\beta_u\gamma_u$. Поэтому
$$\gamma^{\prime 2}=\frac{m_\pi^2}{m_K^2}(\gamma_u+1)^2$$ и
$$(\beta^\prime\gamma^\prime)^2=\gamma^{\prime 2}-1=\left(\frac{m_\pi}{m_K}
\right)^2\beta_u^2\gamma_u^2=\frac{m_\pi^2}{m_K^2}(\gamma_u^2-1).$$
Отсюда
$$\frac{m_\pi^2}{m_K^2}(\gamma_u+1)^2-1=\frac{m_\pi^2}{m_K^2}
(\gamma_u^2-1),\;\;
1=\frac{m_\pi^2}{m_K^2}\Biggl[\gamma_u^2+2\gamma_u+1-
\gamma_u^2+1\Biggr]=\frac{2m_\pi^2}{m_K^2}(\gamma_u+1)$$
и, следовательно, $$\gamma_u=\frac{m_K^2}{2m_\pi^2}-1.$$

\vspace*{5mm}
\noindent $\bullet$  $\bullet$  $\bullet$
 
\noindent
For the relative velocity we have got in the problem {\bf 4.3}
$$\beta_u^2=\frac{(\vec{\beta}_1-\vec{\beta}_2)^2-(\vec{\beta}_1
\times\vec{\beta}_2)^2}{(1-\vec{\beta}_1\cdot\vec{\beta}_2)^2}.$$
Since the decay is symmetric, $\beta_1=\beta_2$ and we will have
$$\beta_u^2=\frac{2\beta^2_1(1-\cos{\theta})-\beta_1^2\sin^2{\theta}}
{(1-\beta^2_1\cos{\theta})^2}.$$
But
$2(1-cos{\theta})-\sin^2{\theta}=2(1-\cos{\theta})-(1-\cos{\theta})
(1+\cos{\theta})=(1-\cos{\theta})^2$. Consequently
$$\beta_u^2=\beta^2_1\frac{(1-\cos{\theta})^2}{(1-\beta^2_1\cos
{\theta})^2}.$$
After the decay, the $\pi$-meson has the energy $E_1=\frac{E}{2}$ 
and momentum $p_1=\sqrt{E_1^2-m_\pi^2}=\frac{E}{2}\sqrt{1-\frac{4m_\pi^2}
{E^2}}$ ($c=1$ is assumed). Therefore
$\beta_1=\frac{p_1}{E_1}=\sqrt{1-\frac{4m_\pi^2}{E^2}}$.
But $\cos{\theta}$ is uniquely determined by $E$. Indeed, the laws of 
conservation of energy $m_K\gamma=2m_\pi\gamma_1$ and momentum
$m_K\beta\gamma=2m_\pi\beta_1\cos{\theta}$ imply
$$\cos{\theta}=\frac{\beta}{\beta_1}.$$ \noindent Hence we get
$\beta_u^2=\frac{(\beta_1-\beta)^2}{(1-\beta_1\beta)^2}$.
Consequently $1-\beta_u^2=\frac{(1-\beta_1^2)(1-\beta^2)}{(1-\beta_1\beta)^2}$,
или $\gamma_u=\gamma_1\gamma(1-\beta_1\beta)$. On the other hand,
$$\gamma_1(1-\beta_1\beta)=\gamma_1(1-\beta_1^2\cos{\theta})=
\frac{1}{m_\pi^2\gamma_1}p_1\cdot
p_2,$$ where $p_1$, $p_2$ are 4-momenta of the pions. From the conservation 
of 4-momentum $p_K=p_1+p_2$ we get
$m_K^2=(p_1+p_2)^2=2m_\pi^2+2p_1\cdot p_2$ and, therefore, 
$$p_1\cdot p_2=\frac{m_K^2-2m_\pi^2}{2}.$$ 
\noindent Finally,
$$\gamma_u=\frac{1}{m_\pi^2}\,p_1\cdot
p_2=\frac{m_K^2}{2m_\pi^2}-1$$ and
$$\beta_u=\sqrt{1-\frac{1}{\gamma_u^2}}=\frac{\sqrt{1-
\frac{4m_\pi^2}{m_K^2}}}{1-\frac{2m_\pi^2}{m_K^2}}$$
does not depend on $E$.

More simply, this result can be obtained by noting that
$\gamma_u=\gamma_{\vec{V}_1\ominus\vec{V}_2}=\gamma_1\gamma_2(1-
\vec{\beta}_1\cdot\vec{\beta}_2)$
and $$\frac{m_K^2-2m_\pi^2}{2}=p_1\cdot
p_2=E_1E_2-\vec{p}_1\cdot\vec{p}_2=m_\pi^2\gamma_1\gamma_2
(1-\vec{\beta}_1\cdot\vec{\beta}_2).$$ Consequently,
$\gamma_u=\frac{m_K^2-2m_\pi^2}{2m_\pi^2}$ even for
asymmetric decay. 

Another simple solution: consider the decay in the rest frame of one of the 
pions. In this system, before the decay, $K^0$ meson is moving with some 
velocity $V^\prime$, and the second pion after the decay has a velocity 
$u$. The conservation of energy requires $m_K\gamma^\prime=m_\pi+m_\pi
\gamma_u$ (recall that $c=1$) and conservation of momentum
$m_K\beta^\prime\gamma^\prime=m_\pi\beta_u\gamma_u$. Consequently,
$$\gamma^{\prime 2}=\frac{m_\pi^2}{m_K^2}(\gamma_u+1)^2$$ and
$$(\beta^\prime\gamma^\prime)^2=\gamma^{\prime 2}-1=\left(\frac{m_\pi}{m_K}
\right)^2\beta_u^2\gamma_u^2=\frac{m_\pi^2}{m_K^2}(\gamma_u^2-1).$$
Hence
$$\frac{m_\pi^2}{m_K^2}(\gamma_u+1)^2-1=\frac{m_\pi^2}{m_K^2}
(\gamma_u^2-1),\;\;
1=\frac{m_\pi^2}{m_K^2}\Biggl[\gamma_u^2+2\gamma_u+1-
\gamma_u^2+1\Biggr]=\frac{2m_\pi^2}{m_K^2}(\gamma_u+1)$$
and, consequently, $$\gamma_u=\frac{m_K^2}{2m_\pi^2}-1.$$

\subsection{}
Дифференцируя $u\cdot u=c^2$, сразу получаем $a\cdot u=0$. Найдем
компоненты $a_\mu$. $$a_0=\frac{du_0}{d\tau}=c\gamma\frac{d\gamma}{dt}.$$ 
Но
$$\frac{d\gamma}{dt}=\beta\dot{\beta}\gamma^3,$$ т.к.
$\gamma=(1-\beta^2)^{-1/2}$. Здесь
$\dot{\beta}=\frac{d\beta}{dt}$ - известная функция
по условию. Следовательно, $a_0=c\gamma^4\beta\dot{\beta}$.
$$\vec{a}=\frac{d\vec{u}}{d\tau}=\gamma\frac{d(\vec{V}\gamma)}
{dt}=\gamma
c\,\frac{d(\vec{\beta}\gamma)}{dt}=\gamma
c(\dot{\vec{\beta}}\gamma+\vec{\beta}\beta\dot{\beta}\gamma^3).$$
Т.к. движение прямолинейное, $\vec{\beta}=\beta\vec{n}$ и
$\dot{\vec{\beta}}=\dot{\beta}\vec{n}$, где $\vec{n}$ - единичный
вектор. Поэтому $$\vec{a}=\gamma^2
c\dot{\beta}\vec{n}(1+\beta^2\gamma^2)=c\gamma^4\dot{\beta}\vec{n}.$$
Следовательно,
$a_\mu=(c\gamma^4\beta\dot{\beta},\,c\gamma^4\dot{\beta}\vec{n})$
и $$a\cdot
a=(c\gamma^4\dot{\beta})^2(\beta^2-1)=-c^2\gamma^6\dot{\beta}^2.$$

\vspace*{5mm}
\noindent $\bullet$  $\bullet$  $\bullet$
 
\noindent
Differentiating $u\cdot u=c^2$ , we immediately get $a\cdot u=0$. Let us find
the components of $a_\mu$. 
$$a_0=\frac{du_0}{d\tau}=c\gamma\frac{d\gamma}{dt}.$$ 
But
$$\frac{d\gamma}{dt}=\beta\dot{\beta}\gamma^3,$$ because
$\gamma=(1-\beta^2)^{-1/2}$. Here $\dot{\beta}=\frac{d\beta}{dt}$ is a known 
function by the condition. Consequently, $a_0=c\gamma^4\beta\dot{\beta}$.
$$\vec{a}=\frac{d\vec{u}}{d\tau}=\gamma\frac{d(\vec{V}\gamma)}
{dt}=\gamma
c\,\frac{d(\vec{\beta}\gamma)}{dt}=\gamma
c(\dot{\vec{\beta}}\gamma+\vec{\beta}\beta\dot{\beta}\gamma^3).$$
Because the motion is rectilinear, $\vec{\beta}=\beta\vec{n}$ and
$\dot{\vec{\beta}}=\dot{\beta}\vec{n}$, where $\vec{n}$ is the unit vector 
in the direction of motion. Therefore $$\vec{a}=\gamma^2
c\dot{\beta}\vec{n}(1+\beta^2\gamma^2)=c\gamma^4\dot{\beta}\vec{n}.$$
Consequently,
$a_\mu=(c\gamma^4\beta\dot{\beta},\,c\gamma^4\dot{\beta}\vec{n})$
and $$a\cdot
a=(c\gamma^4\dot{\beta})^2(\beta^2-1)=-c^2\gamma^6\dot{\beta}^2.$$

\section*{Контрольная работа 6}
\setcounter{section}{6}
\setcounter{subsection}{0}
\subsection{}
4-импульсы фотонов 
$$p_1=(\frac{h\nu_1}{c},\,\frac{h\nu_1}{c}\;\vec{n})\;\; \mbox{и} \;\;
p_2=(\frac{h\nu_2}{c},\,-\frac{h\nu_2}{c}\;\vec{n}),$$ 
где $\vec{n}$ -- единичный 
вектор. $$M^2c^2=(p_1+p_2)^2=\frac{h^2}{c^2}[(\nu_1+\nu_2)^2-
(\nu_1-\nu_2)^2]=4\,\frac{h^2}{c^2}\,\nu_1\nu_2.$$ Поэтому масса системы 
$$M=\frac{2h}{c^2}\sqrt{\nu_1\nu_2},$$ а скорость 
$$\vec{V}=\frac{\vec{p}c^2}{E}=c\,\frac{\nu_1-\nu_2}{\nu_1+\nu_2}
\,\vec{n}.$$

\vspace*{5mm}
\noindent $\bullet$  $\bullet$  $\bullet$
 
\noindent
4-momenta of photons are 
$$p_1=(\frac{h\nu_1}{c},\;\frac{h\nu_1}{c}\,\vec{n})\;\;\; \mbox{and} 
\;\;\; p_2=(\frac{h\nu_2}{c},\,-\frac{h\nu_2}{c}\;\vec{n}),$$ 
where $\vec{n}$ is the unit vector. 
$$M^2c^2=(p_1+p_2)^2=\frac{h^2}{c^2}[(\nu_1+\nu_2)^2-
(\nu_1-\nu_2)^2]=4\,\frac{h^2}{c^2}\,\nu_1\nu_2.$$ Therefore, the mass of
the system is 
$$M=\frac{2h}{c^2}\sqrt{\nu_1\nu_2},$$ and its velocity equals to 
$$\vec{V}=\frac{\vec{p}c^2}{E}=c\,\frac{\nu_1-\nu_2}{\nu_1+\nu_2}
\,\vec{n}.$$

\subsection{}
Рассмотрим распад $\pi^0\to 2\gamma$. Из сохранения 4-импульса $p=k_1+k_2$ 
получаем (положили $c=1$) $m_{\pi^0}^2=2k_1\cdot k_2=2E_1E_2 
(1-\cos{\theta})=4E_1E_2\sin^2{\frac{\theta}{2}}$ и 
$$\sin^2{\frac{\theta}{2}}=
\frac{m_{\pi^0}^2}{4E_1E_2}=\frac{m_{\pi^0}^2}{4E_1(E-E_1)}.$$ Когда 
$E_1=\frac{E}{2}=E_2$, 
$E_1(E-E_1)$ достигает максимума и, следовательно, минимальный угол разлета
определяется из $$\sin^2{\frac{\theta_{min}}{2}}=\frac{m_{\pi^0}^2}{E^2}=
\frac{1}{\gamma_{\pi^0}^2},$$ или $\theta_{min}=2\arcsin{\frac{1}{\gamma_
{\pi^0}}}$.
Этот угол тем меньше, чем больше $\gamma_{\pi^0}$. Т.е. в распаде $\phi\to
\pi^+\pi^-\pi^0$ надо обеспечить максимальную энергию $\pi^0$-мезона. 
Рассмотрим $\pi^+\pi^-$ систему как одну частицу 4-импульсом $p^*$. Тогда 
$p_\phi-p_{\pi^0}=p^*$ и $(p_\phi-p_{\pi^0})^2=m_\phi^2-2m_\phi 
E+m_{\pi^0}^2=
m^{*2}$. Отсюда $$E=\frac{m_\phi^2+m_{\pi^0}^2-m^{*2}}{2m_\phi}$$ и она 
максимальна, когда $m^*$ минимальна. Но в системе центра масс $\pi^+\pi^-$ 
составной частицы $m^{*2}=(E_+^\prime+E_-^\prime,\vec{0})^2=4(m_{\pi^\pm}^2+
p^{\prime 2})$. Отсюда видно, что минимум достигается при $p^\prime=0$ и 
$m_{min}^{*2}=4m_{\pi^\pm}^2$. Поэтому
$$E_{max}=\frac{m_\phi^2+m_{\pi^0}^2-4m_{\pi^\pm}^2}{2m_\phi}$$ и самый 
минимальный угол разлета $\gamma$-квантов будет $$\theta_{min}=2\arcsin{
\frac{2m_{\pi^0}m_\phi}{m_\phi^2+m_{\pi^0}^2-4m_{\pi^\pm}^2}}.$$

\vspace*{5mm}
\noindent $\bullet$  $\bullet$  $\bullet$
 
\noindent
Consider the decay $\pi^0\to 2\gamma$.  From the 4-momentum conservation,
$p=k_1+k_2$, we get ($c=1$ is assumed) $m_{\pi^0}^2=2k_1\cdot k_2=2E_1E_2 
(1-\cos{\theta})=4E_1E_2\sin^2{\frac{\theta}{2}}$ and 
$$\sin^2{\frac{\theta}{2}}=
\frac{m_{\pi^0}^2}{4E_1E_2}=\frac{m_{\pi^0}^2}{4E_1(E-E_1)}.$$ When
$E_1=\frac{E}{2}=E_2$, the function $E_1(E-E_1)$ reaches its maximum and, 
consequently, the minimum opening angle of the two gammas is determined by
$$\sin^2{\frac{\theta_{min}}{2}}=\frac{m_{\pi^0}^2}{E^2}=
\frac{1}{\gamma_{\pi^0}^2},$$ or $\theta_{min}=2\arcsin{\frac{1}{\gamma_
{\pi^0}}}$. The larger is $\gamma_{\pi^0}$, the smaller is this angle.
Consequently, in the decay $\phi\to\pi^+\pi^-\pi^0$ we should maximize the 
energy of the  $\pi^0$-meson. Consider the $\pi^+\pi^-$ subsystem as a single 
particle with 4-momentum $p^*$. Then $p_\phi-p_{\pi^0}=p^*$ and 
$(p_\phi-p_{\pi^0})^2=m_\phi^2-2m_\phi E+m_{\pi^0}^2=m^{*2}$. Hence 
$$E=\frac{m_\phi^2+m_{\pi^0}^2-m^{*2}}{2m_\phi}$$ and it is maximal when $m^*$
is minimal. But in the center of mass system of the $\pi^+\pi^-$ composite 
particle, $m^{*2}=(E_+^\prime+E_-^\prime,\vec{0})^2=4(m_{\pi^\pm}^2+
p^{\prime 2})$. This shows that the minimum is achieved when $p^\prime=0$
and $m_{min}^{*2}=4m_{\pi^\pm}^2$. Therefore,
$$E_{max}=\frac{m_\phi^2+m_{\pi^0}^2-4m_{\pi^\pm}^2}{2m_\phi}$$
and the smallest opening angle of the two gammas is
$$\theta_{min}=2\arcsin{
\frac{2m_{\pi^0}m_\phi}{m_\phi^2+m_{\pi^0}^2-4m_{\pi^\pm}^2}}.$$

\subsection{}
Рассмотрим электрон и ядро до столкновения как одну частицу с массой $M_1$, а 
после столкновения -- как одну частицу с массой $M_2$. Причем $M_1^2=(p_e+
p_A)^2=m^2+M^2+2ME$, где $p_A$ -- 4-импульс неподвижного ядра. Из $p_1-k=p_2$,
где $k$ 4-импульс фотона, а $p_{1,2}$ 4-импульсы системы электрон-ядро до и 
после столкновения, получим в системе центра масс $M_1^2-2M_1E_\gamma^\prime=
M_2^2$. Т.е. в этой системе энергия тормозного фотона $E_\gamma^\prime=\frac
{M_1^2-M_2^2}{2M_1}$. В Л-системе будем иметь $E_\gamma=\gamma_1E_\gamma^
\prime(1+\beta_1\cos{\theta^\prime})$, где $\beta_1,\gamma_1$ относятся к 
системе электрон-ядро до столкновения. Видно, что $E_\gamma$ максимальна когда
$\theta^\prime=0$ и $M_2$
минимальна. Но $M_{2,min}=m+M$, т.к. в система покоя электрон-ядро (после
столкновения) $M_2=\sqrt{m^2+p^{\prime 2}}+\sqrt{M^2+p^{\prime 2}}$, где 
$p^\prime$ -- величина импульса частиц в этой системе. Поэтому
$$E_{\gamma,max}=\gamma_1\frac{M_1^2-(m+M)^2}{2M_1}(1+\beta_1).$$
Но $M_1^2-(m+M)^2=m^2+M^2+2ME-m^2-M^2-2mM=2M(E-m)$ и
$$\frac{\gamma_1(1+\beta_1)}{M_1}=\frac{\gamma_1^2(1+\beta_1)}{M_1\gamma_1}=
\frac{1}{E+M}\,\frac{1}{1-\beta_1},$$
где $M_1\gamma_1=E+M$ есть энергия системы электрон-ядро до столкновения, а
$\beta_1=\frac{p}{E+M}$ -- скорость центра масс этой системы. Следовательно,
$E_{\gamma,max}=\frac{M(E-m)}{E+M-p}$. Если восстановить $c$ (выше мы 
предполагали $c=1$):
$$E_{\gamma,max}=\frac{M(E-mc^2)}{\frac{E}{c^2}+M-\frac{p}{c}}=
\frac{mMc^2(\gamma-1)}{m\gamma+M-m\beta\gamma}.$$
Окончательно $$E_{\gamma,max}=mc^2(\gamma-1)\left [1+\frac{m}{M}\gamma(1-\beta)
\right ]^{-1},\;
{\mbox{где}} \; \gamma=\left [1-\frac{V^2}{c^2}\right ]^{-1/2}.$$
Т.к. $\frac{m}{M}\ll 1$, то $E_{\gamma,max}$ практически равна кинетической 
энергии электрона.

Ответ можно записать еще в таком виде
$$E_{\gamma,max}=\frac{Mc^2(E-mc^2)}{E-pc+Mc^2}=\frac{mc^2(\gamma-1)}
{1+\frac{m}{M}\sqrt{\frac{1-\beta}{1+\beta}}}.$$

\vspace*{5mm}
\noindent $\bullet$  $\bullet$  $\bullet$
 
\noindent
Consider the electron and the nucleus before the collision as a single 
composite particle with mass $M_1$, and after the collision - as a single 
particle with mass $M_2$. At that $M_1^2=(p_e+p_A)^2=m^2+M^2+2ME$,
where $p_A$ is 4-momentum of the stationary nucleus. From $p_1-k=p_2$,
where $k$ is 4-momentum of the photon, and $p_{1,2}$ are 4-momenta of the 
electron-nucleus system before and after the collision, we obtain in the 
center of mass frame:  $M_1^2-2M_1E_\gamma^\prime=M_2^2$. Hence, in this 
system, the energy of the bremsstrahlung photon is $E_\gamma^\prime=\frac
{M_1^2-M_2^2}{2M_1}$. In the L-system, we have $E_\gamma=\gamma_1E_\gamma^
\prime(1+\beta_1\cos{\theta^\prime})$, where $\beta_1,\gamma_1$ refer to the
system of electron-nucleus before the collision. It is clear that 
$E_\gamma$ is maximal when $\theta^\prime=0$ and $M_2$ is minimal.
But $M_{2,min}=m+M$, because In the rest frame of the electron-nucleus system
(after the collision) $M_2=\sqrt{m^2+p^{\prime 2}}+\sqrt{M^2+p^{\prime 2}}$,
where $p^\prime$ is the magnitude of momentum of the particles in this system. 
Therefore, $$E_{\gamma,max}=\gamma_1\frac{M_1^2-(m+M)^2}{2M_1}(1+\beta_1).$$
But $M_1^2-(m+M)^2=m^2+M^2+2ME-m^2-M^2-2mM=2M(E-m)$ and
$$\frac{\gamma_1(1+\beta_1)}{M_1}=\frac{\gamma_1^2(1+\beta_1)}{M_1\gamma_1}=
\frac{1}{E+M}\,\frac{1}{1-\beta_1},$$
where $M_1\gamma_1=E+M$ is the energy of the electron-nucleus system before 
the collision, and $\beta_1=\frac{p}{E+M}$ is the velocity of the center of 
mass of this system. Consequently, $E_{\gamma,max}=\frac{M(E-m)}{E+M-p}$.
If we restore the light velocity $c$ (above we have assumed $c=1$), we get
$$E_{\gamma,max}=\frac{M(E-mc^2)}{\frac{E}{c^2}+M-\frac{p}{c}}=
\frac{mMc^2(\gamma-1)}{m\gamma+M-m\beta\gamma}.$$
Finally, $$E_{\gamma,max}=mc^2(\gamma-1)\left [1+\frac{m}{M}\gamma(1-\beta)
\right ]^{-1},\;
{\mbox{where}} \; \gamma=\left [1-\frac{V^2}{c^2}\right ]^{-1/2}.$$
Because $\frac{m}{M}\ll 1$, $E_{\gamma,max}$ is almost equal to the 
kinetic energy of the electron.

The answer can also be rewritten in the following form
$$E_{\gamma,max}=\frac{Mc^2(E-mc^2)}{E-pc+Mc^2}=\frac{mc^2(\gamma-1)}
{1+\frac{m}{M}\sqrt{\frac{1-\beta}{1+\beta}}}.$$

\subsection{}
Рассмотрим распад $A\to B^+B^-$ в системе покоя частицы $A$. Пусть в этой 
системе $B^+$ имеет скорость $\beta_B^\prime$ и вылетает под углом $\theta^
\prime$ относительно скорости $A$ в лабораторной системе. Тогда для скорости 
$B^+$ в Л-системе будем иметь 
$$\beta_{1,x}=\frac{\beta_B^\prime\cos{\theta^\prime}+\beta_A}{1+\beta_B
^\prime\beta_A\cos{\theta^\prime}} \;\;\; \mbox{и} \;\;\; 
\beta_{1,y}=\frac{\beta_B^\prime\sin{\theta^\prime}}{\gamma_A(1+\beta_B
^\prime\beta_A\cos{\theta^\prime})}. $$
А для скорости $B^-$:
$$\beta_{2,x}=\frac{-\beta_B^\prime\cos{\theta^\prime}+\beta_A}{1-\beta_B
^\prime\beta_A\cos{\theta^\prime}} \;\;\; \mbox{и} \;\;\; 
\beta_{1,y}=\frac{-\beta_B^\prime\sin{\theta^\prime}}{\gamma_A(1-\beta_B
^\prime\beta_A\cos{\theta^\prime})}. $$
\begin{figure}[htb]
\centerline{\epsfig{figure=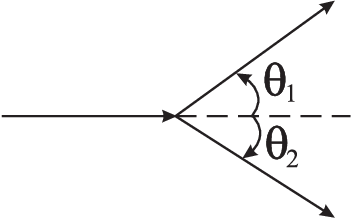}}
\end{figure}

\noindent Поэтому $B^+$ и $B^-$ в Л-системе вылетают под углами $\theta_1$ и
$\theta_2$ ($\theta_1$ отсчитывается против часовой стрелки, а $\theta_2$ -- 
по часовой), удовлетворяющими
$$\tg{\theta_1}=\frac{\beta_B^\prime\sin{\theta^\prime}}{\gamma_A (\beta_A+
\beta_B^\prime\cos{\theta^\prime})} \;\; \mbox{и} \;
\tg{\theta_2}=\frac{\beta_B^\prime\sin{\theta^\prime}}{\gamma_A (\beta_A-
\beta_B^\prime\cos{\theta^\prime})}.$$
Угол разлета $\theta=\theta_1+\theta_2$ и
$$\tg{\theta}=\frac{\tg{\theta_1}+\tg{\theta_2}}{1-\tg{\theta_1}
\tg{\theta_2}}=\frac{2\beta_A\beta^\prime_B\gamma_A\sin{\theta^\prime}}
{\gamma_A^2(\beta_A^2-\beta^{\prime 2}_B\cos^2{\theta^\prime})-
\beta^{\prime 2}_B\sin^2{\theta^\prime}}.$$
Но $\gamma_A^2(\beta_A^2-\beta^{\prime 2}_B\cos^2{\theta^\prime})-
\beta^{\prime 2}_B\sin^2{\theta^\prime}=\gamma_A^2(\beta_A^2-\beta^
{\prime 2}_B)+\beta^{\prime 2}_B(\gamma_A^2-1)\sin^2{\theta^\prime}=
\gamma_A^2 [\beta_A^2-\beta^{\prime 2}_B+\beta_A^2\beta^{\prime 2}_B
\sin^2{\theta^\prime}]$, т.к. $\gamma_A^2-1=\gamma_A^2\beta_A^2$. Поэтому
$$\tg{\theta}=\frac{2\beta_A\beta^\prime_B}{\gamma_A}\frac{\sin{\theta
^\prime}}{\beta_A^2-\beta^{\prime 2}_B+\beta_A^2\beta^{\prime 2}_B
\sin^2{\theta^\prime}}.$$
Функция
$$f(x)=\frac{x}{\beta_A^2-\beta^{\prime 2}_B+\beta_A^2\beta^{\prime 2}_B
x^2}$$
показана на рисунке. 
\begin{figure}[htb]
\centerline{\epsfig{figure=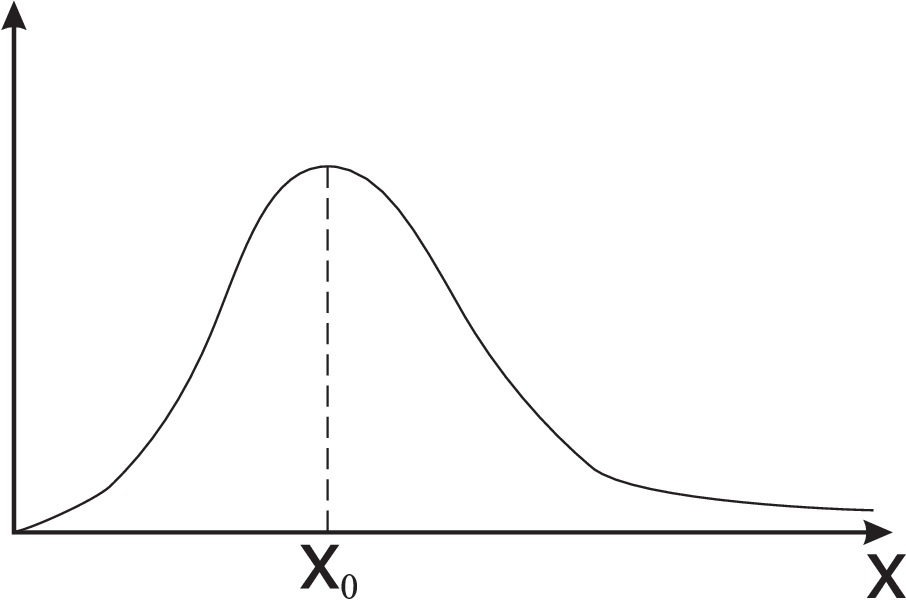,height=3cm}}
\end{figure}

\noindent Она имеет максимум в точке 
$$x_0=\sqrt{\frac{1}{\beta^{\prime 2}_B}-\frac{1}{\beta_A^2}}.$$
Заметим, что в нашем случае $\beta_A>\beta^{\prime}_B$, т.к. энергия $\phi$-мезона
очень большая. Кроме того $x=\sin{\theta^\prime}$ и, следовательно, должны 
иметь $x_0^2\le 1$, или $$\beta_A^2\le\frac{\beta^{\prime 2}_B}{1- \beta^{
\prime 2}_B},$$ что дает $\beta_A\le \beta^{\prime}_B\gamma^\prime_B=\frac{1}
{m_B}p^\prime$, где $p^\prime$ есть импульс $B^\pm$ в системе покоя частицы 
$A$. Но в этой системе закон сохранения энергии дает $m_A=2\sqrt{p^{\prime 2}+
m_B^2}$. Отсюда $p^\prime=m_B\sqrt{\frac{m_A^2}{4m_B^2}-1}$ 
и получаем условие
$$\beta_A\le\sqrt{\frac{m_A^2}{4m_B^2}-1}.$$ В случае распада $\phi\to \pi^+
\pi^-$ условие $x_0^2\le 1$ не ограничивает $\beta_\phi$, так как $\sqrt{\frac{
m_\phi^2}{4m_\pi^2}-1}>1$. Когда $\theta^\prime$ меняется от $0$ до $\pi$,
$\tg{\theta}$ меняется так
\begin{figure}[htb]
\centerline{\epsfig{figure=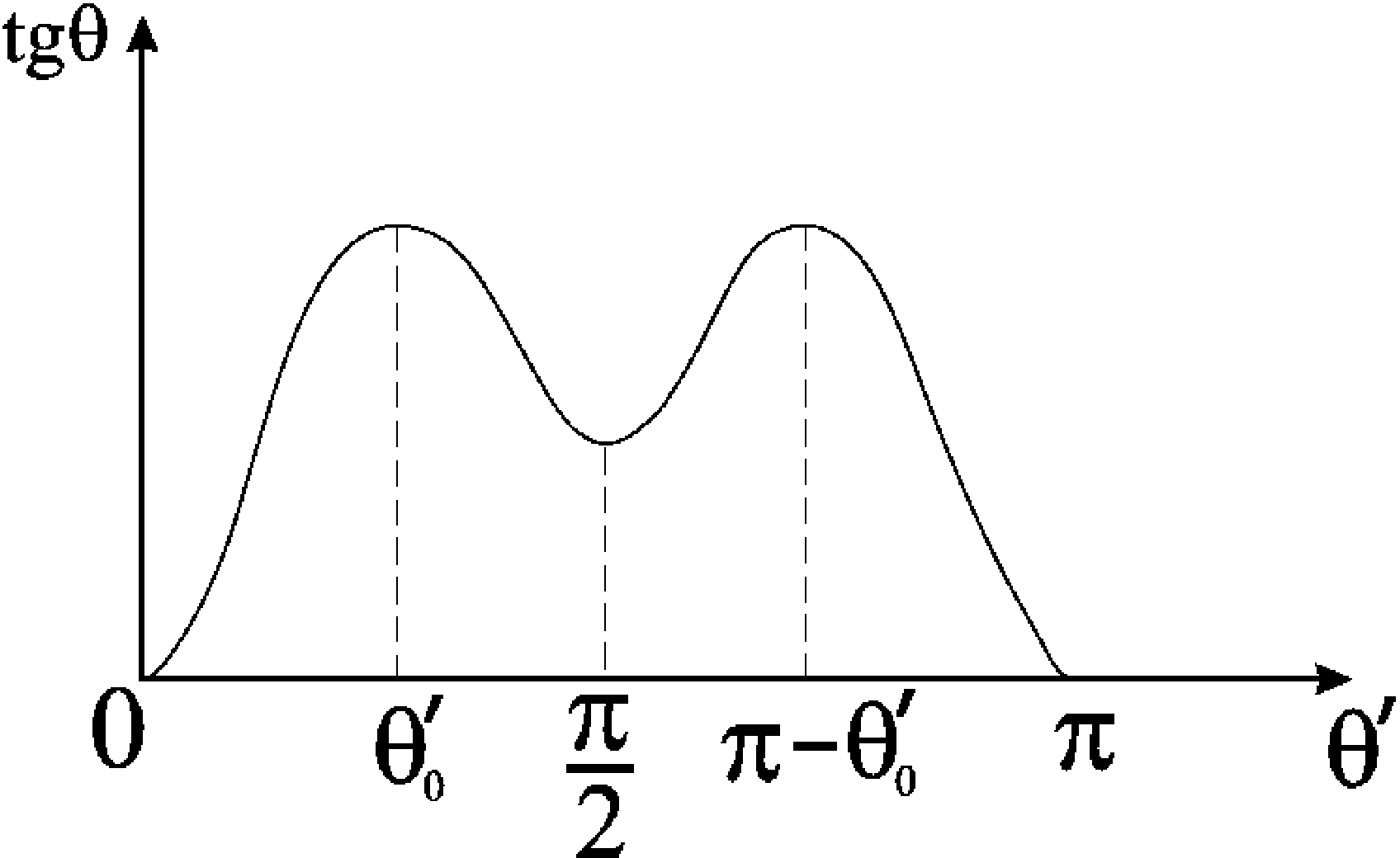,height=4cm}}
\end{figure}

\noindent
Максимальному углу разлета соответствует $\theta_0^\prime$ и $$\sin{\theta_0
^\prime}=\sqrt{\frac{1}{\beta_\pi^{\prime 2}}-\frac{1}{\beta_\phi^2}}.$$
Следовательно,
$$\tg{\theta_{max}}=\frac{2\beta_\phi\beta_\pi^\prime}{\gamma_\phi}
\frac{\sqrt{\beta_\phi^2-\beta_\pi^{\prime 2}}/\beta_\pi^\prime\beta_\phi}
{2(\beta_\phi^2-\beta_\pi^{\prime 2})}=\frac{1}{\gamma_\phi\sqrt{\beta_\phi^2
-\beta_\pi^{\prime 2}}}.$$
Так как $\gamma_\phi\gg 1$, то
$$\theta^{(\pi)}_{max}\approx\frac{1}{\gamma_\phi\sqrt{1-\beta_\pi^{\prime 2}
}}=\frac{\gamma_\pi^\prime}{\gamma_\phi}.$$
Но для распада $\phi\to K^+ K^-$,  $\sqrt{\frac{m_\phi^2}{4m_K^2}-1}<1$ и для
больших энергии $\phi$-мезона, когда $\beta_\phi>\sqrt{\frac{m_\phi^2}{4m_K^2}
-1}$, максимум $x_0$ лежит вне физической области (т.е. не достигается ни при 
каком $\theta^\prime$). В этом случае $\tg{\theta}$ меняется так
\begin{figure}[htb]
\centerline{\epsfig{figure=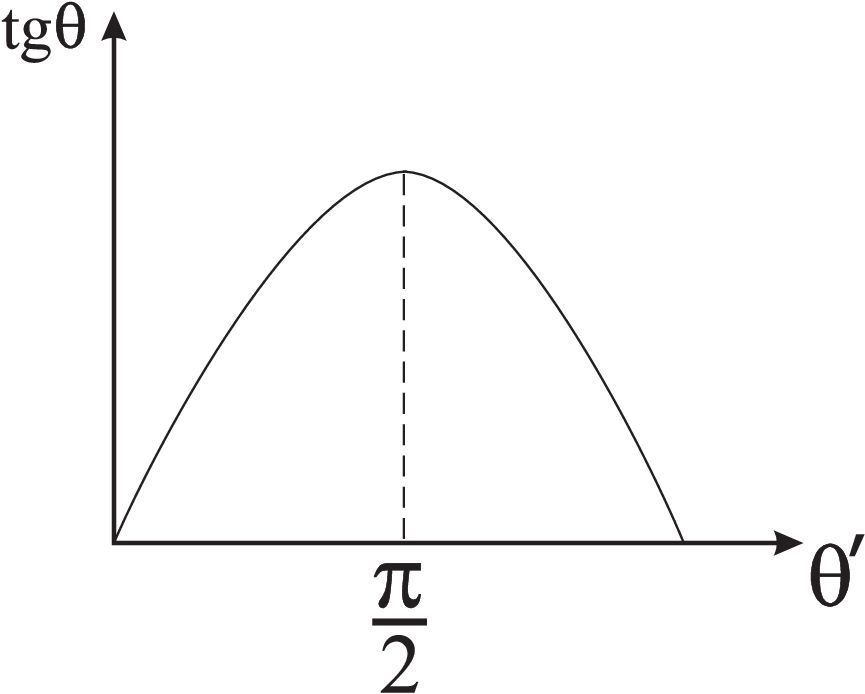,height=4cm}}
\end{figure}

\noindent
и максимальный угол разлета соответствует $\theta^\prime=\frac{\pi}{2}$, т.е.
симметричному разлету. Тогда
$$\tg{\theta_{max}}=\frac{2\beta_\phi\beta_K^\prime}{\gamma_\phi}
\frac{1}{\beta_\phi^2-\beta_K^{\prime 2}+\beta_\phi^2\beta_K^{\prime 2}}
\approx \frac{2\beta_K^\prime}{\gamma_\phi}.$$
Следовательно, $\theta^{(K)}_{max}\approx \frac{2\beta_K^\prime}{\gamma_\phi}$.
Поэтому в пределе очень больших энергии $\phi$-мезона
$$\frac{\theta_{max}(\phi\to K^+ K^-)}{\theta_{max}(\phi\to \pi^+ \pi^-)}
\approx \frac{2\beta_K^\prime}{\gamma_\pi^\prime}.$$
Как уже нашли, $\gamma_B^{\prime 2}\beta_B^{\prime 2}=\gamma_B^{\prime 2}-1=
\frac{m_A^2}{4m_B^2}-1$. Т.е. $\gamma_B^\prime=\frac{m_A}{2m_B}$ и
$\beta_B^\prime=\sqrt{1-\frac{1}{\gamma_B^{\prime 2}}}=\sqrt{1-\frac{4m_B^2}
{m_A^2}}$. Поэтому, окончательно,
$$\frac{\theta_{max}(\phi\to K^+ K^-)}{\theta_{max}(\phi\to \pi^+ \pi^-)}
\approx \frac{4m_\pi}{m_\phi}\sqrt{1-\frac{4m_K^2}{m_\phi^2}}\approx 0.14.$$

\vspace*{5mm}
\noindent $\bullet$  $\bullet$  $\bullet$
 
\noindent
Consider the decay  $A\to B^+B^-$ in the rest frame of the particle $A$. 
Suppose in this system the particle $B^+$ has a velocity $\beta_B^\prime$
and flies at an angle $\theta^\prime$ relative to the direction of motion 
of $A$ in the laboratory frame. Then the velocity of $B^+$ in the L-frame 
will be
$$\beta_{1,x}=\frac{\beta_B^\prime\cos{\theta^\prime}+\beta_A}{1+\beta_B
^\prime\beta_A\cos{\theta^\prime}} \;\;\; \mbox{and} \;\;\; 
\beta_{1,y}=\frac{\beta_B^\prime\sin{\theta^\prime}}{\gamma_A(1+\beta_B
^\prime\beta_A\cos{\theta^\prime})}. $$
While the velocity of $B^-$:
$$\beta_{2,x}=\frac{-\beta_B^\prime\cos{\theta^\prime}+\beta_A}{1-\beta_B
^\prime\beta_A\cos{\theta^\prime}} \;\;\; \mbox{and} \;\;\; 
\beta_{1,y}=\frac{-\beta_B^\prime\sin{\theta^\prime}}{\gamma_A(1-\beta_B
^\prime\beta_A\cos{\theta^\prime})}. $$
\begin{figure}[htb]
\centerline{\epsfig{figure=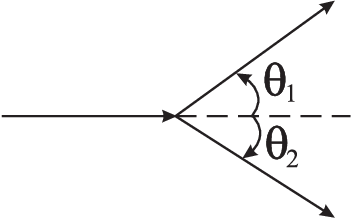}}
\end{figure}

\noindent Hence $B^+$ and $B^-$ are emitted in the L-system at angles 
$\theta_1$ and $\theta_2$  ($\theta_1$ is reckoned counterclockwise, and 
$\theta_2$ is counted  clockwise) satisfying 
$$\tg{\theta_1}=\frac{\beta_B^\prime\sin{\theta^\prime}}{\gamma_A (\beta_A+
\beta_B^\prime\cos{\theta^\prime})} \;\; \mbox{and} \;
\tg{\theta_2}=\frac{\beta_B^\prime\sin{\theta^\prime}}{\gamma_A (\beta_A-
\beta_B^\prime\cos{\theta^\prime})}.$$
The opening angle is $\theta=\theta_1+\theta_2$ and
$$\tg{\theta}=\frac{\tg{\theta_1}+\tg{\theta_2}}{1-\tg{\theta_1}
\tg{\theta_2}}=\frac{2\beta_A\beta^\prime_B\gamma_A\sin{\theta^\prime}}
{\gamma_A^2(\beta_A^2-\beta^{\prime 2}_B\cos^2{\theta^\prime})-
\beta^{\prime 2}_B\sin^2{\theta^\prime}}.$$
But $\gamma_A^2(\beta_A^2-\beta^{\prime 2}_B\cos^2{\theta^\prime})-
\beta^{\prime 2}_B\sin^2{\theta^\prime}=\gamma_A^2(\beta_A^2-\beta^
{\prime 2}_B)+\beta^{\prime 2}_B(\gamma_A^2-1)\sin^2{\theta^\prime}=
\gamma_A^2 [\beta_A^2-\beta^{\prime 2}_B+\beta_A^2\beta^{\prime 2}_B
\sin^2{\theta^\prime}]$, because $\gamma_A^2-1=\gamma_A^2\beta_A^2$. 
Therefore,
$$\tg{\theta}=\frac{2\beta_A\beta^\prime_B}{\gamma_A}\frac{\sin{\theta
^\prime}}{\beta_A^2-\beta^{\prime 2}_B+\beta_A^2\beta^{\prime 2}_B
\sin^2{\theta^\prime}}.$$
the function
$$f(x)=\frac{x}{\beta_A^2-\beta^{\prime 2}_B+\beta_A^2\beta^{\prime 2}_B
x^2}$$
is shown in the figure below. 
\begin{figure}[htb]
\centerline{\epsfig{figure=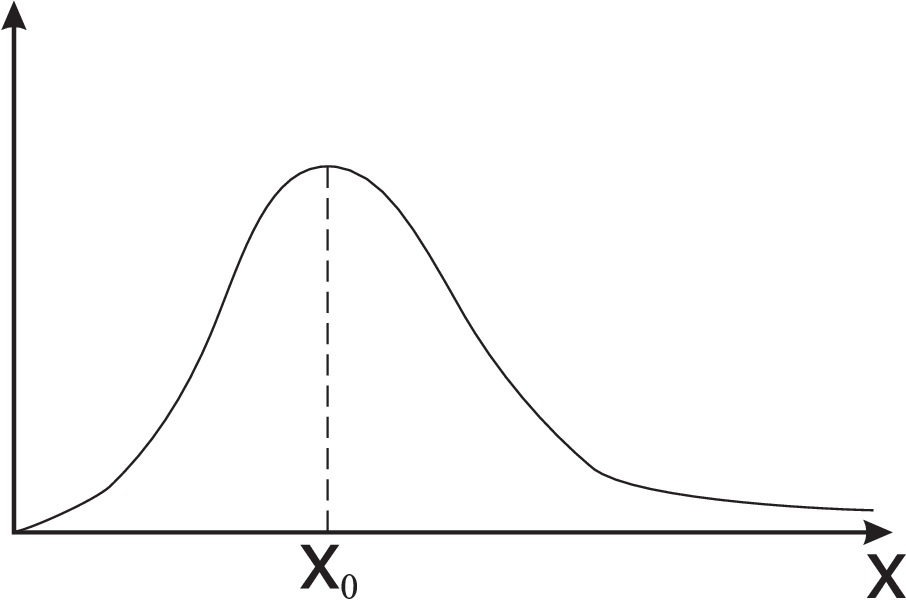,height=3cm}}
\end{figure}

\noindent It has the maximum at the point 
$$x_0=\sqrt{\frac{1}{\beta^{\prime 2}_B}-\frac{1}{\beta_A^2}}.$$
Note that in our case $\beta_A>\beta^{\prime}_B$, since the energy of 
the $\phi$-meson is very large. In addition $x=\sin{\theta^\prime}$ and, 
therefore, we should have $x_0^2\le 1$, or $$\beta_A^2\le
\frac{\beta^{\prime 2}_B}{1- \beta^{\prime 2}_B},$$ 
which gives $\beta_A\le \beta^{\prime}_B\gamma^\prime_B=
\frac{1}{m_B}p^\prime$, where $p^\prime$ is the magnitude of the momentum of 
$B^\pm$ in the rest frame of the particle $A$. But in this system, the energy
conservation law gives $m_A=2\sqrt{p^{\prime 2}+m_B^2}$.
Hence $p^\prime=m_B\sqrt{\frac{m_A^2}{4m_B^2}-1}$ and we obtain the condition
$$\beta_A\le\sqrt{\frac{m_A^2}{4m_B^2}-1}.$$  In the case of the $\phi\to 
\pi^+\pi^-$ decay, the condition $x_0^2\le 1$ does not restrict $\beta_\phi$,
because $\sqrt{\frac{m_\phi^2}{4m_\pi^2}-1}>1$. When $\theta^\prime$ varies 
from $0$ to $\pi$, $\tg{\theta}$ is changing as follows
\begin{figure}[htb]
\centerline{\epsfig{figure=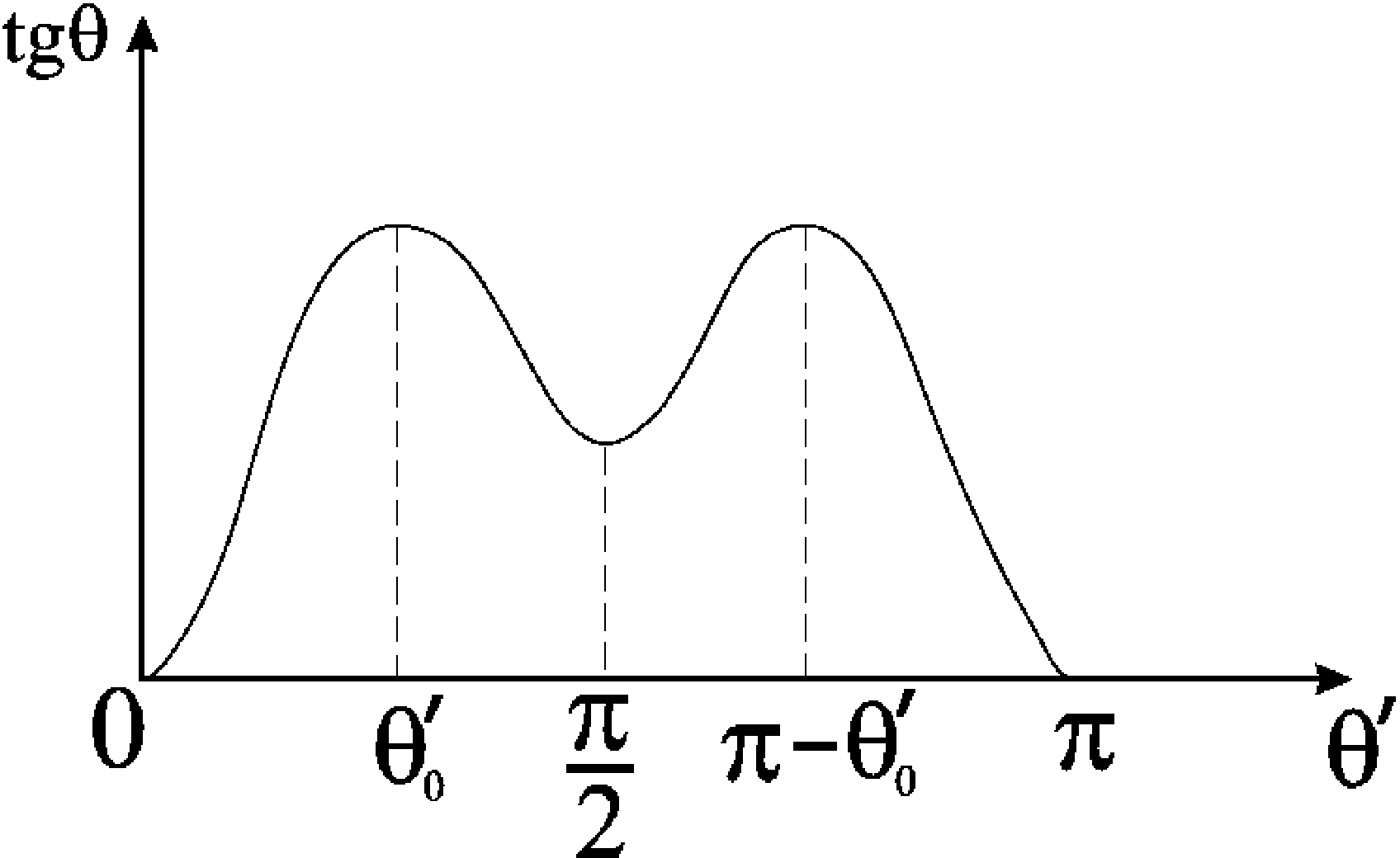,height=4cm}}
\end{figure}

\noindent
The maximum opening angle corresponds to $\theta_0^\prime$ with 
$$\sin{\theta_0^\prime}=\sqrt{\frac{1}{\beta_\pi^{\prime 2}}-
\frac{1}{\beta_\phi^2}}.$$
Hence,
$$\tg{\theta_{max}}=\frac{2\beta_\phi\beta_\pi^\prime}{\gamma_\phi}
\frac{\sqrt{\beta_\phi^2-\beta_\pi^{\prime 2}}/\beta_\pi^\prime\beta_\phi}
{2(\beta_\phi^2-\beta_\pi^{\prime 2})}=\frac{1}{\gamma_\phi\sqrt{\beta_\phi^2
-\beta_\pi^{\prime 2}}}.$$
Since $\gamma_\phi\gg 1$, then
$$\theta^{(\pi)}_{max}\approx\frac{1}{\gamma_\phi\sqrt{1-\beta_\pi^{\prime 2}
}}=\frac{\gamma_\pi^\prime}{\gamma_\phi}.$$
But for the decay $\phi\to K^+ K^-$,  $\sqrt{\frac{m_\phi^2}{4m_K^2}-1}<1$ 
and for very high-energy $\phi$-meson, when $\beta_\phi>\sqrt{\frac{m_\phi^2}
{4m_K^2}-1}$, the maximum of $x_0$ lies outside the physical domain (that is, 
it is not attained at some $\theta^\prime$). In this case, $\tg{\theta}$
is changing as shown in the figure below
\begin{figure}[htb]
\centerline{\epsfig{figure=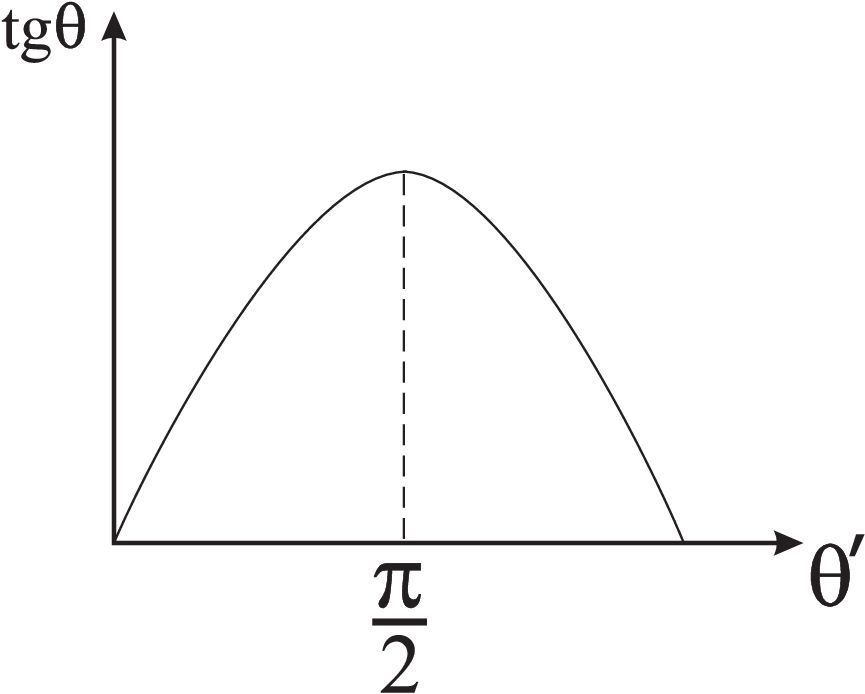,height=4cm}}
\end{figure}
and, hence, the maximum opening angle corresponds to $\theta^\prime=
\frac{\pi}{2}$, that is, to the symmetric decay. Then
$$\tg{\theta_{max}}=\frac{2\beta_\phi\beta_K^\prime}{\gamma_\phi}
\frac{1}{\beta_\phi^2-\beta_K^{\prime 2}+\beta_\phi^2\beta_K^{\prime 2}}
\approx \frac{2\beta_K^\prime}{\gamma_\phi}.$$
Consequently, $\theta^{(K)}_{max}\approx \frac{2\beta_K^\prime}{\gamma_\phi}$.
Therefore, in the limit of very high-energy $\phi $-meson,
$$\frac{\theta_{max}(\phi\to K^+ K^-)}{\theta_{max}(\phi\to \pi^+ \pi^-)}
\approx \frac{2\beta_K^\prime}{\gamma_\pi^\prime}.$$
As was already found, $\gamma_B^{\prime 2}\beta_B^{\prime 2}=
\gamma_B^{\prime 2}-1=\frac{m_A^2}{4m_B^2}-1$. That is, $\gamma_B^\prime=
\frac{m_A}{2m_B}$ and $\beta_B^\prime=\sqrt{1-\frac{1}{\gamma_B^{\prime 2}}}
=\sqrt{1-\frac{4m_B^2} {m_A^2}}$. Therefore, finally,
$$\frac{\theta_{max}(\phi\to K^+ K^-)}{\theta_{max}(\phi\to \pi^+ \pi^-)}
\approx \frac{4m_\pi}{m_\phi}\sqrt{1-\frac{4m_K^2}{m_\phi^2}}\approx 0.14.$$

\subsection{}
Закон сохранения 4-импульса $p=q_0+q_++q_-$ дает (предполагаем $c=1$) $(p-
q_0)^2=m_\omega^2+m_{\pi^0}^2-2m_\omega E_0=(q_++q_-)^2=2m_{\pi^\pm}^2+2(E_+
E_--p_+p_-\cos{\theta})$. Отсюда угол разлета между $\pi^+$ и $\pi^-$ 
$$\cos{\theta}=\frac{2E_1E_--m_\omega^2-m_{\pi^0}^2+2m_{\pi^\pm}^2
+2m_\omega (m_\omega-E_1-E_-)}{2\sqrt{E_1^2-m_{\pi^\pm}^2}\sqrt{E_-^2-
m_{\pi^\pm}^2}},$$
так как $E_+=E_1$ и $E_0=m_\omega-E_+-E_-$. Но $\cos^2{\theta}\le 1$, что дает
условие $4(E_1^2-m_{\pi^\pm}^2)(E_-^2-m_{\pi^\pm}^2)\ge
[2E_1E_--m_\omega^2-m_{\pi^0}^2+2m_{\pi^\pm}^2+2m_\omega 
(m_\omega-E_1-E_-)]^2$. После некоторой алгебры, получим $4E_-^2(m_\omega^2-
2m_\omega E_1+m_{\pi^\pm}^2)+4E_-(E_1-m_\omega)(m_\omega^2-2m_\omega E_1-
m_{\pi^0}^2+2m_{\pi^\pm}^2)+(m_\omega^2-2m_\omega E_1-m_{\pi^0}^2+
2m_{\pi^\pm}^2)^2-4m_{\pi^\pm}^2(E_1^2-m_{\pi^\pm}^2)\le 0$. Удобно перейти
к безразмерным величинам 
$$x_1=\frac{E_1}{m_\omega},\;\; x=\frac{E_-}{m_\omega},\;\;
\mu=\frac{m_{\pi^\pm}}{m_\omega}\;\;\mbox{и} \;\; \mu_0=\frac{m_{\pi^0}}
{m_\omega}.$$
Тогда $4x^2(1-2x_1+\mu^2)-4x(1-x_1)(1-2x_1-\mu_0^2+2\mu^2)+
(1-2x_1-\mu_0^2+2\mu^2)^2+4\mu^2(x_1^2-\mu^2)\le 0$. Но $16(1-x_1)^2
(1-2x_1-\mu_0^2+2\mu^2)^2-16(1-2x_1+\mu^2)(1-2x_1-\mu_0^2+2\mu^2)^2-
64\mu^2(x_1^2-\mu^2)(1-2x_1+\mu^2)=16(x_1^2-\mu^2)[(1-2x_1-\mu_0^2+2\mu^2)^2-
4\mu^2(1-2x_1+\mu^2)]$ и $(1-2x_1-\mu_0^2+2\mu^2)^2-4\mu^2(1-2x_1+\mu^2)=
(1-2x_1-\mu_0^2)^2+4\mu^4+4\mu^2(1-2x_1-\mu_0^2)-4\mu^2(1-2x_1+\mu^2)=
(1-2x_1-\mu_0^2)^2-4\mu^2\mu_0^2=(1-2x_1-\mu_0^2-2\mu\mu_0)
(1-2x_1-\mu_0^2+2\mu\mu_0)$. Поэтому, окончательно, $x_{min}\le x \le x_{max}$,
где
$$x_{min}=\frac{(1-x_1)(1-2x_1-\mu_0^2+2\mu^2)-\sqrt{D}}{2(1-2x_1+\mu^2)},$$
$$x_{max}=\frac{(1-x_1)(1-2x_1-\mu_0^2+2\mu^2)+\sqrt{D}}{2(1-2x_1+\mu^2)},$$
и $$D=(x_1^2-\mu^2)(1-2x_1-\mu_0^2-2\mu\mu_0)(1-2x_1-\mu_0^2+2\mu\mu_0).$$

\vspace*{5mm}
\noindent $\bullet$  $\bullet$  $\bullet$
 
\noindent
The conservation of 4-momentum $p=q_0+q_++q_-$ gives ($c=1$ is assumed) $(p-
q_0)^2=m_\omega^2+m_{\pi^0}^2-2m_\omega E_0=(q_++q_-)^2=2m_{\pi^\pm}^2+2(E_+
E_--p_+p_-\cos{\theta})$. Hence, the opening angle between $\pi^+$ and 
$\pi^-$ is 
$$\cos{\theta}=\frac{2E_1E_--m_\omega^2-m_{\pi^0}^2+2m_{\pi^\pm}^2
+2m_\omega (m_\omega-E_1-E_-)}{2\sqrt{E_1^2-m_{\pi^\pm}^2}\sqrt{E_-^2-
m_{\pi^\pm}^2}},$$
since $E_+=E_1$ and $E_0=m_\omega-E_+-E_-$. But $\cos^2{\theta}\le 1$, which 
gives the condition $4(E_1^2-m_{\pi^\pm}^2)(E_-^2-m_{\pi^\pm}^2)\ge
[2E_1E_--m_\omega^2-m_{\pi^0}^2+2m_{\pi^\pm}^2+2m_\omega 
(m_\omega-E_1-E_-)]^2$. After some algebra, we get $4E_-^2(m_\omega^2-
2m_\omega E_1+m_{\pi^\pm}^2)+4E_-(E_1-m_\omega)(m_\omega^2-2m_\omega E_1-
m_{\pi^0}^2+2m_{\pi^\pm}^2)+(m_\omega^2-2m_\omega E_1-m_{\pi^0}^2+
2m_{\pi^\pm}^2)^2-4m_{\pi^\pm}^2(E_1^2-m_{\pi^\pm}^2)\le 0$. It is convenient 
to use dimensionless quantities
$$x_1=\frac{E_1}{m_\omega},\;\; x=\frac{E_-}{m_\omega},\;\;
\mu=\frac{m_{\pi^\pm}}{m_\omega}\;\;\mbox{и} \;\; \mu_0=\frac{m_{\pi^0}}
{m_\omega}.$$
Then $4x^2(1-2x_1+\mu^2)-4x(1-x_1)(1-2x_1-\mu_0^2+2\mu^2)+
(1-2x_1-\mu_0^2+2\mu^2)^2+4\mu^2(x_1^2-\mu^2)\le 0$. But $16(1-x_1)^2
(1-2x_1-\mu_0^2+2\mu^2)^2-16(1-2x_1+\mu^2)(1-2x_1-\mu_0^2+2\mu^2)^2-
64\mu^2(x_1^2-\mu^2)(1-2x_1+\mu^2)=16(x_1^2-\mu^2)[(1-2x_1-\mu_0^2+2\mu^2)^2-
4\mu^2(1-2x_1+\mu^2)]$ and $(1-2x_1-\mu_0^2+2\mu^2)^2-4\mu^2(1-2x_1+\mu^2)=
(1-2x_1-\mu_0^2)^2+4\mu^4+4\mu^2(1-2x_1-\mu_0^2)-4\mu^2(1-2x_1+\mu^2)=
(1-2x_1-\mu_0^2)^2-4\mu^2\mu_0^2=(1-2x_1-\mu_0^2-2\mu\mu_0)
(1-2x_1-\mu_0^2+2\mu\mu_0)$. Therefore, finally, $x_{min}\le x \le x_{max}$,
where
$$x_{min}=\frac{(1-x_1)(1-2x_1-\mu_0^2+2\mu^2)-\sqrt{D}}{2(1-2x_1+\mu^2)},$$
$$x_{max}=\frac{(1-x_1)(1-2x_1-\mu_0^2+2\mu^2)+\sqrt{D}}{2(1-2x_1+\mu^2)},$$
and $$D=(x_1^2-\mu^2)(1-2x_1-\mu_0^2-2\mu\mu_0)(1-2x_1-\mu_0^2+2\mu\mu_0).$$

\section*{Контрольная работа 7}
\setcounter{section}{7}
\setcounter{subsection}{0}
\subsection{}
Если собственная длина желоба $2L$, координаты событий падения шаров в системе
$S$ равны $t_L=\frac{L}{u}$, $x_L=-L$ и $t_R=\frac{L}{u}$, $x_R=L$. 
В системе наблюдателя ($S^\prime$) будем иметь
$$t^\prime_L=\gamma_V\left (\frac{L}{u}+\frac{V}{c^2}L\right )\;\;
\mbox{и} \;\; t^\prime_R=\gamma_V\left (\frac{L}{u}-\frac{V}{c^2}L
\right ),$$
т.е. правый шар падает раньше. Но после его падения состояние тока в сечении 
$AA$ не может измениться мгновенно: информация о том, что короткое замыкание 
исчезло дойдет не раньше, чем через время $\tau^\prime=\frac{2L/\gamma_V}
{c-V}$. 

\begin{figure}[htb]
\centerline{\epsfig{figure=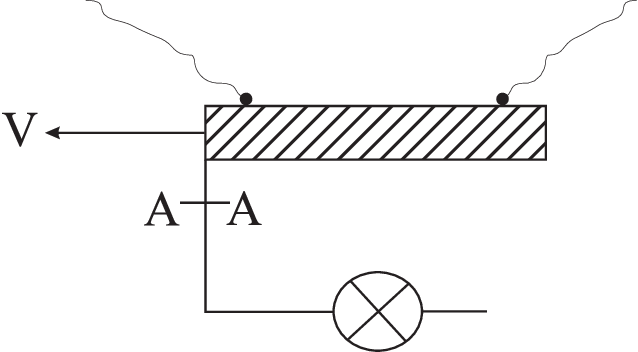,height=4cm}}
\end{figure}
\noindent Но 
$$t^\prime_R+\tau^\prime=L\gamma_V\left [\frac{1}{u}-\frac{V}{c^2}+
2\frac{1}{c}\left (1+\frac{V}{c}\right )\right ]=L\gamma_V\left [\frac{1}{u}
+\frac{V}{c^2}+\frac{2}{c}\right ]=t^\prime_L+2\gamma_V\frac{L}{c}>
t^\prime_L.$$
Следовательно, к тому времени когда в системе $S^\prime$ сечение $AA$ 
``узнает'', что короткое замыкание исчезло, левый шар тоже окажется упавшим и 
лампочка не зажгется.

\vspace*{5mm}
\noindent $\bullet$  $\bullet$  $\bullet$
 
\noindent
In the system $S$, if the proper length of the gutter is $2L$, the coordinates 
of the events of falling of left and right spheres are  
$t_L=\frac{L}{u}$, $x_L=-L$ and $t_R=\frac{L}{u}$, $x_R=L$.
In the observer's system ($S^\prime$), we will have
$$t^\prime_L=\gamma_V\left (\frac{L}{u}+\frac{V}{c^2}L\right )\;\;
\mbox{and} \;\; t^\prime_R=\gamma_V\left (\frac{L}{u}-\frac{V}{c^2}L
\right ),$$
that is, the right sphere falls earlier. But after its fall, the state of 
current in the cross-section $AA$ can not change instantaneously: information 
that the electric element is no longer short circuited arrives at $AA$ not 
earlier than after the time $\tau^\prime=\frac{2L/\gamma_V}{c-V}$.

\begin{figure}[htb]
\centerline{\epsfig{figure=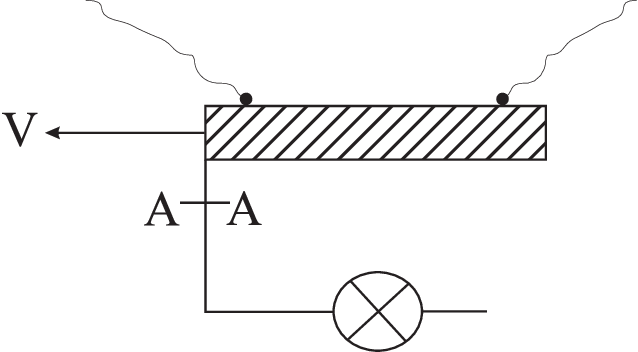,height=4cm}}
\end{figure}
\noindent But 
$$t^\prime_R+\tau^\prime=L\gamma_V\left [\frac{1}{u}-\frac{V}{c^2}+
2\frac{1}{c}\left (1+\frac{V}{c}\right )\right ]=L\gamma_V\left [\frac{1}{u}
+\frac{V}{c^2}+\frac{2}{c}\right ]=t^\prime_L+2\gamma_V\frac{L}{c}>
t^\prime_L.$$
Consequently, in the system $S^\prime$, by the time when the cross-section 
$AA$ will become aware that a short circuit has disappeared, the left sphere, 
too, is already fallen down and the bulb will not light up.

\subsection{}
Пусть сигнал догоняет ракету в момент, когда скорость ракеты равна $V$. В 
сопутствующей инерциальной СО частота сигнала будет $\nu^\prime=\gamma\nu(1-
\beta)$. После отражения, в этой системе частота не поменяется, а импульс 
фотона поменяет знак. Поэтому в Л-системе частота отраженного сигнала будет
$$\nu_{\mbox{отр}}=\gamma\nu^\prime(1-\beta)=\nu\gamma^2(1-\beta)^2=\frac{1-
\beta}{1+\beta}\,\nu.$$
Следовательно, $\nu_{\mbox{отр}}=Z\nu$, где 
$$Z=\frac{1-\beta}{1+\beta}=\frac{2}{1+\beta}-1.$$ 
Посмотрим как меняется $Z$ со временем. 
$$\frac{dZ}{dt}=-\frac{2}
{(1+\beta)^2}\frac{d\beta}{dt}.$$ Но 
$$\frac{d\beta}{dt}=\frac{g}{c\gamma^3}$$
(ускорение ракеты в Л-системе $\gamma^3$-раз меньше). Поэтому
$$\frac{dZ}{dt}=-\frac{2g}{c\gamma^3}\frac{1}{(1+\beta)^2}=-\frac{2g}{c}
(1-\beta)\sqrt{\frac{1-\beta}{1+\beta}}=-\frac{2g}{c}(1-\beta)\sqrt{Z}.$$
Перейдем к запаздывающему времени $t_-=t-\frac{x}{c}=\frac{x_-}{c}$ (вспомните 
переменные светового конуса). Тогда $dt_-=dt-\frac{dx}{c}=(1-\beta)dt$ и 
уравнение для $Z$ перепишется так $$\frac{dZ}{dt_-}=-\frac{2g}{c}\sqrt{Z},$$ 
или $$\frac{dZ}{2\sqrt{Z}}=-\frac{g}{c}dt_-.$$ 
Интегрируя и принимая во внимание, что
в начальный момент $t=0,\, x=0$ и, следовательно, $t_-=0,\, Z(0)=1$ (т.к. 
$\beta(0)=0$), получаем $\sqrt{Z}-1=-\frac{g}{c}t_-$. Сигнал прошел расстояние 
$x$ за время $t-\tau$. Поэтому $x=c(t-\tau)$ и $t_-=t-\frac{x}{c}=\tau$.
Окончательно, $\sqrt{Z}=1-\frac{g\tau}{c}$ и $Z=\left (1-\frac{g\tau}{c}
\right)^2$. Следовательно, $$\nu_{\mbox{отр}}=\left (1-\frac{g\tau}{c}\right)^2
\nu.$$
 
\vspace*{5mm}
\noindent $\bullet$  $\bullet$  $\bullet$
 
\noindent
Let the signal reaches the rocket at a time when the velocity of the rocket is 
equal to $V$. In a co-moving inertial reference frame, the frequency of the 
signal will be  $\nu^\prime=\gamma\nu(1-\beta)$. After the reflection, in this
system, the frequency is not changed, while the photon momentum changes sign.
Therefore, in the L-system, frequency of the reflected signal will be
$$\nu_{\mbox{ref}}=\gamma\nu^\prime(1-\beta)=\nu\gamma^2(1-\beta)^2=\frac{1-
\beta}{1+\beta}\,\nu.$$
Hence, $\nu_{\mbox{ref}}=Z\nu$, where 
$$Z=\frac{1-\beta}{1+\beta}=\frac{2}{1+\beta}-1.$$ 
Let's see how $Z$ varies with time. 
$$\frac{dZ}{dt}=-\frac{2}
{(1+\beta)^2}\frac{d\beta}{dt}.$$ But 
$$\frac{d\beta}{dt}=\frac{g}{c\gamma^3}$$
(acceleration of the rocket in the L-system is $\gamma^3$-times less). 
Therefore,
$$\frac{dZ}{dt}=-\frac{2g}{c\gamma^3}\frac{1}{(1+\beta)^2}=-\frac{2g}{c}
(1-\beta)\sqrt{\frac{1-\beta}{1+\beta}}=-\frac{2g}{c}(1-\beta)\sqrt{Z}.$$
Let us introduce the retarded time $t_-=t-\frac{x}{c}=\frac{x_-}{c}$
(recall the definition of the light-cone variables). Then $dt_-=dt-\frac{dx}
{c}=(1-\beta)dt$ and the equation for $Z$ can be rewritten as 
$$\frac{dZ}{dt_-}=-\frac{2g}{c}\sqrt{Z},$$ or $$\frac{dZ}{2\sqrt{Z}}=-
\frac{g}{c}dt_-.$$ 
Integrating and taking into account that $t=0,\, x=0$ at the initial time,
and, consequently, $t_-=0,\, Z(0)=1$ (because $\beta(0)=0$), we get 
$\sqrt{Z}-1=-\frac{g}{c}t_-$. The signal covered the distance $x$ during the
time $t-\tau$. Therefore, $x=c(t-\tau) $ and $t_-= t-\frac{x}{c} =\tau $.
Finally,  $\sqrt{Z}=1-\frac{g\tau}{c}$ and $Z=\left (1-\frac{g\tau}{c}
\right)^2$. Consequently, 
$$\nu_{\mbox{ref}}=\left (1-\frac{g\tau}{c}\right)^2\nu.$$

\subsection{}
\noindent $ $

\begin{figure}[htb]
\centerline{\epsfig{figure=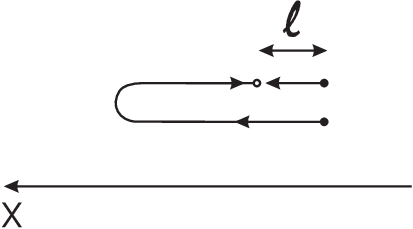}}
\end{figure}
Импульсы электрона и протона перед столкновением будут 
$p_{e,x}=p_{e,0}-e{\cal{E}} t$ и $p_{p,x}=e{\cal{E}} t$. 
Следовательно, $p_{e,x}+p_{p,x}=
p_0$, где $p_0=p_{e,0}$ -- начальный импульс электрона. Работа электрических 
сил определяется только конечными и начальными положениями частиц. Поэтому 
перед столкновением энергий частиц будут $E_e=E_0-e{\cal{E}} l$ и $E_p=m_p
c^2+e{\cal{E}} l$. Т.е. $E_e+E_p=E_0+m_pc^2$. Следовательно, суммарный 
4-импульс перед столкновением будет (положили $c=1$) $(E_0+m_p,p_0,0,0)$. Из 
сохранения 4-импульса следует $(E_0+m_p)^2-p_0^2=(p^\prime_e+p^\prime_p+q_
{e^+}+q_{e^-})^2$, где $p^\prime_e,\,p^\prime_p$ -- 4-импульсы электрона и 
протона после столкновения, а $q_{e^+},\,q_{e^-}$ -- 4-импульсы новых частиц. 
Но $(E_0+m_p)^2-p_0^2=E_0^2-p_0^2+2E_0m_p+m_p^2=m_e^2+m_p^2+2E_0m_p$ и 
$(p^\prime_e+p^\prime_p+q_{e^+}+q_{e^-})^2\ge(m_e+m_p+2m_e)^2=9m_e^2+m_p^2+
6m_em_p$ (инвариантная масса $(p^\prime_e+p^\prime_p+q_{e^+}+q_{e^-})^2$ 
минимальна, когда в системе центра масс все частицы неподвижны). 
Следовательно, $m_e^2+m_p^2+2E_0m_p\ge 9m_e^2+m_p^2+6m_em_p$ и (восстановили 
$c$) $$E_0\ge m_ec^2\left (3+4\frac{m_e}{m_p}\right ).$$

\vspace*{5mm}
\noindent $\bullet$  $\bullet$  $\bullet$
 
\begin{figure}[htb]
\centerline{\epsfig{figure=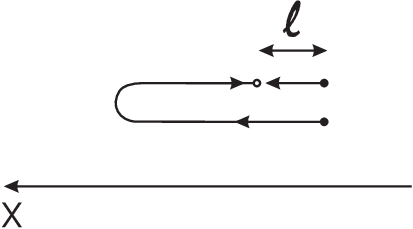}}
\end{figure}
The electron and proton momenta before the collision will be
$p_{e,x}=p_{e,0}-e{\cal{E}} t$ and $p_{p,x}=e{\cal{E}} t$.
Therefore, $p_{e,x}+p_{p,x}=p_0$, where $p_0=p_{e,0}$ is the initial momentum 
of the electron. Work done by electrical  force is determined solely by the 
final and initial positions. Therefore, before the collision, energies of the 
particles will be $E_e=E_0-e{\cal{E}} l$ and $E_p=m_p c^2+e{\cal{E}} l$.
That is, $E_e+E_p=E_0+m_pc^2$. Consequently, the total 4-momentum before the 
collision will be (we have put $c=1$) $(E_0+m_p,p_0,0,0)$. Conservation of the
4-momentum implies  $(E_0+m_p)^2-p_0^2=(p^\prime_e+p^\prime_p+q_{e^+}+
q_{e^-})^2$, where  $p^\prime_e,\,p^\prime_p$ are 4-momenta of the electron 
and proton after the collision, and  $q_{e^+},\,q_{e^-}$  are 4-momenta of the 
new particles. But $(E_0+m_p)^2-p_0^2=E_0^2-p_0^2+2E_0m_p+m_p^2=m_e^2+m_p^2+
2E_0m_p$ and $(p^\prime_e+p^\prime_p+q_{e^+}+q_{e^-})^2\ge(m_e+m_p+2m_e)^2=
9m_e^2+m_p^2+6m_em_p$ (the invariant mass $(p^\prime_e+p^\prime_p+q_{e^+}+
q_{e^-})^2$ is minimal when in the center of mass system all particles are at
rest). Hence $m_e^2+m_p^2+2E_0m_p\ge 9m_e^2+m_p^2+6m_em_p$ and (we have 
restored $c$) $$E_0\ge m_ec^2\left (3+4\frac{m_e}{m_p}\right ).$$

\subsection{}
Сперва некоторые соотношения для ракеты. Так как в системе земли ускорение 
ракеты
$a=\frac{g}{\gamma^3}$, то $\dot{\beta}\equiv\frac{d\beta}{dt}=\frac{g}{c
\gamma^3}$. Но $\dot{\gamma}=\gamma^3\beta\dot{\beta}$ и 
$$\frac{d}{dt}(\beta
\gamma)=\dot{\beta}\gamma+\beta\dot{\gamma}=\dot{\beta}\gamma(1+\beta^2
\gamma^2)=\dot{\beta}\gamma^3=\frac{g}{c}.$$ Следовательно, $\beta\gamma=
\frac{gt}{c}$ и $\gamma=\sqrt{1+\beta^2\gamma^2}=\sqrt{1+\frac{g^2t^2}{c^2}}$.
За время $\tau$ ракета улетит на расстояние 
$$L=\int\limits_0^\tau V(t)dt=c\int\limits_0^\tau\beta dt=c\int\limits_0^\beta
\beta\frac{c\gamma^3}{g}d\beta=\frac{c^2}{2g}\int\limits_0^{\beta^2}
(1-x)^{-3/2}dx=\frac{c^2}{g}(\gamma-1).$$
При этом собственное время ракеты
$$t^\prime=\int\limits_0^\tau\frac{dt}{\gamma}=\int\limits_0^\tau\frac{dt}
{\sqrt{1+\frac{g^2t^2}{c^2}}}=\frac{c}{g}\,\mathrm{arcsh}{\frac{g\tau}{c}}.$$ 
Следовательно,
$$\tau=\frac{c}{g}\sh{\left (\frac{gt^\prime}{c}\right )} \;\; \mbox{и} \;\;
\gamma(t)=\sqrt{1+\frac{g^2t^2}{c^2}}=\ch{\left (\frac{gt^\prime}{c}
\right )},\;\; \beta^2(t)=1-\frac{1}{\gamma^2}=\th^2{\left (\frac{gt^\prime}
{c}\right )},$$
т.е. $$\beta(t)=\th{\left (\frac{gt^\prime}{c}\right )}.$$
Координаты в системе ракеты $S^\prime$ совпадают с координатами в мгновенной 
сопутствующей системе $\tilde{S^\prime}$. Но в момент $t=0$ начала систем $S$ и
$\tilde{S^\prime}$ не совпадают и часы $\tilde{S^\prime}$ показывают не нуль, а
некоторое время 
$$t_0^\prime=t^\prime-\frac{\tau}{\gamma}=t^\prime-\frac{c}{g}\th{\left 
(\frac{gt^\prime}{c}\right )}.$$
Поэтому рассмотрим еще две инерциальные системы: $\tilde{S^\prime}_1$ который 
в момент $t=0$ находился в той же точке с координатой 
$$x_0=L-V\tau=\frac{c^2}{g}\left [ \ch{\left (\frac{gt^\prime}{c}\right )}-
1\right ]-\frac{c^2}{g}\th{\left (\frac{gt^\prime}{c}\right )}
\sh{\left (\frac{gt^\prime}{c}\right )}=\frac{c^2}{g}\frac{1}{\ch{\left 
(\frac{gt^\prime}{c}\right )}}- \frac{c^2}{g},$$
где находился начало системы $\tilde{S^\prime}$, но часы $\tilde{S^\prime}_1$
при этом показывали $t_1^\prime=0$. Вторая система $S_1$ неподвижна 
относительно Земли, но ее начало находится в точке с координатой $x_0$ (в 
системе Земли). Следовательно, в момент $t=t_1^\prime=0$ начала $S_1$ и 
$\tilde{S^\prime}_1$ совпадают. Поэтому стандартные формулы преобразования 
Лоренца применимы для этих систем без модификации и получаем $x_1=\gamma(
\tilde{x^\prime}_1+V\tilde{t^\prime}_1)$ и $t_1=\gamma\left (\tilde{t^\prime}
_1+\frac{V}{c^2}\tilde{x^\prime}_1\right )$. Но $x=x_0+x_1$ (заметим, что 
$x_0<0$) и $t=t_1$. С другой стороны, $\tilde{x^\prime}_1= \tilde{x^\prime}=
x^\prime$ и $\tilde{t^\prime}_1=\tilde{t^\prime}-\tilde{t^\prime}_0=t^\prime-
\tilde{t^\prime}_0=\frac{c}{g}\th{\left (\frac{gt^\prime}{c}\right )}$.
Поэтому
$$x_1=\ch{\left (\frac{gt^\prime}{c}\right )}\left [x^\prime+\frac{c^2}{g}
\th^2{\left (\frac{gt^\prime}{c}\right )}\right ]$$
и
$$x=x_1+x_0=-\frac{c^2}{g}+\ch{\left (\frac{gt^\prime}{c}\right )}\left [
x^\prime+\frac{c^2}{g}\th^2{\left (\frac{gt^\prime}{c}\right )}+\frac{c^2}{g}
\frac{1}{\ch^2{\left (\frac{gt^\prime}{c}\right )}}\right ]=$$
$$\hspace*{-75mm}=
\ch{\left (\frac{gt^\prime}{c}\right )}\left [x^\prime+\frac{c^2}{g}\right ]-
\frac{c^2}{g}.$$
Кроме того
$$t=t_1=\ch{\left (\frac{gt^\prime}{c}\right )}\left [\frac{c}{g}
\th{\left (\frac{gt^\prime}{c}\right )}+\frac{x^\prime}{c}
\th{\left (\frac{gt^\prime}{c}\right )}\right ]=\frac{c}{g}
\sh{\left (\frac{gt^\prime}{c}\right )}\left [1+\frac{g}{c^2}x^\prime
\right ].$$
Окончательно получаем, что координаты события $(ct^\prime,x^\prime,y^\prime,
z^\prime)$ в системе ракеты и координаты $(ct,x,y,z)$ в системе Земли связаны
соотношениями
$$t=\frac{c}{g}\sh{\left (\frac{gt^\prime}{c}\right )}\left [1+\frac{g}{c^2}
x^\prime \right ],\; x=\ch{\left (\frac{gt^\prime}{c}\right )}\left 
[x^\prime+\frac{c^2}{g}\right ]-\frac{c^2}{g},\; y=y^\prime,\; z=z^\prime.$$

\vspace*{5mm}
\noindent $\bullet$  $\bullet$  $\bullet$
 
\noindent
First, some relations for the rocket. Since in the earth's frame the rocket
acceleration is $a=\frac{g}{\gamma^3}$, then $\dot{\beta}\equiv\frac{d\beta}
{dt}=\frac{g}{c\gamma^3}$. But $\dot{\gamma}=\gamma^3\beta\dot{\beta}$ and 
$$\frac{d}{dt}(\beta
\gamma)=\dot{\beta}\gamma+\beta\dot{\gamma}=\dot{\beta}\gamma(1+\beta^2
\gamma^2)=\dot{\beta}\gamma^3=\frac{g}{c}.$$ Consequently, $\beta\gamma=
\frac{gt}{c}$ и $\gamma=\sqrt{1+\beta^2\gamma^2}=\sqrt{1+\frac{g^2t^2}{c^2}}$.
During the time span $\tau$, the rocket have flied the distance 
$$L=\int\limits_0^\tau V(t)dt=c\int\limits_0^\tau\beta dt=c\int\limits_0^\beta
\beta\frac{c\gamma^3}{g}d\beta=\frac{c^2}{2g}\int\limits_0^{\beta^2}
(1-x)^{-3/2}dx=\frac{c^2}{g}(\gamma-1).$$
At that, the proper time elapsed in the rocket equals to
$$t^\prime=\int\limits_0^\tau\frac{dt}{\gamma}=\int\limits_0^\tau\frac{dt}
{\sqrt{1+\frac{g^2t^2}{c^2}}}=\frac{c}{g}\,\mathrm{arcsh}{\frac{g\tau}{c}}.$$ 
Hence,
$$\tau=\frac{c}{g}\sh{\left (\frac{gt^\prime}{c}\right )} \; \mbox{and} \;
\gamma(t)=\sqrt{1+\frac{g^2t^2}{c^2}}=\ch{\left (\frac{gt^\prime}{c}
\right )},\; \beta^2(t)=1-\frac{1}{\gamma^2}=\th^2{\left (\frac{gt^\prime}
{c}\right )},$$
that is, $$\beta(t)=\th{\left (\frac{gt^\prime}{c}\right )}.$$
Coordinates in the rocket's frame $S^\prime$ coincide with the coordinates in 
the instan\-tan\-ous comoving frame $\tilde{S^\prime}$. However, at the time 
$t=0$, the origins of the systems $S$ and $\tilde{S^\prime}$ do not coincide 
and the clock $\tilde{S^\prime}$ does not read zero, but some other time
$$t_0^\prime=t^\prime-\frac{\tau}{\gamma}=t^\prime-\frac{c}{g}\th{\left 
(\frac{gt^\prime}{c}\right )}.$$
Therefore, we consider two other inertial reference frames. At the time $t=0$,
the origin of the first frame $\tilde{S^\prime}_1$ has the same $x$-coordinate
$$x_0=L-V\tau=\frac{c^2}{g}\left [ \ch{\left (\frac{gt^\prime}{c}\right )}-
1\right ]-\frac{c^2}{g}\th{\left (\frac{gt^\prime}{c}\right )}
\sh{\left (\frac{gt^\prime}{c}\right )}=\frac{c^2}{g}\frac{1}{\ch{\left 
(\frac{gt^\prime}{c}\right )}}- \frac{c^2}{g},$$
as the origin of the frame $\tilde{S^\prime}$, but the clocks of $\tilde
{S^\prime}_1$ read at that the time $t_1^\prime=0$. The second frame $S_1$ 
is at rest with respect to the Earth, but its origin is located at the point
$x_0$ (in the Earth's frame). Therefore, at the moment $t=t_1^\prime=0$, the
origins of $S_1$ and $\tilde{S^\prime}_1$ coincide and, hence, the 
coordinates in these systems are related by standard Lorentz transformations
$x_1=\gamma(\tilde{x^\prime}_1+V\tilde{t^\prime}_1)$ and $t_1=\gamma\left 
(\tilde{t^\prime}_1+\frac{V}{c^2}\tilde{x^\prime}_1\right )$. But $x=x_0+x_1$ 
(note that $x_0<0$) and $t=t_1$. On the other hand, $\tilde{x^\prime}_1= 
\tilde{x^\prime}=x^\prime$ and $\tilde{t^\prime}_1=\tilde{t^\prime}-
\tilde{t^\prime}_0=t^\prime-\tilde{t^\prime}_0=\frac{c}{g}
\th{\left (\frac{gt^\prime}{c}\right )}$.
Consequently,
$$x_1=\ch{\left (\frac{gt^\prime}{c}\right )}\left [x^\prime+\frac{c^2}{g}
\th^2{\left (\frac{gt^\prime}{c}\right )}\right ]$$
and
$$x=x_1+x_0=-\frac{c^2}{g}+\ch{\left (\frac{gt^\prime}{c}\right )}\left [
x^\prime+\frac{c^2}{g}\th^2{\left (\frac{gt^\prime}{c}\right )}+\frac{c^2}{g}
\frac{1}{\ch^2{\left (\frac{gt^\prime}{c}\right )}}\right ]=$$
$$\hspace*{-75mm}=
\ch{\left (\frac{gt^\prime}{c}\right )}\left [x^\prime+\frac{c^2}{g}\right ]-
\frac{c^2}{g}.$$
Moreover,
$$t=t_1=\ch{\left (\frac{gt^\prime}{c}\right )}\left [\frac{c}{g}
\th{\left (\frac{gt^\prime}{c}\right )}+\frac{x^\prime}{c}
\th{\left (\frac{gt^\prime}{c}\right )}\right ]=\frac{c}{g}
\sh{\left (\frac{gt^\prime}{c}\right )}\left [1+\frac{g}{c^2}x^\prime
\right ].$$
Finally, we find that the coordinates of the event $(ct^\prime,x^\prime,
y^\prime,z^\prime)$ in the rocket's frame and the coordinates $(ct,x,y,z)$ of 
the same event in the Earth frame are related as follows
$$t=\frac{c}{g}\sh{\left (\frac{gt^\prime}{c}\right )}\left [1+\frac{g}{c^2}
x^\prime \right ],\; x=\ch{\left (\frac{gt^\prime}{c}\right )}\left 
[x^\prime+\frac{c^2}{g}\right ]-\frac{c^2}{g},\; y=y^\prime,\; z=z^\prime.$$

\subsection{}
Получим уравнение движения для зеркала. За время $dt$ на зеркало упадет 
участок луча лазера длиной $dl=(c-V)dt$. Лазер этот участок излучит за время 
$dt_0=\frac{dl}{c}=(1-\beta)dt$. Следовательно, на зеркало упало световой 
энергии ${\cal{E}}=Ndt_0=N(1-\beta)dt$, где $N$ -- мощность лазера. Если из 
этой энергии ${\cal{E}}^\prime$ отразилась, то законы сохранения энергии и 
импульса будут иметь вид ${\cal{E}}+E={\cal{E}}^\prime+E+dE$ и 
$\frac{{\cal{E}}}{c}+
p=-\frac{{\cal{E}}^\prime}{c}+p+dp$. Отсюда $d(E+cp)=2{\cal{E}}=
2N(1-\beta)dt$. Но $E=mc^2\gamma$ и $p=mc\beta\gamma$. Вводя запаздывающее 
время $t_-=t-\frac{x}{c}$ и замечая, что $dt_-=(1-\beta)dt$, получаем 
$$\frac{d}{dt_-}\left [\gamma(1+\beta)\right ]=\frac{2N}{mc^2}.$$
Отсюда
$$\gamma(1+\beta)=\sqrt{\frac{1+\beta}{1-\beta}}=\frac{2N}{mc^2}\,t_-+1,$$
т.к. в начальный момент $t_-=0$ и $\beta=0$. Следовательно,
$$\beta=\frac{\left (\frac{2N}{mc^2}t_-+1\right )^2-1}
{\left (\frac{2N}{mc^2}t_-+1\right )^2+1}.$$
Зеркало улетит на расстояние $$x=\int\limits_0^tc\beta dt=c\int\limits_0^{t_-}
\frac{\beta}{1-\beta}dt_-.$$ Но
$$\frac{\beta}{1-\beta}=\frac{\left (\frac{2N}{mc^2}\,t_-
+1\right )^2-1}{2}$$
и
$$\frac{x}{c}=\frac{1}{2}\int\limits_0^{t_-}\left [\left 
(\frac{2N}{mc^2}\,t_- +1\right )^2-1\right ]dt_-=\frac{mc^2}{12N}\left [
\left (\frac{2N}{mc^2}\,t_-+1\right )^3-1\right ]-\frac{t_-}{2}.$$
Отсюда
$$2\frac{x}{c}+t_-=t+\frac{x}{c}=\frac{mc^2}{6N}\left [\left (\frac{2N}
{mc^2}\,t_-+1\right )^3-1\right ],$$
или
$$6\frac{N}{mc^2}\,t_+=\left (\frac{2N}{mc^2}\,t_-+1\right )^3-1,$$
где $t_\pm=t\pm\frac{x}{c}$. Но
$$\frac{Nx}{mc^3}=\frac{9\cdot 10^9\cdot 10^{15}}{1\cdot 27 \cdot 10^{24}}=
\frac{1}{3}.$$
Следовательно, 
$$2\left (\frac{ct}{x}+1\right )=\left [\frac{2}{3}\left (\frac{ct}{x}-1
\right )+1\right ]^3-1.$$
Пусть
$$\frac{2}{3}\left (\frac{ct}{x}-1\right )+1=\tau.$$
Для $\tau$ будем иметь уравнение $3\tau+2=\tau^3$. Или $3(\tau+1)=\tau^3+1=
(\tau+1)(\tau^2-\tau+1)$. Следовательно, $(\tau+1)(\tau^2-\tau-2)=(\tau+1)^2
(\tau-2)=0$. Устраивает только $\tau=2$, т.к. $\tau>0$. Поэтому 
$$\frac{ct}{x}=
\frac{3\tau-1}{2}=\frac{5}{2}\;\; \mbox{и} \;\; t=\frac{5}{2}\frac{x}{c}=
\frac{5}{2}\,\frac{10^{15}}{3\cdot 10^8}=\frac{5}{6}10^7~\mbox{сек}.$$ 
Таким образом, зеркало улетит на расстояние $10^{15}~\mbox{м}$ за 
$\frac{5}{6}10^7~\mbox{сек}$. 

\vspace*{5mm}
\noindent $\bullet$  $\bullet$  $\bullet$
 
\noindent
Let's derive the equation of motion for the mirror. Over the time $dt$, 
a portion of the laser beam with the length $dl=(c-V)dt$ will fall at the 
mirror. The laser will emit this portion of the beam during the time
$dt_0=\frac{dl}{c}=(1-\beta)dt$. Therefore, the light energy which fell at the 
mirror equals to ${\cal{E}}=Ndt_0=N(1-\beta)dt$, where $N$ is the power of the 
laser. If the part of this energy, ${\cal{E}}^\prime$, was reflected, the 
conservation laws of energy and momentum will have the forms 
${\cal{E}}+E={\cal{E}}^\prime+E+dE$ and $\frac{{\cal{E}}}{c}+
p=-\frac{{\cal{E}}^\prime}{c}+p+dp$, respectively. Hence $d(E+cp)=2{\cal{E}}=
2N(1-\beta)dt$. But $E=mc^2\gamma$ and $p=mc\beta\gamma$. Introducing the
retarded time $t_-=t-\frac{x}{c}$ and noting that $dt_-=(1-\beta)dt$, we get 
$$\frac{d}{dt_-}\left [\gamma(1+\beta)\right ]=\frac{2N}{mc^2}.$$
Therefore,
$$\gamma(1+\beta)=\sqrt{\frac{1+\beta}{1-\beta}}=\frac{2N}{mc^2}\,t_-+1,$$
because at the initial moment $t_-=0$ and $\beta=0$. Hence,
$$\beta=\frac{\left (\frac{2N}{mc^2}t_-+1\right )^2-1}
{\left (\frac{2N}{mc^2}t_-+1\right )^2+1}.$$
The mirror will fly a distance $$x=\int\limits_0^tc\beta dt=c\int\limits_0^
{t_-}\frac{\beta}{1-\beta}dt_-.$$ But
$$\frac{\beta}{1-\beta}=\frac{\left (\frac{2N}{mc^2}\,t_-+1\right )^2-1}{2}$$
and
$$\frac{x}{c}=\frac{1}{2}\int\limits_0^{t_-}\left [\left 
(\frac{2N}{mc^2}\,t_- +1\right )^2-1\right ]dt_-=\frac{mc^2}{12N}\left [
\left (\frac{2N}{mc^2}\,t_-+1\right )^3-1\right ]-\frac{t_-}{2}.$$
Hence
$$2\frac{x}{c}+t_-=t+\frac{x}{c}=\frac{mc^2}{6N}\left [\left (\frac{2N}
{mc^2}\,t_-+1\right )^3-1\right ],$$
or
$$6\frac{N}{mc^2}\,t_+=\left (\frac{2N}{mc^2}\,t_-+1\right )^3-1,$$
where $t_\pm=t\pm\frac{x}{c}$. But
$$\frac{Nx}{mc^3}=\frac{9\cdot 10^9\cdot 10^{15}}{1\cdot 27 \cdot 10^{24}}=
\frac{1}{3}.$$
Consequently, 
$$2\left (\frac{ct}{x}+1\right )=\left [\frac{2}{3}\left (\frac{ct}{x}-1
\right )+1\right ]^3-1.$$
Let
$$\frac{2}{3}\left (\frac{ct}{x}-1\right )+1=\tau.$$
For $\tau$ we get the equation $3\tau+2=\tau^3$. Or $3(\tau+1)=\tau^3+1=
(\tau+1)(\tau^2-\tau+1)$. Therefore, $(\tau+1)(\tau^2-\tau-2)=(\tau+1)^2
(\tau-2)=0$. Only the root $\tau=2$ is appropriate, because $\tau>0$. 
Therefore, 
$$\frac{ct}{x}=
\frac{3\tau-1}{2}=\frac{5}{2}\;\; \mbox{and} \;\; t=\frac{5}{2}\frac{x}{c}=
\frac{5}{2}\,\frac{10^{15}}{3\cdot 10^8}=\frac{5}{6}10^7~\mbox{sec}.$$ 
Thus, the mirror will cover the distance $10^{15}~\mbox{m}$ in 
$\frac{5}{6}10^7~\mbox{sec}$. 

\section*{Контрольная работа 8}
\setcounter{section}{8}
\setcounter{subsection}{0}
\subsection{}
\begin{figure}[htb]
\centerline{\epsfig{figure=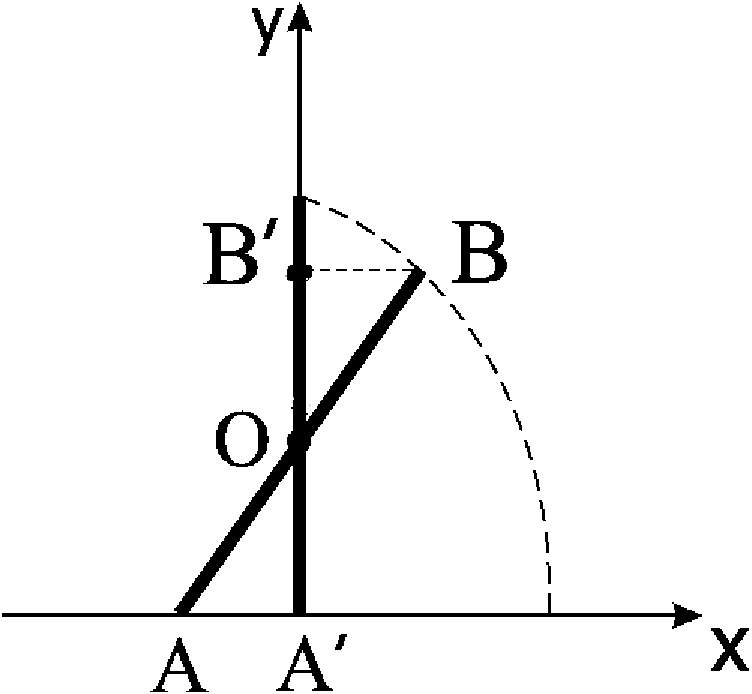,height=7cm}}
\end{figure}
Пусть координаты верхнего конца в некоторый момент времени $x$ и $y$. \\
В $x$-направлении на стержень сила не действует, поэтому центр масс стержня 
$O$ будет падать вертикально и всегда будет находится на оси $y$. Т.к. 
стержень однородный, $AO=OB$ и, следовательно, $A^\prime O=OB^\prime=
\frac{y}{2}$. Тогда из прямоугольного треугольника $OB^\prime B$ получаем 
$x^2+\frac{y^2}{4}=\frac{l^2}{4}$. Это и есть уравнение траектории верхнего 
конца и представляет собой эллипс.

\vspace*{5mm}
\noindent $\bullet$  $\bullet$  $\bullet$
 
\begin{figure}[htb]
\centerline{\epsfig{figure=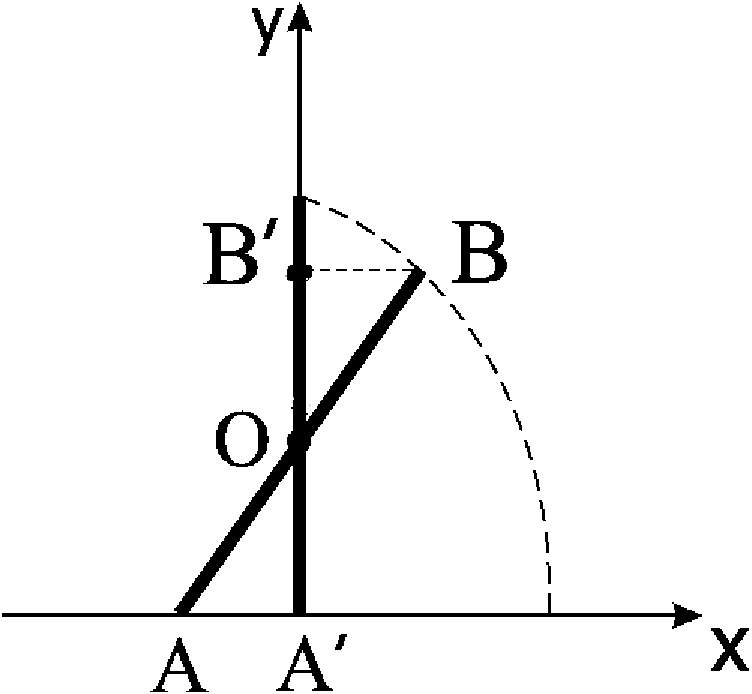,height=7cm}}
\end{figure}
Let the coordinates of the upper end of the rod are $x$ and $y$ at some time.
In the $x$-direction, no force is acting on the rod. Therefore, the center of 
mass of the rod $O$ will fall down vertically and will always be located on 
the $y$-axis. Since the rod is homogeneous, $AO=OB$, and hence, $A^\prime O=
OB^\prime=\frac{y}{2}$. Then we get from the right-angled triangle 
$OB^\prime B$ the relation $x^2+\frac{y^2}{4}=\frac{l^2}{4}$. This is just 
the equation of the trajectory of the upper end of the rod and it is 
an ellipse.

\subsection{}
Электрон притягивается к плоскости с силой $\frac{e^2}{(2z)^2}$, где $z$ -- 
расстояние до плоскости. Следовательно, уравнение движения будет $m\ddot{z}=
-\frac{e^2}{4z^2}$. Умножаем обе части на $\dot{z}$,
$$m\dot{z}\ddot{z}=\frac{m}{2}\frac{d}{dt}(\dot{z}^2)=-\frac{e^2}{4z^2}
\dot{z}=\frac{e^2}{4}\frac{d}{dt}\left (\frac{1}{z}\right ).$$
Отсюда $\frac{m\dot{z}^2}{2}-\frac{e^2}{4z}=\mathrm{const}$. Но вначале $z=H$ и
$\dot{z}=0$. Следовательно, $\mathrm{const}=-
\frac{e^2}{4H}$ и 
$$\frac{m\dot{z}^2}{2}=\frac{e^2}{4}\left (\frac{1}{z}-\frac{1}{H}\right ).$$
Поэтому (заметим, что $\dot{z}<0$, т.к. ось $z$ у нас направлена вертикально 
вверх)
$$\frac{dz}{dt}=-\sqrt{\frac{e^2}{2m}\left (\frac{1}{z}-\frac{1}{H}
\right )}.$$
Разделяем переменные и интегрируем
$$T=-\int\limits_H^0\frac{dz}{\sqrt{\frac{e^2}{2m}\left (\frac{1}{z}-
\frac{1}{H}\right )}}=\int\limits_0^H\frac{dz}{\sqrt{\frac{e^2}{2m}\left 
(\frac{1}{z}-\frac{1}{H}\right )}}.$$
Сделаем подстановку
$$\sqrt{\frac{1}{z}-\frac{1}{H}}=\sqrt{\frac{1}{H}}\,\tg{\varphi},$$
т.е.
$$\frac{1}{z}=\frac{1}{H}(1+\tg^2{\varphi})=\frac{1}{H\cos^2{\varphi}}\;\;
\mbox{и}\;\;z=H\cos^2{\varphi}.$$
Получаем
$$T=2\frac{H}{l}\,\sqrt{2mH}\int\limits_0^{\pi/2}\cos^2{\varphi}\,d\varphi=
\frac{\pi H}{l}\,\sqrt{\frac{mH}{2}}.$$

\clearpage
\noindent $\bullet$  $\bullet$  $\bullet$
 
\noindent
The electron is attracted to the plane with a force $\frac{e^2}{(2z)^2}$, 
where $z$ is the distance from the plane. Hence, the equation of motion 
of the electron is  $m\ddot{z}=-\frac{e^2}{4z^2}$. Multiply both sides by 
$\dot{z}$,
$$m\dot{z}\ddot{z}=\frac{m}{2}\frac{d}{dt}(\dot{z}^2)=-\frac{e^2}{4z^2}
\dot{z}=\frac{e^2}{4}\frac{d}{dt}\left (\frac{1}{z}\right ).$$
Hence $\frac{m\dot{z}^2}{2}-\frac{e^2}{4z}=\mathrm{const}$.
But at the initial moment $z=H$ and $\dot{z}=0$. Therefore,
$\mathrm{const}=-\frac{e^2}{4H}$ and 
$$\frac{m\dot{z}^2}{2}=\frac{e^2}{4}\left (\frac{1}{z}-\frac{1}{H}\right ).$$
Consequently (note that $\dot{z}<0$, because the $z$-axis is directed upward)
$$\frac{dz}{dt}=-\sqrt{\frac{e^2}{2m}\left (\frac{1}{z}-\frac{1}{H}
\right )}.$$
We can separate the variables and integrate
$$T=-\int\limits_H^0\frac{dz}{\sqrt{\frac{e^2}{2m}\left (\frac{1}{z}-
\frac{1}{H}\right )}}=\int\limits_0^H\frac{dz}{\sqrt{\frac{e^2}{2m}\left 
(\frac{1}{z}-\frac{1}{H}\right )}}.$$
If we make the substitution
$$\sqrt{\frac{1}{z}-\frac{1}{H}}=\sqrt{\frac{1}{H}}\,\tg{\varphi},$$
that is,
$$\frac{1}{z}=\frac{1}{H}(1+\tg^2{\varphi})=\frac{1}{H\cos^2{\varphi}}\;\;
\mbox{and}\;\;z=H\cos^2{\varphi},$$
we get
$$T=2\frac{H}{l}\,\sqrt{2mH}\int\limits_0^{\pi/2}\cos^2{\varphi}\,d\varphi=
\frac{\pi H}{l}\,\sqrt{\frac{mH}{2}}.$$

\subsection{}
Закон сохранения энергии
$$\frac{m\dot{x}^2}{2}+U_0\tg^2{\left (\frac{x}{a}\right )}=E.$$
Отсюда
$$\frac{dx}{dt}=\sqrt{\frac{2}{m}\left (E-U_0\tg^2{\left (\frac{x}{a}\right )}
\right )}$$
и (выбираем начало отсчета так, что в момент $t=0$ координата $x=0$)
$$t=\sqrt{\frac{m}{2E}}\int\limits_0^x\frac{dy}{\sqrt{1-\frac{U_0}{E}
\tg^2{\left (\frac{y}{a}\right )}}}=\sqrt{\frac{m}{2E}}\int\limits_0^x
\frac{\cos{\left (\frac{y}{a}\right )}\,dy}{\sqrt{1-\frac{E+U_0}{E}
\sin^2{\left (\frac{y}{a}\right ) }}}.$$
Сделаем подстановку
$$\sqrt{\frac{E+U_0}{E}}\sin{\left (\frac{y}{a}\right ) }=\sin{\varphi}.$$
Тогда 
$$\frac{1}{a}\sqrt{\frac{E+U_0}{E}}\cos{\left (\frac{y}{a}\right ) }\,dy=
\cos{\varphi}\;d\varphi $$
и
$$t=a\sqrt{\frac{m}{2(E+U_0)}}\int\limits
_0^{\varphi_{max}}\frac{\cos{\varphi}\,d\varphi}{\cos{\varphi}}=
a\sqrt{\frac{m}{2(E+U_0)}}\,\varphi_{max},$$
где
$$\sin{\varphi_{max}}=\sqrt{\frac{E+U_0}{E}}\sin{\left (\frac{x}{a}\right 
) }. $$
Следовательно,
$$\sqrt{\frac{E+U_0}{E}}\sin{\left (\frac{x}{a}\right ) }=\sin{\left (
\sqrt{\frac{2(E+U_0)}{m}}\,\frac{t}{a} \right )}$$
и, окончательно,
$$x=a\arcsin{\left [\sqrt{\frac{E}{E+U_0}}\sin{\left (
\sqrt{\frac{2(E+U_0)}{m}}\,\frac{t}{a} \right )}\right ]}.$$ 

\vspace*{5mm}
\noindent $\bullet$  $\bullet$  $\bullet$
 
\noindent
The energy is conserved
$$\frac{m\dot{x}^2}{2}+U_0\tg^2{\left (\frac{x}{a}\right )}=E.$$
Therefore,
$$\frac{dx}{dt}=\sqrt{\frac{2}{m}\left (E-U_0\tg^2{\left (\frac{x}{a}\right )}
\right )}$$
and (the origin of the coordinate system is chosen in such a way that at the
initial moment $t=0$ we have $x=0$)
$$t=\sqrt{\frac{m}{2E}}\int\limits_0^x\frac{dy}{\sqrt{1-\frac{U_0}{E}
\tg^2{\left (\frac{y}{a}\right )}}}=\sqrt{\frac{m}{2E}}\int\limits_0^x
\frac{\cos{\left (\frac{y}{a}\right )}\,dy}{\sqrt{1-\frac{E+U_0}{E}
\sin^2{\left (\frac{y}{a}\right ) }}}.$$
Let us substitute 
$$\sqrt{\frac{E+U_0}{E}}\sin{\left (\frac{y}{a}\right ) }=\sin{\varphi}.$$
Then
$$\frac{1}{a}\sqrt{\frac{E+U_0}{E}}\cos{\left (\frac{y}{a}\right ) }\,dy=
\cos{\varphi}\;d\varphi $$
and
$$t=a\sqrt{\frac{m}{2(E+U_0)}}\int\limits
_0^{\varphi_{max}}\frac{\cos{\varphi}\,d\varphi}{\cos{\varphi}}=
a\sqrt{\frac{m}{2(E+U_0)}}\,\varphi_{max},$$
where
$$\sin{\varphi_{max}}=\sqrt{\frac{E+U_0}{E}}\sin{\left (\frac{x}{a}\right 
) }. $$
Consequently,
$$\sqrt{\frac{E+U_0}{E}}\sin{\left (\frac{x}{a}\right ) }=\sin{\left (
\sqrt{\frac{2(E+U_0)}{m}}\,\frac{t}{a} \right )}$$
and, finally,
$$x=a\arcsin{\left [\sqrt{\frac{E}{E+U_0}}\sin{\left (
\sqrt{\frac{2(E+U_0)}{m}}\,\frac{t}{a} \right )}\right ]}.$$

\subsection{}
Рассмотрим сначала движение шарика вниз. Ось $x$ направим вниз. Изменение
кинетической энергии  шарика
$$\Delta \frac{mV^2}{2}=mg\Delta x-\alpha V^2 \Delta x.$$
Но $\Delta \frac{mV^2}{2}=mV\Delta V$ и получаем $mV\frac{dV}{dx}=
mg-\alpha V^2$. Отсюда $\frac{mVdV}{mg-\alpha V^2}=dx$
и
$$H=\int\limits_0^{V_{max}}\frac{\frac{1}{2}mdV^2}{mg-\alpha V^2}=
-\frac{m}{2\alpha}\ln{\left ( 1-\frac{\alpha V_{max}^2}{mg}\right )}.$$
Это определяет скорость шарика перед ударом
$$V_{max}^2=\frac{mg}{\alpha}\left (1-e^{-\frac{2\alpha H}{m}}\right ).$$
Т.к. удар упругий, шарик отскочит с такой же скоростью. Рассмотрим теперь
движение вверх. Ось $x$ направим вверх. Будем иметь
$$\Delta \frac{mV^2}{2}=-mg\Delta x-\alpha V^2\Delta x.$$
Или $mV\frac{dV}{dx}=-(mg+\alpha V^2)$. Отсюда высота
подъема
$$h=\int\limits_0^h dx=-\int\limits_{V_{max}}^0\frac{\frac{1}{2}mdV^2}
{mg+\alpha V^2}=\frac{m}{2\alpha}\ln{\left ( 1+\frac{\alpha V^2_{max}}
{mg}\right )}.$$
Подставляя $V^2_{max}$, окончательно получаем
$$h=\frac{m}{2\alpha}\ln{\left [2-e^{-\frac{2\alpha H}{m}}\right ]}.$$

\vspace*{5mm}
\noindent $\bullet$  $\bullet$  $\bullet$
 
\noindent
We first consider the downward motion of the ball assuming that the $x$-axis 
is directed downward. The change of the kinetic energy of the ball is 
determined by the energy balance equation
$$\Delta \frac{mV^2}{2}=mg\Delta x-\alpha V^2 \Delta x.$$ 
But $\Delta \frac{mV^2}{2}=mV\Delta V$ and we get $mV\frac{dV}{dx}=
mg-\alpha V^2$. Hence $\frac{mVdV}{mg-\alpha V^2}=dx$ 
and
$$H=\int\limits_0^{V_{max}}\frac{\frac{1}{2}mdV^2}{mg-\alpha V^2}=
-\frac{m}{2\alpha}\ln{\left ( 1-\frac{\alpha V_{max}^2}{mg}\right )}.$$
This determines the velocity of the ball before the impact
$$V_{max}^2=\frac{mg}{\alpha}\left (1-e^{-\frac{2\alpha H}{m}}\right ).$$
Because the impact is elastic, the ball will bounce with the same velocity. 
We now consider the upward movement with the $x$ axis directed upwards. 
We will have
$$\Delta \frac{mV^2}{2}=-mg\Delta x-\alpha V^2\Delta x.$$ 
Or $mV\frac{dV}{dx}=-(mg+\alpha V^2)$. Consequently, the lifting height of 
the ball equals to
$$h=\int\limits_0^h dx=-\int\limits_{V_{max}}^0\frac{\frac{1}{2}mdV^2}
{mg+\alpha V^2}=\frac{m}{2\alpha}\ln{\left ( 1+\frac{\alpha V^2_{max}}
{mg}\right )}.$$
Substituting $V^2_{max}$, we finally get
$$h=\frac{m}{2\alpha}\ln{\left [2-e^{-\frac{2\alpha H}{m}}\right ]}.$$

\subsection{}
Заметим, что $\frac{d\vec{L}}{dt}=\vec{r}\times\frac{d\vec{p}}{dt}$, т.к.
$\frac{d\vec{r}}{dt}\times\vec{p}=\vec{V}\times\vec{p}=0$. Но
$$\frac{d\vec{p}}{dt}=\frac{e}{c}\,\vec{V}\times\vec{B}=\frac{eg}{c}\frac{
\vec{V}\times\vec{r}}{r^3}.$$ 
Следовательно, $$\frac{d\vec{L}}{dt}=\frac{eg}{cr^3}\,\vec{r}\times(
\vec{V}\times\vec{r}).$$ Используя $\vec{r}\times(\vec{V}\times\vec{r})=
r^2\vec{V}-(\vec{r}\cdot\vec{V})\vec{r}$, получаем
$$\frac{d\vec{L}}{dt}=\frac{eg}{c}\left (\frac{\dot{\vec{r}}}{r}-\frac{
\vec{r}\cdot\dot{\vec{r}}}{r^3}\vec{r}\right ).$$
Но $\vec{r}\cdot\dot{\vec{r}}=\frac{1}{2}\frac{d}{dt}(\vec{r}^{\,2})=
\frac{1}{2}\frac{d}{dt}(r^2)=r\dot{r}$. Таким образом
$$\frac{d\vec{L}}{dt}=\frac{eg}{c}\left (\frac{\dot{\vec{r}}}{r}-
\frac{\dot{r}}{r^2}\vec{r}\right )=\frac{eg}{c}\frac{d}{dt}\left 
(\frac{\vec{r}}{r}\right ).$$
или 
$$\frac{d}{dt}\left (\vec{L}-\frac{eg}{c}\frac{\vec{r}}{r}\right )=0\;\;
\mbox{и}\;\;\vec{L}-\frac{eg}{c}\frac{\vec{r}}{r}= \vec{L}_0-\frac{eg}{c}
\frac{\vec{r}_0}{r_0},$$ где
$L_0$ и $r_0$ есть значения момента импульса и радиус-вектора в начальный 
момент времени. Окончательно,
$$\vec{L}=\vec{L}_0+\frac{eg}{c}\left (\frac{\vec{r}}{r}-\frac{\vec{r}_0}
{r_0}\right ).$$

\vspace*{5mm}
\noindent $\bullet$  $\bullet$  $\bullet$
 
\noindent
Note that $\frac{d\vec{L}}{dt}=\vec{r}\times\frac{d\vec{p}}{dt}$, because
$\frac{d\vec{r}}{dt}\times\vec{p}=\vec{V}\times\vec{p}=0$. But
$$\frac{d\vec{p}}{dt}=\frac{e}{c}\,\vec{V}\times\vec{B}=\frac{eg}{c}\frac{
\vec{V}\times\vec{r}}{r^3}.$$ 
Therefore, $$\frac{d\vec{L}}{dt}=\frac{eg}{cr^3}\,\vec{r}\times(
\vec{V}\times\vec{r}).$$ Using $\vec{r}\times(\vec{V}\times\vec{r})=
r^2\vec{V}-(\vec{r}\cdot\vec{V})\vec{r}$, we get
$$\frac{d\vec{L}}{dt}=\frac{eg}{c}\left (\frac{\dot{\vec{r}}}{r}-\frac{
\vec{r}\cdot\dot{\vec{r}}}{r^3}\vec{r}\right ).$$
However, $\vec{r}\cdot\dot{\vec{r}}=\frac{1}{2}\frac{d}{dt}(\vec{r}^{\,2})=
\frac{1}{2}\frac{d}{dt}(r^2)=r\dot{r}$. Therefore,
$$\frac{d\vec{L}}{dt}=\frac{eg}{c}\left (\frac{\dot{\vec{r}}}{r}-
\frac{\dot{r}}{r^2}\vec{r}\right )=\frac{eg}{c}\frac{d}{dt}\left 
(\frac{\vec{r}}{r}\right ).$$
Or
$$\frac{d}{dt}\left (\vec{L}-\frac{eg}{c}\frac{\vec{r}}{r}\right )=0\;\;
\mbox{и}\;\;\vec{L}-\frac{eg}{c}\frac{\vec{r}}{r}= \vec{L}_0-\frac{eg}{c}
\frac{\vec{r}_0}{r_0},$$ where
$L_0$ and $r_0$ are the values of the angular momentum and radius vector at 
the initial time. Finally,
$$\vec{L}=\vec{L}_0+\frac{eg}{c}\left (\frac{\vec{r}}{r}-\frac{\vec{r}_0}
{r_0}\right ).$$

\subsection{}
Согласно решению задачи 5.29 из \cite{1},
$$V_x=\frac{e{\cal{E}} c^2 t}{\sqrt{E_0^2+(e{\cal{E}} ct)^2}}\;\;
\mbox{и}\;\; V_y=\frac{E_0V_0}{\sqrt{E_0^2+(e{\cal{E}} ct)^2}}.$$
Отсюда
$$x=\int\limits_0^tV_x\,dt=\frac{1}{2e{\cal{E}}}\int\limits_0^t\frac
{d(e{\cal{E}} ct)^2}{\sqrt{E_0^2+(e{\cal{E}} ct)^2}}=
\frac{E_0}{e{\cal{E}}}\left (
\sqrt{1+\left (\frac{e{\cal{E}} ct}{E_0}\right )^2}-1\right )$$
и
$$y=\int\limits_0^tV_y\,dt=\frac{E_0V_0}{e{\cal{E}}c}\int\limits_0^t\frac
{d(e{\cal{E}} ct)}{\sqrt{E_0^2+(e{\cal{E}} ct)^2}}=\frac{E_0V_0}{e{\cal{E}}c}
\,\mathrm{arcsh}{\frac{e{\cal{E}} ct}{E_0}}.$$
Из второго уравнения 
$$\frac{e{\cal{E}} ct}{E_0}=\sh{\frac{e{\cal{E}}cy}{E_0V_0}}$$
подставим в первом:
$$x=\frac{E_0}{e{\cal{E}}}\left (\ch{\frac{e{\cal{E}}cy}{E_0V_0}}-1\right ).$$
Это и есть уравнение траектории.

\vspace*{5mm}
\noindent $\bullet$  $\bullet$  $\bullet$
 
\noindent
according to the solution of the problem 5.29 from \cite{1},
$$V_x=\frac{e{\cal{E}} c^2 t}{\sqrt{E_0^2+(e{\cal{E}} ct)^2}}\;\;
\mbox{и}\;\; V_y=\frac{E_0V_0}{\sqrt{E_0^2+(e{\cal{E}} ct)^2}}.$$
Hence
$$x=\int\limits_0^tV_x\,dt=\frac{1}{2e{\cal{E}}}\int\limits_0^t\frac
{d(e{\cal{E}} ct)^2}{\sqrt{E_0^2+(e{\cal{E}} ct)^2}}=
\frac{E_0}{e{\cal{E}}}\left (
\sqrt{1+\left (\frac{e{\cal{E}} ct}{E_0}\right )^2}-1\right )$$
and
$$y=\int\limits_0^tV_y\,dt=\frac{E_0V_0}{e{\cal{E}}c}\int\limits_0^t\frac
{d(e{\cal{E}} ct)}{\sqrt{E_0^2+(e{\cal{E}} ct)^2}}=\frac{E_0V_0}{e{\cal{E}}c}
\,\mathrm{arcsh}{\frac{e{\cal{E}} ct}{E_0}}.$$
From the second equation, we can determine
$$\frac{e{\cal{E}} ct}{E_0}=\sh{\frac{e{\cal{E}}cy}{E_0V_0}}$$
and substitute this into the first equation. As a result we get the 
equation of the trajectory:
$$x=\frac{E_0}{e{\cal{E}}}\left (\ch{\frac{e{\cal{E}}cy}{E_0V_0}}-1\right ).$$

\section*{Контрольная работа 9}
\setcounter{section}{9}
\setcounter{subsection}{0}
\subsection{}
Пусть поршни сдвинулись справа на $x$. Объем воздуха между поршнями станет 
$V^\prime=V-S_1x+S_2x=V-(S_1-S_2)x$, а давление 
$$p^\prime=\frac{pV}{V^\prime}=\frac{p}{1-\frac{1}{V}(S_1-S_2)x}\approx
p+\frac{p}{V}(S_1-S_2)x.$$
Возвращающая сила будет 
$$F=S_1p^\prime+S_2p-S_1p-S_2p^\prime=(S_1-S_2)(p^\prime-p)\approx 
\frac{p}{V}(S_1-S_2)^2x.$$
Следовательно, эффективный коэффициент жесткости $k=\frac{p}{V}(S_1-S_2)^2$
и частота 
$$\omega=\sqrt{\frac{k}{m}}=\sqrt{\frac{p}{mV}}\,(S_1-S_2).$$

\vspace*{5mm}
\noindent $\bullet$  $\bullet$  $\bullet$
 
\noindent
Let the Pistons moved to the right by $x$. The volume of air between the 
pistons becomes $V^\prime=V-S_1x+S_2x=V-(S_1-S_2)x$, and the pressure
$$p^\prime=\frac{pV}{V^\prime}=\frac{p}{1-\frac{1}{V}(S_1-S_2)x}\approx
p+\frac{p}{V}(S_1-S_2)x.$$
The restoring force equals to
$$F=S_1p^\prime+S_2p-S_1p-S_2p^\prime=(S_1-S_2)(p^\prime-p)\approx 
\frac{p}{V}(S_1-S_2)^2x.$$
Consequently, the effective spring constant is $k=\frac{p}{V}(S_1-S_2)^2$
and we get for the frequency
$$\omega=\sqrt{\frac{k}{m}}=\sqrt{\frac{p}{mV}}\,(S_1-S_2).$$

\subsection{}
В пределе бесконечно легкой сгибаемости каната, между частями $AB$ и $BC$ нет 
упругого взаимодействия. 

\begin{figure}[htb]
\centerline{\epsfig{figure=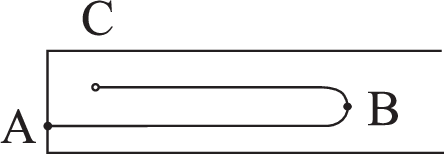,height=2cm}}
\end{figure}
\noindent Поэтому, после окончания первоначального импульса 
силы, который сообщает верхней половине скорость $V$, в точке $A$ сила 
натяжения исчезнет ($AB$ покоится и со стороны $BC$ на него не действует сила.
Поэтому и в точке $A$ не может действовать на $AB$ никакая сила). Это 
означает, что импульс каната сохраняется. Если $C$ отстоит от конца трубки 
на $x$, то $AB=\frac{1}{2}(L+x)$ (это следует из
$L=x+2BC$). Поэтому скорость точки изгиба 2-раза меньше, чем $\dot{x}$ и когда
она равна $u$, верхняя половина движется со скоростью $2u$. Следовательно,
$\frac{1}{2}(L-x)\frac{M}{L}\,2u=\frac{M}{2}\,V_0$. Отсюда $x=\left (1-
\frac{V_0}{2u}\right )L$ и $$AB=L\left (1-\frac{V_0}{4u}\right ).$$

Поведение реальной веревки может быть несколько другим. Смотрите обсуждение 
проблены падающей цепи в \cite{22A,22B,22C,22D,22E,22F,22}.

\clearpage
\noindent $\bullet$  $\bullet$  $\bullet$
 
\noindent
In the limit of infinite flexibility of the rope, there are no elastic 
interactions between $AB$ and $BC$ parts of the rope.

\begin{figure}[htb]
\centerline{\epsfig{figure=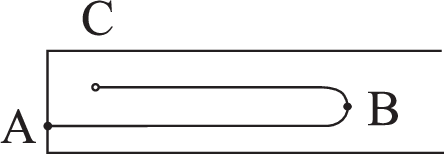,height=2cm}}
\end{figure}
\noindent Consequently, after the initial impulse, which gives the top half of 
the rope its initial velocity $V$, The tension at the point $A$ will disappear 
($AB$ is at rest and $BC$ does not exert a force on it. Therefore, at the 
point $A$ there is no force acting on $AB$ too). This implies that the 
momentum of the rope is conserved. If the distance between $C$ and the 
left-end of the tube is $x$, then $AB=\frac{1}{2}(L+x)$ (this follows from
$L=x+2BC$). Therefore, the velocity of the inflection point is two times less 
than $\dot{x}$ and when it is equal to $u$, the top half of the rope moves 
with the velocity $2u$. Consequently,
$\frac{1}{2}(L-x)\frac{M}{L}\,2u=\frac{M}{2}\,V_0$. Hence $x=\left (1-
\frac{V_0}{2u}\right )L$ and $$AB=L\left (1-\frac{V_0}{4u}\right ).$$

Real ropes may exhibit quite a different behaviour. See the discussion of 
the falling chain problem in \cite{22A,22B,22C,22D,22E,22F,22}.

\subsection{}
\noindent $ $

\begin{figure}[htb]
\centerline{\epsfig{figure=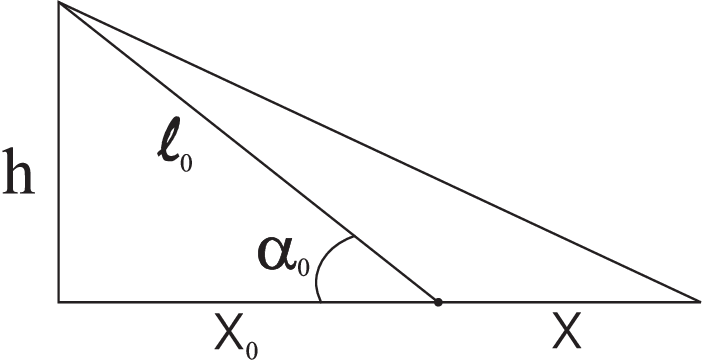,height=4cm}}
\end{figure}
\noindent Пусть тело сместилось на $x$. Новая длина пружины
$$l=\sqrt{h^2+(x_0+x)^2}\approx \sqrt{l_0^2+2x_0x}=l_0\sqrt{1+2\cos{
\alpha_0}\;\frac{x}{l_0}}\approx l_0\left (1+\cos{\alpha_0}\;\frac{x}
{l_0}\right ),$$
т.к. $x\ll x_0$ и $x_0=l_0\cos{\alpha_0}$. Сила натяжения пружины $F=k\Delta 
l=k(l-l_0)\approx k\cos{\alpha_0}\, x$. Ее проекция на горизонтальное 
направление $F_x=F\cos{\alpha}\approx k\cos^2{\alpha_0}\, x$. Следовательно,
эффективный коэффициент жесткости для горизонтального направления 
$k_{\mbox{эфф}}= k\cos^2{\alpha_0}$ и частота
$$\omega=\sqrt{\frac{k_{\mbox{эфф}}}{m}}=\sqrt{\frac{k}{m}}\,
\cos{\alpha_0}.$$

\vspace*{5mm}
\noindent $\bullet$  $\bullet$  $\bullet$
 
\begin{figure}[htb]
\centerline{\epsfig{figure=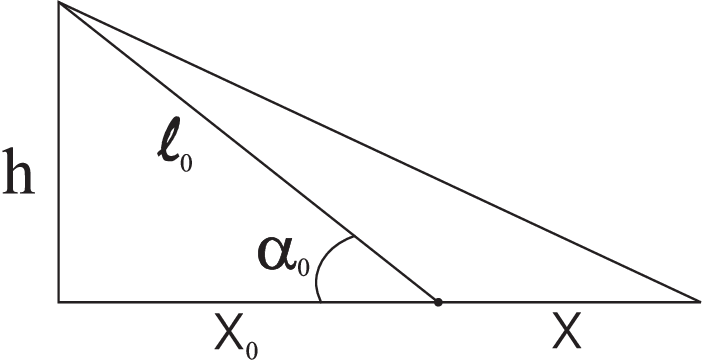,height=4cm}}
\end{figure}
\noindent  Let the body has been displaced  by $x$. The new length of the 
spring is
$$l=\sqrt{h^2+(x_0+x)^2}\approx \sqrt{l_0^2+2x_0x}=l_0\sqrt{1+2\cos{
\alpha_0}\;\frac{x}{l_0}}\approx l_0\left (1+\cos{\alpha_0}\;\frac{x}
{l_0}\right ),$$
because $x\ll x_0$ and $x_0=l_0\cos{\alpha_0}$. The force exerted by the spring
equals to $F=k\Delta l=k(l-l_0)\approx k\cos{\alpha_0}\, x$. Its projection on 
the horizontal direction is  $F_x=F\cos{\alpha}\approx k\cos^2{\alpha_0}\, x$. 
Therefore, the effective stiffness of the spring in the horizontal direction 
is given by the coefficient $k_{\mbox{eff}}= k\cos^2{\alpha_0}$
and the frequency of small oscillations is given by
$$\omega=\sqrt{\frac{k_{\mbox{eff}}}{m}}=\sqrt{\frac{k}{m}}\,
\cos{\alpha_0}.$$

\subsection{}
Период движения
$$T=\sqrt{2m}\int\limits_{x_1}^{x_2}\frac{dx}{\sqrt{E-U(x)}}=2\sqrt{2m}
\frac{\partial}{\partial E}\int\limits_{x_1}^{x_2}\sqrt{E-U(x)}\,dx.$$
Изменение периода
$$\Delta T=2\sqrt{2m}\frac{\partial}{\partial E}\Delta I,$$
где
$$\Delta I=\int\limits_{x_1+\delta x_1}^{x_2+\delta x_2}\sqrt{E-U-
\delta U}\,dx-\int\limits_{x_1}^{x_2}\sqrt{E-U(x)}\,dx.$$
Но
$$\sqrt{E-U-\delta U}\approx \sqrt{E-U}\left (1-\frac{\delta U}{2(E-U)}
\right )$$
и
$$\int\limits_{x_1+\delta x_1}^{x_2+\delta x_2}
\sqrt{E-U}\,dx-\int\limits_{x_1}^{x_2}\sqrt{E-U}\,dx\approx 0,$$
так как
$$\int\limits_{x_2}^{x_2+\delta x_2}\sqrt{E-U}\,dx \;\; \mbox{и} \;\;
\int\limits_{x_1+\delta x_1}^{x_1}\sqrt{E-U}\,dx$$
величины второго порядка малости (вблизи точек поворота $\sqrt{E-U}\approx 0$).
Следовательно, 
$$\Delta I\approx -\frac{1}{2}\int\limits_{x_1}^{x_2}\frac{\delta U}
{\sqrt{E-U}}\,dx\approx -\frac{m\beta}{8}\int\limits_{x_1}^{x_2}
\frac{x^4dx}{\sqrt{E-\frac{m\omega^2x^2}{2}}}.$$
После подстановки $y=\sqrt{\frac{m\omega^2}{2E}}\,x$, будем иметь
$$\Delta I\approx -\frac{\beta E^2}{\sqrt{2m}\,m\,\omega^5}
\int\limits_{-1}^1\frac{y^4dy}{\sqrt{1-y^2}}=-\frac{2\beta E^2}
{\sqrt{2m}\,m\,\omega^5}\int\limits_0^1\frac{y^4dy}{\sqrt{1-y^2}}.$$
Если сделать еще одну подстановку $y=\sin{\theta}$, получим
$$\Delta I\approx -\frac{2\beta E^2}{\sqrt{2m}\,m\,\omega^5}
\int\limits_0^{\pi/2}\sin^4{\theta}\,d\theta.$$
Но 
$$\sin^4{\theta}=\frac{1}{4}(1-\cos{2\theta})^2=\frac{1}{4}\left [
1-2\cos{2\theta}+\frac{1}{2}(1+\cos{4\theta})\right ]$$
и
$$\int\limits_0^{\pi/2}\cos{2\theta}\,d\theta=\int\limits_0^{\pi/2}
\cos{4\theta}\,d\theta=0.$$
Поэтому
$$\int\limits_0^{\pi/2}\sin^4{\theta}\,d\theta=\frac{\pi}{2}\,
\frac{1}{4}\,\left (1+\frac{1}{2}\right )=\frac{3\pi}{16}$$
и
$$\Delta I\approx -\frac{3\beta E^2\pi}{8\sqrt{2m}\,m\,\omega^5}.$$
Тогда
$$\Delta T= 2\sqrt{2m}\frac{\partial}{\partial E}\Delta I\approx
-\frac{3\pi\beta E}{2m\omega^5}.$$
Т.е. новый период
$$T=\frac{2\pi}{\omega}-\frac{3\pi\beta E}{2m\omega^5}=\frac{2\pi}{\omega}
\left (1-\frac{3}{4}\frac{\beta E}{m\omega^4}\right ).$$
Поправка мала, если $\frac{\beta E}{m\omega^4}\ll 1$.

\vspace*{5mm}
\noindent $\bullet$  $\bullet$  $\bullet$
 
\noindent
The oscillation period is given by
$$T=\sqrt{2m}\int\limits_{x_1}^{x_2}\frac{dx}{\sqrt{E-U(x)}}=2\sqrt{2m}
\frac{\partial}{\partial E}\int\limits_{x_1}^{x_2}\sqrt{E-U(x)}\,dx.$$
The change of the period is
$$\Delta T=2\sqrt{2m}\frac{\partial}{\partial E}\Delta I,$$
where
$$\Delta I=\int\limits_{x_1+\delta x_1}^{x_2+\delta x_2}\sqrt{E-U-
\delta U}\,dx-\int\limits_{x_1}^{x_2}\sqrt{E-U(x)}\,dx.$$
but
$$\sqrt{E-U-\delta U}\approx \sqrt{E-U}\left (1-\frac{\delta U}{2(E-U)}
\right )$$
and
$$\int\limits_{x_1+\delta x_1}^{x_2+\delta x_2}
\sqrt{E-U}\,dx-\int\limits_{x_1}^{x_2}\sqrt{E-U}\,dx\approx 0,$$
because the magnitudes of the integrals
$$\int\limits_{x_2}^{x_2+\delta x_2}\sqrt{E-U}\,dx \;\; \mbox{and} \;\;
\int\limits_{x_1+\delta x_1}^{x_1}\sqrt{E-U}\,dx$$
are of the second order (in the vicinity of the turning points 
$\sqrt{E-U}\approx 0$).
Consequently, 
$$\Delta I\approx -\frac{1}{2}\int\limits_{x_1}^{x_2}\frac{\delta U}
{\sqrt{E-U}}\,dx\approx -\frac{m\beta}{8}\int\limits_{x_1}^{x_2}
\frac{x^4dx}{\sqrt{E-\frac{m\omega^2x^2}{2}}}.$$
After the substitution $y=\sqrt{\frac{m\omega^2}{2E}}\,x$, we will have
$$\Delta I\approx -\frac{\beta E^2}{\sqrt{2m}\,m\,\omega^5}
\int\limits_{-1}^1\frac{y^4dy}{\sqrt{1-y^2}}=-\frac{2\beta E^2}
{\sqrt{2m}\,m\,\omega^5}\int\limits_0^1\frac{y^4dy}{\sqrt{1-y^2}}.$$
If we make one more substitution $y=\sin{\theta}$, we get
$$\Delta I\approx -\frac{2\beta E^2}{\sqrt{2m}\,m\,\omega^5}
\int\limits_0^{\pi/2}\sin^4{\theta}\,d\theta.$$
But
$$\sin^4{\theta}=\frac{1}{4}(1-\cos{2\theta})^2=\frac{1}{4}\left [
1-2\cos{2\theta}+\frac{1}{2}(1+\cos{4\theta})\right ]$$
and
$$\int\limits_0^{\pi/2}\cos{2\theta}\,d\theta=\int\limits_0^{\pi/2}
\cos{4\theta}\,d\theta=0.$$
therefore,
$$\int\limits_0^{\pi/2}\sin^4{\theta}\,d\theta=\frac{\pi}{2}\,
\frac{1}{4}\,\left (1+\frac{1}{2}\right )=\frac{3\pi}{16}$$
and
$$\Delta I\approx -\frac{3\beta E^2\pi}{8\sqrt{2m}\,m\,\omega^5}.$$
Then
$$\Delta T= 2\sqrt{2m}\frac{\partial}{\partial E}\Delta I\approx
-\frac{3\pi\beta E}{2m\omega^5}.$$
Therefore, the new period is
$$T=\frac{2\pi}{\omega}-\frac{3\pi\beta E}{2m\omega^5}=\frac{2\pi}{\omega}
\left (1-\frac{3}{4}\frac{\beta E}{m\omega^4}\right ).$$
The correction is small if $\frac{\beta E}{m\omega^4}\ll 1$.

\subsection{} 
Пусть энергия фотона $E$. Т.к. после реакции продукты покоятся, импульс 
электрона по величине равен импульсу фотона $E$ (положили $c=1$), а энергия 
электрона $E_e=\sqrt{m^2+E^2}$, где $m$ -- масса электрона. Закон сохранения
энергии дает $E+\sqrt{m^2+E^2}=3m$. Отсюда (восстановили $c$) 
$$E=\frac{4}{3}mc^2.$$

\vspace*{5mm}
\noindent $\bullet$  $\bullet$  $\bullet$
 
\noindent
Let the photon's energy is $E$. Since after the reaction all reaction products 
are at rest, the electron momentum equals in magnitude to the photon momentum 
$E$ (we have assumed $c=1$), and its energy is $E_e=\sqrt{m^2+E^2}$, where $m$ 
is the electron mass. The energy conservation law gives $E+\sqrt{m^2+E^2}=3m$. 
Hence (we have restored $c$)
$$E=\frac{4}{3}mc^2.$$

\section*{Контрольная работа 10}
\setcounter{section}{10}
\setcounter{subsection}{0}
\subsection{}
\begin{figure}[htb]
\centerline{\epsfig{figure=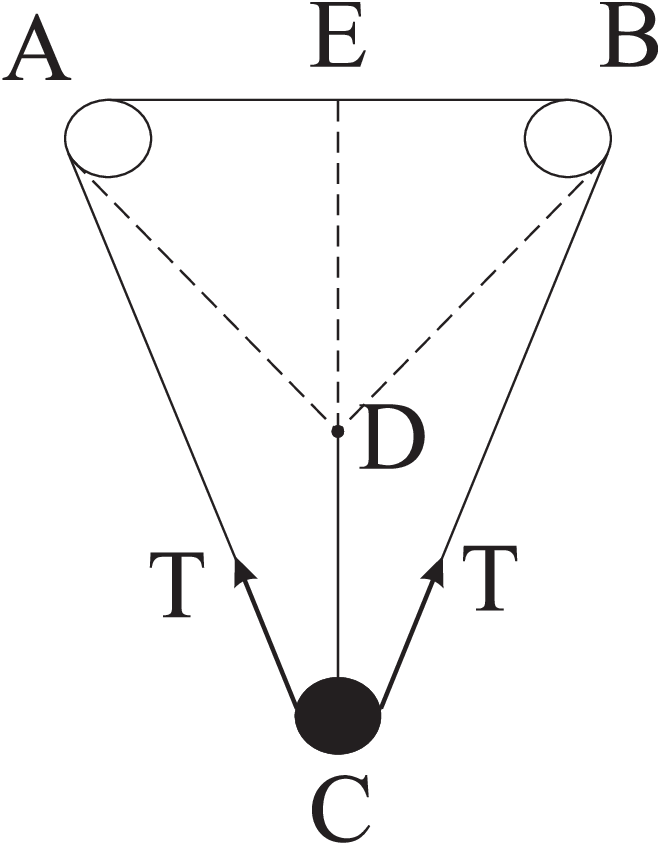,height=5cm}}
\end{figure}\noindent
Пусть сместили из положения равновесия на $CD = x$. Т.к.
$AE = \frac{l}{2}$ и $ED = \frac{\sqrt{3}}{2}\,l$, удлинение нити
будет $$\Delta l = 2 \sqrt{\left( \frac{l}{2}\right)^{2}+\left( \frac{
\sqrt{3}}{2}l+x\right)^{2}}+l-2l = l \sqrt{1+\left( \sqrt{3}+2
\frac{x}{l}\right)^{2}}-l\approx l+\sqrt{3}\,x.$$ Возвращающая
сила будет $F=2T\cos(\alpha+\Delta\alpha)-mg$, где
$\alpha+\Delta\alpha=\angle ECB$, $T=k\Delta l=k(l+\sqrt{3}\,x)$. 
В положение равновесия $2kl\cos30^{\circ}=mg$. Поэтому
$$k=\frac{mg}{2l\cos{30^{\circ}}}=\frac{mg}{\sqrt{3}\,l}\;\;\mbox{и} \;\;
T=\frac{mg}{\sqrt{3}\,l}(l+\sqrt{3}\,x).$$ Далее,
$$\cos{(\alpha+\Delta\alpha)}=\frac{\frac{\sqrt{3}}{2}\,l+x}{\sqrt{(\frac
{\sqrt{3}}{2}\,l+x)^{2}+(\frac{l}{2})^{2}}}\approx\frac{\sqrt{3}+2\frac{x}{l}}
{2}\,\frac{1}{\sqrt{1+\sqrt{3}\,\frac{x}{l}}}\approx\frac{\sqrt{3}}{2}+
\frac{x}{4l}.$$
Поэтому возвращающая сила равна
$$F=2\frac{mg}{\sqrt{3}\,l}(l+\sqrt{3}\,x)\left(\frac{\sqrt{3}}{2}+
\frac{x}{4l}\right)-mg=\frac{2mg}{\sqrt{3}\,l}\left(\frac{\sqrt{3}}{2}\,l+
\frac{7}{4}x\right)-mg=\frac{7mg}{2\sqrt{3}\,l}\;x.$$
Следовательно, эффективный коэффициент жесткости
$$k_{\mbox{эфф}}=\frac{7mg}{2\sqrt{3}\,l}$$ и частота малых колебаний равна
$$\omega=\sqrt{\frac{k_{\mbox{эфф}}}{m}}=\sqrt{\frac{7g}{2\sqrt{3}\,l}}.$$

\vspace*{5mm}
\noindent $\bullet$  $\bullet$  $\bullet$
 
\begin{figure}[htb]
\centerline{\epsfig{figure=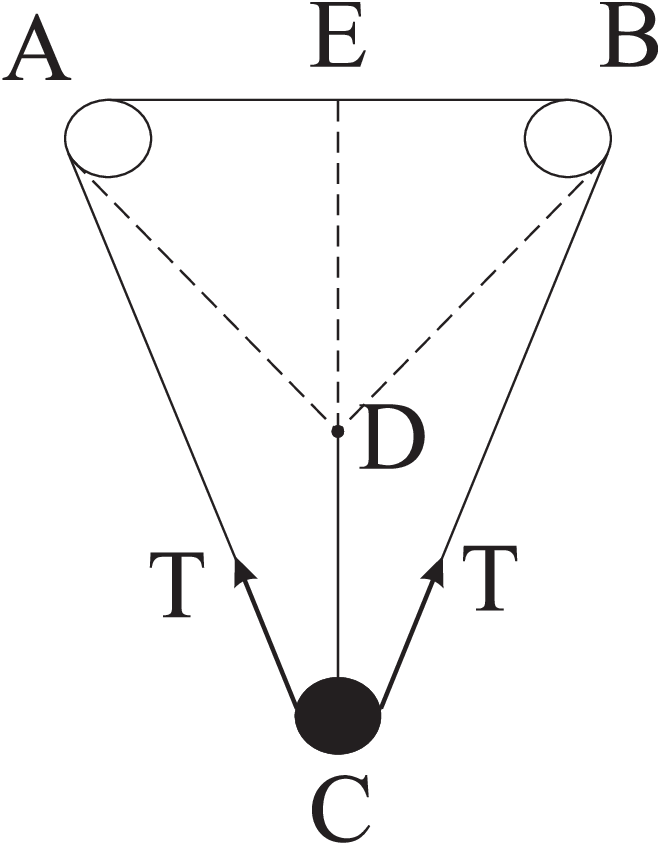,height=5cm}}
\end{figure}

\noindent
Suppose the ball is slightly displaced from the equilibrium position by 
$CD = x$. Since 
$AE = \frac{l}{2}$ and $ED = \frac{\sqrt{3}}{2}\,l$, the string will be
elongated by 
$$\Delta l = 2 \sqrt{\left( \frac{l}{2}\right)^{2}+\left( \frac{
\sqrt{3}}{2}l+x\right)^{2}}+l-2l = l \sqrt{1+\left( \sqrt{3}+2
\frac{x}{l}\right)^{2}}-l\approx l+\sqrt{3}\,x.$$ The magnitude of the
restoring force will be $F=2T\cos(\alpha+\Delta\alpha)-mg$, where
$\alpha+\Delta\alpha=\angle ECB$, $T=k\Delta l=k(l+\sqrt{3}\,x)$. 
In the equilibrium state $2kl\cos30^{\circ}=mg$. Hence
$$k=\frac{mg}{2l\cos{30^{\circ}}}=\frac{mg}{\sqrt{3}\,l}\;\;\mbox{and} \;\;
T=\frac{mg}{\sqrt{3}\,l}(l+\sqrt{3}\,x).$$ Next we have,
$$\cos{(\alpha+\Delta\alpha)}=\frac{\frac{\sqrt{3}}{2}\,l+x}{\sqrt{(\frac
{\sqrt{3}}{2}\,l+x)^{2}+(\frac{l}{2})^{2}}}\approx\frac{\sqrt{3}+2\frac{x}{l}}
{2}\,\frac{1}{\sqrt{1+\sqrt{3}\,\frac{x}{l}}}\approx\frac{\sqrt{3}}{2}+
\frac{x}{4l}.$$
Therefore, the restoring force  is
$$F=2\frac{mg}{\sqrt{3}\,l}(l+\sqrt{3}\,x)\left(\frac{\sqrt{3}}{2}+
\frac{x}{4l}\right)-mg=\frac{2mg}{\sqrt{3}\,l}\left(\frac{\sqrt{3}}{2}\,l+
\frac{7}{4}x\right)-mg=\frac{7mg}{2\sqrt{3}\,l}\;x.$$
Consequently, the effective spring constant emerges
$$k_{\mbox{eff}}=\frac{7mg}{2\sqrt{3}\,l}$$ and the frequency of small 
vertical oscillations is
$$\omega=\sqrt{\frac{k_{\mbox{eff}}}{m}}=\sqrt{\frac{7g}{2\sqrt{3}\,l}}.$$

\subsection{}
Переданная энергия $E=\frac{m}{2}|z|^{2}$, где
$$z=\frac{F_{0}}{m}e^{i\omega
t}\int\limits_{-\infty}^{t}e^{-\frac{t'^{2}}{\tau^{2}}}e^{-i\omega
t'}dt', \; t\rightarrow\infty.$$ 
Но $|e^{i\omega t}|=1$, поэтому
$$|z|^{2}=\frac{F_{0}^{2}}{m^{2}}\;\left|\int\limits_{-\infty}^{\infty}
e^{-\frac{t'^{2}}{\tau^{2}}-i\omega
t'}dt'\right|^{2}.$$ 
Но $$\frac{t'^{2}}{\tau^{2}}+i\omega
t'=\frac{t'^{2}}{\tau^{2}}+2i\omega\,\frac{t'}{\tau}\,\frac{\tau}{2}+
\left(\frac{i\omega\tau}{2}\right)^{2}-\left(\frac{i\omega\tau}{2}
\right)^{2}=\left(\frac{t'}{\tau}+i\frac{\omega\tau}{2}\right)^{2}+
\frac{\omega^{2}\tau^{2}}{4}.$$ 
Поэтому
$$\int\limits_{-\infty}^{\infty}e^{-\frac{t'^{2}}{\tau^{2}}-i\omega
t'}dt'=e^{-\frac{\omega^{2}\tau^{2}}{4}}\int\limits_{-\infty}^{\infty}
e^{-(\frac{t'}{\tau}+i\frac{\omega\tau}{2})^{2}}dt'=\tau
e^{-\frac{\omega^{2}\tau^{2}}{4}}\int\limits_{-\infty}^{\infty}e^{-z^{2}}dz=
\sqrt{\pi}\,\tau\, e^{-\frac{\omega^{2}\tau^{2}}{4}}.$$ 
Следовательно, передается энергия 
$$E(\tau)=\frac{\pi F_{0}^{2}}{2m}\,\tau^{2}\,e^{-\frac{\omega^{2}
\tau^{2}}{2}}.$$ 
Найдем максимум функции $f(x)=xe^{-\alpha x}$. $f'(x)=e^{-\alpha
x}-\alpha xe^{-\alpha x}=0$ дает $x=\frac{1}{\alpha}$.
Следовательно, максимально эффективно энергия передается при
$$\tau^{2}=\frac{2}{\omega^{2}}\;\;\mbox{и}\;\;
E_{max}=\frac{\pi F_{0}^{2}}{2m}\,\frac{2}{\omega^{2}}\,e^{-1}=
\frac{\pi F_{0}^{2}}{m\omega^{2}\,e}.$$

\vspace*{5mm}
\noindent $\bullet$  $\bullet$  $\bullet$
 
\noindent
The transferred energy is $E=\frac{m}{2}|z|^{2}$, where
$$z=\frac{F_{0}}{m}e^{i\omega
t}\int\limits_{-\infty}^{t}e^{-\frac{t'^{2}}{\tau^{2}}}e^{-i\omega
t'}dt', \; t\rightarrow\infty.$$ 
But $|e^{i\omega t}|=1$, therefore,
$$|z|^{2}=\frac{F_{0}^{2}}{m^{2}}\;\left|\int\limits_{-\infty}^{\infty}
e^{-\frac{t'^{2}}{\tau^{2}}-i\omega
t'}dt'\right|^{2}.$$ 
Moreover, $$\frac{t'^{2}}{\tau^{2}}+i\omega
t'=\frac{t'^{2}}{\tau^{2}}+2i\omega\,\frac{t'}{\tau}\,\frac{\tau}{2}+
\left(\frac{i\omega\tau}{2}\right)^{2}-\left(\frac{i\omega\tau}{2}
\right)^{2}=\left(\frac{t'}{\tau}+i\frac{\omega\tau}{2}\right)^{2}+
\frac{\omega^{2}\tau^{2}}{4}.$$ 
Hence
$$\int\limits_{-\infty}^{\infty}e^{-\frac{t'^{2}}{\tau^{2}}-i\omega
t'}dt'=e^{-\frac{\omega^{2}\tau^{2}}{4}}\int\limits_{-\infty}^{\infty}
e^{-(\frac{t'}{\tau}+i\frac{\omega\tau}{2})^{2}}dt'=\tau
e^{-\frac{\omega^{2}\tau^{2}}{4}}\int\limits_{-\infty}^{\infty}e^{-z^{2}}dz=
\sqrt{\pi}\,\tau\, e^{-\frac{\omega^{2}\tau^{2}}{4}}.$$ 
Consequently, we get for the transferred energy 
$$E(\tau)=\frac{\pi F_{0}^{2}}{2m}\,\tau^{2}\,e^{-\frac{\omega^{2}
\tau^{2}}{2}}.$$ 
Let us find the maximum of the function $f(x)=xe^{-\alpha x}$. The  
condition $f'(x)=e^{-\alpha x}-\alpha xe^{-\alpha x}=0$ gives 
$x=\frac{1}{\alpha}$.
Therefore, the maximally effective energy transfer takes place at
$$\tau^{2}=\frac{2}{\omega^{2}}\;\;\mbox{and}\;\;
E_{max}=\frac{\pi F_{0}^{2}}{2m}\,\frac{2}{\omega^{2}}\,e^{-1}=
\frac{\pi F_{0}^{2}}{m\omega^{2}\,e}.$$

\subsection{}
Уравнение движения
$$\ddot{x}+\omega^{2}x+\frac{\alpha}{m}\dot{x}=\frac{F(t)}{m}.$$
Обозначим $\frac{\alpha}{m}=2\beta$. Надо
решить уравнение
$$\ddot{x}+2\beta\dot{x}+\omega^{2}x=\frac{F(t)}{m}.$$ Пусть
$z=\dot{x}+\lambda_{1}x$ и подберем $\lambda_{2}$ так, чтобы было
$\dot{z}+\lambda_{2}z=\ddot{x}+2\beta\dot{x}+\omega^{2}x$. Но
$\dot{z}+\lambda_{2}z=\ddot{x}+\lambda_{1}\dot{x}+\lambda_{2}\dot{x}+
\lambda_{1}\lambda_{2}\,x$.
Т.е. $$\left\{
\begin{array}{ll}
 2\beta=\lambda_{1}+\lambda_{2}\\
\omega^{2}=\lambda_{1}\lambda_{2}\\
\end{array}\right.$$
и $\lambda_{1}, \lambda_{2}$ есть решения уравнения
$\lambda^{2}-2\beta\lambda+\omega^{2}=0$. Т.е.
$$\lambda_{1,2}=\beta\pm\sqrt{\beta^{2}-\omega^{2}}=\beta\pm
i\Omega,\;\; \mbox{где} \;\Omega=\sqrt{\omega^{2}-\beta^{2}}.$$ Будем
предполагать, что $\omega>\beta$, т.е. трение маленькое. Пусть
$\lambda_{1}=\beta+i\Omega$ и $\lambda_{2}=\beta-i\Omega$. Тогда
$z=\dot{x}+\beta x+i\Omega x$ и уравнение будет
$$\dot{z}+(\beta-i\Omega)z=\frac{F(t)}{m}.$$ Решение однородного
уравнения пропорционально $e^{-(\beta-i\Omega)t}$. Поэтому возьмем
$z=A(t)e^{-(\beta-i\Omega)t}$. Для A(t) получим уравнение
$$\dot{A}(t)=\frac{F(t)}{m}e^{(\beta-i\Omega)t}.$$ Окончательно
$$z=e^{-(\beta-i\Omega)t}\left\{z_{0}+\int\limits_{0}^{t}\frac{F(t')}{m}
e^{(\beta-i\Omega)t'}dt'\right\}.$$

\vspace*{5mm}
\noindent $\bullet$  $\bullet$  $\bullet$
 
\noindent
The equation of motion is
$$\ddot{x}+\omega^{2}x+\frac{\alpha}{m}\dot{x}=\frac{F(t)}{m}.$$
Let us denote $\frac{\alpha}{m}=2\beta$. We have to solve the equation
$$\ddot{x}+2\beta\dot{x}+\omega^{2}x=\frac{F(t)}{m}.$$ Let
$z=\dot{x}+\lambda_{1}x$ and choose $\lambda_{2}$ so that
$\dot{z}+\lambda_{2}z=\ddot{x}+2\beta\dot{x}+\omega^{2}x$. But
$\dot{z}+\lambda_{2}z=\ddot{x}+\lambda_{1}\dot{x}+\lambda_{2}\dot{x}+
\lambda_{1}\lambda_{2}\,x$.
That is, $$\left\{
\begin{array}{ll}
 2\beta=\lambda_{1}+\lambda_{2}\\
\omega^{2}=\lambda_{1}\lambda_{2}\\
\end{array}\right.$$
and $\lambda_{1}, \lambda_{2}$ are solutions of the equation 
$\lambda^{2}-2\beta\lambda+\omega^{2}=0$. That is,
$$\lambda_{1,2}=\beta\pm\sqrt{\beta^{2}-\omega^{2}}=\beta\pm
i\Omega,\;\; \mbox{where} \;\;\; \Omega=\sqrt{\omega^{2}-\beta^{2}}.$$ 
We will assume that the friction is small and, therefore, $\omega>\beta$. 
Let $\lambda_{1}=\beta+i\Omega$ and $\lambda_{2}=\beta-i\Omega$. Then
$z=\dot{x}+\beta x+i\Omega x$ and the equation takes the form 
$$\dot{z}+(\beta-i\Omega)z=\frac{F(t)}{m}.$$ Solution of the homogeneous 
equation is proportional to $e^{-(\beta-i\Omega)t}$. Therefore, we will take
$z=A(t)e^{-(\beta-i\Omega)t}$. For A(t) we get the equation
$$\dot{A}(t)=\frac{F(t)}{m}e^{(\beta-i\Omega)t}.$$ Finally,
$$z=e^{-(\beta-i\Omega)t}\left\{z_{0}+\int\limits_{0}^{t}\frac{F(t')}{m}
e^{(\beta-i\Omega)t'}dt'\right\}.$$

\vspace*{5mm}
\subsection{}
\noindent $ $

\begin{figure}[htb]
\centerline{\epsfig{figure=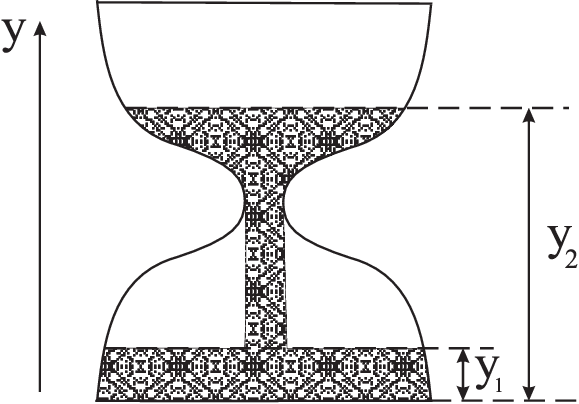,height=4cm}}
\end{figure}
\noindent Пусть M -- масса песочных часов (включая песок). Положение
центра масс определяется уравнением
$$My_{\mbox{ц.м.}}=\int\limits_{0}^{y_{1}(t)}y\rho
S(y)dy+\int\limits_{h}^{y_{2}(t)}y\rho S(y)dy+C,$$ где
$h$ -- высота узкого места часов, $\rho$ -- плотность песка, $S(y)$ есть 
поперечное сечение часов на
высоте $y$ и $C$ соответствует вкладу в $My_{\mbox{ц.м.}}$ от каркаса
песочных часов и от той части песка, которая в полете. Т.к. поток
стационарный, $C$ не зависит от времени. Заметим, что
$$\frac{d}{dt}\int\limits_{0}^{f(t)}F(x)dx=\lim\limits_{\Delta
t\rightarrow 0}\frac{1}{\Delta t}\left[\int\limits_{0}^{f(t+\Delta
t)}F(x)dx-\int\limits_{0}^{f(t)}F(x)dx\right]=$$ $$=\lim\limits_{\Delta
t\rightarrow 0}\frac{1}{\Delta
t}\int\limits_{f(t)}^{f(t)+f'(t)\Delta t}F(x)dx=f'(t)F[f(t)],$$
т.к. 
$$f(t+\Delta t)\approx f(t)+f'(t)\Delta t\;\;\mbox{и} \;\;
\int\limits_{a}^{a+\Delta a}F(x)dx\approx F(a)\Delta a.$$ 
Поэтому будем иметь
$$M\dot{y}_{\mbox{ц.м.}}=\rho\left[\dot{y_{1}}y_{1}S(y_{1})+
\dot{y_{2}}y_{2}S(y_{2})\right].$$
Если поток песка равен $\mu$, то $\mu=\rho
S(y_{1})\,\dot{y_{1}}=-\rho S(y_{2})\,\dot{y_{2}}$, т.к. $\rho
S(y_{1})\Delta y_{1}=\rho S(y_{2})(-\Delta y_{2})$ есть количество
песка, которое переносится из верхней половины в нижнюю за время
$\Delta t$. Следовательно, $M\dot{y}_{\mbox{ц.м.}}= \mu(y_{1}-y_{2})$. В
начальный момент времени $\dot{y}_{\mbox{ц.м.}}=-\frac{\mu}{M}H$, $H$ -
высота часов (вернее, высота верхнего уровня песка при $t=0$). Эту
скорость центр масс приобретает в начальный период установления
стационарного потока. Дифференцируя еще раз, имеем
$M\ddot{y}_{\mbox{ц.м.}}=\mu(\dot{y_{1}}-\dot{y_{2}})$. Но
$$\dot{y_{1}}=\frac{\mu}{\rho S(y_{1})}\;\;\mbox{и} \;\;
\dot{y_{2}}=-\frac{\mu}{\rho S(y_{2})}.$$ 
Следовательно,
$$M\ddot{y}_{\mbox{ц.м.}}=\frac{\mu^{2}}{\rho}\left[\frac{1}{S(y_{1})}+
\frac{1}{S(y_{2})}\right]>0.$$
Т.е. ускорение направлено вверх, несмотря на то что центр масс
опускается вниз. С другой стороны, $M\ddot{y}_{\mbox{ц.м.}}=F-Mg$, где
$F$ - показание весов. Поэтому $F>Mg$ всегда, пока существует
стационарный поток песка.

\vspace*{5mm}
\noindent $\bullet$  $\bullet$  $\bullet$
 
\begin{figure}[htb]
\centerline{\epsfig{figure=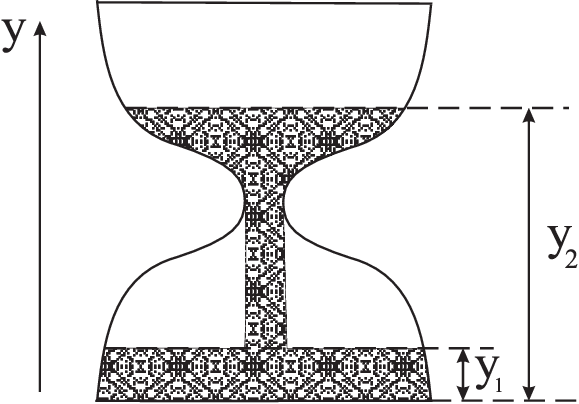,height=4cm}}
\end{figure}\noindent
Let $M$ be the mass of the hourglass (the mass of sand included). The center
of mass position is determined by equation
$$My_{\mbox{cm}}=\int\limits_{0}^{y_{1}(t)}y\rho
S(y)dy+\int\limits_{h}^{y_{2}(t)}y\rho S(y)dy+C,$$ where
$h$ is the height of the hourglass bottleneck, $\rho$ is the density of sand, 
$S(y)$ is hourglass cross-section at a height $y$ and $C$ corresponds to the 
contribution in $My_{\mbox{cm}}$ from the hourglass frame and the part of the 
sand, which is in flight. Since the flow is steady-state, $C$ does not depend 
on time. Note that
$$\frac{d}{dt}\int\limits_{0}^{f(t)}F(x)dx=\lim\limits_{\Delta
t\rightarrow 0}\frac{1}{\Delta t}\left[\int\limits_{0}^{f(t+\Delta
t)}F(x)dx-\int\limits_{0}^{f(t)}F(x)dx\right]=$$ $$=\lim\limits_{\Delta
t\rightarrow 0}\frac{1}{\Delta
t}\int\limits_{f(t)}^{f(t)+f'(t)\Delta t}F(x)dx=f'(t)F[f(t)],$$
because 
$$f(t+\Delta t)\approx f(t)+f'(t)\Delta t\;\;\mbox{and} \;\;
\int\limits_{a}^{a+\Delta a}F(x)dx\approx F(a)\Delta a.$$ 
Therefore, we have
$$M\dot{y}_{\mbox{cm}}=\rho\left[\dot{y_{1}}y_{1}S(y_{1})+
\dot{y_{2}}y_{2}S(y_{2})\right].$$
If the sand flux is $\mu$, then $\mu=\rho
S(y_{1})\,\dot{y_{1}}=-\rho S(y_{2})\,\dot{y_{2}}$, because $\rho
S(y_{1})\Delta y_{1}=\rho S(y_{2})(-\Delta y_{2})$ is the amount of sand 
which is transferred from the upper half to the lower one during the time
$\Delta t$. Consequently, $M\dot{y}_{\mbox{cm}}= \mu(y_{1}-y_{2})$. At the
initial moment $\dot{y}_{\mbox{cm}}=-\frac{\mu}{M}H$, $H$ being the height
of the hourglass (or rather, the height of the top level of the sand at $t=0$).
The center of mass gains this initial velocity in the brief period of 
establishing the steady flow. Differentiating again, we have
$M\ddot{y}_{\mbox{ц.м.}}=\mu(\dot{y_{1}}-\dot{y_{2}})$. But
$$\dot{y_{1}}=\frac{\mu}{\rho S(y_{1})}\;\;\mbox{and} \;\;
\dot{y_{2}}=-\frac{\mu}{\rho S(y_{2})}.$$ 
Therefore,
$$M\ddot{y}_{\mbox{cm}}=\frac{\mu^{2}}{\rho}\left[\frac{1}{S(y_{1})}+
\frac{1}{S(y_{2})}\right]>0.$$
That is, the center of mass acceleration is directed upward despite the fact 
that the center of mass descends. On the other hand, $M\ddot{y}_{\mbox{cm}}=
F-Mg$, where $F$ is the reading of the balance. Therefore, $F>Mg$ always 
as long as there is a steady flow of sand.

\subsection{}
Энергия $E=mc^{2}\gamma+\frac{1}{2}m\omega^{2}x^{2}$
сохраняется. В начальный момент $x=A$ и $\gamma=1$, поэтому
$E=mc^{2}+\frac{1}{2}m\omega^{2}A^{2}$. Следовательно,
$$\gamma=\frac{E}{mc^{2}}-\frac{\omega^{2}}{2c^{2}}\,x^{2}=
1+\frac{\omega^{2}}{2c^{2}}(A^{2}-x^{2}).$$
Период колебаний
$$T=4\int\limits_{0}^{A}\frac{dx}{V}=\frac{4}{c}\int\limits_{0}^{A}
\frac{dx}{\beta}.$$
Но
$\beta=\sqrt{1-\frac{1}{\gamma^{2}}}=\frac{\sqrt{\gamma^{2}-1}}{\gamma}$.
Следовательно,
$$T=\frac{4}{c}\int\limits_{0}^{A}\frac{\gamma}{\sqrt{\gamma^{2}-1}}\,dx=
\frac{8c}{\omega^{2}}\frac{d}{dA^{2}}\int\limits_{0}^{A}
\sqrt{\gamma^{2}-1}\,dx.$$
Т.к. $$y=\frac{\omega^{2}}{2c^{2}}(A^{2}-x^{2})\ll 1,$$ то
$$\sqrt{\gamma^{2}-1}=\sqrt{(1+y)^{2}-1}\approx\sqrt{2y\left(1+
\frac{y}{2}\right)}\approx\sqrt{2y}\left(1+\frac{y}{4}\right).$$
Поэтому $T=T_{0}+T_{1}$, где
$$T_{0}=\frac{8c}{\omega^{2}}\,\frac{d}{dA^{2}}\int\limits_{0}^{A}\sqrt{\frac{
\omega^{2}}{c^{2}}(A^{2}-x^{2})}\;dx=\frac{8}{\omega}\,\frac{d}{dA^{2}}\int
\limits_{0}^{A}\sqrt{A^{2}-x^{2}}\,dx.$$
Сделаем подстановку $x=A \sin\varphi$. Тогда
$$\int\limits_{0}^{A}\sqrt{A^{2}-x^{2}}\,dx=\int\limits_{0}^{\frac{\pi}{2}}
A^{2}\cos^{2}\varphi\,d\varphi=\frac{A^{2}}{2}\int\limits_{0}^{\frac{\pi}{2}}
(1+\cos2\varphi)\,d\varphi=\frac{\pi A^{2}}{4}$$ и $T_{0}=\frac{2\pi}{\omega}$.
$$T_{1}=\frac{\omega}{c^{2}}\,\frac{d}{dA^{2}}\int\limits_{0}^{A}(A^{2}-x^{2})^
{3/2}dx=\frac{\omega}{c^{2}}\,\frac{d}{dA^{2}}\left[A^{4}\int
\limits_{0}^{\frac{\pi}{2}}\cos^{4}\varphi\, d\varphi\right].$$ Но
$$\cos^{4}\varphi=\left[\frac{1}{2}(1+\cos2\varphi)\right]^{2}=\frac{1}{4}
\left[1+2\cos2\varphi+\frac{1}{2}(1+\cos4\varphi)\right]$$
и $$\int\limits_{0}^{\frac{\pi}{2}}\cos^{4}\varphi\,
d\varphi=\frac{\pi}{2}\,\frac{1}{4}\left(1+\frac{1}{2}\right)=
\frac{3}{16}\pi.$$
Далее, $$\frac{dA^{4}}{dA^{2}}=2A^{2}\;\;\mbox{и}\;\;
T_{1}=\frac{3\pi}{8}\,\frac{\omega A^{2}}{c^{2}}.$$ Окончательно,
$$T=\frac{2\pi}{\omega}\left[1+\frac{3}{16}\frac{\omega^{2}A^{2}}
{c^{2}}\right].$$
Замечание:
$$\frac{d}{dA^{2}}\int\limits_{0}^{A}\sqrt{\gamma^{2}-1}\,dx=
\int\limits_{0}^{A}\left(\frac{\partial}{\partial
A^{2}}\,\sqrt{\gamma^{2}-1}\right)dx+\left .\frac{dA}{dA^{2}}\,
\sqrt{\gamma^{2}(x,A)-1}\,\right|_{x=A}.$$
Но второй член зануляется, т.к. $\gamma(x,A)=1$ при $x=A$.

\vspace*{5mm}
\noindent $\bullet$  $\bullet$  $\bullet$
 
\noindent
The energy $E=mc^{2}\gamma+\frac{1}{2}m\omega^{2}x^{2}$
is conserved. Initially, $x=A$ and $\gamma=1$, that imply
$E=mc^{2}+\frac{1}{2}m\omega^{2}A^{2}$. Consequently,
$$\gamma=\frac{E}{mc^{2}}-\frac{\omega^{2}}{2c^{2}}\,x^{2}=
1+\frac{\omega^{2}}{2c^{2}}(A^{2}-x^{2}).$$
The period of oscillations is
$$T=4\int\limits_{0}^{A}\frac{dx}{V}=\frac{4}{c}\int\limits_{0}^{A}
\frac{dx}{\beta}.$$
But
$\beta=\sqrt{1-\frac{1}{\gamma^{2}}}=\frac{\sqrt{\gamma^{2}-1}}{\gamma}$.
Therefore,
$$T=\frac{4}{c}\int\limits_{0}^{A}\frac{\gamma}{\sqrt{\gamma^{2}-1}}\,dx=
\frac{8c}{\omega^{2}}\frac{d}{dA^{2}}\int\limits_{0}^{A}
\sqrt{\gamma^{2}-1}\,dx.$$
Since $$y=\frac{\omega^{2}}{2c^{2}}(A^{2}-x^{2})\ll 1,$$ we have
$$\sqrt{\gamma^{2}-1}=\sqrt{(1+y)^{2}-1}\approx\sqrt{2y\left(1+
\frac{y}{2}\right)}\approx\sqrt{2y}\left(1+\frac{y}{4}\right).$$
Hence $T=T_{0}+T_{1}$, where
$$T_{0}=\frac{8c}{\omega^{2}}\,\frac{d}{dA^{2}}\int\limits_{0}^{A}\sqrt{\frac{
\omega^{2}}{c^{2}}(A^{2}-x^{2})}\;dx=\frac{8}{\omega}\,\frac{d}{dA^{2}}\int
\limits_{0}^{A}\sqrt{A^{2}-x^{2}}\,dx.$$
Let us make the substitution $x=A \sin\varphi$. Then
$$\int\limits_{0}^{A}\sqrt{A^{2}-x^{2}}\,dx=\int\limits_{0}^{\frac{\pi}{2}}
A^{2}\cos^{2}\varphi\,d\varphi=\frac{A^{2}}{2}\int\limits_{0}^{\frac{\pi}{2}}
(1+\cos2\varphi)\,d\varphi=\frac{\pi A^{2}}{4}$$ and 
$T_{0}=\frac{2\pi}{\omega}$.
$$T_{1}=\frac{\omega}{c^{2}}\,\frac{d}{dA^{2}}\int\limits_{0}^{A}(A^{2}-x^{2})^
{3/2}dx=\frac{\omega}{c^{2}}\,\frac{d}{dA^{2}}\left[A^{4}\int
\limits_{0}^{\frac{\pi}{2}}\cos^{4}\varphi\, d\varphi\right].$$ But
$$\cos^{4}\varphi=\left[\frac{1}{2}(1+\cos2\varphi)\right]^{2}=\frac{1}{4}
\left[1+2\cos2\varphi+\frac{1}{2}(1+\cos4\varphi)\right]$$
and $$\int\limits_{0}^{\frac{\pi}{2}}\cos^{4}\varphi\,
d\varphi=\frac{\pi}{2}\,\frac{1}{4}\left(1+\frac{1}{2}\right)=
\frac{3}{16}\pi.$$
Further, we have $$\frac{dA^{4}}{dA^{2}}=2A^{2}\;\;\mbox{and}\;\;
T_{1}=\frac{3\pi}{8}\,\frac{\omega A^{2}}{c^{2}}.$$ Finally,
$$T=\frac{2\pi}{\omega}\left[1+\frac{3}{16}\frac{\omega^{2}A^{2}}
{c^{2}}\right].$$
Note:
$$\frac{d}{dA^{2}}\int\limits_{0}^{A}\sqrt{\gamma^{2}-1}\,dx=
\int\limits_{0}^{A}\left(\frac{\partial}{\partial
A^{2}}\,\sqrt{\gamma^{2}-1}\right)dx+\left .\frac{dA}{dA^{2}}\,
\sqrt{\gamma^{2}(x,A)-1}\,\right|_{x=A}.$$
But the second term vanishes because $\gamma(x,A)=1$ when $x=A$.

\section*{Контрольная работа 11}
\setcounter{section}{11}
\setcounter{subsection}{0}
\subsection{}
Пусть $x_{1}$ и $x_{2}$ смещения шариков из положения равновесия
(т. к. колебания маленькие, вертикальными смешениями
пренебрегаем). Уравнения движения будут
$ m\ddot{x_{1}} = -m\omega^{2}x_{1}+k(x_{2}-x_{1})$ и
$ m\ddot{x_{2}} = -m\omega^{2}x_{2}-k(x_{2}-x_{1})$ или
$\ddot{x_{1}}+\left(\omega^{2}+\frac{k}{m}\right)x_{1}=\frac{k}{m}x_{2}$,
$\ddot{x_{2}}+\left(\omega^{2}+\frac{k}{m}\right)x_{2}=\frac{k}{m}x_{1}$.
Вводя новые переменные $X=x_{1}+x_{2}$ и $x=x_{1}-x_{2}$, получаем
$\ddot{X}+\omega^{2}X=0$ и $\ddot{x}+\left(\omega^{2}+2\frac{k}{m}
\right)x=0$. В начальный момент времени $x_{1}(0)=A,\; x_{2}(0)=0,\;
\dot{x_{1}}=\dot{x_{2}}=0$. Т.е. $X(0)=x(0)=A$ и
$\dot{X_{0}}=\dot{x_{0}}=0$. Поэтому $X(t)=A\cos{\omega t}$ и
$x(t)=A\cos{\sqrt{\omega^{2}+2\frac{k}{m}}\,t}$. Но $x_{1}(t)=(X+x)/2,\;
x_{2}=(X-x)/2$. Следовательно,
$$x_{1}=\frac{A}{2}\left [\cos{\omega t}+\cos{\sqrt{\omega^{2}+
2\frac{k}{m}}\,t}
\right ]$$ и
$$x_{2}(t)=\frac{A}{2}\left [\cos{\omega t}-\cos{\sqrt{\omega^{2}+
2\frac{k}{m}}\,t}\right ].$$ Но
$$\sqrt{\omega^{2}+2\frac{k}{m}}\approx\omega+\frac{k}{m\omega}$$ и
$$\cos{\alpha}+\cos{\beta}=2\cos{\frac{\alpha+\beta}{2}}\cos{\frac{
\alpha-\beta}{2}},\;
\cos{\alpha}-\cos{\beta}=2\sin{\frac{\alpha+\beta}{2}}\sin{\frac{
\beta-\alpha}{2}}.$$
Поэтому 
$$x_{1}(t)\approx A\cos{\omega t}\,\cos{\frac{kt}{2m\omega}}, \;\;
x_{2}(t)=A\sin{\omega t}\,\sin{\frac{kt}{2m\omega}}.$$ 
Это показывает, что когда $\frac{kt}{2m\omega}=\frac{\pi}{2}$, колебания 
первого шарика затухают, а колебания второго максимальны. Следовательно, через
время  $$t=\frac{\pi m \omega}{k}=T\frac{m\omega^2}{2k}$$
колебания полностью перейдут ко второму маятнику. Здесь $T=\frac{2\pi}{\omega}$
есть период колебаний, когда шарики не связаны.

\vspace*{5mm}
\noindent $\bullet$  $\bullet$  $\bullet$
 
\noindent
Let $x_{1}$ and $x_{2}$ are horizontal displacements of the balls from the 
equilibrium state (since the oscillations are small, we neglect vertical
displacements). The equations of motion are
$ m\ddot{x_{1}} = -m\omega^{2}x_{1}+k(x_{2}-x_{1})$ and
$ m\ddot{x_{2}} = -m\omega^{2}x_{2}-k(x_{2}-x_{1})$ or
$\ddot{x_{1}}+\left(\omega^{2}+\frac{k}{m}\right)x_{1}=\frac{k}{m}x_{2}$,
$\ddot{x_{2}}+\left(\omega^{2}+\frac{k}{m}\right)x_{2}=\frac{k}{m}x_{1}$.
Introducing new variables $X=x_{1}+x_{2}$ and $x=x_{1}-x_{2}$, we get
$\ddot{X}+\omega^{2}X=0$ and $\ddot{x}+\left(\omega^{2}+2\frac{k}{m}
\right)x=0$. At the initial moment $x_{1}(0)=A,\; x_{2}(0)=0,\;
\dot{x_{1}}=\dot{x_{2}}=0$. That is, $X(0)=x(0)=A$ and
$\dot{X_{0}}=\dot{x_{0}}=0$. Hence $X(t)=A\cos{\omega t}$ and
$x(t)=A\cos{\sqrt{\omega^{2}+2\frac{k}{m}}\,t}$. But $x_{1}(t)=(X+x)/2,\;
x_{2}=(X-x)/2$. Consequently,
$$x_{1}=\frac{A}{2}\left [\cos{\omega t}+\cos{\sqrt{\omega^{2}+
2\frac{k}{m}}\,t}
\right ]$$ and
$$x_{2}(t)=\frac{A}{2}\left [\cos{\omega t}-\cos{\sqrt{\omega^{2}+
2\frac{k}{m}}\,t}\right ].$$ However,
$$\sqrt{\omega^{2}+2\frac{k}{m}}\approx\omega+\frac{k}{m\omega}$$ and
$$\cos{\alpha}+\cos{\beta}=2\cos{\frac{\alpha+\beta}{2}}\cos{\frac{
\alpha-\beta}{2}},\;
\cos{\alpha}-\cos{\beta}=2\sin{\frac{\alpha+\beta}{2}}\sin{\frac{
\beta-\alpha}{2}}.$$
Therefore, 
$$x_{1}(t)\approx A\cos{\omega t}\,\cos{\frac{kt}{2m\omega}}, \;\;
x_{2}(t)=A\sin{\omega t}\,\sin{\frac{kt}{2m\omega}}.$$ 
This shows that when $\frac{kt}{2m\omega}=\frac{\pi}{2}$, the oscillations
of the first ball decay, and the oscillations of the second ball are maximal. 
Hence it takes the time $$t=\frac{\pi m \omega}{k}=T\frac{m\omega^2}{2k}$$
the oscillations to completely pass to the second pendulum. Here, 
$T=\frac{2\pi}{\omega}$ is the period of oscillations when the balls are not 
related.

\subsection{}
Перейдем к цилиндрическим координатам. Тогда
$V^{2}=\dot{\rho}^{2}+\rho^{2}\dot{\varphi}^{2}+\dot{z}^{2}$ и
$x\dot{y}-y\dot{x}=\rho\cos{\varphi}\,(\dot{\rho}\sin{\varphi}+
\rho\dot{\varphi}\cos{\varphi})-\rho\sin{\varphi}\,(\dot{\rho}\cos{\varphi}
-\rho\dot{\varphi}\sin{\varphi})=\rho^{2}\dot{\varphi}$.
Поэтому
$${\cal{L}}=\frac{m}{2}(\dot{\rho}^{2}+\rho^{2}\dot{\varphi}^{2}+\dot{z}^{2})
+\frac{qB}{2c}\rho^{2}\dot{\varphi}$$
и обобщенный импульс, соответствующий координате $\varphi$,
будет
$$p_{\varphi}=\frac{\partial {\cal{L}}}{\partial \dot{\varphi}}=m\rho^{2}
\dot{\varphi}+\frac{qB}{2c}\rho^{2}.$$
$\oint p_{\varphi}\,d\varphi$ адиабатический инвариант. Но когда электрон 
вращается в магнитном поле $B$, его угловая скорость $\omega=\dot{\varphi}$
определяется из $m\omega^{2}\rho=\frac{q}{c}\,VB=\frac{qB}{c}\omega\rho$. 
Т.е. $\dot{\varphi}=\frac{qB}{mc}$. Поэтому
$$\oint p_{\varphi}\,d\varphi=\oint\left (\frac{qB}{c}\rho^{2}+
\frac{qB}{2c}\rho^{2}\right )d\varphi=3\pi\frac{qB}{c}\rho^{2}.$$
Следовательно, произведение  $B\rho^{2}$ является инвариантом и когда поле
увеличится в два раза, радиус орбиты электрона
уменьшится $\sqrt{2}$ раза. Заметим, что инвариантность $\rho^{2}B$
означает, что поток магнитной индукции сквозь орбиту электрона
является адиабатическим инвариантом.

\vspace*{5mm}
\noindent $\bullet$  $\bullet$  $\bullet$
 
\noindent
Let us pass to cylindrical coordinates. Then
$V^{2}=\dot{\rho}^{2}+\rho^{2}\dot{\varphi}^{2}+\dot{z}^{2}$ и
$x\dot{y}-y\dot{x}=\rho\cos{\varphi}\,(\dot{\rho}\sin{\varphi}+
\rho\dot{\varphi}\cos{\varphi})-\rho\sin{\varphi}\,(\dot{\rho}\cos{\varphi}
-\rho\dot{\varphi}\sin{\varphi})=\rho^{2}\dot{\varphi}$.
Therefore,
$${\cal{L}}=\frac{m}{2}(\dot{\rho}^{2}+\rho^{2}\dot{\varphi}^{2}+\dot{z}^{2})
+\frac{qB}{2c}\rho^{2}\dot{\varphi}$$
and the generalized momentum corresponding to the $\varphi$ coordinate is
$$p_{\varphi}=\frac{\partial {\cal{L}}}{\partial \dot{\varphi}}=m\rho^{2}
\dot{\varphi}+\frac{qB}{2c}\rho^{2}.$$
$\oint p_{\varphi}\,d\varphi$ is an adiabatic invariant. But when the electron
rotates in a magnetic field $B$, its angular velocity $\omega=\dot{\varphi}$
is determined by $m\omega^{2}\rho=\frac{q}{c}\,VB=\frac{qB}{c}\omega\rho$.
That is, $\dot{\varphi}=\frac{qB}{mc}$. Therefore,
$$\oint p_{\varphi}\,d\varphi=\oint\left (\frac{qB}{c}\rho^{2}+
\frac{qB}{2c}\rho^{2}\right )d\varphi=3\pi\frac{qB}{c}\rho^{2}.$$
Consequently, the product $B\rho^{2}$ is invariant and when the field
strength is doubled, the radius of the electron's orbit will be
reduced by $\sqrt{2}$ times. Note that the invariance of $\rho^{2}B$
means that the magnetic flux through the electron's orbit is an adiabatic 
invariant.

\clearpage
\subsection{}
\begin{figure}[htb]
\centerline{\epsfig{figure=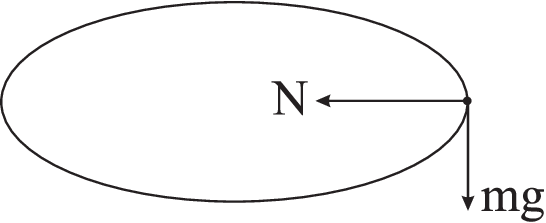,height=2cm}}
\end{figure}
\noindent
На кольцо действует как сила притяжения $mg$, так и сила реакции $N$. Поэтому 
сила трения $F=\mu\sqrt{N^{2}+m^{2}g^{2}}$. Но $N$ сообщает кольцу
центростремительное ускорение. Поэтому  $mV^{2}/R=N$ и
$$F=\mu mg\sqrt{1+\frac{V^4}{g^2R^2}}.$$
Следовательно, скорость кольца уменьшается по закону
$$m\frac{dV}{dt}=-F=-\mu mg\sqrt{1+\frac{V^4}{g^2R^2}}$$
и кольцо остановится через время
$$T=\frac{1}{\mu g}\int\limits_0^{V_0}\frac{dV}{\sqrt{1+\frac{V^4}{g^2R^2}}}
\approx \frac{1}{\mu g}\int\limits_0^{V_0}\left (1-\frac{V^4}{2g^2R^2}\right )
dV=\frac{1}{\mu g}\left [V_0-\frac{V_0^5}{10 g^2 R^2}\right ].$$
Окончательно 
$$T=\frac{V_0}{\mu g}\left [1-\frac{V_0^4}{10 g^2 R^2}\right ].$$

\vspace*{5mm}
\noindent $\bullet$  $\bullet$  $\bullet$
 
\begin{figure}[htb]
\centerline{\epsfig{figure=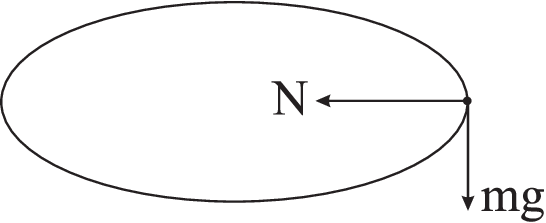,height=2cm}}
\end{figure}

\noindent
Both the force of gravity $mg$ and the reaction force $N$ are acting on the 
ring. Therefore, the friction force is $F=\mu\sqrt{N^{2}+m^{2}g^{2}}$. But 
the reaction force  $N$ gives the ring a centripetal acceleration. Therefore,
$mV^{2}/R=N$ and $$F=\mu mg\sqrt{1+\frac{V^4}{g^2R^2}}.$$
Consequently, the velocity of the ring decreases in accordance with the law
$$m\frac{dV}{dt}=-F=-\mu mg\sqrt{1+\frac{V^4}{g^2R^2}}$$
and the ring stops after the time
$$T=\frac{1}{\mu g}\int\limits_0^{V_0}\frac{dV}{\sqrt{1+\frac{V^4}{g^2R^2}}}
\approx \frac{1}{\mu g}\int\limits_0^{V_0}\left (1-\frac{V^4}{2g^2R^2}\right )
dV=\frac{1}{\mu g}\left [V_0-\frac{V_0^5}{10 g^2 R^2}\right ].$$
Finally,
$$T=\frac{V_0}{\mu g}\left [1-\frac{V_0^4}{10 g^2 R^2}\right ].$$

\subsection{}
\begin{figure}[htb]
\centerline{\epsfig{figure=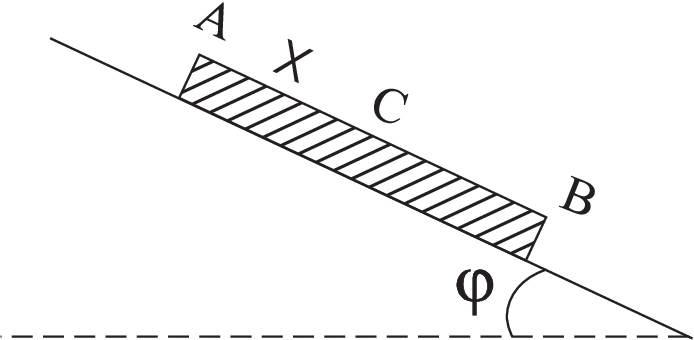,height=4cm}}
\end{figure}
\noindent
Рассмотрим процесс теплового расширения свинцового листа. Пусть при этом точка 
$C$ на растоянии $x$ от верхнего края неподвижна относительно крыши. Тогда 
участок $AC$ при расширении движется вверх, а участок $CB$ вниз. Это определяет
направление сил трения действующих на них. Так как ускорения центр масс $AC$ и 
$CB$ очень маленькие, можно считать, что действующие на них силы уравновешены.
Для $AC$ это дает уравнение
$$\frac{m}{l}xg\sin{\varphi}+\mu\frac{m}{l}xg\cos{\varphi}-T=0,$$ 
где $T$ есть сила с которой $CB$ действует на $AC$. Аналогично для $CB$ будем 
иметь 
$$\frac{m}{l}(l-x)g\sin{\varphi}-\mu\frac{m}{l}(l-x)g\cos{\varphi}+T=0.$$ 
Складывая эти уравнения, получаем
$$\frac{m}{l}lg\sin{\varphi}-\mu\frac{m}{l}(l-2x)g\cos{\varphi}=0.$$ Отсюда
$$x=\frac{l}{2}\left (1-\frac{\tg{\varphi}}{\mu}\right ).$$
если температура увеличивается на $\Delta t$, длина листа станет 
$l^\prime=(1+\alpha\Delta t)l$. 

\begin{figure}[htb]
\centerline{\epsfig{figure=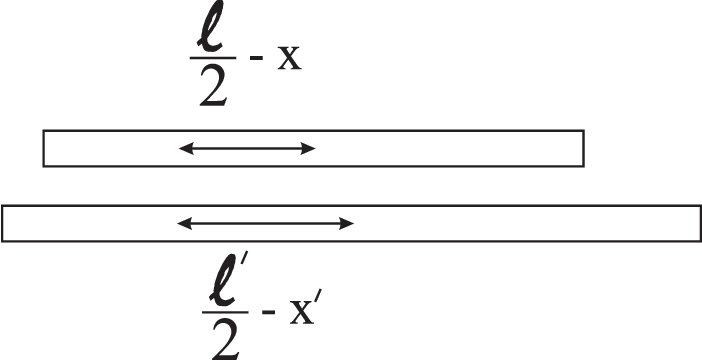,height=4cm}}
\end{figure}\noindent
При этом его центр масс сместится, как показывает рисунок. Смешение равно
$\Delta=\left (\frac{l^\prime}{2}-x^\prime\right )-\left(\frac{l}{2}-x
\right )$, где $x^\prime=\frac{l^\prime}{2}\left(1-\frac{\tg{\varphi}}{\mu}
\right )$. Следовательно,
$$\Delta=\frac{\tg{\varphi}}{2\mu}(l^\prime-l)=\alpha\Delta t \,l\,
\frac{\tg{\varphi}}{2\mu}.$$
Когда температура понижается, участок $AC$ движется вниз, а участок $CB$ 
вверх. Поэтому уравнения будут (мы используем те же самие обозначения $x$ и 
$T$, хотя численно они могут быть другими):
$$\frac{m}{l}xg\sin{\varphi}-\mu\frac{m}{l}xg\cos{\varphi}+T=0\;
\mbox{и}\; \frac{m}{l}(l-x)g\sin{\varphi}+\mu\frac{m}{l}(l-x)g
\cos{\varphi}-T=0.$$
Отсюда $\frac{2x}{l}=1+\frac{\tg{\varphi}}{\mu} $ и $x=\frac{l}{2}\left (1+
\frac{\tg{\varphi}}{\mu}\right )$. Смещение центра масс вниз будет
$$\Delta^\prime=\left (x-\frac{l}{2}\right )-\left(x^\prime-\frac{l^\prime}
{2}\right)=\frac{\tg{\varphi}}{2\mu}(l-l^\prime)=\alpha\Delta t \,l\,
\frac{\tg{\varphi}}{2\mu}=\Delta.$$
Следовательно, за одни сутки центр масс смещается вниз на $$2\Delta=\alpha l
\,\Delta t\,\frac{\tg{\varphi}}{\mu}.$$ За N суток смешение составит
$$\alpha Nl\,\Delta t\,\frac{\tg{\varphi}}{\mu}.$$

\vspace*{5mm}
\noindent $\bullet$  $\bullet$  $\bullet$
 
\begin{figure}[htb]
\centerline{\epsfig{figure=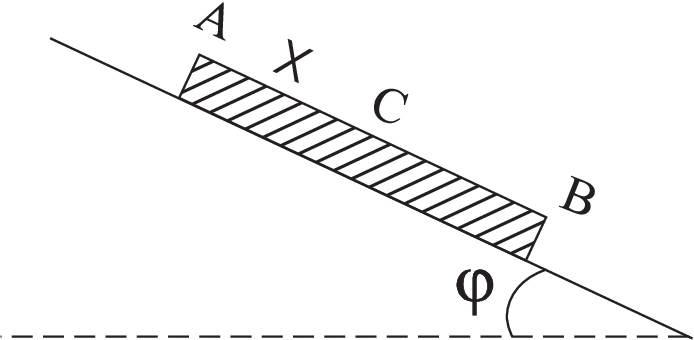,height=4cm}}
\end{figure}
\noindent
Consider the process of thermal expansion of a lead sheet. Let the point
$C$ at a distance $x$ from the top edge remains fixed with respect to the 
roof during the expansion. Then the part $AC$ of the sheet moves upward and 
the part $CB$ moves downward during the expansion. This determines directions 
of the frictional forces acting on them. Since accelerations of the centers
of mass of $AC$ and $CB$ are very small, we can assume that the forces acting 
on them are balanced. For $AC$ this condition is given by the equation
$$\frac{m}{l}xg\sin{\varphi}+\mu\frac{m}{l}xg\cos{\varphi}-T=0,$$ 
where $T$ where is the force with which the $CB$ part acts on $AC$. 
Similarly, for $CB$ we will have
$$\frac{m}{l}(l-x)g\sin{\varphi}-\mu\frac{m}{l}(l-x)g\cos{\varphi}+T=0.$$ 
Adding these equations, we obtain
$$\frac{m}{l}lg\sin{\varphi}-\mu\frac{m}{l}(l-2x)g\cos{\varphi}=0.$$ Hence
$$x=\frac{l}{2}\left (1-\frac{\tg{\varphi}}{\mu}\right ).$$
if the temperature is increased by $\Delta t$, the length of the leaf becomes 
$l^\prime=(1+\alpha\Delta t)l$. 

\begin{figure}[htb]
\centerline{\epsfig{figure=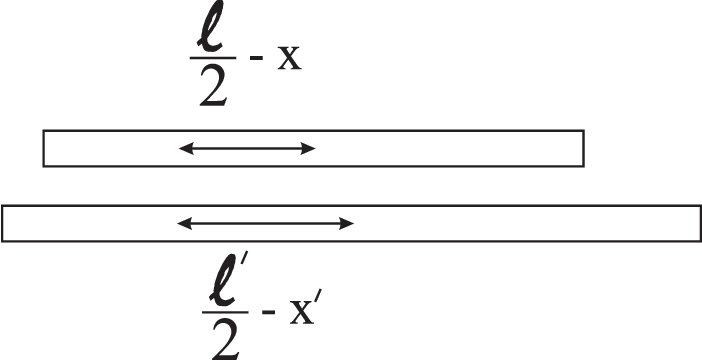,height=4cm}}
\end{figure}\noindent At that its center of mass shifts, as is shown in the 
Figure. The shift equals
$\Delta=\left (\frac{l^\prime}{2}-x^\prime\right )-\left(\frac{l}{2}-x
\right )$, where $x^\prime=\frac{l^\prime}{2}\left(1-\frac{\tg{\varphi}}{\mu}
\right )$. Therefore,
$$\Delta=\frac{\tg{\varphi}}{2\mu}(l^\prime-l)=\alpha\Delta t \,l\,
\frac{\tg{\varphi}}{2\mu}.$$
When the temperature drops, the part $AC$ moves down and the part $CB$ moves
up. Therefore, the corresponding equations are (we use the same notations
$x$ and $T$, although numerically they may be different):
$$\frac{m}{l}xg\sin{\varphi}-\mu\frac{m}{l}xg\cos{\varphi}+T=0\,
\mbox{and}\; \frac{m}{l}(l-x)g\sin{\varphi}+\mu\frac{m}{l}(l-x)g
\cos{\varphi}-T=0.$$
Hence $\frac{2x}{l}=1+\frac{\tg{\varphi}}{\mu} $ and $x=\frac{l}{2}\left (1+
\frac{\tg{\varphi}}{\mu}\right )$. The downward displacement of the center of 
mass equals
$$\Delta^\prime=\left (x-\frac{l}{2}\right )-\left(x^\prime-\frac{l^\prime}
{2}\right)=\frac{\tg{\varphi}}{2\mu}(l-l^\prime)=\alpha\Delta t \,l\,
\frac{\tg{\varphi}}{2\mu}=\Delta.$$
Consequently, one day downward displacement of the center of mass is 
$$2\Delta=\alpha l \,\Delta t\,\frac{\tg{\varphi}}{\mu}.$$ For N days, the 
shift will be
$$\alpha Nl\,\Delta t\,\frac{\tg{\varphi}}{\mu}.$$

\subsection{}
Энергия и импульс подсистемы $(\nu_\tau\gamma)$ определяются из
законов сохранения энергии и импульса при распаде
$\tau^-\to(\pi^{-}\pi^{0})+(\nu_{\tau}\gamma)$ и равны $E=\frac{m_{\tau}}{2}$,
$p=\sqrt{E^{2}-\left (\frac{m_{\tau}}{3}\right )^{2}}=\frac{\sqrt{5}}{6}
m_{\tau}$ (положили $c=1$). Дальше рассмотрим ``распад'' $ (\nu_{\tau}\gamma)
\to \nu_{\tau}+\gamma$. С учетом $(p_{\nu\gamma}-p_\gamma)^{2}=p_{\nu}^{2}=0$,
$p^{2}_{\nu\gamma}=m^{2}_{\nu\gamma}=\frac{m_{\tau}^{2}}{9}$ 
и $p_{\nu\gamma}\cdot p_{\gamma}=EE_{\gamma}-\vec{p}\cdot
\vec{p}_{\gamma}=E_{\gamma}(E-p\cos{\theta})$, получаем
$$\frac{m_{\tau}^{2}}{9}-2E_{\gamma}(E-p\cos{\theta})=0.$$
Поэтому
$$E_{\gamma}=\frac{m_{\tau}^{2}}{18(E-p\cos{\theta})}$$
и, так как $E^{2}-p^{2}=m_{\nu\gamma}^{2}=\frac{1}{9}m_{\tau}^{2}$,
$$E_{\gamma,min}=\frac{m_{\tau}^{2}}{18(E+p)}=\frac{1}{2}(E-p),\;\;
E_{\gamma,max}=\frac{m_{\tau}^{2}}{18(E-p)}=\frac{1}{2}(E+p).$$
Следовательно (восстановили $c$),
$$\frac{m_{\tau}c^{2}}{4}\left (1-\frac{\sqrt{5}}{3}\right )\leq E_\gamma\leq
\frac{m_{\tau}c^{2}}{4}\left (1+\frac{\sqrt{5}}{3}\right ).$$

\vspace*{5mm}
\noindent $\bullet$  $\bullet$  $\bullet$
 
\noindent
The energy and momentum of the $(\nu_\tau\gamma)$ subsystem is determined
from the energy-momentum conservation laws for the decay
$\tau^-\to(\pi^{-}\pi^{0})+(\nu_{\tau}\gamma)$ that give 
$E=\frac{m_{\tau}}{2}$, $p=\sqrt{E^{2}-\left (\frac{m_{\tau}}{3}\right )^{2}}
=\frac{\sqrt{5}}{6} m_{\tau}$ (we have assumed  $c=1$). Let us consider 
further the ``decay'' $ (\nu_{\tau}\gamma)\to \nu_{\tau}+\gamma$. Given 
$(p_{\nu\gamma}-p_\gamma)^{2}=p_{\nu}^{2}=0$, $p^{2}_{\nu\gamma}=m^{2}_{\nu
\gamma}=\frac{m_{\tau}^{2}}{9}$ and $p_{\nu\gamma}\cdot 
p_{\gamma}=EE_{\gamma}-\vec{p}\cdot \vec{p}_{\gamma}=E_{\gamma}(E-p
\cos{\theta})$, we obtain
$$\frac{m_{\tau}^{2}}{9}-2E_{\gamma}(E-p\cos{\theta})=0.$$
Hence
$$E_{\gamma}=\frac{m_{\tau}^{2}}{18(E-p\cos{\theta})}$$
and, since $E^{2}-p^{2}=m_{\nu\gamma}^{2}=\frac{1}{9}m_{\tau}^{2}$, we get
$$E_{\gamma,min}=\frac{m_{\tau}^{2}}{18(E+p)}=\frac{1}{2}(E-p),\;\;
E_{\gamma,max}=\frac{m_{\tau}^{2}}{18(E-p)}=\frac{1}{2}(E+p).$$
Consequently, (we have restored $c$),
$$\frac{m_{\tau}c^{2}}{4}\left (1-\frac{\sqrt{5}}{3}\right )\leq E_\gamma\leq
\frac{m_{\tau}c^{2}}{4}\left (1+\frac{\sqrt{5}}{3}\right ).$$

\section*{Контрольная работа 12}
\setcounter{section}{12}
\setcounter{subsection}{0}
\subsection{}
\begin{figure}[htb]
\centerline{\epsfig{figure=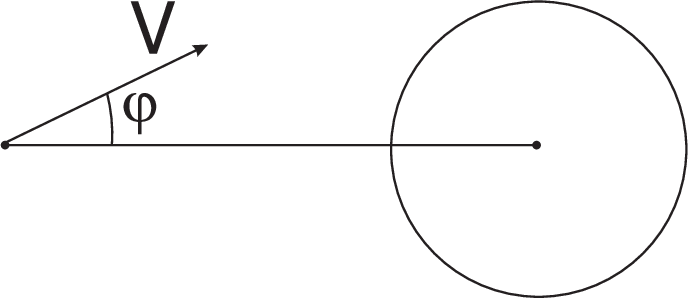,height=4cm}}
\end{figure}\noindent
Пусть космический аппарат находиться на расстоянии $r$ от центра
Земли и его скорость $\vec{V}$составляет угол
$\varphi $ с направлением на центр. Аппарат упадет на Землю, если
$r_{\min } = \frac{p}{1 + e} \le R$, $R$ -- радиус Земли. Но $$e =
\sqrt {1 + \frac{2EL^2}{m\alpha ^2}} = \sqrt {1 + \frac{2Ep}{\alpha
}} .$$ Поэтому получаем $$\left( {\frac{p}{R} - 1} \right)^2
=\frac{p^2}{R^2} - 2\frac{p}{R} + 1 \le e^2 = 1 + \frac{2Ep}{\alpha
}.$$ Или $$\frac{p}{R^2} \le \frac{2}{\alpha }\left( {E + \frac{\alpha
}{R}} \right).$$ Учтем здесь $$E = \frac{mV^2}{2} - \frac{\alpha }{r}\;\;
\mbox{и}\;\; p = \frac{L^2}{m\alpha } = \frac{mV^2}{\alpha }\sin ^2{\varphi}
\;r^2,$$ будем иметь $$mV^2\left( {\frac{r^2}{R^2}\sin ^2\varphi - 1}
\right) \le 2\frac{\alpha }{rR}(r - R).$$ Это неравенство будет
удовлетворяться при любом значении $V$, только если $$\sin ^2{\varphi}
\,\frac{r^2}{R^2} - 1 \le 0,\;\; \mbox{или} \;\;\sin \varphi \le 
\frac{r}{R}.$$ 
Т.е. вектор скорости аппарата должен лежать внутри телесного угла, с
которым Земля видна из аппарата.

\vspace*{5mm}
\noindent $\bullet$  $\bullet$  $\bullet$
 
\begin{figure}[htb]
\centerline{\epsfig{figure=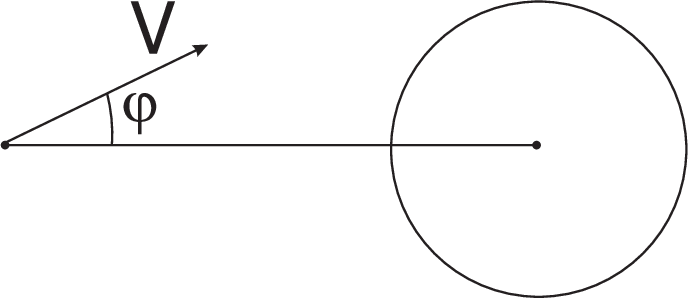,height=4cm}}
\end{figure}\noindent
Let the spacecraft is at a distance $r$ from the center of the Earth and its 
velocity $\vec{V}$ makes an angle $\varphi $ with the direction to the center.
The spacecraft will fall to the Earth, if
$r_{\min } = \frac{p}{1 + e} \le R$, $R$ being the Earth radius. But $$e =
\sqrt {1 + \frac{2EL^2}{m\alpha ^2}} = \sqrt {1 + \frac{2Ep}{\alpha
}} .$$ Therefore, we get $$\left( {\frac{p}{R} - 1} \right)^2
=\frac{p^2}{R^2} - 2\frac{p}{R} + 1 \le e^2 = 1 + \frac{2Ep}{\alpha
}.$$ Or $$\frac{p}{R^2} \le \frac{2}{\alpha }\left( {E + \frac{\alpha
}{R}} \right).$$ Taking into account $$E = \frac{mV^2}{2} - \frac{\alpha }{r}
\;\;\mbox{and}\;\; p = \frac{L^2}{m\alpha } = \frac{mV^2}{\alpha }\sin ^2
{\varphi}\;r^2,$$ we get $$mV^2\left( {\frac{r^2}{R^2}\sin ^2\varphi - 1}
\right) \le 2\frac{\alpha }{rR}(r - R).$$ This inequality will be
satisfied for any value of $V$, only if $$\sin ^2{\varphi}
\,\frac{r^2}{R^2} - 1 \le 0,\;\; \mbox{or} \;\;\sin \varphi \le 
\frac{r}{R}.$$ 
That is, the velocity vector of the spacecraft must lie within the solid angle 
with that the Earth is visible from the spacecraft.

\subsection{} 
Из равенств $$E = \frac{m}{2}(\dot {r}^2 + r^2\dot {\varphi}^2) 
+ U(r)$$ и $L = mr^2\dot {\varphi }$, получаем $$\frac{E}{L^2} =
\frac{1}{2mr^4}\left( {\frac{dr}{d\varphi }} \right)^2 +
\frac{1}{2mr^2} + \frac{U(r)}{L^2}.$$ Но $$\frac{1}{r^4}\left(
{\frac{dr}{d\varphi }} \right)^2 = \left( {\frac{ds}{d\varphi }}
\right)^2,\;\; \mbox{где}\;\; s = \frac{1}{r}.$$ Поэтому $$U = E -
\frac{L^2}{2m}\left[ {\left( {\frac{ds}{d\varphi }} \right)^2 + s^2}
\right].$$ Подставляя здесь $$s = \frac{1}{p}\left[ 1 + e\cos{
(\varepsilon \varphi )} \right],$$ и выражая ответ через $r$, получаем 
$$U(r) = E -\frac{L^2\varepsilon^2}{2mp^2}(e^2 - 1) - \frac{L^2(1 - 
\varepsilon ^2)}{2mr^2}-\frac{L^2\varepsilon ^2}{mp}\,\frac{1}{r}.$$

\vspace*{5mm}
\noindent $\bullet$  $\bullet$  $\bullet$
 
\noindent
From the equalities $$E = \frac{m}{2}(\dot {r}^2 + r^2\dot {\varphi}^2) 
+ U(r)$$ and $L = mr^2\dot {\varphi }$, we obtain $$\frac{E}{L^2} =
\frac{1}{2mr^4}\left( {\frac{dr}{d\varphi }} \right)^2 +
\frac{1}{2mr^2} + \frac{U(r)}{L^2}.$$ But $$\frac{1}{r^4}\left(
{\frac{dr}{d\varphi }} \right)^2 = \left( {\frac{ds}{d\varphi }}
\right)^2,\;\; \mbox{where}\;\; s = \frac{1}{r}.$$ Therefore, $$U = E -
\frac{L^2}{2m}\left[ {\left( {\frac{ds}{d\varphi }} \right)^2 + s^2}
\right].$$ Substituting here  $$s = \frac{1}{p}\left[ 1 + e\cos{
(\varepsilon \varphi )} \right],$$ 
and expressing the answer in terms of $r$, we get $$U(r) = E -
\frac{L^2\varepsilon^2}{2mp^2}(e^2 - 1) - \frac{L^2(1 - \varepsilon ^2)}
{2mr^2}-\frac{L^2\varepsilon ^2}{mp}\,\frac{1}{r}.$$

\subsection{} 
\begin{figure}[htb]
\centerline{\epsfig{figure=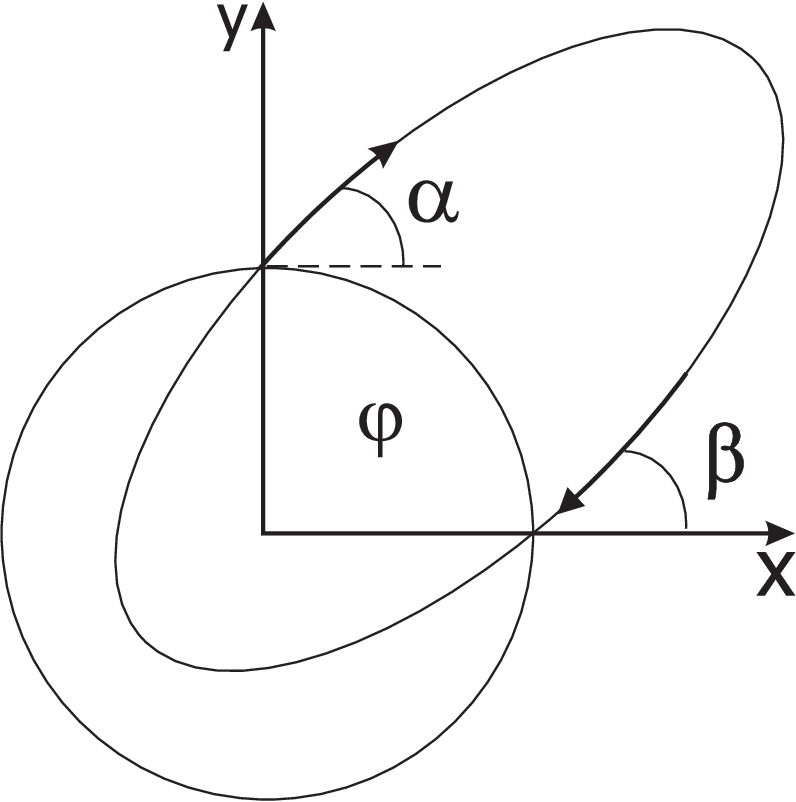,height=5cm}}
\end{figure}\noindent
Угол $\varphi $ отсчитываем от оси $x$. При движении
ракеты $\dot{\varphi}<0$, поэтому в качестве вектора Гамильтона
берем $$\vec{u}=\vec{V}+\frac{\tilde{\alpha}}{L}\,\vec
{e}_\varphi \, .$$ 
Здесь $\tilde {\alpha } = GmM$ (так обозначили, т.к. $\alpha $ у нас 
обозначает угол). Сохранение его $x$-проекции дает $$V\cos \alpha -
\frac{\tilde {\alpha }}{L} = - V\cos{\beta}$$ (на полюсе и на
экваторе величина скорости ракеты будет одно и тоже, из сохранения
энергии). Сохранение $y$-проекции дает $$V\sin \alpha = \frac{\tilde
{\alpha }}{L} - V\sin \beta .$$ Следовательно, $$\sin{\alpha} + \sin
{\beta} = \frac{\tilde {\alpha }}{VL} = \cos {\alpha} + \cos {\beta} .$$
Или $$2\sin {\frac{\alpha + \beta }{2}}\cos {\frac{\alpha - \beta }{2}} =
2\cos{ \frac{\alpha + \beta }{2}}\cos {\frac{\alpha - \beta }{2}}.$$
Отсюда $\tg{\frac{\alpha + \beta }{2}} = 1$, и $\beta = \frac{\pi }{2}
- \alpha $. Тогда $$\sin{\alpha} + \sin{\beta} = \sin{\alpha} + \cos
{\alpha} = \frac{\tilde {\alpha }}{VL} = \frac{\tilde {\alpha
}}{mV^2R\cos{\alpha}}.$$ Следовательно,
$$V^2 = \frac{\tilde {\alpha }}{mR\cos{\alpha}\,(\cos{\alpha} + \sin
{\alpha} )}$$ минимальна, когда максимальна $\cos{\alpha} (\cos{\alpha} +
\sin{\alpha} )$. Но $$\frac{d}{d\alpha }\left[ {\cos{\alpha}\,(\cos
{\alpha} + \sin{\alpha} )} \right] = - 2\cos{\alpha} \sin{\alpha} + \cos
{2\alpha} = \cos {2\alpha} - \sin {2\alpha} = 0$$ дает $\tg{2\alpha} = 1$ и
$\alpha = \frac{\pi }{8}$. Следовательно, оптимальный угол запуска
ракеты составляет $\frac{\pi }{8}$.

\clearpage
\noindent $\bullet$  $\bullet$  $\bullet$ 

\begin{figure}[htb]
\centerline{\epsfig{figure=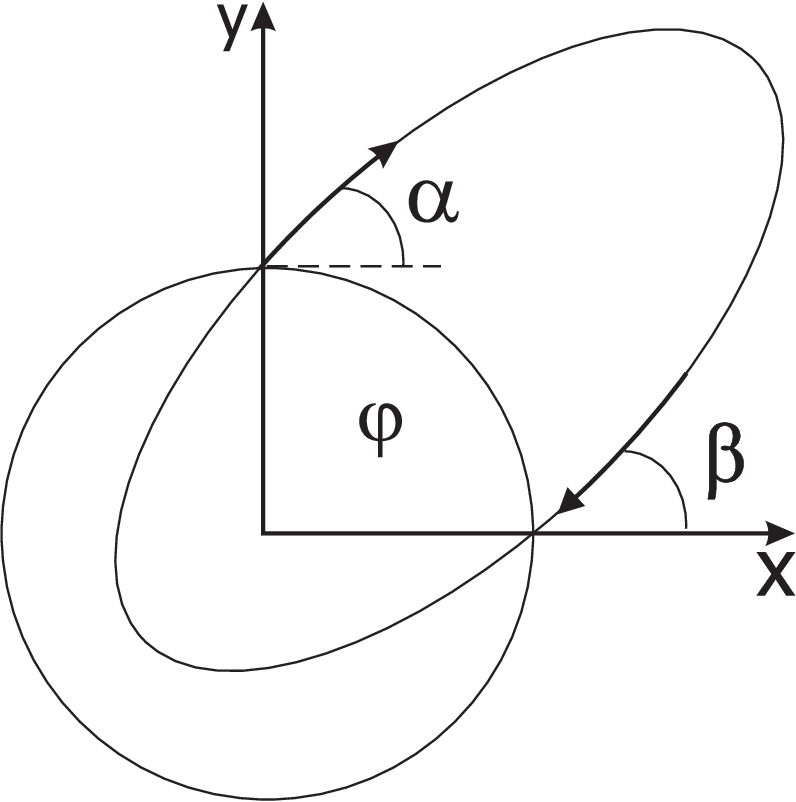,height=5cm}}
\end{figure}\noindent
Angle $\varphi$ is measured counterclockwise from the axis $x$. When the
rocket moves, $\dot{\varphi}<0$. Therefore, as the Hamilton vector we will
take $$\vec{u}=\vec{V}+\frac{\tilde{\alpha}}{L}\,\vec
{e}_\varphi \, .$$ 
Here $\tilde {\alpha } = GmM$ (so designated because the $\alpha$ we had
already used to denote the angle). Conservation of its $x$-projection gives 
$$V\cos \alpha -
\frac{\tilde {\alpha }}{L} = - V\cos{\beta}$$
(at the pole and the equator, the rocket velocities are the same, 
according to the energy conservation law). Conservation of the  
$y$-projection gives $$V\sin \alpha = \frac{\tilde
{\alpha }}{L} - V\sin \beta .$$ Therefore, $$\sin{\alpha} + \sin
{\beta} = \frac{\tilde {\alpha }}{VL} = \cos {\alpha} + \cos {\beta} .$$
Or $$2\sin {\frac{\alpha + \beta }{2}}\cos {\frac{\alpha - \beta }{2}} =
2\cos{ \frac{\alpha + \beta }{2}}\cos {\frac{\alpha - \beta }{2}}.$$
Hence $\tg{\frac{\alpha + \beta }{2}} = 1$, and $\beta = \frac{\pi }{2}
- \alpha $. Then $$\sin{\alpha} + \sin{\beta} = \sin{\alpha} + \cos
{\alpha} = \frac{\tilde {\alpha }}{VL} = \frac{\tilde {\alpha
}}{mV^2R\cos{\alpha}}.$$ Therefore,
$$V^2 = \frac{\tilde {\alpha }}{mR\cos{\alpha}\,(\cos{\alpha} + \sin
{\alpha} )}.$$ It is minimal, when $\cos{\alpha} (\cos{\alpha} +
\sin{\alpha} )$ is maximal. But $$\frac{d}{d\alpha }\left[ {\cos{\alpha}\,
(\cos{\alpha} + \sin{\alpha} )} \right] = - 2\cos{\alpha} \sin{\alpha} + \cos
{2\alpha} = \cos {2\alpha} - \sin {2\alpha} = 0$$ gives $\tg{2\alpha} = 1$ and
$\alpha = \frac{\pi }{8}$. Consequently, the optimal launching angle for the 
missile is $\frac{\pi}{8}$.

\subsection{}
Момент импульса $L = mr^2\dot {\varphi }$. Но $\frac{1}{2}r^2d\varphi $ 
есть площадь сектора $OAB$. Следовательно,
$\frac{1}{2}\int {r^2d\varphi } = \frac{L}{2m}\int {dt} $ означает,
что $\frac{LT}{2m} = $площадь эллипса =$\pi ab$, где $T$- период
обращения. 

\begin{figure}[htb]
\centerline{\epsfig{figure=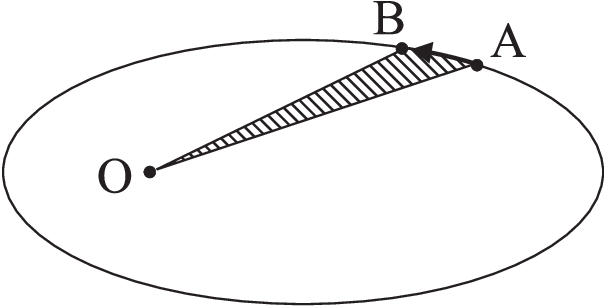,height=3cm}}
\end{figure}
\noindent Но из $e = \frac{\sqrt {a^2 - b^2} }{a}$ следует, что $$b
= a\sqrt {1 - e^2} = a\sqrt { - \frac{2EL^2}{m\alpha ^2}} = L\sqrt
{\frac{a}{m\alpha }} ,$$ т.к. $$e = \sqrt {1 + \frac{2EL^2}{m\alpha
^2}}\;\;\mbox{и}\;\; E = - \frac{\alpha }{2a}.$$ Поэтому получаем 
$$T = 2\pi\sqrt {\frac{m}{\alpha }}\, a\,\sqrt a.$$ При рассмотрении задачи 
двух тел (Солнце с массой $M$ и планета с массой $m)$, в предыдущих
формулах мы должны сделать замену $m \to \mu = \frac{mM}{m + M}$ и
$\alpha = GmM$. Поэтому будем иметь $$T = 2\pi \sqrt {\frac{a}{G(m +
M)}} \; a.$$ Следовательно, $$\frac{T_1^2 }{T_2^2 } = \frac{m_2 +
M}{m_1 + M}\;\frac{a_1^3 }{a_2^3 }.$$ Третий закон Кеплера $$\frac{T_1^2
}{T_2^2 } = \frac{a_1^3 }{a_2^3 }$$ получится, только если пренебречь
массами планет по сравнению с массой Солнца.

\vspace*{5mm}
\noindent $\bullet$  $\bullet$  $\bullet$
 
\noindent
For the angular momentum we have $L = mr^2\dot {\varphi }$. But 
$\frac{1}{2}r^2d \varphi $ is the area of the sector $OAB$. Consequently,
$\frac{1}{2}\int {r^2d\varphi } = \frac{L}{2m}\int {dt}$ means
that $\frac{LT}{2m} = $ area of the ellipse = $\pi ab$, where $T$ is the 
orbital period.

\begin{figure}[htb]
\centerline{\epsfig{figure=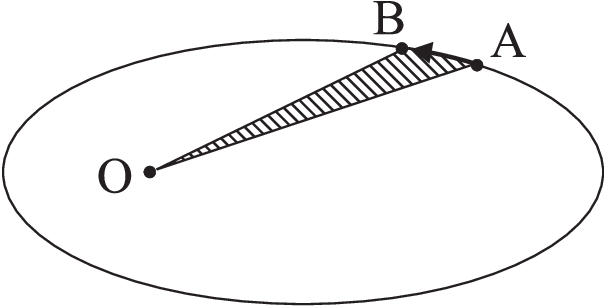,height=3cm}}
\end{figure}
\noindent But it follows from $e = \frac{\sqrt {a^2 - b^2} }{a}$ that 
$$b= a\sqrt {1 - e^2} = a\sqrt { - \frac{2EL^2}{m\alpha ^2}} = L\sqrt
{\frac{a}{m\alpha }} ,$$ because $$e = \sqrt {1 + \frac{2EL^2}{m\alpha
^2}}\;\;\mbox{and}\;\; E = - \frac{\alpha }{2a}.$$ Therefore, we get 
$$T = 2\pi\sqrt {\frac{m}{\alpha }}\, a\,\sqrt a.$$ When considering the 
two-body problem (the Sun with mass $M$ and the planet with a mass $m$), in 
the previous formulas we must make the substitutions $m \to \mu = \frac{mM}
{m + M}$ and $\alpha = GmM$. Therefore, we will have 
$$T = 2\pi \sqrt {\frac{a}{G(m +M)}} \; a.$$
Consequently, $$\frac{T_1^2 }{T_2^2 } = \frac{m_2 +
M}{m_1 + M}\;\frac{a_1^3 }{a_2^3 }.$$
Kepler's third law  $$\frac{T_1^2}{T_2^2 } = \frac{a_1^3 }{a_2^3 }$$  
is obtained only if we neglect the masses of the planets compared to the mass 
of the Sun.

\subsection{} 
Так как $$E = - \frac{\alpha }{2a} = - \frac{\alpha }{r_{\min } +
r_{\max } },$$ энергии спутников равны $$E_1 = - \frac{\alpha }{2R} =
\frac{mV_1^2 }{2} - \frac{\alpha }{R}\;\;\mbox{и}\;\; E_2 = - \frac{\alpha
}{8R} = \frac{mV_2^2 }{2} - \frac{\alpha }{R},$$ где $V_1 $ и $V_2$ 
их скорости в перигее и $\alpha = GmM$. Отсюда $$V_1 = \sqrt
{\frac{\alpha }{mR}} \;\;\mbox{и} \;\; V_2 = \sqrt{\frac{7\alpha}{4mR}}.$$
Из сохранения импульса $2mV = mV_1 + mV_2 $, находим скорость после
стыковки $$V = \frac{1}{2}(V_1 + V_2 ) = \frac{\sqrt 7 + 2}{4}\sqrt
{\frac{\alpha }{mR}}.$$ Поэтому полная энергия после стыковки будет
$$E = \frac{2mV^2}{2} - \frac{2\alpha }{R} = \frac{(\sqrt 7 +
2)^2}{16}\frac{\alpha }{R} - \frac{2\alpha }{R} = - \frac{21 -
4\sqrt 7 }{16}\,\frac{\alpha }{R}.$$ С другой стороны $$E = -
\frac{2\alpha }{R + r},$$ где $r $-- новый апогей. Поэтому получаем
$$r = \frac{11 + 4\sqrt 7 }{21 - 4\sqrt 7 }\,R = \frac{343 + 128\sqrt 7
}{329}\,R \approx 2.07R.$$

\vspace*{5mm}
\noindent $\bullet$  $\bullet$  $\bullet$
 
\noindent
Since $$E = - \frac{\alpha }{2a} = - \frac{\alpha }{r_{\min } +
r_{\max } },$$ the satellites energies are $$E_1 = - \frac{\alpha }{2R} =
\frac{mV_1^2 }{2} - \frac{\alpha }{R}\;\;\mbox{and}\;\; E_2 = - \frac{\alpha
}{8R} = \frac{mV_2^2 }{2} - \frac{\alpha }{R},$$ where $V_1 $ and $V_2$ 
are their perigee velocities and $\alpha = GmM$. Therefore, 
$$V_1 = \sqrt{\frac{\alpha }{mR}} \;\;\;\mbox{and} \;\;\;
V_2 = \sqrt{\frac{7\alpha}{4mR}}.$$
From the conservation of momentum $2mV = mV_1 + mV_2 $, we find the speed 
after the docking $$V = \frac{1}{2}(V_1 + V_2 ) = \frac{\sqrt 7 + 2}{4}\sqrt
{\frac{\alpha }{mR}}.$$ Therefore, the total energy after the docking will be
$$E = \frac{2mV^2}{2} - \frac{2\alpha }{R} = \frac{(\sqrt 7 +
2)^2}{16}\frac{\alpha }{R} - \frac{2\alpha }{R} = - \frac{21 -
4\sqrt 7 }{16}\,\frac{\alpha }{R}.$$ On the other hand $$E = -
\frac{2\alpha }{R + r},$$ where $r$ is the new apogee. Therefore, we get
$$r = \frac{11 + 4\sqrt 7 }{21 - 4\sqrt 7 }\,R = \frac{343 + 128\sqrt 7
}{329}\,R \approx 2.07R.$$

\subsection{} 
При круговом движении $$\left| {\frac{d\vec {p}}{dr}} \right| =
\frac{p\Delta \varphi }{\Delta t} = \frac{pV}{R}.$$ Поэтому должно
быть $$\frac{pV}{R} = \left| { - \frac{dU}{dr}} \right|_{r = R} =
\frac{\alpha \beta }{R^{\beta + 1}}.$$ Но $L = pR$ (при круговом
движении). Следовательно, $$\frac{LV}{R^2} = \frac{\alpha \beta
}{R^{\beta + 1}}\;\;\mbox{и} \;\; V = \frac{\alpha \beta }{LR^{\beta - 1}}.$$
Ограничение на $L$ возникает, если $\beta = 1$. Тогда $\frac{\alpha
}{L} = V < c$ дает $$L > \frac{\alpha }{c}.$$ Если $\beta \ne 1$,
круговые орбиты возможны для всех значений момента импульса.

Если $\beta = 1$, радиус круговой орбиты находится так: из $\frac{pV}{R} =
\frac{\alpha }{R^2}$ имеем $$\frac{mV^2}{\sqrt {1 - \frac{V^2}{c^2}} } =
\frac{\alpha }{R}\;\;\mbox{и} \;\; R = \frac{\alpha }{mV^2}\sqrt {1 - 
\frac{V^2}{c^2}}.$$
Подставляя сюда $V = \frac{\alpha }{L}$, будем иметь $$R = \frac{L^2}{m\alpha
}\sqrt {1 - \frac{\alpha ^2}{L^2c^2}}.$$

\vspace*{5mm}
\noindent $\bullet$  $\bullet$  $\bullet$
 
\noindent
When moving on a circle, we have $$\left| {\frac{d\vec {p}}{dr}} \right| =
\frac{p\Delta \varphi }{\Delta t} = \frac{pV}{R}.$$ Therefore, the following 
relation must hold $$\frac{pV}{R} = \left| { - \frac{dU}{dr}} \right|_{r = R}=
\frac{\alpha \beta }{R^{\beta + 1}}.$$ But $L = pR$ (for circular motion). 
Consequently, $$\frac{LV}{R^2} = \frac{\alpha \beta
}{R^{\beta + 1}}\;\;\mbox{and} \;\; V = \frac{\alpha \beta }{LR^{\beta - 1}}.$$
Restriction on $L$ occurs if $\beta = 1$. Then $\frac{\alpha
}{L} = V < c$ gives $$L > \frac{\alpha }{c}.$$ If $\beta \ne 1$,
circular orbits are possible for all values of angular momentum.

If $\beta = 1$, the radius of the circular orbit can be found as follows:
from $\frac{pV}{R} =\frac{\alpha }{R^2}$, we have $$\frac{mV^2}{\sqrt {1 - 
\frac{V^2}{c^2}} } =\frac{\alpha }{R}\;\;\mbox{and} \;\; R = \frac{\alpha }
{mV^2}\sqrt {1 - \frac{V^2}{c^2}}.$$
Substituting here $V = \frac{\alpha }{L}$, we obtain $$R = \frac{L^2}{m\alpha
}\sqrt {1 - \frac{\alpha ^2}{L^2c^2}}.$$

\section*{Контрольная работа 13}
\setcounter{section}{13}
\setcounter{subsection}{0}
\subsection{}
Сила, которая действует на электрон, равна (в системе CGSE) 
$$\vec{F}=\frac{e}{c}\;\vec{V}\times \vec{B}=\frac{eg}{cr^3}\,\vec{V}\times 
\vec{r}.$$ 
Изменение импульса
$$\Delta\vec{p}=\int\limits_{-\infty}^\infty \vec{F}\,dt=\frac{eg}{c}\int
\limits_{-\infty}^\infty\frac{\vec{V}\times{\vec{r}}}{r^3}\,dt.$$
Так как электрон быстрый, при вычислении интеграла можно считать, что электрон 
движется по прямой с неизменной скоростью. 

\begin{figure}[htb]
\centerline{\epsfig{figure=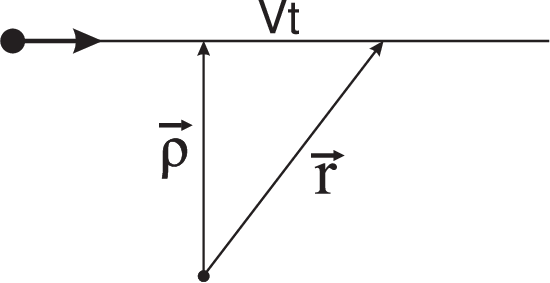,height=4cm}}
\end{figure}\noindent
Тогда $\vec{r}=\vec{\rho}+\vec{V}t$
и $\vec{V}\times\vec{r}=\vec{V}\times \vec{\rho}$. Кроме того,
$r=\sqrt{\rho^2+V^2t^2}$. Поэтому
$$ \Delta\vec{p}\approx \frac{eg}{c}\vec{V}\times\vec{\rho}\int\limits_
{-\infty}^\infty\frac{dt}{(\rho^2+V^2t^2)^{3/2}}=\left.\frac{eg}{c}\frac
{\vec{V}\times{\vec{\rho}}}{\rho^2}\; \frac{t}{\sqrt{\rho^2+V^2t^2}}
\right|_ {-\infty}^{\infty}=\frac{2eg}{c}\frac{\vec{V}\times{\vec{\rho}}}
{V\rho^2}.$$
Импульс электрона после рассеяния
$\vec{q}=\vec{p}+\Delta{\vec{p}}$, причем $q\approx p$, так как
$\Delta \vec{p}\perp \vec{p}$. Пусть $\theta$ -- угол
рассеяния. Тогда 
$$| \vec{q}\times\vec{p} |=p^2\sin{\theta}=|\Delta\vec{p}\times \vec{p}|=
\frac{2eg}{cV\rho^2}\,|\vec{p}\times(\vec{V}\times\vec{\rho})|.$$ 
Но
$\vec{p}\times(\vec{V}\times{\vec{\rho}})=\vec{V}\,(\vec{p}\cdot
\vec{\rho}) - \vec{\rho}\,(\vec{p}\cdot \vec{V})= -\vec{\rho}\,pV$, так как
$\vec{p} \perp \vec{\rho}$ и $\vec{p} \parallel \vec{V}$. Поэтому
получаем $$p^2\sin{\theta}\approx\frac{2eg}{cV\rho^2}\,{\rho}pV\;\;\mbox{и}
\;\;\theta\approx\sin{\theta}\approx\frac{2eg}{c\rho p}.$$

\clearpage
\noindent $\bullet$  $\bullet$  $\bullet$
 
\noindent
The force which acts on the electron is (in the CGSE system)
$$\vec{F}=\frac{e}{c}\;\vec{V}\times \vec{B}=\frac{eg}{cr^3}\,\vec{V}\times 
\vec{r}.$$ 
Momentum change caused by this force equals to
$$\Delta\vec{p}=\int\limits_{-\infty}^\infty \vec{F}\,dt=\frac{eg}{c}\int
\limits_{-\infty}^\infty\frac{\vec{V}\times{\vec{r}}}{r^3}\,dt.$$
Since the electron is fast moving, while calculating the integral, we can 
assume that the electron moves in a straight line with a constant speed.

\begin{figure}[htb]
\centerline{\epsfig{figure=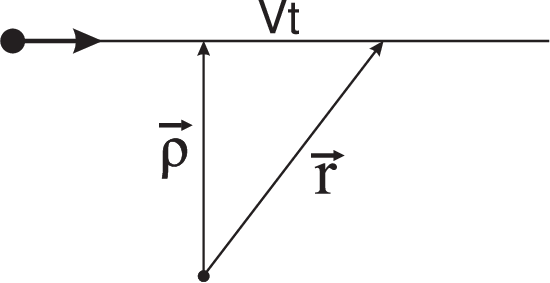,height=4cm}}
\end{figure}\noindent
Then $\vec{r}=\vec{\rho}+\vec{V}t$
and $\vec{V}\times\vec{r}=\vec{V}\times \vec{\rho}$. In addition,
$r=\sqrt{\rho^2+V^2t^2}$. Therefore,
$$ \Delta\vec{p}\approx \frac{eg}{c}\vec{V}\times\vec{\rho}\int\limits_
{-\infty}^\infty\frac{dt}{(\rho^2+V^2t^2)^{3/2}}=\left.\frac{eg}{c}\frac
{\vec{V}\times{\vec{\rho}}}{\rho^2}\; \frac{t}{\sqrt{\rho^2+V^2t^2}}
\right|_ {-\infty}^{\infty}=\frac{2eg}{c}\frac{\vec{V}\times{\vec{\rho}}}
{V\rho^2}.$$
The momentum of the electron after the scattering is
$\vec{q}=\vec{p}+\Delta{\vec{p}}$, wherein $q\approx p$, because
$\Delta \vec{p}\perp \vec{p}$. Let $\theta$ be the  scattering angle. Then 
$$| \vec{q}\times\vec{p} |=p^2\sin{\theta}=|\Delta\vec{p}\times \vec{p}|=
\frac{2eg}{cV\rho^2}\,|\vec{p}\times(\vec{V}\times\vec{\rho})|.$$ 
But
$\vec{p}\times(\vec{V}\times{\vec{\rho}})=\vec{V}\,(\vec{p}\cdot
\vec{\rho}) - \vec{\rho}\,(\vec{p}\cdot \vec{V})= -\vec{\rho}\,pV$, since
$\vec{p} \perp \vec{\rho}$ and $\vec{p} \parallel \vec{V}$. Hence we get
$$p^2\sin{\theta}\approx\frac{2eg}{cV\rho^2}\,{\rho}pV\;\;\mbox{and}
\;\;\theta\approx\sin{\theta}\approx\frac{2eg}{c\rho p}.$$

\subsection{}
Так как момент импульса $L=mr^2|\dot{\varphi}|$, то
энергия частицы $$E=\frac{m}{2}(\dot{r}^2+r^2{\dot{\varphi}}^2)+U(r)=
\frac{m\dot{r}^2}{2}+\frac{L^2}{2mr^2}+U(r).$$ Отсюда ( для
$\dot{r}<0$)
$$dt=-\left[\frac{2}{m}[E-U(r)]-\frac{L^2}{m^2r^2}\right]^{-1/2}dr.$$ С другой
стороны, из $\frac{d\varphi}{dt}=\frac{L}{mr^2}$ получаем ( заметим,
что $\dot{\varphi}>0$ и $|\dot{\varphi}|=\dot{\varphi}$ )
$$\varphi_0=-\int\limits_{\infty}^{r_{min}}
\frac{\frac{L}{2mr^2}dr}{\sqrt{\frac{2}{m}[E-U(r)]-\frac{L^2}{m^2r^2}}}=
\int\limits_{r_{min}}^{\infty}
\frac{\rho\frac{dr}{r^2}}{\sqrt{1-\frac{\rho^2}{r^2}-\frac{U(r)}{T}}},$$
так как $E=T=\frac{m{V_{\infty}}^2}{2}$ и $L=m{\rho}V_{\infty}$ (см. рисунок).

\begin{figure}[htb]
\centerline{\epsfig{figure=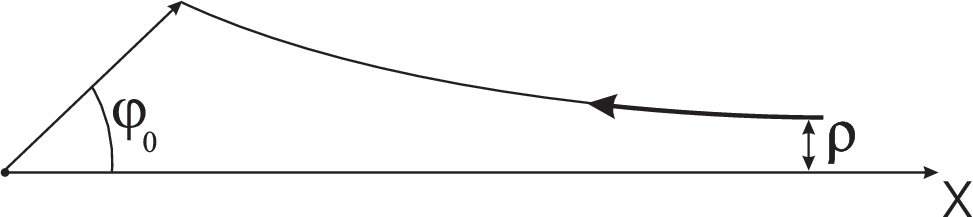,height=2cm}}
\end{figure}\noindent
Это справедливо для любого центрального потенциала. В случае
$U(r)=\frac{\alpha}{r^2}$ будем иметь
$$\varphi_0=\int\limits_{r_{min}}^{\infty}
\frac{\rho\frac{dr}{r^2}}{\sqrt{1-(\rho^2+\frac{\alpha}{T})\frac{1}
{r^2}}},$$ причем $1-(\rho^2+\frac{\alpha}{T})\frac{1}{r^2_{min}}=0$ 
(при $r=r_{min}\,$, $\dot{r}=0$). Сделаем подстановку $s=\frac{1}{r}$.
Тогда $$\varphi_0=\int\limits_0^{s_0}
\frac{\rho{ds}}{\sqrt{1-(\rho^2+\frac{\alpha}{T})s^2}},\;\;
s_0=\frac{1}{\sqrt{\rho^2+\frac{\alpha}{T}}}.$$ Но
$$\int\frac{dx}{\sqrt{1-x^2}}=\arcsin{x}\;\;\mbox{и}\;\;
\varphi_0=\left.\frac{\rho}{\sqrt{\rho^2+\frac{\alpha}{T}}}
\arcsin{\left(\sqrt{\rho^2+\frac{\alpha}{T}}\,s\right)}\right|_0^{s_0}=
\frac{\pi{\rho}}{2\sqrt{\rho^2+\frac{\alpha}{T}}}.$$ 

\begin{figure}[htb]
\centerline{\epsfig{figure=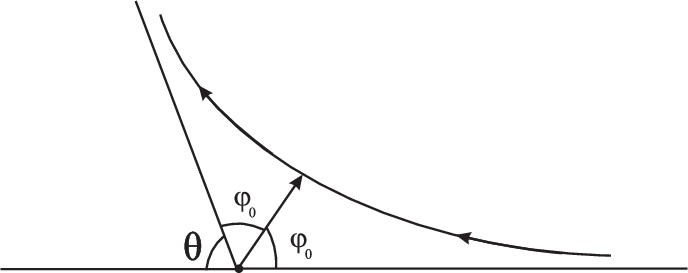,height=4cm}}
\end{figure}\noindent
Угол рассеяния $\theta$ и
$\varphi_0$ связаны так $\theta=\pi-2\varphi_0$. Поэтому
$$\theta=\pi-\frac{\pi{\rho}}{\sqrt{\rho^2+\frac{\alpha}{T}}}=
\pi\left[1-\frac{\rho}{\sqrt{\rho^2+\frac{\alpha}{T}}}\right].$$
Если $\theta=\frac{\pi}{2}$, прицельный параметр будет определяться
из $$\frac{\rho}{\sqrt{\rho^2+\frac{\alpha}{T}}}=\frac{1}{2}.$$ Отсюда
$\rho^2=\frac{\alpha}{3T}$ и сечение рассеяния на угол, больший
$90^{\circ}$, будет $$\sigma=\pi\rho^2=\frac{\alpha\pi}{3T}.$$

\vspace*{5mm}
\noindent $\bullet$  $\bullet$  $\bullet$
 
\noindent
Since the angular momentum is $L=mr^2|\dot{\varphi}|$, the energy of the
particle equals to $$E=\frac{m}{2}(\dot{r}^2+r^2{\dot{\varphi}}^2)+U(r)=
\frac{m\dot{r}^2}{2}+\frac{L^2}{2mr^2}+U(r).$$ Hence (for $\dot{r}<0$)
$$dt=-\left[\frac{2}{m}[E-U(r)]-\frac{L^2}{m^2r^2}\right]^{-1/2}dr.$$ On the
other hand, from $\frac{d\varphi}{dt}=\frac{L}{mr^2}$ we get (note that
$\dot{\varphi}>0$ and $|\dot{\varphi}|=\dot{\varphi}$ )
$$\varphi_0=-\int\limits_{\infty}^{r_{min}}
\frac{\frac{L}{2mr^2}dr}{\sqrt{\frac{2}{m}[E-U(r)]-\frac{L^2}{m^2r^2}}}=
\int\limits_{r_{min}}^{\infty}
\frac{\rho\frac{dr}{r^2}}{\sqrt{1-\frac{\rho^2}{r^2}-\frac{U(r)}{T}}},$$
because $E=T=\frac{m{V_{\infty}}^2}{2}$ and $L=m{\rho}V_{\infty}$ (see the 
figure below).

\begin{figure}[htb]
\centerline{\epsfig{figure=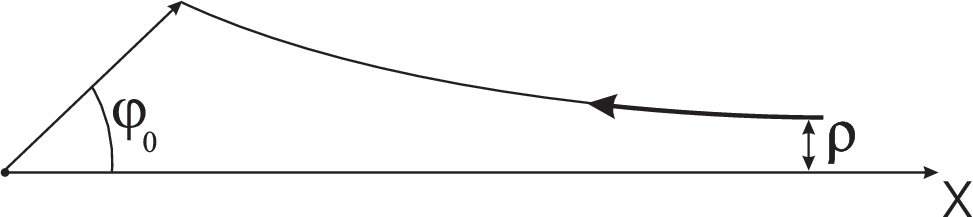,height=2cm}}
\end{figure}\noindent
This is true for any central potential. In the case of
$U(r)=\frac{\alpha}{r^2}$, we have
$$\varphi_0=\int\limits_{r_{min}}^{\infty}
\frac{\rho\frac{dr}{r^2}}{\sqrt{1-(\rho^2+\frac{\alpha}{T})\frac{1}
{r^2}}},$$ wherein $1-(\rho^2+\frac{\alpha}{T})\frac{1}{r^2_{min}}=0$ 
(at $r=r_{min}\,$, we have $\dot{r}=0$). Let us make the substitution 
$s=\frac{1}{r}$. Then $$\varphi_0=\int\limits_0^{s_0}
\frac{\rho{ds}}{\sqrt{1-(\rho^2+\frac{\alpha}{T})s^2}},\;\;
s_0=\frac{1}{\sqrt{\rho^2+\frac{\alpha}{T}}}.$$ But
$$\int\frac{dx}{\sqrt{1-x^2}}=\arcsin{x}\;\;\mbox{and}\;\;
\varphi_0=\left.\frac{\rho}{\sqrt{\rho^2+\frac{\alpha}{T}}}
\arcsin{\left(\sqrt{\rho^2+\frac{\alpha}{T}}\,s\right)}\right|_0^{s_0}=
\frac{\pi{\rho}}{2\sqrt{\rho^2+\frac{\alpha}{T}}}.$$ 

\noindent The scattering angle $\theta$ and the angle
$\varphi_0$ are related as follows $\theta=\pi-2\varphi_0$. 
\begin{figure}[htb]
\centerline{\epsfig{figure=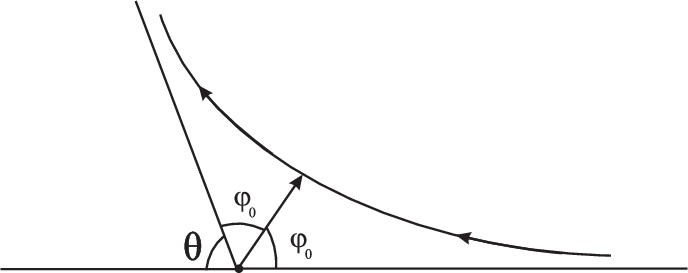,height=4cm}}
\end{figure}

\noindent Therefore,
$$\theta=\pi-\frac{\pi{\rho}}{\sqrt{\rho^2+\frac{\alpha}{T}}}=
\pi\left[1-\frac{\rho}{\sqrt{\rho^2+\frac{\alpha}{T}}}\right].$$

\noindent If $\theta=\frac{\pi}{2}$, the impact parameter is determined
from $$\frac{\rho}{\sqrt{\rho^2+\frac{\alpha}{T}}}=\frac{1}{2}.$$ Hence
$\rho^2=\frac{\alpha}{3T}$ and the scattering cross section at angles 
greater than $90^{\circ}$ is equal to 
$$\sigma=\pi\rho^2=\frac{\alpha\pi}{3T}.$$

\subsection{}
\begin{figure}[htb]
\centerline{\epsfig{figure=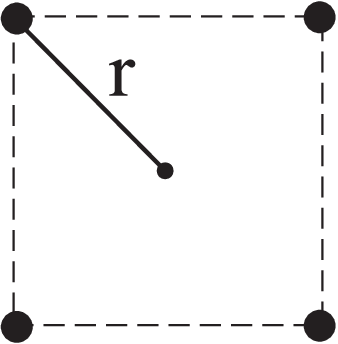,height=4cm}}
\end{figure}
Из симметрии задачи ясно, что звезды всегда будут в
вершинах некоторого квадрата. Пусть расстояние от звезд до центра
этого квадрата равно $r$. Тогда сторона квадрата будет
$a=\sqrt{2}\,r$. Сила притяжения, действующая на звезду со стороны
других звезд, направлена к центру и по величине равна
$$F=G\frac{M^2}{(2r)^2}+2G\frac{M^2}{a^2}\cos45^{\circ}
=\frac{GM^2}{r^2}\,\frac{1+2\sqrt{2}}{4}.$$ То есть можно считать, что
каждая звезда движется в центральном поле притяжения
$$U(r)=-\frac{\alpha}{r},\;\; \alpha=GM^2\,\frac{1+2\sqrt{2}}{4}$$ с центром в
центре первоначального квадрата. По формуле Резерфорда
$$\tan\frac{\theta}{2}=\frac{\alpha}{MV^2b}$$ находим угол рассеяния звезд
$$\theta=2\arctan{\left[\frac{GM}{bV^2}\,\frac{1+2\sqrt{2}}{4}\right]}.$$

\vspace*{5mm}
\noindent $\bullet$  $\bullet$  $\bullet$
 
\begin{figure}[htb]
\centerline{\epsfig{figure=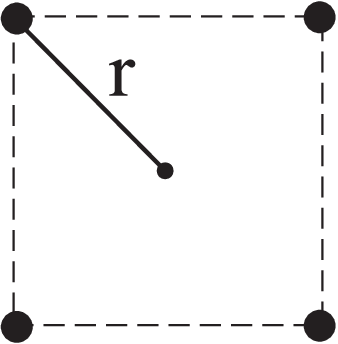,height=4cm}}
\end{figure}
It is clear from the symmetry of the problem that the stars will always be
at the vertexes of some square. Let the distance from the star to the center 
of this square is equal to $r$. Then the side of the square will be
$a=\sqrt{2}\,r$. An  attractive force acting on the star by other stars is 
directed to the center and is equal in magnitude to
$$F=G\frac{M^2}{(2r)^2}+2G\frac{M^2}{a^2}\cos45^{\circ}
=\frac{GM^2}{r^2}\,\frac{1+2\sqrt{2}}{4}.$$ That is, we can assume that
each star moves in a central field 
$$U(r)=-\frac{\alpha}{r},\;\; \alpha=GM^2\,\frac{1+2\sqrt{2}}{4}$$ 
with the center of attraction that coincides to the center of the original 
square. According to the Rutherford formula 
$$\tan\frac{\theta}{2}=\frac{\alpha}{MV^2b},$$
we find the scattering angle of the stars
$$\theta=2\arctan{\left[\frac{GM}{bV^2}\,\frac{1+2\sqrt{2}}{4}\right]}.$$

\subsection{}
Докажем, что в оптимальной ситуации центр тяжести любой верхней части 
конструкции находится на одной вертикали с правым краем пластины, на которой 
она опирается. 

\begin{figure}[htb]
\centerline{\epsfig{figure=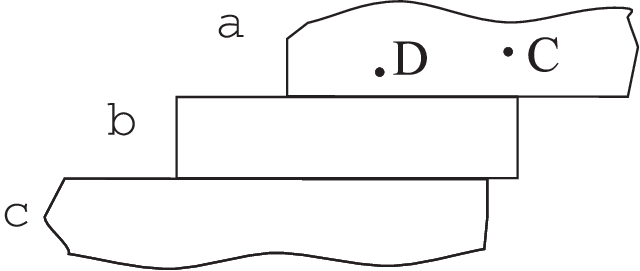,height=2cm}}
\end{figure}\noindent
Пусть это не так и центр
тяжести $C$ верхней части $a$ находится левее края пластины $b$.
Тогда часть $a$ можно сдвинуть вправо скажем на $x$ и она все равно
устойчиво будет опираться на пластину $b$. При этом центр тяжести
$D$ системы $b+a$ тоже сдвинется вправо, но на меньшую величину
$y=\frac{M}{M+m}x$, где $M$ -- масса части $a$ и $m$ -- масса
пластины. Если теперь систему $b+a$ сдвинем влево на $y$, ее центр
масс $D$ окажется в прежней точке и $b+a$ будет снова устойчиво
опираться на нижнюю часть $c$. После этих манипуляций пролет
увеличится на $x-y=\frac{m}{M+m}x$. Это доказывает сделанное
утверждение. 
\begin{figure}[htb]
\centerline{\epsfig{figure=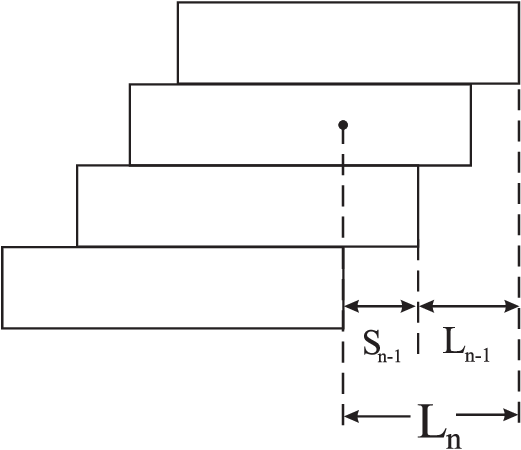,height=5cm}}
\end{figure}\noindent

\noindent
Следовательно, максимальный пролет $L_n$ удовлетворяет
рекуррентному соотношению $L_n=S_{n-1}+L_{n-1}$, где $S_{n-1}$ есть
расстояние по горизонтали от правого нижнего края верхней
конструкции из $(n-1)$ пластин до ее центра масс ( см. рисунок ).

С другой стороны, для центра масс конструкции из $n$ пластин, когда
центр масс верхней части из $(n-1)$ пластин проходит через правый
край нижней пластины, имеем
$$S_n=\frac{m}{m+(n-1)m}~\frac{a}{2}=\frac{a}{2}~\frac{1}{n}.$$ 
Поэтому
$$L_n=S_{n-1}+L_{n-1}=S_{n-1}+S_{n-2}+L_{n-2}=\ldots=S_{n-1}+S_{n-2}+
\ldots+S_1=$$ $$=\frac{a}{2}\left(1+\frac{1}{2}+\frac{1}{3}+\ldots+
\frac{1}{n-1}\right)\approx\frac{a}{2}\ln{n}$$ 
при $n\gg 1$. Тогда $L_n=3a$ дает $\ln{n}=6$ и $n=e^6\approx 400$.

Более общий случай, когда пластины не обязательно укладываются одна на другой и могут 
быть использованы в качестве противовеса, рассматривается в \cite{23A,23B,23C}.

\vspace*{5mm}
\noindent $\bullet$  $\bullet$  $\bullet$
 
\noindent
Let us prove that in the optimal situation the center of gravity of any 
top-level construction is located on the vertical line through the right edge 
of the plate on which it is based.

\begin{figure}[htb]
\centerline{\epsfig{figure=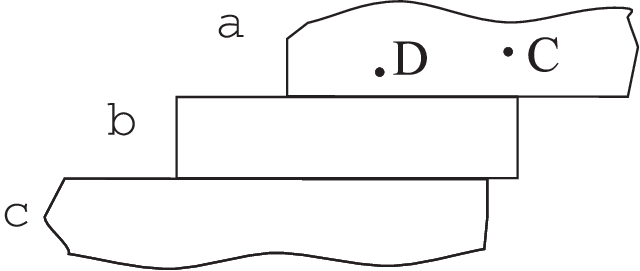,height=2cm}}
\end{figure}\noindent
Suppose this is not true and the center of gravity $C$ of the top part $a$ 
lies to the left of the edge of the plate $b$. Then the part $a$ can be 
shifted to the right say by $x$ and it will be still stably based on 
the plate $b$. At that the center of gravity $D$ of the system $b+a$ also 
moves to the right but by a smaller amount $y=\frac{M}{M+m}x$, where $M$ is 
the mass of the part $a$ and $m$ is the mass of the plate. If the system $b+a$ 
is now shifted to the left by $y$, its center of mass $D$ will return to the 
same point and $b+a$ will be again stably based on the lower part $c$. 
However, after these manipulations the overhang increases by  
$x-y=\frac{m}{M+m}x$. This proves the assertion. Hence, the maximum overhang 
$L_n$ satisfies the recurrence relation $L_n=S_{n-1}+L_{n-1}$, where $S_{n-1}$ 
is the horizontal distance from the bottom right edge of the upper 
construction of the $(n-1)$-plate stack to its center of mass (see the figure).

\begin{figure}[htb]
\centerline{\epsfig{figure=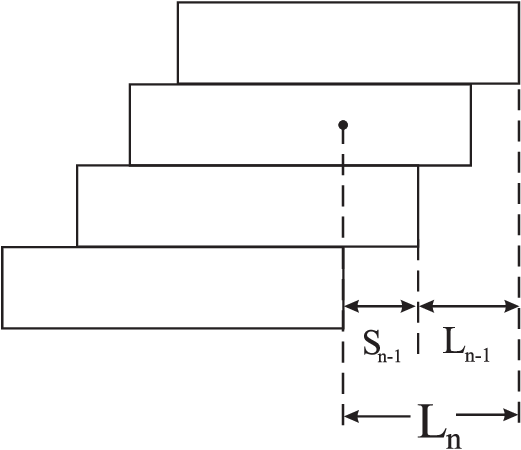,height=5.0cm}}
\end{figure}\noindent
On the other hand, for the center of mass of the $n$-plate stack, when the
center of mass of its upper part, consisting of $(n-1)$ plates, passes through 
the right edge of the bottom plate, we have
$$S_n=\frac{m}{m+(n-1)m}~\frac{a}{2}=\frac{a}{2}~\frac{1}{n}.$$
Therefore,
$$L_n=S_{n-1}+L_{n-1}=S_{n-1}+S_{n-2}+L_{n-2}=\ldots=S_{n-1}+S_{n-2}+
\ldots+S_1=$$ $$=\frac{a}{2}\left(1+\frac{1}{2}+\frac{1}{3}+\ldots+
\frac{1}{n-1}\right)\approx\frac{a}{2}\ln{n}$$
for $n\gg 1$. Then $L_n=3a$ gives $\ln{n}=6$ and $n=e^6\approx 400$.

More general case, when plates are not necessarily placed one on another and
can be used as counterweights, is considered in \cite{23A,23B,23C}.

\subsection{}
\begin{figure}[htb]
\centerline{\epsfig{figure=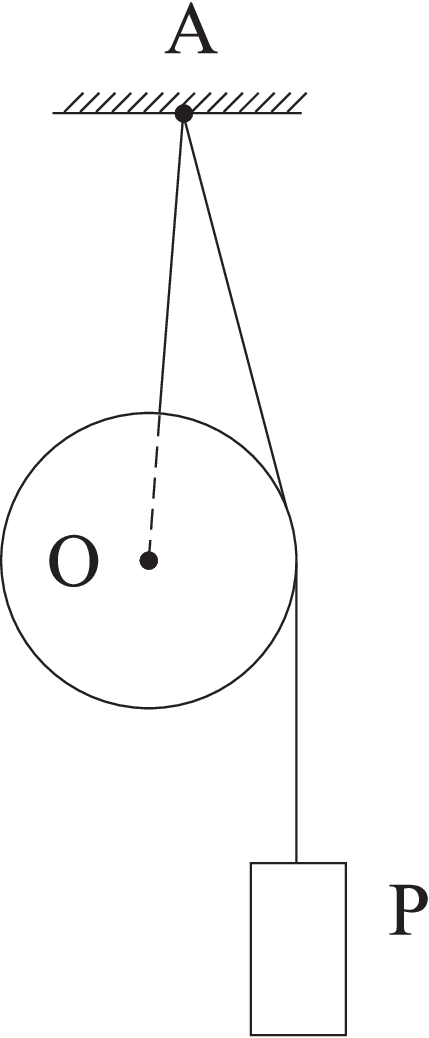,height=4cm}}
\end{figure}\noindent
При равновесии центр тяжести системы находится на одной
вертикали с точкой подвеса $A$. Пусть прямая $AO$ образует угол
$\varphi$ с вертикалью и начало системы отсчета находится в точке
$A$. Тогда в равновесии
$$0=x_{\mbox{ц.м.}}=\frac{1}{P+Q}[-Ql\sin{\varphi}+P(R-l\sin{\varphi})]$$
(центр шара находится влево от вертикали на расстоянии $l\sin{\varphi}$ и гиря 
находится вправо от вертикали на расстоянии $R-l\sin{\varphi}$). 
Следовательно, $PR=(Q+P)l\sin{\varphi}$ и
$$\sin{\varphi}=\frac{R}{l}\frac{P}{Q+P}.$$

\vspace*{5mm}
\noindent $\bullet$  $\bullet$  $\bullet$ 

\begin{figure}[htb]
\centerline{\epsfig{figure=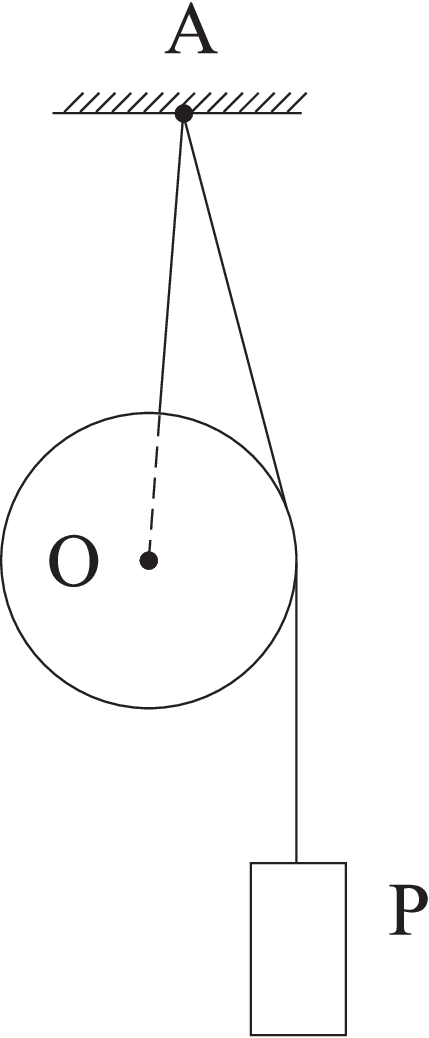,height=4cm}}
\end{figure}\noindent
At equilibrium, the center of gravity of the system lies on the vertical line
through the suspension point $A$. Let the line $AO$ makes an angle
$\varphi$ with the vertical and the origin of the reference frame is located 
at the point $A$. Then, in equilibrium,
$$0=x_{\mbox{cm}}=\frac{1}{P+Q}[-Ql\sin{\varphi}+P(R-l\sin{\varphi})]$$
(the center of the ball is on the left side of the vertical at the distance 
$l\sin{\varphi}$ and the weight is on the right side at the distance
$R-l\sin{\varphi}$). 
Hence, $PR=(Q+P)l\sin{\varphi}$ and
$$\sin{\varphi}=\frac{R}{l}\frac{P}{Q+P}.$$

\clearpage
\section*{Контрольная работа 14}
\setcounter{section}{14}
\setcounter{subsection}{0}
\subsection{}
\begin{figure}[htb]
\centerline{\epsfig{figure=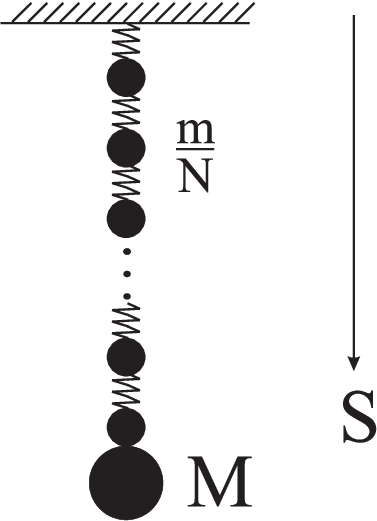,height=4cm}}
\end{figure}\noindent
Заменим канат на систему $N$ грузиков массы $\frac{m}{N}$ каждый и
$N$ невесомых пружин длины $\frac{l}{N}$ каждый. Жесткость
маленькой пружинки будет в $N$-раз больше, чем жесткость каната
(т.к. у нее длина в $N$-раз меньше). Поэтому уравнение движения
$i$-го грузика будет
$$\frac{m}{N}\ddot{S}_i=\frac{m}{N}g-Nk\left(S_i-S_{i-1}-\frac{l}{N}\right)+
Nk\left(S_{i+1}-S_i-\frac{l}{N}\right),\;\;i<N,$$ где $S_i$ - его координата. 
Это уравнение можно записать так
$$\ddot{S}_i=g+\frac{kN^2}{m}\Biggl[S_{i+1}-S_i-(S_i-S_{i-1})\Biggr].$$
Нас интересует предел $N\to\infty$ поэтому введем непрерывную
переменную $x$ через $x=i\frac{l}{N}$ и будем рассматривать $S_i$
как функцию $x$: $S_i\equiv S(x,t)$. Тогда
$$S_{i+1}-S_i=S(x+dx,t)-S(x,t)=\frac{\partial S(x,t)}{\partial
x}dx,$$ где $dx=\frac{l}{N}$, и
$$S_i-S_{i-1}=S(x,t)-S(x-dx,t)=\frac{\partial S(x-dx,t)}{\partial
x}dx.$$ Поэтому, в пределе $N\to\infty$, $S_{i+1}-S_i-(S_i-S_{i-1})$ 
можно заменить на
$$\Biggl[\frac{\partial S(x,t)}{\partial
x}-\frac{\partial S(x-dx,t)}{\partial x}\Biggr]dx=\frac{\partial^2
S(x,t)}{\partial x^2}(dx)^2=\frac{\partial^2 S}{\partial
x^2}\frac{l^2}{N^2}$$ и уравнение движения перепишется так
$$\frac{\partial^2 S(x,t)}{\partial
t^2}-\frac{kl^2}{m}\,\frac{\partial^2 S(x,t)}{\partial x^2}=g.$$ Но
это для $i<N$, т.е. $x<l$. Для последнего грузика вместе с телом
$M$, уравнение движения будет
$$\left(M+\frac{m}{N}\right)\ddot{S}_N=\left(M+\frac{m}{N}\right)g-Nk
\left(x_N-x_{N-1}-\frac{l}{N}\right).$$
Для перехода к непрерывному пределу, заметим что
$$N(x_N-x_{N-1})\to N[S(l,t)-S(l-dx,t)]\to N\frac{\partial
S(l-dx,t)}{\partial x}dx\to l\frac{\partial S(l,t)}{\partial x}.$$
Следовательно, в непрерывном пределе получим уравнение
$$\frac{\partial^2 S(l,t)}{\partial
t^2}=g-\frac{kl}{M}\Biggl[\frac{\partial S(l,t)}{\partial
l}-1\Biggr].$$ Кроме этого будем иметь граничное условие
$S(0,t)=0$, т.к. чем ближе точка подвеса, тем меньше растягивается
соответствующий участок каната. Пусть
$S(x,t)=S_0(x)+A(x)\cos{\omega t}$, где $S_0(x)$ соответствует положению
равновесия. Т.е. 
$$-\frac{kl^2}{m}\,\frac{d^2S_0(x)}{dx^2}=g\;\;\mbox{и}\;\;
0=g-\frac{kl}{M}\Biggl[\frac{dS_0(l)}{dl}-1\Biggr].$$ Подстановка
$S(x,t)$ в уравнения движения дает
$$A\omega^2+\frac{kl^2}{m}\,\frac{d^2A}{dx^2}=0,$$ 
и, при $x=l$,
$$\omega^2A(l)=\frac{kl}{M}\,\frac{dA(l)}{dl}.$$ Поэтому
$$A(x)=a_1\cos{\left(\sqrt{\frac{m\omega^2}{k}}\,\frac{x}{l}\right)}+
a_2\sin{\left(\sqrt{\frac{m\omega^2}{k}}\,\frac{x}{l}\right)},$$ где
$a_1$, $a_2$ - некоторые константы. Условие $S(0,t)=0$ дает
$A(0)=0$ и $a_1=0$. Тогда как
$\omega^2A(l)=\frac{kl}{M}\,\frac{dA(l)}{dl}$ условие означает
$$\omega^2\sin{\left[\sqrt{\frac{m}{k}}\omega\right]}=\frac{kl}{M}
\sqrt{\frac{m}{k}}\,\omega\,\frac{1}{l}\cos{\left[\sqrt{\frac{m}{k}}
\omega\right]}.$$
Или
$$\sqrt{\frac{m}{k}}\,\omega\,\mathrm{\tg}{\left[\sqrt{\frac{m}{k}}
\omega\right]}=
\frac{m}{M}.$$
Это трансцендентное уравнение определяет частоту вертикальных
колебаний $\omega$. Рассмотрим несколько предельных случаев. Если $m\to
0$,
$$\sqrt{\frac{m}{k}}\,\omega\,\mathrm{\tg}{\left[\sqrt{\frac{m}{k}}
\omega\right]}\to
\frac{m}{k}\,\omega^2$$
и получим $\omega^2=\frac{k}{M}$. Если $M\to 0$, то
$\mathrm{\tg}{\left[\sqrt{\frac{m}{k}}\omega\right]}=\infty$ и
$\sqrt{\frac{m}{k}}\,\omega=\frac{\pi}{2}+n\pi$, или
$$\omega=(2n+1)\,\frac{\pi}{2}\,\sqrt{\frac{k}{m}}, \;\; n=0,1,2,\ldots$$ 
Пусть теперь $\frac{m}{M}<<1$. Используя $\mathrm{\tg}x\approx
x+\frac{1}{3}x^3$, получим
$$\frac{m}{k}\,\omega^2\left(1+\frac{1}{3}\,\frac{m}{k}\,\omega^2\right)
\approx\frac{\frac{m}{k}\,\omega^2}{1-\frac{1}{3}\,\frac{m}{k}\,\omega^2}
=\frac{m}{M}.$$
Решая относительно $\omega^2$, будем иметь
$$\omega^2=\frac{k}{M+\frac{1}{3}m}\approx\frac{k}{M}\left(1-\frac{1}{3}\,
\frac{m}{M}\right).$$

\vspace*{5mm}
\noindent $\bullet$  $\bullet$  $\bullet$
 
\begin{figure}[htb]
\centerline{\epsfig{figure=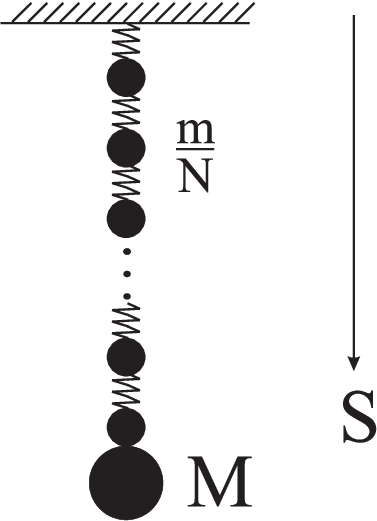,height=4cm}}
\end{figure}\noindent
Replace the rope by a system of $N$ weights of mass $\frac{m}{N}$
each connected by $N$ weightless springs of length $\frac{l}{N}$ each.
Stiffness of the small springs will be $N$-times greater than the stiffness 
of the rope (Because they have $N$-times less length). Therefore, the 
equation of motion of the $i$-th weight will be
$$\frac{m}{N}\ddot{S}_i=\frac{m}{N}g-Nk\left(S_i-S_{i-1}-\frac{l}{N}\right)+
Nk\left(S_{i+1}-S_i-\frac{l}{N}\right),\;\;i<N,$$ where $S_i$ is its 
coordinate. This equation can be rewritten as follows
$$\ddot{S}_i=g+\frac{kN^2}{m}\Biggl[S_{i+1}-S_i-(S_i-S_{i-1})\Biggr].$$
We are interested in the limit $N\to\infty$, so we introduce a continuous
variable $x$ through $x=i\frac{l}{N}$ and consider $S_i$ as a function of
$x$: $S_i\equiv S(x,t)$. Then
$$S_{i+1}-S_i=S(x+dx,t)-S(x,t)=\frac{\partial S(x,t)}{\partial
x}dx,$$ where $dx=\frac{l}{N}$, and
$$S_i-S_{i-1}=S(x,t)-S(x-dx,t)=\frac{\partial S(x-dx,t)}{\partial
x}dx.$$ Therefore, in the limit $N\to\infty$, $S_{i+1}-S_i-(S_i-S_{i-1})$ 
can be replaced by 
$$\Biggl[\frac{\partial S(x,t)}{\partial
x}-\frac{\partial S(x-dx,t)}{\partial x}\Biggr]dx=\frac{\partial^2
S(x,t)}{\partial x^2}(dx)^2=\frac{\partial^2 S}{\partial
x^2}\frac{l^2}{N^2}$$ and the equation of motion takes the form
$$\frac{\partial^2 S(x,t)}{\partial
t^2}-\frac{kl^2}{m}\,\frac{\partial^2 S(x,t)}{\partial x^2}=g.$$ But
this is true only for $i<N$, that is, $x<l$. For the last weight, together 
with the body $M$, the equation of motion is
$$\left(M+\frac{m}{N}\right)\ddot{S}_N=\left(M+\frac{m}{N}\right)g-Nk
\left(x_N-x_{N-1}-\frac{l}{N}\right).$$
To get the continuum limit, we note that
$$N(x_N-x_{N-1})\to N[S(l,t)-S(l-dx,t)]\to N\frac{\partial
S(l-dx,t)}{\partial x}dx\to l\frac{\partial S(l,t)}{\partial x}.$$
Consequently, in the continuum limit we obtain the equation
$$\frac{\partial^2 S(l,t)}{\partial
t^2}=g-\frac{kl}{M}\Biggl[\frac{\partial S(l,t)}{\partial
l}-1\Biggr].$$ In addition, we have the boundary condition $S(0,t)=0$,
since the closer the point of suspension, the lower is the stretching of the
appropriate section of the rope. Let $S(x,t)=S_0(x)+A(x)\cos{\omega t}$,
where $S_0(x)$ corresponds to the position at equilibrium. That is,
$$-\frac{kl^2}{m}\,\frac{d^2S_0(x)}{dx^2}=g\;\;\mbox{и}\;\;
0=g-\frac{kl}{M}\Biggl[\frac{dS_0(l)}{dl}-1\Biggr].$$ Substitution of $S(x,t)$
in the equations of motion gives
$$A\omega^2+\frac{kl^2}{m}\,\frac{d^2A}{dx^2}=0,$$ 
and, at $x=l$,
$$\omega^2A(l)=\frac{kl}{M}\,\frac{dA(l)}{dl}.$$ Therefore,
$$A(x)=a_1\cos{\left(\sqrt{\frac{m\omega^2}{k}}\,\frac{x}{l}\right)}+
a_2\sin{\left(\sqrt{\frac{m\omega^2}{k}}\,\frac{x}{l}\right)},$$ where
$a_1$, $a_2$ are some constants. The condition $S(0,t)=0$ gives
$A(0)=0$ and $a_1=0$.  Whereas
$\omega^2A(l)=\frac{kl}{M}\,\frac{dA(l)}{dl}$ condition implies
$$\omega^2\sin{\left[\sqrt{\frac{m}{k}}\omega\right]}=\frac{kl}{M}
\sqrt{\frac{m}{k}}\,\omega\,\frac{1}{l}\cos{\left[\sqrt{\frac{m}{k}}
\omega\right]}.$$
Or
$$\sqrt{\frac{m}{k}}\,\omega\,\mathrm{\tg}{\left[\sqrt{\frac{m}{k}}
\omega\right]}=
\frac{m}{M}.$$
This transcendental equation determines the frequency of vertical
oscillations $\omega$. Let us consider several limiting cases. 
If $m\to 0$, then
$$\sqrt{\frac{m}{k}}\,\omega\,\mathrm{\tg}{\left[\sqrt{\frac{m}{k}}
\omega\right]}\to
\frac{m}{k}\,\omega^2$$
and we get $\omega^2=\frac{k}{M}$. If $M\to 0$, then
$\mathrm{\tg}{\left[\sqrt{\frac{m}{k}}\omega\right]}=\infty$ and
$\sqrt{\frac{m}{k}}\,\omega=\frac{\pi}{2}+n\pi$, or
$$\omega=(2n+1)\,\frac{\pi}{2}\,\sqrt{\frac{k}{m}}, \;\; n=0,1,2,\ldots$$ 
Now let $\frac{m}{M}<<1$. Using $\mathrm{\tg}x\approx x+\frac{1}{3}x^3$, 
we get 
$$\frac{m}{k}\,\omega^2\left(1+\frac{1}{3}\,\frac{m}{k}\,\omega^2\right)
\approx\frac{\frac{m}{k}\,\omega^2}{1-\frac{1}{3}\,\frac{m}{k}\,\omega^2}
=\frac{m}{M}.$$
Solving with respect to $\omega^2$, we obtain
$$\omega^2=\frac{k}{M+\frac{1}{3}m}\approx\frac{k}{M}\left(1-\frac{1}{3}\,
\frac{m}{M}\right).$$

\subsection{}
Пусть кольцо радиуса $r$ повернулось на маленький угол $\theta$.
Тогда оно поднимется на высоту
$$h=l-\sqrt{l^2-r^2\theta^2}\approx\frac{r^2\theta^2}{2l}.$$
Заметим, что
$$\dot{h}=\frac{r^2\theta\dot{\theta}}{l}=\frac{r\theta}{l}r
\dot{\theta}<<r\dot{\theta},$$
поэтому энергией поступательного движения $\frac{m\dot{h}^2}{2}$
можно пренебречь по сравнению с энергией вращательного движения
$\frac{I\dot{\theta}^2}{2}=\frac{m}{2}r^2\dot{\theta}^2$. Полная
энергия кольца
$$E=\frac{m}{2}r^2\dot{\theta}^2+mgh=\frac{1}{2}mr^2\dot{\theta}^2+
\frac{1}{2}mg\frac{r^2}{l}\theta^2.$$
Поэтому
$0=\frac{dE}{dt}=mr^2\dot{\theta}\left(\ddot{\theta}+\frac{g}{l}\theta\right)$
дает уравнение движения $\ddot{\theta}+\frac{g}{l}\theta=0$.
Следовательно, частота крутильных колебаний
$\omega=\sqrt{\frac{g}{l}}$.

\vspace*{5mm}
\noindent $\bullet$  $\bullet$  $\bullet$
 
\noindent
Let the ring of radius $r$ has turned by a small angle $\theta$.
Then it rises to a height
$$h=l-\sqrt{l^2-r^2\theta^2}\approx\frac{r^2\theta^2}{2l}.$$
Note that
$$\dot{h}=\frac{r^2\theta\dot{\theta}}{l}=\frac{r\theta}{l}r
\dot{\theta}<<r\dot{\theta}.$$
Therefore, the energy of translational motion $\frac{m\dot{h}^2}{2}$
can be neglected compared to the energy of rotational motion
$\frac{I\dot{\theta}^2}{2}=\frac{m}{2}r^2\dot{\theta}^2$. The total energy 
of the ring is
$$E=\frac{m}{2}r^2\dot{\theta}^2+mgh=\frac{1}{2}mr^2\dot{\theta}^2+
\frac{1}{2}mg\frac{r^2}{l}\theta^2.$$
The conservation of energy,
$0=\frac{dE}{dt}=mr^2\dot{\theta}\left(\ddot{\theta}+\frac{g}{l}\theta\right)$,
gives the equation of motion $\ddot{\theta}+\frac{g}{l}\theta=0$.
Consequently, the frequency of torsional oscillations is
$\omega=\sqrt{\frac{g}{l}}$.

\subsection{}
Из-за цилиндрической симметрии, $z$-компонента углового момента
$L_z=m\rho^2\dot{\varphi}$ сохраняется. Энергия частицы
$$E=\frac{m}{2}\left(\dot{\rho}^2+\rho^2\dot{\varphi}^2\right)+\frac{m}{2}
\dot{z}^2+mgz=\frac{m}{2}(1+z^{\prime\, 2})\dot{\rho}^2
+ \frac{L_z^2}{2m\rho^2}+mgz,$$ где $z^\prime=\frac{dz}{d\rho}$. Энергия
сохраняется, поэтому
$$0=\frac{dE}{dt}=\dot{\rho}\Biggl[m\ddot{\rho}(1+z^{\prime\, 2})+
m\dot{\rho}^2z^\prime z^{\prime \prime}-
\frac{L_z^2}{4m\rho^3}+mgz^\prime\Biggr]$$
и получим уравнение движения
$$m(1+z^{\prime\, 2})\ddot{\rho}+m\dot{\rho}^2z^\prime z^{\prime \prime}
-\frac{L_z^2}{4m\rho^3}+mgz^\prime=0.$$
Пусть $\rho=r+x$, где $r$ -- радиус равновесной круговой орбиты, а
$x$ -- возмущение. Тогда 
$$\frac{L_z^2}{4mr^3}=mgz^\prime(r),\;\;
z^\prime(\rho)\approx z^\prime(r)+z^{\prime \prime}(r)x, \;\;
\frac{1}{\rho^3}=\frac{1}{(r+x)^3}\approx\frac{1}{r^3}\Biggl(1-\frac{3x}{r}
\Biggr),$$
и, с точностью до линейных по $x$ членов, уравнение движения
станет
$$m\Biggl(1+z^{\prime\, 2}(r)\Biggr)\ddot{x}+\frac{L_z^2}{4mr^3}\,\frac{3x}{r}
+mgz^{\prime \prime}(r)x=0.$$
Или, если учтем $\frac{L_z^2}{4mr^3}=mgz^\prime(r)$,
$$\Biggl(1+z^{\prime\, 2}(r)\Biggr)\ddot{x}+g\Biggl[z^{\prime \prime}(r)+
3\frac{z^\prime(r)}{r}\Biggr]x=0.$$
Если $z^{\prime \prime}(r)+3\frac{z^\prime(r)}{r}>0$, будем иметь осцилляции 
около равновесной орбиты с частотой
$$\omega=\sqrt{g\frac{z^{\prime \prime}(r)+3\frac{z^\prime(r)}{r}}
{1+z^{\prime\, 2}(r)}}.$$
 Если $z^{\prime \prime}(r)+3\frac{z^\prime(r)}{r}<0$, возмущение $x(t)$ будет
экспоненциально расти и орбита не будет стабильной. Если
$z^{\prime\prime}(r)+3\frac{z^\prime(r)}{r}=0$, возмущение будет расти линейно 
со временем, и орбита тоже не будет стабильной. Таким образом,
условием устойчивого равновесия является
$z^{\prime\prime}(r)+3\frac{z^\prime(r)}{r}>0$. Из $\rho^2-z^2=R^2$ следует, 
что $zz^\prime=\rho$. Т.е. $z^\prime=\frac{\rho}{z}$. Тогда
$$z^{\prime \prime}=\frac{1}{z}-\frac{\rho z^\prime}{z^2}=\frac{1}{z}-
\frac{\rho^2}{z^3}.$$ Поэтому
$$z^{\prime \prime}(r)+3\frac{z^\prime(r)}{r}=\frac{1}{z}-\frac{r^2}{z^3}+
3\frac{1}{z}=
\frac{1}{z}\Biggl(4-\frac{r^2}{z^2}\Biggr).$$
Но $z>0$ (иначе вертикальной компоненты силы реакции опоры не
будет). Следовательно, орбита устойчива, если
$$4>\frac{r^2}{z^2}=\frac{r^2}{r^2-R^2}.$$ Отсюда $3r^2>4R^2$ и
$$r>\frac{2}{\sqrt{3}}R.$$

\vspace*{5mm}
\noindent $\bullet$  $\bullet$  $\bullet$
 
\noindent
Because of the cylindrical symmetry, $z$-component of the angular momentum
$L_z=m\rho^2\dot{\varphi}$ is conserved. The particle energy is
$$E=\frac{m}{2}\left(\dot{\rho}^2+\rho^2\dot{\varphi}^2\right)+\frac{m}{2}
\dot{z}^2+mgz=\frac{m}{2}(1+z^{\prime\, 2})\dot{\rho}^2
+ \frac{L_z^2}{2m\rho^2}+mgz,$$ where $z^\prime=\frac{dz}{d\rho}$. It is also 
conserved. Hence
$$0=\frac{dE}{dt}=\dot{\rho}\Biggl[m\ddot{\rho}(1+z^{\prime\, 2})+
m\dot{\rho}^2z^\prime z^{\prime \prime}-
\frac{L_z^2}{4m\rho^3}+mgz^\prime\Biggr]$$
and we get the following equation of motion
$$m(1+z^{\prime\, 2})\ddot{\rho}+m\dot{\rho}^2z^\prime z^{\prime \prime}
-\frac{L_z^2}{4m\rho^3}+mgz^\prime=0.$$
Let $\rho=r+x$, where $r$ is  the radius of the equilibrium circular orbit 
and $x$ is a perturbation. Then
$$\frac{L_z^2}{4mr^3}=mgz^\prime(r),\;\;
z^\prime(\rho)\approx z^\prime(r)+z^{\prime \prime}(r)x, \;\;
\frac{1}{\rho^3}=\frac{1}{(r+x)^3}\approx\frac{1}{r^3}\Biggl(1-\frac{3x}{r}
\Biggr),$$
and, up to linear in $x$ terms, the equation of motion becomes
$$m\Biggl(1+z^{\prime\, 2}(r)\Biggr)\ddot{x}+\frac{L_z^2}{4mr^3}\,\frac{3x}{r}
+mgz^{\prime \prime}(r)x=0.$$
Or, if we take into account $\frac{L_z^2}{4mr^3}=mgz^\prime(r)$,
$$\Biggl(1+z^{\prime\, 2}(r)\Biggr)\ddot{x}+g\Biggl[z^{\prime \prime}(r)+
3\frac{z^\prime(r)}{r}\Biggr]x=0.$$
If $z^{\prime \prime}(r)+3\frac{z^\prime(r)}{r}>0$, we have oscillations 
around the equilibrium orbit with frequency 
$$\omega=\sqrt{g\frac{z^{\prime \prime}(r)+3\frac{z^\prime(r)}{r}}
{1+z^{\prime\, 2}(r)}}.$$
If $z^{\prime \prime}(r)+3\frac{z^\prime(r)}{r}<0$, the perturbation $x(t)$ 
will grow exponentially and the orbit will be not stable. If
$z^{\prime\prime}(r)+3\frac{z^\prime(r)}{r}=0$, the perturbation will grow 
linearly with time, and the orbit also will be not stable. Thus, the condition 
for stable equilibrium is $z^{\prime\prime}(r)+3\frac{z^\prime(r)}{r}>0$. 
It follows from $\rho^2-z^2=R^2$ that $zz^\prime=\rho$. That is, $z^\prime=
\frac{\rho}{z}$. Then
$$z^{\prime \prime}=\frac{1}{z}-\frac{\rho z^\prime}{z^2}=\frac{1}{z}-
\frac{\rho^2}{z^3}.$$ Therefore,
$$z^{\prime \prime}(r)+3\frac{z^\prime(r)}{r}=\frac{1}{z}-\frac{r^2}{z^3}+
3\frac{1}{z}=
\frac{1}{z}\Biggl(4-\frac{r^2}{z^2}\Biggr).$$
But $z>0$ (otherwise we will not have the vertical component of the ground 
reaction force). Consequently, the orbit is stable if
$$4>\frac{r^2}{z^2}=\frac{r^2}{r^2-R^2}.$$ Hence $3r^2>4R^2$ and
$$r>\frac{2}{\sqrt{3}}R.$$

\subsection{}
Стержень начнет вращаться относительно точки $O$ с угловым
ускорением $\dot{\omega}=\frac{mg(l/2)}{I}$, где $l$-длина
стержня, а
$I=\frac{ml^2}{12}+m\Bigl(\frac{l}{2}\Bigr)^2=\frac{1}{3}ml^2$ -- его
момент  инерции относительно $O$. Следовательно,
$\dot{\omega}=\frac{3}{2}\,\frac{g}{l}$. Тогда центр масс стержня будет
двигаться с линейным ускорением
$a=\dot{\omega}\frac{l}{2}=\frac{3}{4}g$. С другой стороны,
$a=\frac{mg-T}{m}$. Следовательно, $T=m(g-a)=\frac{1}{4}mg$.
\begin{figure}[htb]
\centerline{\epsfig{figure=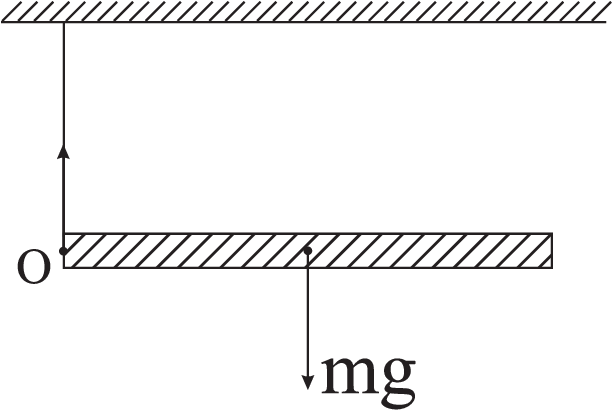,height=4cm}}
\end{figure}

\noindent $\bullet$  $\bullet$  $\bullet$
%\begin{figure}[htb]
%\centerline{\epsfig{figure=fig59.eps,height=4cm}}
%\end{figure}\noindent 

\noindent The rod will begin to rotate about the point $O$ with an angular 
acceleration $\dot{\omega}=\frac{mg(l/2)}{I}$, where $l$ is the length
of the rod and
$I=\frac{ml^2}{12}+m\Bigl(\frac{l}{2}\Bigr)^2=\frac{1}{3}ml^2$ is its
moment of inertia about $O$. Consequently,
$\dot{\omega}=\frac{3}{2}\,\frac{g}{l}$. Then the center of mass of the rod 
will move with a linear acceleration
$a=\dot{\omega}\frac{l}{2}=\frac{3}{4}g$. On the other hand,
$a=\frac{mg-T}{m}$. Hence, $T=m(g-a)=\frac{1}{4}mg$.

\subsection{}
\begin{figure}[htb]
\centerline{\epsfig{figure=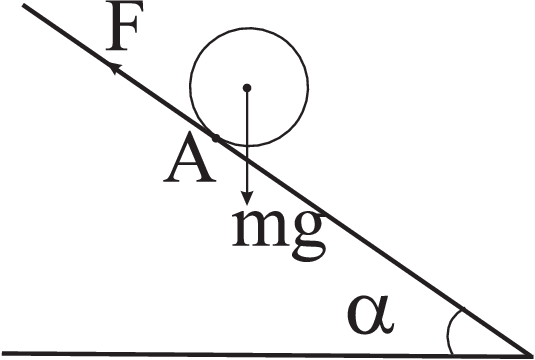,height=4cm}}
\end{figure}\noindent
Сила трения $F$ раскручивает цилиндр с угловым ускорением
$\dot{\omega}=\frac{FR}{I}=\frac{2F}{mR}$. В точке касания
$A$ скорость равна нулю. Поэтому $V-\omega R=0$, где $V$ есть скорость
центра масс цилиндра. Тогда $a=\dot{\omega}R=\frac{2F}{m}$. С
другой стороны, $ma=mg\sin{\alpha}-F$. Следовательно,
$F=\frac{1}{3}mg\sin{\alpha}$. Но сила трения не может
превысить максимальное значение $\mu N=\mu mg\cos{\alpha}$.
Следовательно $\frac{1}{3}mg\sin{\alpha}\leq\mu mg\cos{\alpha}$ и
если $\tg{\alpha}\leq 3\mu$, цилиндр будет скатываться без
скольжения. Раз цилиндр скатывается со скольжением, значит $\alpha
> \arctg{(3\mu)}$.

\vspace*{5mm}
\noindent $\bullet$  $\bullet$  $\bullet$
 
\begin{figure}[htb]
\centerline{\epsfig{figure=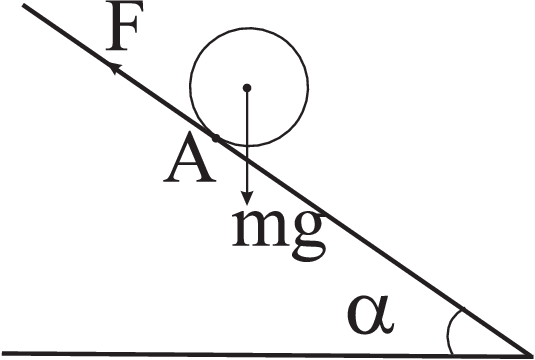,height=4cm}}
\end{figure}\noindent
The friction force $F$ rotates the cylinder with angular acceleration
$\dot{\omega}=\frac{FR}{I}=\frac{2F}{mR}$. At the point of tangency $A$, 
the velocity is zero. Therefore $V-\omega R=0$, where $V$ is the 
instantaneous velocity of the center of mass of the cylinder. Then 
$a=\dot{\omega}R=\frac{2F}{m}$. On the other hand, $ma=mg\sin{\alpha}-F$.
Hence $F=\frac{1}{3}mg\sin{\alpha}$. But the friction force  
cannot exceed the maximum value $\mu N=\mu mg\cos{\alpha}$.
Therefore, $\frac{1}{3}mg\sin{\alpha}\leq\mu mg\cos{\alpha}$ and if
$\tg{\alpha}\leq 3\mu$, the cylinder will roll down without slipping. 
As the cylinder slides, then  $\alpha > \arctg{(3\mu)}$.

\subsection{}
Скорость шариков в системе наблюдателя находим по формуле сложения
скоростей: 
$$V_1=\frac{u+V}{1+\frac{uV}{c^2}}\;\; {\mbox{и}}\;\;
V_2=\frac{V-u}{1-\frac{uV}{c^2}}.$$ Суммарный импульс шариков
(после разрыва цепи) будет
$$mV_1\gamma_1+mV_2\gamma_2=mV_1\gamma_V\gamma_u\Biggl(1+\frac{uV}{c^2}\Biggr)
+ mV_2\gamma_V\gamma_u\Biggl(1-\frac{uV}{c^2}\Biggr)=2mV\gamma_V\gamma_u.$$
Здесь мы воспользовались формулой
$$\gamma_{\vec{V}\oplus\vec{u}}=\gamma_V\gamma_u\Biggl(1+\frac{\vec{u}\cdot
\vec{V}}{c^2}\Biggr).$$
Заметим, что $p_1+p_2\neq 2mV\gamma_{V}=q_1+q_2$ -- сумма импульсов
шариков до разрыва цепи. Дело в том, что сжатая пружина имеет
упругую энергию покоя $E$ и, следовательно, массу
$M=\frac{E}{c^2}$. $E$ находим из закона сохранения энергии в
системе покоя трубки
$$E+2mc^2=2mc^2\gamma_u$$ 
(при этом мы массой не деформированной пружины пренебрегли). Следовательно,
$$M=\frac{E}{c^2}=2m(\gamma_u-1)$$ и импульс сжатой пружины будет
$q=MV\gamma_V=2mV\gamma_V(\gamma_u-1)$. Поэтому суммарный импульс
до разрыва цепи есть
$$q+q_1+q_2=2mV\gamma_V(\gamma_u-1)+2mV\gamma_V=2mV\gamma_V\gamma_u=p_1+p_2.$$

\vspace*{5mm}
\noindent $\bullet$  $\bullet$  $\bullet$
 
\noindent
Velocities of the balls in the observer's frame are given by the velocity 
addition formula:
$$V_1=\frac{u+V}{1+\frac{uV}{c^2}}\;\; {\mbox{and}}\;\;
V_2=\frac{V-u}{1-\frac{uV}{c^2}}.$$ The balls total momentum 
(after breaking of the string) is
$$mV_1\gamma_1+mV_2\gamma_2=mV_1\gamma_V\gamma_u\Biggl(1+\frac{uV}{c^2}\Biggr)
+ mV_2\gamma_V\gamma_u\Biggl(1-\frac{uV}{c^2}\Biggr)=2mV\gamma_V\gamma_u.$$
Here we used the formula
$$\gamma_{\vec{V}\oplus\vec{u}}=\gamma_V\gamma_u\Biggl(1+\frac{\vec{u}\cdot
\vec{V}}{c^2}\Biggr).$$
Note that $p_1+p_2\neq 2mV\gamma_{V}=q_1+q_2$ -- the sum of balls momenta 
before breaking of the string. The point is that a compressed spring has an 
elastic rest energy $E$ and hence the mass $M=\frac{E}{c^2}$. $E$ is found 
from the energy conservation law in the rest frame of the tube
$$E+2mc^2=2mc^2\gamma_u$$ 
(we have neglected a mass of the not deformed spring). Consequently,
$$M=\frac{E}{c^2}=2m(\gamma_u-1)$$ and the momentum of the deformed spring
in the observer's frame will be
$q=MV\gamma_V=2mV\gamma_V(\gamma_u-1)$. Therefore, in this frame the total 
momentum before the breaking of the string is
$$q+q_1+q_2=2mV\gamma_V(\gamma_u-1)+2mV\gamma_V=2mV\gamma_V\gamma_u=p_1+p_2.$$

\section*{Контрольная работа 15}
\setcounter{section}{15}
\setcounter{subsection}{0}
\subsection{}
В системе полуцилиндра сила инерции $-mg\sin{\alpha}$ компенсирует компоненту
силы тяжести вдоль наклонной плоскости. Поэтому задача эквивалентна нахождению
частоты малых колебаний полуцилиндра на гладкой горизонтальной поверхности в
эффективном поле тяжести $g\cos{\alpha}$.

\begin{figure}[htb]
\centerline{\epsfig{figure=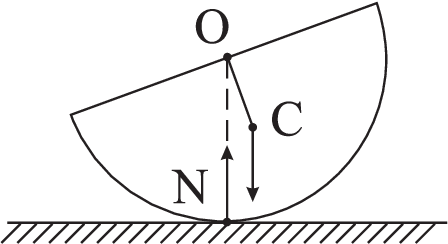,height=4cm}}
\end{figure}\noindent
Так как в горизонтальном направлении сила не 
действует, центр тяжести полуцилиндра движется только по вертикали. При малых 
колебаниях этим смещением по вертикали тоже можно пренебречь и считать, что 
полуцилиндр вращается вокруг центра масс. При наклоне полуцилиндра на угол
$\varphi\ll 1$, сила реакции опоры $N\approx mg\cos{\alpha}$ создает 
возвращающий момент $Nb\sin{\varphi}\approx (mgb\cos{\alpha})\varphi$, где 
$b$ -- расстояние от центра масс $C$ до геометрического центра цилиндра $O$. 
Следовательно, уравнение движения будет $I_c\ddot{\varphi}=-(mgb\cos{\alpha})
\varphi$ и частота малых колебаний 
$$\omega^2=\frac{mgb\cos{\alpha}}{I_c}.$$
Момент инерции полуцилиндра относительно оси цилиндра $I_o$ будет половина 
момента инерции цилиндра (массы $2m$), т.е. $I_o=\frac{1}{2}\,\frac{1}{2}
\,2mR^2=\frac{1}{2}mR^2$. С другой стороны, по теореме Штейнера, 
$I_o=I_c+mb^2$. Отсюда получаем $I_c=\frac{1}{2}m(R^2-2b^2)$. Величину $b$ 
можно найти по второй теореме Гульдена: объем тела вращения равен произведению
длины окружности, описываемой центром тяжести фигуры на ее площадь. Если 
вращать полукруг вокруг своего диаметра,
получится шар. Поэтому, согласно теореме Гульдена, должны иметь $\frac{4}{3}\pi
R^3=\frac{1}{2}\pi R^2 \,2\pi b$. Отсюда $b=\frac{4R}{3\pi}$ и, окончательно,
$$\omega^2=\frac{g\cos{\alpha}}{R}\,\frac{24\pi}{9\pi^2-32}.$$

\vspace*{5mm}
\noindent $\bullet$  $\bullet$  $\bullet$
 
\noindent
In the rest frame of half-cylinder, the inertial force $-mg\sin{\alpha}$ 
compensates the component of gravity force along the inclined plane. 
Therefore, the problem is equivalent to finding the frequency of small 
oscillations of a half-cylinder on a smooth horizontal surface under the
effective field of gravity $g\cos{\alpha}$.

\begin{figure}[htb]
\centerline{\epsfig{figure=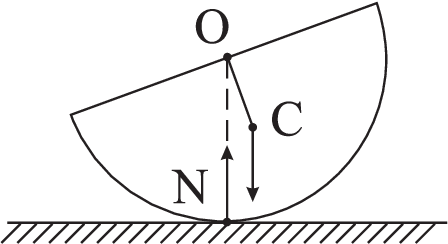,height=4cm}}
\end{figure}\noindent
Since there is no force acting in the horizontal direction, the center of 
mass of the half-cylinder moves only vertically. For small oscillations,
we can neglect this tiny vertical shift too and assume that the
half-cylinder rotates around its center of mass. When tilted at an angle
$\varphi\ll 1$, the ground reaction force  $N\approx mg\cos{\alpha}$
creates a return torque $Nb\sin{\varphi}\approx (mgb\cos{\alpha})\varphi$, 
where $b$ is the distance from the center of mass $C$ to the geometric center 
of the cylinder $O$. Consequently, the equation of motion is
$I_c\ddot{\varphi}=-(mgb\cos{\alpha})\varphi$ and the frequency of small 
oscillations will be
$$\omega^2=\frac{mgb\cos{\alpha}}{I_c}.$$
The Moment of inertia $I_o$ of the half-cylinder about the axis of the 
cylinder is one half of the moment of inertia of the cylinder (of mass $2m$), 
that is, $I_o=\frac{1}{2}\,\frac{1}{2} \,2mR^2=\frac{1}{2}mR^2$.
On the other hand, by Steiner's theorem $I_o=I_c+mb^2$. Hence we get 
$I_c=\frac{1}{2}m(R^2-2b^2)$. Value of $b$ can be found by applying the second 
theorem of Pappus-Guldinus: the volume of a solid of revolution generated by 
rotating a plane figure about an external axis is equal to the product of the 
area of the figure and the distance traveled by its geometric centroid.
If we rotate a semicircle around its diameter, we get a ball. Therefore, by 
theorem of Pappus-Guldinus, we must have $\frac{4}{3}\pi
R^3=\frac{1}{2}\pi R^2 \,2\pi b$. Hence $b=\frac{4R}{3\pi}$ and, finally,
$$\omega^2=\frac{g\cos{\alpha}}{R}\,\frac{24\pi}{9\pi^2-32}.$$

\subsection{}
Выталкивающая сила больше, чем сила тяжести. Тем не менее песочные часы не 
сразу начнут всплывать после переворачивания трубки. Дело в том, что 
выталкивающая сила приложена к геометрическому центру симметричных часов, а 
сила тяжести -- к центру масс. Вначале, когда часы плавают в трубке, центр 
масс находится ниже геометрического центра и это состояние устойчиво 
относительно наклона часов, так как при случайном наклоне возникает 
возвращающий момент. 

\begin{figure}[htb]
\centerline{\epsfig{figure=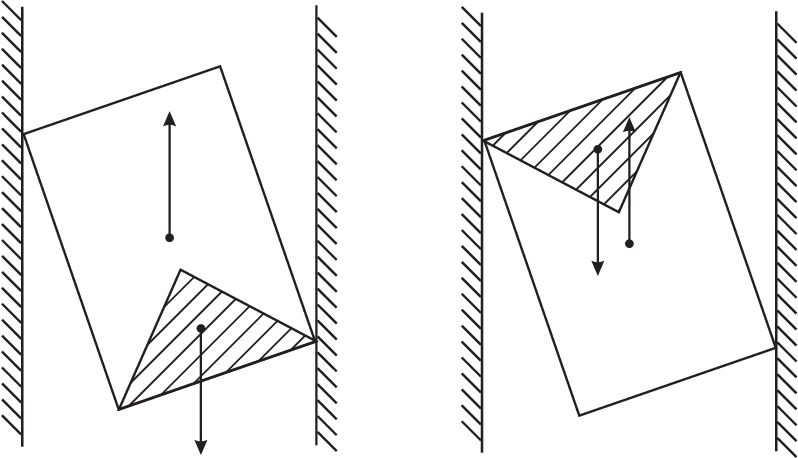,height=4cm}}
\end{figure}\noindent
После переворачивания трубки центр тяжести оказывается выше геометрического 
центра и это состояние не устойчиво относительно случайных наклонов часов. 
В своей попытке развернуться, часы упираются в стенки трубки и давят на них. 
В результате возникает сила трения которая мешает всплытию часов. После 
израсходования примерно половины песка, центр тяжести песочных
часов снова окажется ниже геометрического центра и часы начнут всплывать (на самом 
деле часы начнут всплывать несколько раньше, когда выталкивающая сила превысит 
ослабившее трение).  

\vspace*{5mm}
\noindent $\bullet$  $\bullet$  $\bullet$
 
\noindent
The buoyancy force is greater than the force of gravity. Nevertheless, 
after overturning of the tube, the hourglass will not immediately begin 
to rise. The case is that the buoyancy force is applied to the geometric 
center of the symmetric hourglass and the gravitational force acts trough 
the center of mass. Initially, when the clock is floating in the tube, the 
center of mass is below the geometric center and this state is stable with 
respect to tilting the hourglass, as a random tilt produces a return torque.

\begin{figure}[htb]
\centerline{\epsfig{figure=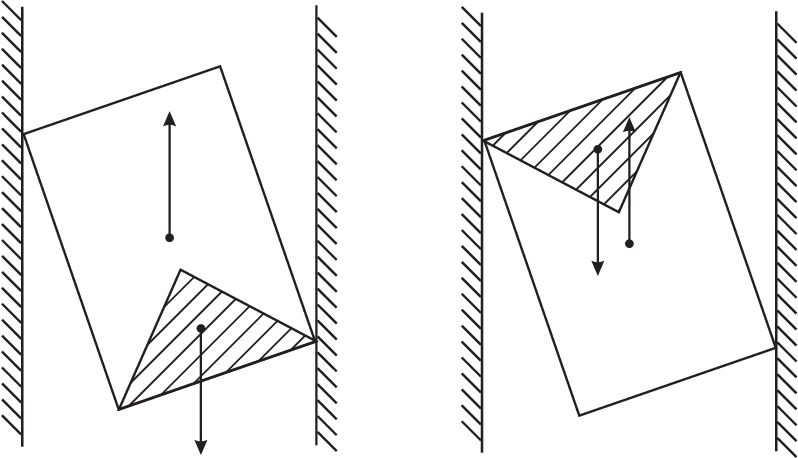,height=4cm}}
\end{figure}\noindent
After the tube is turned over, the center of gravity becomes located above 
the geometric center and this state is not stable under random tilts of the 
hourglass. In its attempt to turn around, the clock is stopped by the wall of 
the tube and it exerts a pressure on the wall. As a result, a friction force 
emerges which prevents the ascent of the hourglass. After about half of the 
sand is spent, the center of gravity of the hourglass will be lowered 
again below the geometric center and the clock will start to rise (in fact, 
it will start to rise a bit earlier when the buoyancy force overcomes the 
decreasing friction).

\subsection{}
Удар рассекает стержень пополам. Пусть каждая половина имеет массу $m$ и длину
$l$. Из-за удара половинка стержня приобретает скорость $V=\frac{1}{m}\int 
Fdt$ и раскручивается относительно центра масс с угловой скоростью $\omega=
\frac{1}{I}\int F\frac{l}{2}dt=\frac{6}{ml}\int Fdt$. Следовательно, конец 
стержня, который опирается на бокал с вином, сразу после удара будет иметь 
скорость $V-\frac{l}{2} \omega=-2V$ и полетит вверх. Поэтому вино не прольется.

Но это рассуждение пренебрегает упругими свойствами стержня. На самом деле 
после удара вдоль стержня распространяется волна поперечной деформации и 
другой конец стержня все равно успевает немного сдвинуться вниз. Поэтому для 
успешной демонстрации опыта надо предусмотреть демпфирование смещений конца стержня
порядка 1~мм \cite{36}.

\vspace*{5mm}
\noindent $\bullet$  $\bullet$  $\bullet$ 

\noindent
The strike cuts the rod in half. Suppose that each half has a mass $m$ and 
a length $l$. Due to the hit, each half of the rod acquires a center of 
mass velocity  $V=\frac{1}{m}\int Fdt$ and begins to rotate about the center 
of mass with angular velocity  $\omega=\frac{1}{I}\int F\frac{l}{2}dt=
\frac{6}{ml}\int Fdt$. Consequently, the end of the rod, which rests on the 
glass of wine, immediately after the impact acquires the velocity
$V-\frac{l}{2} \omega=-2V$ and will fly upwards. Therefore, the wine does 
not spill.

But this argument ignores the elastic properties of the rod. Actually,
after the impact, a wave of transverse strain propagates along the rod and 
the ends of the rod still have a little time to move down. Therefore, for
successful demonstration of the experiment it is  necessary to provide a 
damping of rod-end displacements of the order of 1~mm \cite{36}.

\subsection{}
Из уравнении движения
$$\ddot{\vec{r}}=\frac{eg}{mc}\,\frac{\dot{\vec{r}}\times\vec{r}}{r^3}$$
следует, что квадрат скорости $\dot{\vec{r}}^{\;2}$ и квадрат орбитального 
момента $\vec{L}{\,^2}$ являются интегралами движения. Действительно, 
$0=\dot{\vec{r}}\cdot
\ddot{\vec{r}}=\frac{1}{2}\frac{d}{dt}(\dot{\vec{r}}{\;^2})$. А в задаче 
{\bf 8.5} показали, что вектор пуанкаре 
$$\vec{J}=\vec{L}-\frac{eg}{c}\,\frac{\vec{r}}{r}$$
сохраняется. Но $\vec{r}\cdot\vec{L}=0$, поэтому $\vec{J}{\,^2}=\vec{L}^{\,2}
+\left (\frac{eg}{c}\right )^2$ и, следовательно, $ \vec{L}^{\,2}$ тоже 
сохраняется.

Зависимость $r(t)$ можно найти так. Имеем
$$\frac{d^2\vec{r}^{\;2}}{dt^2}=2\frac{d}{dt}(\dot{\vec{r}}\cdot\vec{r})=
2\dot{\vec{r}}{\;^2}=2V^2,$$
т.к. $\dot{\vec{r}}\cdot\ddot{\vec{r}}=0$. Пусть начало отсчета времени выбрано
так, что при $t=0$ электрон находится на минимальном расстоянии $r_0$ от 
центра поля и, следовательно, $\vec{r}(0)\cdot\dot{\vec{r}}(0)=0$. Тогда 
интегрированием выше приведенного уравнения получаем 
$\vec{r}^{\,2}=V^2t^2+r_0^2$ и $r(t)=\sqrt{V^2t^2+r_0^2}$. Минимальное 
расстояние $r_0$ можно найти из сохранения величин скорости и орбитального 
момента: $L=mV\rho=mVr_0$. Отсюда $r_0=\rho$. 

Далее рассмотрим скалярное произведение $\vec{J}\cdot\vec{r}=Jr\cos{\theta}=
-\frac{eg}{c}r$. Это показывает, что $\cos{\theta}=-\frac{eg}{cJ}$ и угол между
векторами $\vec{r}$ и $\vec{J}$ не меняется с течением времени -- электрон
движется на поверхности конуса с углом раствора $2\theta$, с вершиной в начале 
координат и с осью вдоль вектора Пуанкаре $\vec{J}$. Поэтому удобно ввести 
сферические координаты с осью $z$ вдоль вектора Пуанкаре. Тогда $\vec{r}=r(
\sin{\theta}\cos{\varphi}\vec{i}+\sin{\theta}\sin{\varphi}\vec{j}+
\cos{\theta}\vec{k})$ и $$\dot{\vec{r}}=\dot{r}(\sin{\theta}\cos{\varphi}
\vec{i}+\sin{\theta}\sin{\varphi}\vec{j}+\cos{\theta}\vec{k})+r\dot{\varphi}
(-\sin{\theta}\sin{\varphi}\vec{i}+\sin{\theta}\cos{\varphi}\vec{j}).$$
Поэтому
$$V^2=\dot{\vec{r}}^{\;2}=(\dot{r}^2+r^2\dot{\varphi}^2)\sin^2{\theta}+
\dot{r}^2\cos^2{\theta}=\dot{r}^2+r^2\dot{\varphi}^2\sin^2{\theta}.$$
Отсюда, с учетом $r(t)=\sqrt{V^2t^2+\rho^2}$, получаем
$$\dot{\varphi}=\frac{\sqrt{V^2-\dot{r}^2}}{r\sin{\theta}}=
\frac{V\rho}{(V^2t^2+\rho^2)\sin{\theta}}.$$
Но
$$\sin{\theta}=\sqrt{1-\cos^2{\theta}}=\frac{1}{\sqrt{1+\left (
\frac{eg}{cL}\right )^2}}.$$
Следовательно, 
$$\dot{\varphi}=\frac{V\rho\,\sqrt{1+\left (\frac{eg}{cL}\right )^2}}
{\rho^2+V^2t^2}$$
и, интегрируя в предположении $\varphi(0)=0$, получаем
$$\varphi(t)=\sqrt{1+\left (\frac{eg}{cL}\right )^2}\arctg{\frac{Vt}
{\rho}}.$$ 
Заметим, что при $t\to\pm\infty$ имеем $\dot{r}\to\pm V$, $r\dot{\varphi}
\to 0$ и $$\varphi\to\pm\frac{\lambda \pi}{2},\;\; \mbox{где} \;\;
\lambda= \sqrt{1+\left (\frac{eg}{cL}\right )^2}=
\sqrt{1+\left (\frac{eg}{mcV\rho}\right )^2}. $$
Поэтому скорость электрона в отдаленном прошлом и в отдаленном будущем дается
выражением
$$\dot{\vec{r}}(\pm\infty)=V\sin{\theta}\left [\pm\cos{\left (\frac{
\lambda \pi}{2}\right )}\vec{i}+\sin{\left (\frac{\lambda \pi}{2}\right )}
\vec{j}\right ]\pm V\cos{\theta}\vec{k}.$$
Следовательно, для угла рассеяния $\alpha$ получаем
$$\cos{\alpha}=\frac{\dot{\vec{r}}(\infty)\cdot\dot{\vec{r}}(-\infty)}
{V^2}=-1+\frac{2}{\lambda^2}\sin^2{\left (\frac{\lambda
\pi}{2}\right )}.$$

\vspace*{5mm}
\noindent $\bullet$  $\bullet$  $\bullet$
 
\noindent
It follows from the equation of motion
$$\ddot{\vec{r}}=\frac{eg}{mc}\,\frac{\dot{\vec{r}}\times\vec{r}}{r^3},$$
that the squares of velocity $\dot{\vec{r}}^{\;2}$ and  angular momentum 
$\vec{L}{\,^2}$ are the integrals of motion. Indeed, 
$0=\dot{\vec{r}}\cdot
\ddot{\vec{r}}=\frac{1}{2}\frac{d}{dt}(\dot{\vec{r}}{\;^2})$.
But we have shown in the problem {\bf 8.5} that the Poincar\'{e} vector 
$$\vec{J}=\vec{L}-\frac{eg}{c}\,\frac{\vec{r}}{r}$$ is conserved.
However, $\vec{r}\cdot\vec{L}=0$, so  $\vec{J}{\,^2}=\vec{L}^{\,2}
+\left (\frac{eg}{c}\right )^2$ and, consequently, $ \vec{L}^{\,2}$ is also
conserved.

The $r(t)$ dependence can be found as follows. We have
$$\frac{d^2\vec{r}^{\;2}}{dt^2}=2\frac{d}{dt}(\dot{\vec{r}}\cdot\vec{r})=
2\dot{\vec{r}}{\;^2}=2V^2,$$
since $\dot{\vec{r}}\cdot\ddot{\vec{r}}=0$. Let the time origin is chosen so 
that at $t=0$ the electron is at the minimum distance $r_0$ from the field
center and, therefore, $\vec{r}(0)\cdot\dot{\vec{r}}(0)=0$. Then we get, by
integrating the above given equation, $\vec{r}^{\,2}=V^2t^2+r_0^2$ and 
$r(t)=\sqrt{V^2t^2+r_0^2}$. The minimum distance $r_0$ can be found from the 
fact that the velocity magnitude is unchanged in the magnetic field and  
conservation of angular momentum thus gives $L=mV\rho=mVr_0$. Hence, 
$r_0=\rho$.

Next, consider the scalar product $\vec{J}\cdot\vec{r}=Jr\cos{\theta}=
-\frac{eg}{c}r$. This shows that $\cos{\theta}=-\frac{eg}{cJ}$ and the angle 
between vectors $\vec{r}$ and $\vec{J}$ does not change over time, that is, 
the electron moves on the surface of a cone with opening angle $2\theta$, 
with the vertex at the coordinate origin and with the axis along the 
Poincar\'{e} vector $\vec{J}$. Therefore, it is convenient to introduce
spherical coordinates with the $z$ axis along the Poincar\'{e} vector. Then
$\vec{r}=r(\sin{\theta}\cos{\varphi}\vec{i}+\sin{\theta}\sin{\varphi}\vec{j}+
\cos{\theta}\vec{k})$ and $$\dot{\vec{r}}=\dot{r}(\sin{\theta}\cos{\varphi}
\vec{i}+\sin{\theta}\sin{\varphi}\vec{j}+\cos{\theta}\vec{k})+r\dot{\varphi}
(-\sin{\theta}\sin{\varphi}\vec{i}+\sin{\theta}\cos{\varphi}\vec{j}).$$
Consequently,
$$V^2=\dot{\vec{r}}^{\;2}=(\dot{r}^2+r^2\dot{\varphi}^2)\sin^2{\theta}+
\dot{r}^2\cos^2{\theta}=\dot{r}^2+r^2\dot{\varphi}^2\sin^2{\theta}.$$
Hence, given $r(t)=\sqrt{V^2t^2+\rho^2}$, we obtain
$$\dot{\varphi}=\frac{\sqrt{V^2-\dot{r}^2}}{r\sin{\theta}}=
\frac{V\rho}{(V^2t^2+\rho^2)\sin{\theta}}.$$
But
$$\sin{\theta}=\sqrt{1-\cos^2{\theta}}=\frac{1}{\sqrt{1+\left (
\frac{eg}{cL}\right )^2}}.$$
Therefore, 
$$\dot{\varphi}=\frac{V\rho\,\sqrt{1+\left (\frac{eg}{cL}\right )^2}}
{\rho^2+V^2t^2}$$
and integrating under the  assumption  $\varphi(0)=0$, we get
$$\varphi(t)=\sqrt{1+\left (\frac{eg}{cL}\right )^2}\arctg{\frac{Vt}
{\rho}}.$$ 
Note that for $t\to\pm\infty$ we have $\dot{r}\to\pm V$, $r\dot{\varphi}
\to 0$ and $$\varphi\to\pm\frac{\lambda \pi}{2},\;\; \mbox{where} \;\;
\lambda= \sqrt{1+\left (\frac{eg}{cL}\right )^2}=
\sqrt{1+\left (\frac{eg}{mcV\rho}\right )^2}. $$
Therefore, the electron velocities in the distant past and distant future are 
given by expression
$$\dot{\vec{r}}(\pm\infty)=V\sin{\theta}\left [\pm\cos{\left (\frac{
\lambda \pi}{2}\right )}\vec{i}+\sin{\left (\frac{\lambda \pi}{2}\right )}
\vec{j}\right ]\pm V\cos{\theta}\vec{k}.$$
Consequently, for the scattering angle $\alpha$, we obtain
$$\cos{\alpha}=\frac{\dot{\vec{r}}(\infty)\cdot\dot{\vec{r}}(-\infty)}
{V^2}=-1+\frac{2}{\lambda^2}\sin^2{\left (\frac{\lambda
\pi}{2}\right )}.$$

\subsection{}
Пусть сила трения, которая действует на шар, равна $\vec{F}$. Уравнение 
движения имеют вид $$m\dot{\vec{V}}=\vec{F}\;\;\mbox{и}\;\; 
I\dot{\vec{\omega}}=\vec{R}\times \vec{F},$$ 
где $I=\frac{2}{5}mR^2$ и $\vec{R}$ есть вектор от 
центра шара к точке касания (т.е. радиус-вектор точки касания равен 
$\vec{r}+\vec{R}$). Так как шар не проскальзывает, в точке касания скорость 
на поворотном столе $\vec{\omega}\times (\vec{r}+\vec{R})$ и на поверхности 
шара $\vec{V}+\vec{\omega}\times \vec{R}$ равны друг другу. Дифференцируя 
равенство 
$\vec{\omega}\times (\vec{r}+\vec{R})=\vec{V}+\vec{\omega}\times \vec{R}$,
получаем
$$\vec{\Omega}\times\dot{\vec{r}}=\dot{\vec{V}}+\dot{\vec{\omega}}\times
\vec{R}=\frac{1}{m}\vec{F}+\frac{1}{I}(\vec{R}\times\vec{F})\times\vec{R}.$$
Но $(\vec{R}\times\vec{F})\times\vec{R}=R^2\vec{F}-(\vec{F}\cdot\vec{R})
\vec{R}=R^2\vec{F}$, т.к. $\vec{F}\cdot\vec{R}=0$ (вектор $\vec{R}$ 
перпендикулярен к плоскости поворотного стола). Следовательно,
$$\vec{\Omega}\times\dot{\vec{r}}=\left (\frac{1}{m}+\frac{R^2}{I}
\right )\vec{F}=\frac{7}{2m}\vec{F}.$$
Определяя отсюда $\vec{F}$ и подставляя в уравнение $m\dot{\vec{V}}=\vec{F}$,
получаем $$\dot{\vec{V}}=\frac{2}{7}\vec{\Omega}\times\vec{V}.$$ Или в 
компонентах
$$\dot{V}_x=-\frac{2}{7}\Omega V_y,\;\;\dot{V}_y=\frac{2}{7}\Omega V_x,\;\;
\dot{V}_z=0.$$
Отсюда $$\ddot{V}_x=-\frac{2}{7}\Omega\dot{V}_y=-\left(\frac{2}{7}\Omega
\right)^2V_x.$$ Направим ось $x$ вдоль начальной скорости $\vec{V}_0$. Тогда
решением этого уравнения будет $V_x=V_0\cos{\frac{2}{7}\Omega t}$. 
Следовательно,
$V_y=-\frac{7}{2\Omega}\dot{V}_x=V_0\sin{\frac{2}{7}\Omega t}$. Интегрируя
еще раз, с учетом $x(0)=x_0,\,y(0)=y_0$, получаем
$$x(t)=x_0+\frac{7V_0}{2\Omega}\sin{\frac{2}{7}\Omega t},\;\;
y(t)=y_0+\frac{7V_0}{2\Omega}-\frac{7V_0}{2\Omega}\cos{\frac{2}{7}
\Omega t}.$$
Это показывает, что центр шара движется по окружности радиуса $\frac{7V_0}
{2\Omega}$ с центром в точке с радиус-вектором
$$x_0\,\vec{i}+\left (y_0+\frac{7V_0}{2\Omega}\right )\vec{j}=
\vec{r}_0+\frac{7}{2\Omega^2}\vec{\Omega}\times\vec{V}_0.$$  

\vspace*{5mm}
\noindent $\bullet$  $\bullet$  $\bullet$
 
\noindent
Let the frictional force acting on the ball is $\vec{F}$. Equations of motion 
have the form $$m\dot{\vec{V}}=\vec{F}\;\;\mbox{and}\;\; 
I\dot{\vec{\omega}}=\vec{R}\times \vec{F},$$ 
where $I=\frac{2}{5}mR^2$ and $\vec{R}$ is a vector from the center of the 
sphere to the point of contact (i.e. the radius vector of the point of contact 
is $\vec{r}+\vec{R}$). Since the ball does not slip, at the contact point
the velocities on the turntable, $\vec{\omega}\times (\vec{r}+\vec{R})$, and 
on the surface of the ball, $\vec{V}+\vec{\omega}\times \vec{R}$, are equal 
to each other.  Differentiating the equality 
$\vec{\omega}\times (\vec{r}+\vec{R})=\vec{V}+\vec{\omega}\times \vec{R}$,
we get
$$\vec{\Omega}\times\dot{\vec{r}}=\dot{\vec{V}}+\dot{\vec{\omega}}\times
\vec{R}=\frac{1}{m}\vec{F}+\frac{1}{I}(\vec{R}\times\vec{F})\times\vec{R}.$$
But $(\vec{R}\times\vec{F})\times\vec{R}=R^2\vec{F}-(\vec{F}\cdot\vec{R})
\vec{R}=R^2\vec{F}$, since $\vec{F}\cdot\vec{R}=0$ (vector $\vec{R}$ is
perpendicular to the plane of the turntable). Consequently,
$$\vec{\Omega}\times\dot{\vec{r}}=\left (\frac{1}{m}+\frac{R^2}{I}
\right )\vec{F}=\frac{7}{2m}\vec{F}.$$
Determining $\vec{F}$ from this equation and substituting it into
$m\dot{\vec{V}}=\vec{F}$, we get $$\dot{\vec{V}}=\frac{2}{7}\vec{\Omega}
\times\vec{V}.$$ Or in components
$$\dot{V}_x=-\frac{2}{7}\Omega V_y,\;\;\dot{V}_y=\frac{2}{7}\Omega V_x,\;\;
\dot{V}_z=0.$$
Hence $$\ddot{V}_x=-\frac{2}{7}\Omega\dot{V}_y=-\left(\frac{2}{7}\Omega
\right)^2V_x.$$ 
Let the $x$-axis is along the initial velocity $\vec{V}_0$. Then the solution 
of this equation is $V_x=V_0\cos{\frac{2}{7}\Omega t}$. 
Consequently,
$V_y=-\frac{7}{2\Omega}\dot{V}_x=V_0\sin{\frac{2}{7}\Omega t}$. Integrating
once again, given that $x(0)=x_0,\,y(0)=y_0$, we obtain
$$x(t)=x_0+\frac{7V_0}{2\Omega}\sin{\frac{2}{7}\Omega t},\;\;
y(t)=y_0+\frac{7V_0}{2\Omega}-\frac{7V_0}{2\Omega}\cos{\frac{2}{7}
\Omega t}.$$
This shows that the center of the ball moves on a circle of radius 
$\frac{7V_0}{2\Omega}$, centered at the point with the radius vector
$$x_0\,\vec{i}+\left (y_0+\frac{7V_0}{2\Omega}\right )\vec{j}=
\vec{r}_0+\frac{7}{2\Omega^2}\vec{\Omega}\times\vec{V}_0.$$

\clearpage
\subsection{}
\begin{figure}[htb]
\centerline{\epsfig{figure=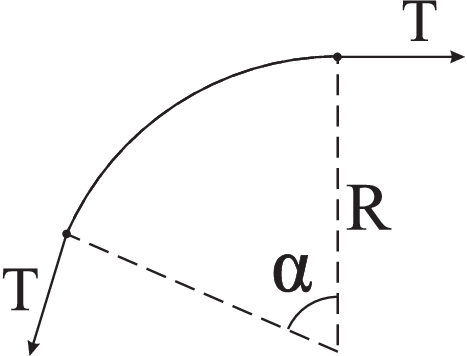,height=2cm}}
\end{figure}

\noindent
Рассмотрим участок проволочки с радиусом кривизны $R$ и длиной $R\alpha\ll 
R$. Равнодействующая сил натяжения $2T\sin{\frac{\alpha}{2}}\approx T\alpha$
уравновешивается силой Ампера $IR\alpha B$. Поэтому получаем $T=IRB$. Но
$R=\frac{p}{eB}$. Следовательно, $T=\frac{IP}{e}$ и $$I=\frac{eT}{p}$$ 
(все величины в системе СИ).

\vspace*{5mm}
\noindent $\bullet$  $\bullet$  $\bullet$
 
\begin{figure}[htb]
\centerline{\epsfig{figure=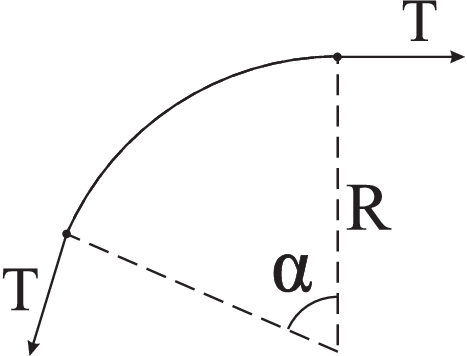,height=2cm}}
\end{figure}

\noindent
Consider a section of the wire with a radius of curvature $R$ and length
$R\alpha\ll R$. The resultant tension force $2T\sin{\frac{\alpha}{2}}
\approx T\alpha$ is balanced by the Ampere force $IR\alpha B$. Therefore,
we have $T=IRB$. But $R=\frac{p}{eB}$. Consequently, $T=\frac{IP}{e}$ and
$$I=\frac{eT}{p}$$ (all quantities are in the SI system).

\section*{Благодарности}
Автор благодарит В.~Яньшоле за составление рисунков. Работа выполнена при 
частичной поддержке грантов НШ-905.2006.2 и  РФФИ 06-02-16192-а.

\end{document}